\documentclass[11pt,letter]{ucthesis}

\def\dsp{\def\baselinestretch{2.0}\large\normalsize}
\dsp

\def\hsp{\def\baselinestretch{1.5}\large\normalsize}

\usepackage[sectionbib]{natbib}
\bibpunct{[}{]}{;}{a}{}{,}

\usepackage{import}
\usepackage{longtable}
\usepackage{lamproof}
\usepackage{mathpartir}
\usepackage{amsmath}
\usepackage{amssymb,amscd}
\usepackage{gastex}
\usepackage{caption}
\usepackage{graphics}
\usepackage{restate}
\usepackage{stmaryrd}
\usepackage{color}
\usepackage{hyperref}
\hypersetup{
  colorlinks,
  citecolor=black,
  filecolor=black,
  linkcolor=black,
  urlcolor=black
}

\newcommand{\comment}[1]{}
\renewcommand{\t}[1]{{\tt #1}}
\newcommand{\mynote}[1]{$\clubsuit$ #1 $\clubsuit$}

\newcommand{\setc}[2]{ \{ #1 ~|~ #2 \}}

\newcommand{\fun}[1]{\mbox{\it #1\/}}
\newcommand{\Iff}[0]{\Leftrightarrow}
\newcommand{\defeq}{\stackrel{\mathrm{def}}{=}}

\newcommand{\dterm}{{\mathcal D}}
\newcommand{\judgestatic}[3]{#1\vdash_{\mathcal{S}} #2\,:\,#3}

\newcommand{\judgehybrid}[3]{#1\vdash^{\t{}} #2\,:\,#3}

\newcommand{\judgehybridc}[4]{#1 \vdash^{\t{}} #2 \,\castinsertsymbol\, #3 \,:\, #4 }

\newcommand{\listof}[1]{\ensuremath{\overline{#1}}}

\newcommand{\proofcase}{ \emph{Case} }

\newcommand{\cal}[1]{\ensuremath{\mathcal{#1}}}

\renewcommand{\t}[1]{{\tt #1}}
\renewcommand{\comment}[1]{ }
\renewcommand{\mynote}[1]{$\clubsuit$ #1 $\clubsuit$}

\renewcommand{\fun}[1]{\mbox{\it #1\/}}
\renewcommand{\Iff}[0]{\Leftrightarrow}
\renewcommand{\defeq}{\stackrel{\mathrm{def}}{=}}
\newcommand{\defiff}{\stackrel{\mathrm{def}}{\Leftrightarrow}}

\newcommand{\myclearpage}{}

\newcommand{\setz}[1]{ \{ #1 \}}

\newcommand{\judgewfsub}[2]{\ensuremath{#1\vdash_{\mathrm{wf}}#2}}
\newcommand{\judgeImp}[3]{\ensuremath{ #1\vdash #2 \Implies #3}}
\newcommand{\judgeTC}[3]{#1\algdash #2 ~\&~ #3}
\newcommand{\lethead}[3]{ \ensuremath{\t{let}~{#1}:{#2}={#3}~\t{in}} }

\newcommand{\judgeC}[4]{\ensuremath{#1\algdash #2\,:\,#3 ~\&~ #4}}

\newcommand{\substT}[3]{[{#1} \mapsto {#2} : {#3}]}

\newcommand{\emptysubst}{[~]}
\newcommand{\emptyenv}{\varnothing}

\newcommand{\existstm}[3]{\exists {#1}:{#2}.~{#3}}
\newcommand{\envfor}[1]{\env_{#1}}
\newcommand{\funcfor}[1]{\mathcal{F}_{#1}}
\newcommand{\lambdahtc}{\lambdah^{\mbox{\tiny HTC}}}
\newcommand{\lambdahr}{\ensuremath{\lambdah^{\mbox{\tiny Recon}}}}

\newcommand{\refpref}[1]{\ref{#1} on page~\pageref{#1}}

\newcommand{\hhref}[2]{\hyperref[#2]{{#1}~\ref{#2}}}
\newcommand{\hhrefpref}[2]{\hyperref[#2]{{#1}~\ref{#2} on page~\pageref{#2}}}
\newcommand{\hhrefppref}[2]{\hyperref[#2]{{#1}~\ref{#2} (page~\pageref{#2})}}

\newenvironment{indentlist}{
  \begin{list}{}{
      
      \setlength{\labelwidth}{35pt}
      \setlength{\leftmargin}{\labelwidth+\labelsep}
    }
}{
  \end{list}
}

\newenvironment{rulelist}{\begin{indentlist}}{\end{indentlist}}
\newenvironment{proofcases}{\begin{indentlist}}{\end{indentlist}}

\newcommand{\hbra}{
\hbox to \textwidth{\vrule width0.3mm height 1.8mm depth-0.3mm
                    \leaders\hrule height1.8mm depth-1.5mm\hfill
                    \vrule width0.3mm height 1.8mm depth-0.3mm}}
\newcommand{\hket}{
\hbox to \textwidth{\vrule width0.3mm height1.5mm
                    \leaders\hrule height0.3mm\hfill
                    \vrule width0.3mm height1.5mm}}

\renewcommand{\hbra}{
\hbox to \columnwidth{\vrule width0.3mm height 1.8mm depth-0.3mm
                    \leaders\hrule height1.8mm depth-1.5mm\hfill
                    \vrule width0.3mm height 1.8mm depth-0.3mm}}
\renewcommand{\hket}{
\hbox to \columnwidth{\vrule width0.3mm height1.5mm
                    \leaders\hrule height0.3mm\hfill
                    \vrule width0.3mm height1.5mm}}

\newcommand{\htbra}{
\hbox to \textwidth{\vrule width0.3mm height 1.8mm depth-0.3mm
                    \leaders\hrule height1.8mm depth-1.5mm\hfill
                    \vrule width0.3mm height 1.8mm depth-0.3mm}}
\newcommand{\htket}{
\hbox to \textwidth{\vrule width0.3mm height1.5mm
                    \leaders\hrule height0.3mm\hfill
                    \vrule width0.3mm height1.5mm}}

\newenvironment{displayfigure}[1]{
  \begin{figure}
  \caption{#1}
  \medskip
  \ \hbra\\[-.8ex]
  \nopagebreak
 }{ 
  \\[-.8ex]
 \nopagebreak
  \hket
  \nopagebreak
 \medskip
  \end{figure}
 }

\newenvironment{displayfigure*}[1]{
  \begin{figure*}[tph!]
   \refstepcounter{figure}
   \caption*{#1}
 \medskip
  \ \htbra\\[-.8ex]
  \nopagebreak
  \footnotesize
 }{ 
  \\[-.8ex]
 \nopagebreak
  \htket
  \nopagebreak
 \medskip
  \end{figure*}
 }

\newcommand{\trulenoname}[2]{
\frac 
	{\strut\begin{array}{@{}c@{}} #1 \end{array}} 
	{\strut\begin{array}{@{}c@{}} #2 \end{array}}
}

\newcommand{\rel}[1]{\raisebox{0ex}[0ex][0ex]{\small\textsc{[#1]}}}
\newcommand{\mydefhead}[2]{\multicolumn{2}{l}{{#1}}&\mbox{\emph{#2}}\\}
\newcommand{\mydefcase}[2]{\qquad& #1 &\mbox{#2}\\}
\newcommand{\headrule}[2]{\multicolumn{3}{@{}l}{\mbox{#2}} & \fbox{#1}\\}

\newcommand{\eg}{\emph{e.g.}}

\newcommand{\ie}{\emph{i.e.}}

\newenvironment{proof}{ {\sc Proof:} }{}

  {
     \normalsize{\sc Proof:}\hspace*{0mm}
}
  {\proofcomplete
\smallskip 
   }

\newcommand{\proofcomplete}{ 
$\square$
}

\newcommand{\qed}{\ensuremath{\square}}

\newcounter{alltheorems}
\numberwithin{alltheorems}{chapter} 

\newtheorem{thm}[alltheorems]{Theorem}
\newtheorem{lem}[alltheorems]{Lemma}
\newtheorem{corollary}[alltheorems]{Corollary}
\newtheorem{requirement}[alltheorems]{Requirement}
\newtheorem{definition}[alltheorems]{Definition}

\newcommand{\e}{\ensuremath{e}}
\newcommand{\tmf}{\ensuremath{f}}
\newcommand{\f}{\ensuremath{f}}
\newcommand{\tmd}{\ensuremath{d}}
\renewcommand{\d}{\ensuremath{d}}
\newcommand{\g}{\ensuremath{g}}
\newcommand{\predp}{\ensuremath{p}}
\newcommand{\predq}{\ensuremath{q}}

\newcommand{\val}{\ensuremath{v}}

\newcommand{\constant}{\ensuremath{k}}
\newcommand{\constructor}{\ensuremath{c}}
\newcommand{\primfun}{\constant}

\newcommand{\tyS}{\ensuremath{S}}
\newcommand{\tyT}{\ensuremath{T}}
\newcommand{\tyU}{\ensuremath{U}}

\newcommand{\lam}[3]{\lambda#1\!:\!#2.\,#3}
\newcommand{\lammulti}[3]{\lambda\listof{#1\!:\!#2}.\,#3}
\newcommand{\plam}[3]{(\lam {#1} {#2} {#3})}

\newcommand{\lamt}[3]{\ensuremath{#1\!:\!#2\rightarrow #3}}
\newcommand{\lamtmulti}[3]{\ensuremath{\listof{#1\!:\!#2}\rightarrow #3}}
\newcommand{\plamt}[3]{(\lamt {#1} {#2} {#3})}
\newcommand{\arrowt}[2]{#1\rightarrow #2}

\newcommand{\predty}[3]{\mbox{\ensuremath{\{#1\!:\!#2~|~#3\}}}}

\newcommand{\predquick}[2]{\mbox{\ensuremath{\{#1~|~#2\}}}}

\newcommand{\app}[2]{#1~#2}
\newcommand{\castl}[2]{\langle{#2}\rangle^{#1}}
\newcommand{\castltwo}[3]{\langle{ {#3} \triangleleft {#2} }\rangle^{#1}}
\newcommand{\cast}[1]{\castl {} {#1}}
\newcommand{\casttwo}[2]{\castltwo {} {#1} {#2}}
\newcommand{\checking}[3]{\ensuremath{\langle {#1}, {#2}, {#3}\rangle}}
\newcommand{\caseexp}[2]{\t{case}~ {#1} ~ \t{of} ~ {#2}}

\newcommand{\patternexp}[2]{{#1} \Mapsto {#2}}

\newcommand{\fix}[1]{\t{fix}_{#1}}
\newcommand{\inl}[1]{{\tt inl}~#1}
\newcommand{\inr}[1]{{\tt inr}~#1}
\newcommand{\fold}[2]{\t{fold}\,[#1]\,#2}
\newcommand{\unfold}[2]{\t{unfold}\,[#1]\,#2}
\renewcommand{\top}{{\tt Top}}
\newcommand{\Nat}{\t{N}}
\newcommand{\Int}{\t{Int}}
\newcommand{\Bool}{\t{Bool}}
\newcommand{\Array}{\t{Array}}
\newcommand{\Unit}{\t{Unit}}
\newcommand{\Matrix}{\t{Matrix}}
\newcommand{\fsucc}{\t{succ}}
\newcommand{\pred}{\t{pred}}
\newcommand{\true}{\t{true}}
\newcommand{\false}{\t{false}}

\newcommand{\dpairt}[3]{#2\!:\!#1\times#3}
\newcommand{\sumtype}[2]{#1+#2}
\newcommand{\rectype}[2]{\mu #1.\,#2}

\newcommand{\castlapp}[3]{\app {\castl {#1} {#2}} {#3}}
\newcommand{\castapp}[2]{\castlapp {} {#1} {#2}}
\newcommand{\castltwoapp}[4]{\app {\castltwo {#1} {#2} {#3}} {#4}}
\newcommand{\casttwoapp}[3]{\castltwoapp {} {#1} {#2} {#3}}
\newcommand{\dpair}[3]{\langle #1=#2,#3\rangle}
\newcommand{\letpl}[5]{\t{let}^{#1}~\langle #2,#3 \rangle = #4~\t{in}~#5}
\newcommand{\letp}[4]{\letpl{}{#1}{#2}{#3}{#4}}
\newcommand{\letexp}[3]{\t{let}~{#1}={#2}~\t{in}~{#3}}
\newcommand{\letexpt}[4]{\t{let}~{#1}\!:\!{#3}={#2}~\t{in}~{#4}}
\newcommand{\ifexp}[3]{\t{if-XXXXXXXXXXXX}~ #1~#2~#3}
\newcommand{\ifexpt}[4]{\t{if}_{#4}~#1~#2~#3}
\newcommand{\arrowd}[3]{\lamt{#2}{#1}{#3}}
\newcommand{\ascriptl}[3]{#2~\t{as}~#3}

\newcommand{\existsty}[2]{\exists #1. \,#2}
\newcommand{\existsbind}[3]{\existsty {#1\!:\!#2} {#3}}
\newcommand{\existsmulti}[3]{\existsty {\listof{#1\!:\!#2}} {#3}}

\newcommand{\pexistsbind}[3]{(\existsbind {#1} {#2} {#3})}

\newcommand{\closingsubst}{\sigma}
\newcommand{\approxfun}{\zeta}
\newcommand{\tyvar}{\alpha}
\newcommand{\ph}{\ensuremath{\psi}}
\newcommand{\tysol}{\pi}
\newcommand{\tmsol}{\rho}

\newcommand{\csub}{\sigma}
\newcommand{\dsub}{\theta}
\newcommand{\dsubpred}[1]{pred(#1)}

\newcommand{\subconset}{C}
\newcommand{\impconset}{P}

\newcommand{\dapp}{\cdot}

\newcommand{\Base}[1]{\ensuremath{\lfloor {#1} \rfloor}}

\newcommand{\red}{\ensuremath{\curvearrowright}}
\newcommand{\lred}{\leadsto}

\newcommand{\meaningf}[1]{\ensuremath{[\![ #1 ]\!]}}
\newcommand{\deltaf}[2]{\delta(#1,#2)}
\newcommand{\subst}[3]{[#1\mapsto\!#2]}

\newcommand{\hole}{\bullet}
\newcommand{\tmctx}{\ensuremath{\mathcal C}}
\newcommand{\tyctx}{\ensuremath{\mathcal D}}
\newcommand{\cbvctx}{\ensuremath{\mathcal E}}
\newcommand{\strictctx}{\ensuremath{\mathcal S}}
\newcommand{\whnfctx}{\ensuremath{\mathcal W}}

\newcommand{\ctxequiv}{\ensuremath{\equiv}}

\newcommand{\checkingpred}[5]{\ensuremath{\langle {\predty {#1} {#2} {#3}}, {#4}, {#5}\rangle}}

\newcommand{\funcasttwo}[6]{
  \lam {#1} {#5} {
      \casttwo {\subst {#1} {\casttwo {#5} {#2} ~ {#1}} {}\, {#3}} {#6}
      ~ 
      ({#4}~ ({\casttwo {#5} {#2} ~ {#1}}))
  }
}

\newcommand{\sword}{\mbox{\,\,$<:$\,\,}}

\newcommand{\Implies}{\Rightarrow}
\newcommand{\algdash}{\Vvdash}

\newcommand{\env}{\Gamma}
\newcommand{\tyenv}{\Gamma}
\newcommand{\envtail}{{\Gamma'}}
\newcommand{\envextra}{{\Gamma''}}
\newcommand{\existsenv}{\Delta}

\newcommand{\judge}[3]{\ensuremath{#1\vdash #2\,:\,#3}}
\newcommand{\plainjudge}[2]{\ensuremath{#1\,:\,#2}}
\newcommand{\judgeT}[2]{\ensuremath{#1\vdash #2}}
\newcommand{\judgeE}[1]{\ensuremath{\vdash #1}}
\newcommand{\judgesub}[3]{\ensuremath{#1\vdash{#2}\sword{#3}}}
\newcommand{\plainjudgesub}[2]{\ensuremath{{#1}\sword{#2}}}

\newcommand{\judgeimp}[3]{\ensuremath{#1\vdash {#2}\Implies #3}}

\newcommand{\njudge}[3]{\ensuremath{#1\not\vdash #2\,:\,#3}}
\newcommand{\njudgeimp}[3]{#1\not\vdash {#2}\Implies #3}
\newcommand{\njudgesub}[3]{#1\not\vdash  #2\sword #3}

\newcommand{\judgeSTLCsym}{\vdash_{\!\!\!\mbox{\sc \tiny stlc}}}
\newcommand{\judgeSTLC}[3]{{#1}\judgeSTLCsym{#2}\,:\,{#3}}

\newenvironment{trules}{
  \begin{mathpar}
} {
  \end{mathpar}
}

\newcommand{\trule}[3]{
  \inferrule*[right=\rel{#1}]{#2}{#3} 
}

\newcommand{\headtrule}[2]{
  \begin{tabular*}{0.98\textwidth}{l@{\extracolsep{\fill}}r}
    \mbox{#2} & \fbox{{#1}} 
  \end{tabular*}
}

\renewcommand{\trulenoname}[2]{
\frac 
	{\strut\begin{array}{@{}c@{}} #1 \end{array}} 
	{\strut\begin{array}{@{}c@{}} #2 \end{array}}
}

\newcommand{\judgeAlg}[3]{\ensuremath{#1\algdash #2\,:\,#3}}
\newcommand{\judgeH}[3]{\ensuremath{#1\vdash #2\,:\,#3}}

\newcommand{\JudgeAlg}[3]{\judgeAlg {\existsenv{#1}} {#2} {#3}}
\newcommand{\JudgeH}[3]{\judgeH {\tyenv{#1}} {#2} {#3}}

\newcommand{\judgeAlgT}[2]{\ensuremath{#1\algdash #2}}
\newcommand{\judgeHT}[2]{\ensuremath{#1\vdash #2}}

\newcommand{\JudgeAlgT}[2]{\judgeAlgT {\existsenv{#1}} {#2}}
\newcommand{\JudgeHT}[2]{\judgeHT {\tyenv{#1}} {#2}}

\newcommand{\judgeAlgsub}[3]{\ensuremath{#1\algdash{#2}\,\sword\,{#3}}}
\newcommand{\judgeHsub}[3]{\ensuremath{#1\vdash{#2}\,\sword\,{#3}}}

\newcommand{\JudgeAlgsub}[3]{\judgeAlgsub {\existsenv{#1}} {#2} {#3}}
\newcommand{\JudgeHsub}[3]{\judgeHsub {\tyenv{#1}} {#2} {#3}}

\newcommand{\judgeinstance}[3]{\ensuremath{#1\vdash{#2}\instanceof{#3}}}

\newcommand{\judgeAlgimp}[3]{\ensuremath{#1\algdash {#2}\Implies #3}}

\newcommand{\self}{\ensuremath{\mathbf{self}}}

\newcommand{\castinsertsymbol}{\hookrightarrow}

\newcommand{\judgesubalg}[4]{#2\algdash^{#1} {#3}\sword{#4}}
\newcommand{\judgeimpalg}[4]{#2\algdash^{#1} {#3}\Implies #4}
\newcommand{\pyes}{\surd}
\newcommand{\pno}{{\times}}
\newcommand{\pmaybe}{{?}}

\newcommand{\judgec}[4]{
  \ensuremath{#1 \algdash #2 \,\castinsertsymbol\, #3 \,:\, #4 }
}
\newcommand{\judgecc}[4]{
  \ensuremath{#1 \algdash #2 \,\castinsertsymbol\, #3 \,\downarrow\, #4 }
}
\newcommand{\judgect}[3]{
  \ensuremath{#1 \algdash #2 \,\castinsertsymbol\, #3  }
}

\newcommand{\lambdah}{\ensuremath{\lambda_\mathcal{H}}}
\newcommand{\lambdas}{\lambda_{\mathcal{S}}}

\newcommand{\bisim}{\ensuremath{\sim}}
\newcommand{\judgesimty}[3]{\ensuremath{#1\vdash #2\,\bisim\,#3}}
\newcommand{\judgesimt}[3]{\ensuremath{#1\vdash #2\,\bisim\,#3}}

\newcommand{\judgesubset}[3]{\ensuremath{#1\vdash {#2} \subseteq {#3}}}
\newcommand{\judgeIn}[3]{
  #1 \vdash #2 \in #3
}

\newcommand{\lambdaex}{\ensuremath{\lambdah^{\mbox{\tiny Comp}}}}

\newcommand{\sage}{{\sc Sage}}

\newcommand{\dynamic}{\t{Dynamic}}

\newcommand{\usepairs}[1]{}
\newcommand{\usesums}[1]{}
\newcommand{\userec}[1]{}

\newcommand{\redctx}{{\cal C}}

\newcommand{\type}{\mbox{\t{*}}}

\newcommand{\unit}{\t{unit}}

\newcommand{\Funty}[2]{#1\rightarrow #2}

\newcommand{\Refine}{\t{Refine}}

\newcommand{\sagefix}{\t{fix}}

\newcommand{\judgesubdb}[4]{#2\vdash_{db}^{#1} {#3}\sword{#4}}

\newcommand{\judgeTPalg}[3]{#2\models_{alg}^{#1} {#3}}
\newcommand{\judgeTP}[2]{#1\models {#2}}
\newcommand{\judgenotTP}[2]{#1\not\models {#2}}

\newenvironment{code} {
  \begin{trivlist} \item \ttfamily \begin{tabular}{l}
} {
  \end{tabular} \end{trivlist}
}

\newcommand{\funtyd}[3]{\lamt {#1} {#2} {#3}}

\newcommand{\STLCsyntax}{
  \[
  \begin{array}{llr}
    \mydefhead {\e,\tmd,\tmf,\g,\predq,\predp ::=\qquad\qquad\qquad\qquad\qquad} {Terms:} 
    \mydefcase \constant                                                         {primitive} 
    \mydefcase x                                                                 {variable} 
    \mydefcase {\lam x \tyS \e}                                                  {abstraction} 
    \mydefcase {\f~\e}                                                           {application} 
    \\
    \mydefhead {\val ::=\qquad\qquad\qquad\qquad\qquad} {Values: ($\subset$ Terms)}
    \mydefcase \constant                                {primitive} 
    \mydefcase {\lam x \tyS \e}                         {abstraction} 
    \\
    \mydefhead {\tyS,\tyT,\tyU::=}    {Types:} 
    \mydefcase {\arrowt \tyS \tyT}      {function type}
    \mydefcase B                        {basic type} 
  \end{array}
  \]
}

\newcommand{\STLCopsem}{
  \[
  \begin{array}{@{}rclr}
    \headrule{$\tmd \red \e$}{Redex \underline{E}valuation}
    \\
    (\lam x \tyS \e) ~ \tmd      & \red &  {\subst x \tmd S}\, \e & \rel{E-$\beta$}
    \\
    \constant ~ \val             & \red &  \deltaf \constant \val & \rel{E-$\delta$}
    \\
    \\
    \\
    \headrule{$\tmd \lred \e$}{Contextual \underline{E}valuation}
    \\
    \qquad\qquad\qquad	\tmctx[\tmd] & \lred & \tmctx[\e] \quad \mbox{if $\tmd \red \e$}
    &\rel{E-Ctx} \\
    \\
    \headrule{\tmctx, \tyctx}{Contexts}
    \\
    \multicolumn{4}{l}{
      \begin{array}{lll}
      \tmctx  & ::= &
      \hole
      ~|~ \app \tmctx \e
      ~|~ \app \e \tmctx
      ~|~ \lam x \tyS \tmctx
    \end{array}
    }
    \\
  \end{array}
  \]
}

\newcommand{\STLCtyperules}{
  {\ssp
  \[
  \begin{array}{ll@{\qquad\qquad\qquad\qquad\qquad\qquad}r}
    \mydefhead {\env::=}     {Environments:} 
    \mydefcase \emptyset     {empty environment}
    \mydefcase {\env,x:\tyT} {environment extension}
  \end{array}
  \]}
  
  \headtrule{$\judgeSTLC \env \e \tyT$}{\underline{STLC} Typing}
  \begin{trules}
    \trule{STLC-Var}{
      (x:\tyT)\in \env
    }{
      \judgeSTLC \env x T
    }

    \trule{STLC-Prim}{
    }{
      \judgeSTLC \env \constant {\fun{ty}(\constant)}
    }
    
    \trule{STLC-Fun}{
      \judgeSTLC {\env,x\!:\!\tyS} \e \tyT
    }{
      \judgeSTLC \env {\lam x \tyS e} {\arrowt \tyS \tyT}
    }
    
    \trule{STLC-App}{
      \judgeSTLC \env \tmf \tyS
      \and
      \judgeSTLC \env \e {\arrowt \tyS \tyT}
    }{
      \judgeSTLC \env {\f ~ \e} \tyT
    }
  \end{trules}
}

\begin{document}


\title{Executable Refinement Types}
\author{Kenneth L. Knowles}
\degreeyear{2014}
\degreemonth{March}
\degree{DOCTOR OF PHILOSOPHY}
\chair{Professor Cormac Flanagan}
\committeememberone{Professor Luca de Alfaro}
\committeemembertwo{Professor Ranjit Jhala}
\numberofmembers{3} 
\deanlineone{Tyrus Miller}
\deanlinetwo{Vice Provost and Dean of Graduate Studies}
\deanlinethree{}
\field{Computer Science}
\campus{Santa Cruz}

\begin{frontmatter}

\maketitle
\copyrightpage
\tableofcontents
\listoffigures

\begin{abstract}
  Precise specifications are integral to effective programming
  practice.  Existing specification disciplines such as structural
  type systems, dynamic contracts, and extended static checking all
  suffer from limitations such as imprecision, false positives, false
  negatives, or excessive manual proof burden. New ways of expressing
  and enforcing program specifications are needed.

  Towards that end, this dissertation introduces \emph{executable
    refinement types} and establishes their metatheory and
  accompanying implementation techniques.  Executable refinement types
  enrich structural type systems with basic types refined by
  semi-decidable predicates. Through the lens of executable refinement
  types, we also address the broader problem of theory and
  implementation for undecidable type systems.

  To establish a firm foundation for the study of executable
  refinement types, this dissertation presents a full formal account
  of their metatheory. Type checking for executable refinement types
  is undecidable.  Nonetheless, they fulfill standard metatheoretical
  correctness criteria including type soundness and extensional
  equivalence.

  To perform type checking for executable refinement types we
  introduce \emph{hybrid type checking}, a type enforcement strategy
  broadly applicable to undecidable type systems. Hybrid type checking
  enforces specifications via static analysis where possible and
  dynamic type casts where necessary. We prove that for any decidable
  approximation of executable refinement types, either: (1) Hybrid
  type checking catches some errors \emph{statically} which the
  decidable approximation would miss, or (2) the decidable
  approximation rejects some correct program which hybrid type
  checking would accept.

  To perform type reconstruction for executable refinement types, we
  radically revise the usual notion of type reconstruction.
  Typeability is undecidable because it subsumes type
  checking. Instead, we propose a more precise definition of type
  reconstruction as a typeability-\emph{preserving} transformation.
  For decidable type systems, our definition coincides with the
  previous.  Using our generalized notion of type reconstruction, we
  demonstrate that type reconstruction for executable refinement types
  is decidable even though type checking is not! We show this by
  providing a syntactic type reconstruction algorithm reminiscent of
  strongest postcondition calculation.

  To enlarge the class of programs for which type checking \emph{is}
  decidable, we formalize the notion of \emph{compositional reasoning}
  for types systems.  Because standard dependent types perform
  non-compositional reasoning, type checking is undecidable even when
  all types appearing in a program fall in a decidable specification
  language.  We present a variant of dependent types which uses
  existential types to achieve compositional reasoning.  Even
  restricted to compositional reasoning, our type system is
  \emph{exact}: It can give any term a type that completely classifies
  that term up to contextual equivalence.  When reasoning
  compositionally, if all the annotations in a program fall into a
  decidable language, then type checking is decidable.  We show this
  with a type checking algorithm for such programs.

  Atop these theoretical foundations we implement \sage{}, a language
  blending executable refinement types, dynamic typing, and
  first-class types. \sage{}'s implementation includes standard
  type-checking machinery, compile-time computation, automatic theorem
  proving, dynamic contract checking, and a database of run-time
  failures which inform the hybrid type checker for future runs.
  Preliminary experiments indicate that \sage{} is effective at
  verifying many common examples statically in a reasonable amount of
  time.  Moreover, every run-time failure in \sage{} can occur at most
  once: From then onwards it becomes a compile time failure.
\end{abstract}

\begin{acknowledgements}
I entered graduate study in search of a community of intellectuals,
and I found it at UC Santa Cruz. I have never encountered such a
spirited and surprising group of scholars, nor any with as broad an
engagement with mathematics, computing, and philosophy, as I found at
UCSC. For my years with them, I am immeasurably grateful.

Particularly for their helpful attention to this work, I thank my
advisor, Cormac Flanagan, and my reading committee members, Luca de
Alfaro and Ranjit Jhala.

Chapters of this dissertation include expository reorganizations of
work by the author and colleagues, updated to unify notation and
theory, engage in more explanation and discussion than possible in a
terse conference paper, provide greater detail in proofs, and survey
subsequent related work.  I would like to thank my coauthors Cormac
Flanagan, Stephen Freund, Aaron Tomb, and Jessica Gronski for their
contributions to \citet{gronski_sage:_2006},
\citet{knowles_type_2007}, \citet{knowles_compositional_2009}, and
\citet{knowles_hybrid_2010}.

Beyond my the intellectual engagement of my colleagues, I also owe
completion of this dissertation to support offered by my family and
friends, so I would like to especially thank my mother, my father, and
Meg.
\end{acknowledgements}\

\begin{displayfigure*}{Summary of Metavariables}
  \label{figure:prelude:metavariables}
  \hsp
  \vspace{-15pt}
  \[
  \begin{array}{l@{\hskip 10em}l}
    \e,\tmd,\,\tmf,\,\g,\,\predq,\,\predp & \mbox{term} \\
    x,\,y,\,z                             & \mbox{variable} \\
    \constant                             & \mbox{primitive constant or function} \\
    \constructor                          & \mbox{constructor supporting pattern matching} \\
    b                                     & \mbox{boolean constant} \\
    m,n                                   & \mbox{natural number or natural number constant} \\
    \delta                                & \mbox{semantic function for constants} \\
    S,\,T,\,U                             & \mbox{type} \\
    B                                     & \mbox{basic type} \\
    Q,R                                   & \mbox{restricted type with limited predicate language} \\
    E,F                                   & \mbox{augmented type with covariant existentials} \\
    a,b                                   & \mbox{three-valued certainty (one of $\pno, \pyes, \pmaybe$)} \\
    \env                                  & \mbox{typing environment} \\
    \existsenv                            & \mbox{typing environment with augmented types} \\
    \theta                                & \mbox{capture-avoiding substitution function} \\
    \sigma,\,\gamma                       & \mbox{closing substitution} \\
    \approxfun                            & \mbox{approximation function to decidable refinements} \\
    \ph                                   & \mbox{term placeholder} \\
    \tysol                                & \mbox{\emph{non}-capture-avoiding type replacement} \\
    \tmsol                                & \mbox{\emph{non}-capture-avoiding placeholder replacement} \\
    \tmctx                                & \mbox{term context} \\
    \tyctx                                & \mbox{type context} \\
    \cbvctx                               & \mbox{call-by-value evaluation context} \\
    \strictctx                            & \mbox{unfolding evaluation context} \\
    \whnfctx                              & \mbox{head-normal form evaluation context} \\
  \end{array}
  \]
  \vspace{-15pt}
\end{displayfigure*}\

\newpage

\begin{displayfigure*}{Summary of Notations}
  \label{figure:prelude:notations}
  \vspace{-15pt}
  \[
  \begin{array}{l@{\hskip 10em}l}
    \lam x T e                               & \mbox{function term} \\
    f~e                                      & \mbox{function application} \\
    \caseexp e {\listof{\patternexp c f}}    & \mbox{case discrimination term} \\
    \casttwo S T                             & \mbox{type cast} \\
    \cast T                                  & \mbox{single-type cast} \\
    \checking {\predty x B \predp} \predq k  & \mbox{cast in progress} \\
    \arrowt S T                              & \mbox{function type} \\
    \lamt x S T                              & \mbox{dependent function type} \\
    \predty x B e                            & \mbox{refinement type} \\
    \existsbind x S T                        & \mbox{existential type} \\
    \Base T, \Base e                         & \mbox{shape of type or term} \\
    \meaningf T                              & \mbox{denotation of type} \\
    \pno, \pyes, \pmaybe                     & \mbox{certainty values of ``no'', ``yes'', ``maybe''} \\
    \subst x e {}                            & \mbox{substitution mapping $x$ to $e$} \\
    \substT x e T                            & \mbox{substitution mapping $x$ to $e$ with type annotation} \\
    \theta\dapp\tyvar                        & \mbox{delayed substitution applied to a type variable} \\
    \theta\dapp\ph                           & \mbox{delayed substitution applied to a placeholder} \\
    \tmctx[e],\,\tyctx[e],\,\cbvctx[e],\,\strictctx[e],\,\whnfctx[e] & \mbox{context application} \\
  \end{array}
  \]
\end{displayfigure*}\

\newpage

\begin{displayfigure*}{Summary of Relations}
  \[
  \begin{array}{l@{\hskip 10em}l}
    e_1 \red e_2                             & \mbox{single-step redex reduction} \\
    e_1 \lred e_2,\, S\lred T                & \mbox{single-step term or type reduction} \\
    e_1 \lred^+ e_2,\, S\lred^+ T            & \mbox{transitive closure of single-step reduction} \\
    e_1 \lred^* e_2,\, S\lred^* T            & \mbox{reflexive-transitive closure of single-step reduction} \\
    e_1 \leftrightsquigarrow e_2             & \mbox{equivalence closure of single-step reduction} \\
    e_1 \ctxequiv e_2                        & \mbox{contextual equivalence of terms} \\
    S \curlywedge T                          & \mbox{wedge product of types} \\
    \sigma \curlywedge \gamma                & \mbox{wedge product of closing substitutions} \\
  \end{array}
  \]
\end{displayfigure*}\

\newpage

\begin{displayfigure*}{Summary of Judgments}
  \label{figure:prelude:judgements}
  \vspace{-15pt}
  \[
  \begin{array}{l@{\hskip 10em}l}
    \judge \env e T                    & \mbox{typing} \\
    \judgeT \env T                     & \mbox{type well-formedness} \\
    \judgeE \env                       & \mbox{typing environment well-formedness} \\
    \judgewfsub \env \dsub             & \mbox{delayed substution well-formedness} \\
    \judgesub \env S T                 & \mbox{subtyping} \\
    \judge {} \sigma \env              & \mbox{closing substitution typing} \\
    \judgesimt \env {e_1} {e_2:T}      & \mbox{extensional equivalence of terms} \\
    \judgesimt {} \sigma {\gamma:\env} & \mbox{extensional equivalence of closing substitutions} \\
    \judgeIn \env e T                  & \mbox{semantic typing} \\
    \judgesubset \env S T              & \mbox{semantic subtyping} \\
    \judgeimp \env {e_1} {e_2}         & \mbox{implication} \\
    \judgeTP \env p                    & \mbox{validity} \\
    \judgec \env d e T                 & \mbox{cast insertion} \\
    \judgecc \env d e T                & \mbox{cast insertion and checking} \\ 
    \judgect \env S T                  & \mbox{cast insertion in types} \\
    \judgesubalg a \env S T            & \mbox{three-valued algorithmic subtyping} \\
    \judgeimpalg a \env S T            & \mbox{three-valued algorithmic implication} \\
    \judgeAlg \env e E                 & \mbox{decidable algorithmic typing} \\
    \judgeAlgT \env R                  & \mbox{decidable algorithmic type well-formedness} \\
    \judgeAlgsub \env R E              & \mbox{decidable algorithmic subtyping} \\
  \end{array}
  \]
\end{displayfigure*}\

\end{frontmatter}



\chapter{Introduction}

\label{chapter:intro}

A program is, of necessity, a completely unambiguous document.  To
manage the enormous detail that this entails,
\citet{parnas_criteria_1972} articulates a modular development
strategy based on \emph{information hiding} where trusted and
precisely specified abstractions of the sections of a program -- now
known as \emph{interfaces} -- reduce the amount of detail that must be
considered at one time, enabling cooperation amongst programmers over
space and time.

In practice, however, interfaces are often only informally or
partially specified. Consider the following increasingly-precise
specifications for the argument to a procedure for inverting a matrix:

\begin{enumerate}
\item
  The argument can be any value expressible in the programming language.
\item
  The argument must be an array of arrays of numbers. 
\item        
  The argument must be a \emph{matrix}, that is, a rectangular (non-ragged)
  array of arrays of  numbers. 
\item
  The argument must be a square matrix. 
\item
  The argument must be a square matrix with nonzero determinant.
\end{enumerate}

All of these specifications may be useful for various purposes, levels
of assurance, and moments in the development of a program.  Simpler
specifications are common during rapid prototyping, at the cost of
proliferation of partial functions (functions which are undefined on
portions of their purported domain). More precise specifications
provide stronger correctness guarantees and better documentation, and
simplify function invocation by reducing or removing the possibility
of passing a ``bad'' input.

The first specification corresponds approximately to what are called
\emph{dynamically typed} (also \emph{unityped} or
\emph{single-sorted}) programming languages. The second item is a
likely specification in \emph{statically typed} languages such as
Java, C++, C\#, OCaml, or Haskell, which automatically distinguish
sorts of program values. The last three specifications remain the
stuff of research, including this dissertation. Often such
specifications, if they are checked at all, are dynamically checked by
ad-hoc hand-written methods as part of a program's execution.

As the precision afforded by a specification language --
its \emph{expressiveness} -- increases, so does the computational
complexity of determining whether a program meets a specification.
Specifications written in less expressive languages, such as
traditional type systems (or no language at all in the case of
unityped languages), can be checked \emph{statically}: There is an
algorithm which, given the text of a program, either accepts that the
program obeys the interface discipline or rejects the program as
ill-formed. In other words, the question of whether a program meets a
specification is decidable.  

More precise specifications, such as those written in formal logics
including recursion/induction, often cannot be checked statically:
They are written in a language expressive enough that no algorithm can
exist that correctly accepts or rejects all programs. This does not
preclude the existence of useful approximate algorithms, but every one
must include false negatives (programs which do not meet their
specification but are accepted) or false positives (programs which do
meet their specification but are rejected) or both.\footnote{The
distinction between whether acceptance or rejection is the positive
case is arbitrary. In program verification spurious errors are usually
considered false positives, while programs that may crash are
considered false negatives.}

\section{Specification Disciplines}

Generally, \emph{static} checking reasons about the text of a program
prior to its execution, thus implicitly reasons about all possible
inputs. \emph{Dynamic} checking avoids this universal quantification
by reasoning only about the values that arise during a particular
execution. Many properties that are difficult to establish statically
are easy or trivial to ensure dynamically, for example that a pointer
is non-null or that a binary tree is balanced.

\subsection{Dynamic Contracts}

Dynamic \emph{contracts} are a programming construct for organizing and
composing hand-written routines that dynamically enforce interface
specifications and for isolating the root cause of a violation during
program execution. Many common specifications can be expressed as
executable contracts,\footnote{Notable classes of properties that are
  not easily checked dynamically are concurrency properties such as
  race-freedom or atomicity and properties requiring (potentially)
  infinite quantification such as associativity or injectivity.} such
as:
\begin{itemize}
\item[-] Subranges: The function \t{squareRoot} requires its argument
  to be a non-negative number.
\item[-] Aliasing restrictions: The function \t{swap} requires that
  its arguments are distinct reference cells.
\item[-] Ordering restrictions: The function \t{binarySearch} requires
  that its argument is a sorted array.
\item[-] Size specifications: The function \t{serializeMatrix} takes
  as input a matrix of size $n$ by $m$, and returns a one-dimensional
  array of size $n\times m$.
\item[-] Arbitrary executable predicates: an interpreter (or code
  generator) for a typed language (or intermediate
  representation~\citep{tarditi_til:_1996}) might naturally require
  that its input be well-typed, \ie, that it satisfies the
  programmer-written (and likely optimized, hence obfuscated) predicate 
  $\t{wellTyped}:\t{Expr}\rightarrow\Bool$.
\end{itemize}

Contracts have been incorporated into many languages, including Turing
\citep{holt_turing_1988}, Anna~\citep{luckham_programming_1990},
Eiffel \citep{meyer_eiffel:_1992},
Sather~\citep{stoutamire_sather_1996}, Blue~\citep{kolling_blue_1996},
Scheme~\citep{findler_contracts_2002},
Spec\# \citep{barnett_spec_2005}, JML~\citep{burdy_overview_2005,
leavens_design_2006}, and Haskell~\citep{hinze_typed_2006,
chitil_practical_2012}. In these languages, functions may have
precondition checks (executed before the body of the function is
evaluated), postcondition checks (executed after the result of the
function has been computed), and invariants (executed when a data
structure is modified, to ensure its integrity).

However, dynamic checking suffers from two limitations. First, it
incurs a computational cost during execution, often running checks
repeatedly and redundantly, and in the worst case is expensive enough
that it is feasible only during testing. For example constraining a
binary tree by the contract $\t{isBinarySearchTree}$ will require a
full traversal of the tree with every insertion, changing the
complexity of the operation from $O(\log n)$ to
$\Omega(n)$.\footnote{The ``solution'' to this problem is to write
  only those contracts which can be efficiently executed, and to make
  sure they are only checked when necessary, via clever programming.}
More seriously, the very property of dynamic checking that makes it
simple -- that specifications are only checked on data values and code
paths of actual executions -- often results in incomplete and late
(possibly post-deployment) detection of defects. For example, consider
the function $\t{abs}$ for computing the absolute value of an integer,
subject to the contract $\arrowt \Int {\t{NonNeg}}$ which states that
$\t{abs}$ requires an integer input and returns a non-negative
result. If we mistakenly simply return the input as the output, then
the function violates its contract for all negative inputs. Yet
executions such as $(\t{abs}~4)$ will leave this defect undiscovered.

\subsection{Refinement Types}
\label{section:intro:refinementtypes}

The term \emph{refinement type}\footnote{The term \emph{refinement
    type} originated with \citet{freeman_refinement_1991} but is now
  used much more broadly.} is not an agreed-upon mathematical term,
but an intuitive label applied to type systems where the basic types
of a language may be ``refined'' to classify a more precise subset of
the terms of the language.  For example, if the type of linked lists
in a structural type system were $\t{List}$, then one may consider
refining this type to carry information about the length of the list,
writing $\t{List}~n$ to describe lists of length $n$. This style of
refinement type, now commonplace, is called an \emph{indexed type}
where $n$ is the index a refinement type.

Dependent ML~\citep{xi_dependent_1999} includes a type system with
indexed types that can express and enforce properties involving linear
inequalities between integer variables, and by this express subranges,
ordering restrictions, and array size constraints. Other refinement
type systems, each enhancing a structural type system with some
decidable logic, include \citet{hayashi_logic_1993},
\citet{denney_refinement_1998}, \citet{xi_dependent_1999},
\citet{xi_imperative_2000}, and \citet{davies_intersection_2000},
\citet{mandelbaum_effective_2003}, and
\citet{dunfield_tridirectional_2004}.

Refinement types improve on dynamic contract checking in terms of
completeness, because once a program is accepted by the type system,
it is certain to satisfy all preconditions and postconditions for all
possible executions, but the specification language is intentionally
restricted to ensure that specifications can always be checked
statically and efficiently so that the analysis can be incorporated
into an iterative workflow.

In addition to limitations to ensure decidability, properties enforced
by indexed types are expressed in an indirect style.  A more direct
way of expressing $\t{List}~n$ using standard set comprehension
notation is $\predty x {\t{List}} {\t{length}(x) = n}$. More complex
types are even less direct, for example the type of lists of length
greater than $k$ in indexed style is
\[
\exists n>k. \, \t{List}~n
\]
where a more direct transliteration of the specification
would be
\[
\predty x {\t{List}} {\t{length}(x) > k}
\]
For more sophisticated data structures and specifications, such as
red-black trees, the indices and methods for expressing the desired
specification are even more complex.

\subsection{Extended Static Checking}

If expressivity is more desirable than efficiency but complete
checking is still desired, as it was for \citet{parnas_technique_1972}
and \citet{dijkstra_discipline_1976}, then \emph{extended static
  checking} may be attempted via a tool such as
PVS~\citep{rushby_subtypes_1998} or
ESC/Java~\citep{flanagan_extended_2002}. Extended static checking
refers to static verification of specifications written in a very
expressive language of preconditions and postconditions.  This
combines the benefits of complete checking and expressive
specifications, but results in large numbers of false positives --
programs where all specifications are satisfied, but the analysis
cannot verify that this is so. To direct the analysis towards
successful verification, the programmer must add many complex
annotations or, in the extreme, provide a manual proof of correctness.

\section{Thesis Overview}

Synthesizing the properties of the specification systems discussed so
far, some desirable properties of a specification system are:

\begin{itemize}
\item[-] No false positives: Programs which meet their
  specification are accepted.

\item[-] No false negatives: Programs which do not meet their
  specification are rejected. (This subsumes complete checking, in
  which a program must meet its specification for \emph{all} inputs or
  it is rejected.)

\item[-] Efficient checking: It can be efficiently and automatically
  determined whether a program meets its specification.  (This
  subsumes decidable checking and often reduces to to efficient
  specification inclusion, in which it can be automatically determined
  whether one specification is strictly more precise than another.)

\item[-] Low proof burden: A human need not
  manually prove complex properties.

\item[-] Specification reconstruction: When some specifications
  are omitted, they can be automatically deduced. (This is similar
  to ``low proof burden'' if one considers programming-supplied
  specifications as a lightweight form of manual proof.)

\item[-] Compositionality: The specification of a program term
  is a sufficient summary for all further analyses.

\item[-] Readable specifications: The specification of a program is a
  self-explanatory mathematical statement about the program.

\item[-] Expressivity: A given set of program values can
  be described in the specification language.

\end{itemize}

The open (and open-ended) problem that this dissertation addresses is
the search for practical combinations of \emph{complete checking} and
\emph{expressive specifications} that nonetheless enjoy as many of
these desirable properties as may be achieved.

In particular, we narrow the search by working with the theoretical
machinery of type systems, while removing any restrictions to ensure
decidability or efficiency of type checking as conventionally
construed. To illustrate the key idea common to these type systems,
consider the following type rule for function application with
subtyping.
\[
\trulenoname{
    \plainjudge \f {\arrowt \tyT \tyU}\qquad
    \plainjudge \e \tyS \qquad
    \plainjudgesub \tyS \tyT
}{
  \plainjudge {\app \tmf \e} \tyU
}
\]
The antecedent $\plainjudge \f {\arrowt \tyT \tyU}$ states that the
program term $\f$ is a function that takes any value of type $\tyT$ as
input and returns a value of type $\tyU$. The antecedent $\plainjudge
\e \tyS$ states that $\e$ is a program term of type $\tyS$. The
question of whether $\f$ may be applied to $\e$ hinges on whether a
member of type $\tyS$ is necessarily also a member of type $\tyT$,
formally posed as ``is $\tyS$ a subtype of $\tyT$?'' and written
$\plainjudgesub \tyS \tyT$.  If the type checker can prove this
subtyping relation, then this application is accepted. Conversely,
if the type checker can prove that this subtyping relation does not
hold, then the program is rejected.  In a conventional, decidable type
system, one of these two cases always holds.

However, for type languages that are sufficiently expressive that type
checking is not decidable, the type checker will encounter situations
where its algorithms can neither prove nor refute the subtype judgment
$\plainjudgesub \tyS \tyT$ (particularly within the time bounds
imposed by interactive compilation). When subtyping is undecidable,
then type checing is undecidable, and type reconstruction as conventionally
defined is undecidable. How shall one proceed?

This dissertation defines such a type system: \emph{executable
  refinement types}. Executable refinement types use the language of
dynamic contracts, so that any predicate that can be written in a
programming language may refine a type. Thus in addition to
traditional types such as $\Int$, $\Bool$, and $\arrowt \Int \Bool$,
the type system includes types such as ${\predty x \Int {x>0}}$ (the
type of positive integers) and $\lamt x \Bool {\predty y \Bool {x
    \Leftrightarrow y}}$ (the type of functions from booleans to
booleans where the return value is $\true$ if and only if the input
value is $\true$).  The techniques gleaned from a close engagement
with executable refinement types are broadly applicable, because
executable refinement types include \emph{all} semidecidable
specifications.

The contributions of this dissertation are organized into chapters as
follows.

\begin{itemize}
\item[-] \hhref{Chapter}{chapter:background} provides references to
  material containing adequate background to contextualize its
  contribution to the field of programming languages.

\item[-] \hhref{Chapter}{chapter:lambdah} presents a core calculus
  $\lambdah$ which will form the theoretical basis of the following 
  chapters and establishes its basic correctness properties using a
  simple, but novel, application of denotational semantics to provide
  a foundation for our refinement type system. In addition to the
  usual soundness criteria, this chapter introduces an original
  variation of the technique of \emph{logical relations} to establish
  the correctness of our notion of ``function'' as well as provide a
  powerful tool for proving program equivalences.

\item[-] \hhref{Chapter}{chapter:htc} introduces \emph{hybrid type
  checking} (HTC), a general technique for combining a static type
  system with dynamic checks, and describes the correctness properties
  any hybrid type checking system should have. We then illustrate
  hybrid type checking for the core calculus of
  \hhref{Chapter}{chapter:lambdah} and prove its correctness using the
  proof tools of the prior chapter. Finally, we compare hybrid type
  checking with static type checking, concluding that any decidable
  restriction always rejects a well-typed program which HTC accepts or
  accepts an ill-typed program which HTC rejects.

\item[-] \hhref{Chapter}{chapter:inference} presents an original
  generalization of the type reconstruction problem that is applicable
  to undecidable type systems, then presents a decision procedure for
  full type reconstruction of $\lambdah$.

\item[-] \hhref{Chapter}{chapter:compositional} examines the special case
  where all specifications are written in a decidable language,
  uncovering a critical flaw in the unrestricted use of dependent
  types in this context. This flaw is resolved by the use of standard
  type-theoretic techniques, and a decision procedure for type
  checking is provided and proved sound.

\item[-] \hhref{Chapter}{chapter:sage} presents \sage{}, an
  implementation of executable refinement types featuring hybrid type
  checking.  In addition to refinement types, \sage{} includes dynamic
  types and first-class computation over types, demonstrating the
  breadth of interesting undecidable type systems to which our
  techniques apply. This chapter presents some preliminary empirical
  validation that hybrid type checking decides most type checking
  queries statically.

\item[-] \hhref{Chapter}{chapter:relatedwork} surveys related work
  developed in parallel or subsequent to the material presented in
  this dissertation, recapitulating citations from substantive
  chapters into coherent research trends.

\item[-] \hhref{Chapter}{chapter:conclusion} concludes and discusses
  future directions.
\end{itemize}  

Chapters~\ref{chapter:intro}-\ref{chapter:lambdah} should be read
prior to any other chapters, but then all of
Chapters~\ref{chapter:htc}-\ref{chapter:sage} may be read mostly
independently before the concluding discussion of
Chapters~\ref{chapter:relatedwork} and~\ref{chapter:conclusion}.

\chapter{Preliminaries}

\label{chapter:background}

This dissertation is best enjoyed with a strong background in
programming languages, especially type systems.  The references and
mathematical content of this chapter are intended to prepare most
readers for the remainder of the dissertation, or at least to allow a
reader to ascertain the degree of overlap between their experience and
the content herein. Please feel free to skip this chapter on first
reading and return to it as needed.

As it does not constitute part of the novel contribution of this
dissertation (except as the particulars of presentation may differ
from others) the material is presented somewhat abruptly and without
extensive motivation or discussion.

\section{Preparatory Readings}
\label{section:background:types}
\label{section:background:tapl}

The fundamentals of programming language theory that form the basic
mathematical setting for this dissertation include: operational
semantics, denotational semantics, axiomatic semantics, type systems
(particularly dependent type systems), and logical relations.

For operational semantics and type systems, \citet{pierce_types_2002}
provides a thorough introduction in a textbook setting, using notation
and terminology similar to that of this dissertation. Its
sequel, \citet{pierce_advanced_2004}, covers advanced topics such as
dependent types and logical relations.

From axiomatic semantics \citep{floyd_assigning_1967,
  hoare_axiomatic_1969, dijkstra_discipline_1976} we draw upon many
  descriptive terms for predicates that may be true of program
  fragments such as \emph{preconditions} (predicates that must hold
  before a procedure is begun), \emph{postconditions} (predicates that
  must hold after a procedure completes), \emph{invariants}
  (predicates that must always hold), \emph{weakest preconditions}
  (the weakest predicate that allows a fragment to complete
  successfully), and \emph{strongest postconditions} (the strongest
  predicate a fragment ensures).
\citet{winskel_formal_1993} is a suitable textbook for a unified
presentation.

\section{The Simply-Typed $\lambda$-Calculus}

The starting point of our metatheoretical work is the simply-typed
$\lambda$-calculus \citep{church_formulation_1940} augmented with
built-in types such as $\Int$ and $\Bool$, and primitives such as
boolean constants ($\true$ and $\false$), integer constants ($1$, $4$,
$-12$, etc), conditionals ($if$), and recursion (in the form of a
fixed-point operator $\t{fix}_T$ for each type $T$) We call this
variant \emph{STLC}.\footnote{This system may also be considered a
  variant of PCF. \citep{scott_type-theoretical_1969}} The material of
this section is thoroughly treated by any introductory text, such as
\citet{pierce_types_2002}, but we present it here because later
developments refer directly to specific details. This section also may
serve to familiarize the reader with our notational conventions.

\subsection{Syntax of STLC}
\label{section:stlc:syntax}

The syntax for STLC described in this section is summarized via
recursive grammars in \hhref{Figure}{figure:stlc:syntax} using our
notation and metavariables. Many metavariables are reserved for terms
and types so that they may be varied for legibility.

\begin{displayfigure}{Syntax of STLC}
  \ssp
  \label{figure:stlc:syntax}
  \STLCsyntax
\end{displayfigure}

\begin{itemize}
\item[-] The type $\arrowt S T$ denotes the type of functions from $S$ to $T$.

\item[-] The type $B$ stands for any of the built-in basic types. We
  freely assume that these include familiar types such as $\Bool$,
  $\Int$, and $\Unit$ (the type with just one element), while
  introducing others as necessary.

\item[-] The term $k$ denotes any of the built-in primitives, which
  may be functions. We freely assume that these include primitives for
  arithmetic (\eg\ $0$, $3$, $+$, $\times$, $>=$), conditionals (\eg\ $\true$,
  $\false$, $\t{if}_T$ for each type $T$), and recursion ($\t{fix}_T$ for
  each type $T$).

\item[-] The term $\lam x S e$ denotes a function taking a parameter
  $x$ of type $S$ and having a result calculated according to the body
  of the function $e$. The particular name chosen for the parameter
  $x$ is of no consequence. For example, the term \mbox{$\lam x \Int
    {x + 3}$} is considered equal to $\lam y \Int {y + 3}$ by
  definition.  Variable binding and renaming ($\alpha$-equivalence) is
  covered thoroughly by the preparatory reading of
  \hhref{Section}{section:background:tapl}.

\item[-] The term $x$ denotes a variable reference -- variable
  references are called \emph{bound} if they occur within the scope of
  a $\lambda$-abstraction of the same variable, and otherwise are
  called \emph{free}. The set of free variables of a term $\e$ is
  written $fv(\e)$. Again, a thorough discussion of free and bound
  variables is relegated to the background reading of
  \hhref{Section}{section:background:tapl}.

\item[-] The term $\f ~ \e$ denotes the application of the term $\f$
  to the term $\e$. Note that $\f$ and $\e$ are simply metavariables
  chosen for legibility; either the function or argument may be any
  term.
\end{itemize}

\subsection{Operational Semantics of STLC}
\label{section:stlc:opsem}

\hhref{Figure}{figure:stlc:opsem} gives formal meaning to STLC terms
via a standard small-step operational semantics. The relation $\tmd
\red \e$ expresses evaluation of \emph{redexes}, short for ``reducible
expressions''. It means that \emph{by definition} we consider $\tmd$
and $\e$ to be equivalent terms, and the notation is chosen to
correspond to the intuition that a program represented by $\tmd$ may
be transformed to a program represented by $\e$ in a ``single step''
of computation. What constitutes a step of computation here is not
suitable for algorithmic analysis, but only as a formal tool for
discussing equivalent programs.

\begin{displayfigure}{Operational Semantics of STLC}
  \ssp
  \label{figure:stlc:opsem}
  \STLCopsem
\end{displayfigure}

The two redex evaluation rules in \hhref{Figure}{figure:stlc:opsem}
concern function application for $\lambda$-abstractions and primitive
functions:

\begin{rulelist}
\item[\rel{E-$\beta$}:] An application of a function $\lam x S \e$ to
  argument $\tmd$ is reduced by replacing the formal parameter $x$ by
  the actual parameter $\tmd$. The notation $\subst x \tmd {}$ denotes
  the standard capture-avoiding substitution of $x$ for $\tmd$; it is
  not a program term but the following operation carried out (by
  induction) on the syntax of $\e$:
  \begin{eqnarray*}
    \subst x \d {}\, k               & \defeq & k \\
    \subst x \d {}\, y               & \defeq & y \qquad \mbox{($y \neq x$)} \\
    \subst x \d {}\, x               & \defeq & \d \\
    \subst x \d {}\, (\f ~ \e)       & \defeq & (\subst x \d {}\, \f) ~ (\subst x \d {}\, \e) \\
    \subst x \d {}\, (\lam y S \e)   & \defeq & \lam y S {\subst x \d {}\, \e} \qquad \mbox{ ($y\neq x$ and $y$ not free in $\tmd$)}
  \end{eqnarray*}
  In the last equation, the bound variable $y$ may always be chosen
  (via $\alpha$-equivalence) to satisfy the side
  condition. Capture-avoiding substitutions may be composed like any
  other (meta-)function and are named using metavariables $\theta$,
  $\sigma$, and $\gamma$.

\item[\rel{E-$\delta$}:] Application of a primitive function is resolved by
  a presumed partial function 
  \[
  \delta:\fun{Prim}\times\fun{Term} \rightharpoonup \fun{Term}
  \]
  which is parameter of the system constrained
  (\hhrefpref{Requirement}{req:stlc:prim}) to be total for appropriately
  typed arguments and to map them to appropriately typed results.

  We assume standard meanings for boolean and arithmetic connectives,
  conditionals, and the fixed point operator. For example, the following
  equations (based on presumed primitives) should all hold for any
  definition of $\delta$.
  \[
  \begin{array}{rcl}
	\deltaf {\t{not}}      \true     &=& \false\\
	\deltaf {\t{not}}      \false    &=& \true\\
	\deltaf {\deltaf + 3}  7         &=& 10\\
	\deltaf {\t{if}_{T}}   \true     &=& \lam x T {\lam y T x}\\
	\deltaf {\t{if}_{T}}   \false    &=& \lam x T {\lam y T y}\\
	\deltaf {\fix{T}}      \e        &=& \e ~ (\fix{T} ~ \e)
  \end{array}
  \]
\end{rulelist}

The single-step evaluation relation $\tmd \lred \e$ denotes redex
evaluation of an arbitrary subterm of $\tmd$. Specifically, redex
evaluation takes place within a context $\tmctx$ that represents a
term with exactly one occurrence of the symbol $\hole$ (a ``hole'')
where a term may be placed, such as $\plam x S {\f ~ \hole} ~ \tmd$.
The notation $\tmctx[e]$ denotes the resulting term when the $\e$ is
placed in the unique hole in $\tmctx$. This operation is \emph{not}
capture-avoiding; $\e$ may have free variables that become bound once
it is placed in $\tmctx$. The reflexive-transitive closure of $\lred$
is written $\lred^*$ and may be read intuitively as ``running the
program for an arbitrary number of steps''.

\subsection{The STLC Type System}

\hhref{Figure}{figure:stlc:typerules} presents the typing rules for STLC
in full, along with the definition of typing environments $\env$ which
bind variables to their declared types. Bindings in the environments
are ordered but environments may be freely treated as partial
functions from variables to types or sets of bindings when order is
insignificant.  The typing judgment
\[
\judgeSTLC \env e T
\]
reads ``Assuming the bindings in $\env$, the term $\e$ may be assigned
STLC type $T$.'' As with the syntax and semantics, these typing rules
are completely standard, and the following narrative descriptions are
provided as a reference for the notation.

\begin{displayfigure}{Typing Rules for STLC}
  \label{figure:stlc:typerules}
  \vspace{-5pt}
  \STLCtyperules
  \vspace{-35pt}
\end{displayfigure}

\begin{rulelist}
\item[\rel{STLC-Var}:] A variable $x$ is assigned its bound type in the
  environment $\env$.
  
\item[\rel{STLC-Prim}:] A primitive $\constant$ is given a type by the
  function $ty$, which is constrained to maintain type soundness along
  with the $\delta$ function. Any definition of $ty$ includes such
  mappings as:
\[
\ssp
\begin{array}[t]{r@{~:~}l}
  \true      & \Bool \\
  \false 	   & \Bool \\
  \Iff	   & \arrowt \Bool {\arrowt \Bool \Bool} \\
  \t{not}	   & \arrowt \Bool \Bool \\
  n          & \Int \\
  +          & \arrowt \Int {\arrowt \Int \Int} \\
  \t{if}_{T} & \arrowt \Bool {\arrowt T {\arrowt T T}}\\
  \fix{T}	   & \arrowt {(\arrowt{T}{T})} T
\end{array}
\]

Primitives with function types may be called \emph{primitive
  functions} and primitives of basic type may be called
\emph{primitive constants}.

\item[\rel{STLC-Fun}:] For a $\lambda$-abstraction $\plam x S e$ the body $\e$ of the function
is checked in an environment extended by $x$ bound with type
$S$. 

\item[\rel{STLC-App}:] For an application $(\f~\e)$, if the term $\f$ in
function position has the function type $\arrowt S T$ and is applied
to an argument of type $\tyS$, then the resulting type of the
application is $T$.

\end{rulelist}

The soundness of this calculus is well-established and we do not
present the proofs here, but introduce the requirement on primitives
upon which this system is parameterized, and state the classical
results.

\begin{requirement}[Types of STLC Primitives]
  \label{req:stlc:prim}
  \

  For each $\constant\in\fun{Prim}$, if $\constant$ is a primitive
  function then it cannot get stuck and its behavior is compatible
  with its type, \ie\ if $\judgeSTLC \emptyset {\app \constant v} T$
  then $\deltaf \constant v$ is defined and $\judgeSTLC \emptyset
  {\deltaf \constant v} T$.
\end{requirement}

The \emph{preservation} lemma proved upon
\hhref{Requirement}{req:stlc:prim} ensures that for any well-typed
term $\e$ that in any evaluation sequence proceeding
\[
\e \lred \e_1 \lred \e_2 \lred \cdots 
\]
each $\e_i$ is well-typed with the same type as $\e$. 

\begin{lem}[Preservation for STLC]
  For any term $\e$, if $\judgeSTLC \env \e T$ and $\e \lred \e'$,
  then $\judgeSTLC \env {\e'} T$.
\end{lem}

Since terms that reduce to or from each other are definitionally
equivalent, this indicates that typing is well-defined for the full
equivalence class. However, preservation may be vacuous if no such
reduction sequence exists, such as for terms like $3 ~ (4 ~ - ~ +)$
(deliberately constructed to be nonsensical). If the type system
assigned this term type $\Bool$, preservation would still hold, so the
\emph{progress} lemma proves that the only well-typed terms with empty
evaluation sequences are those in the known set of values.

\begin{lem}[Progress for STLC]
  For any term $\e$, if $\judgeSTLC \emptyset e T$ then either $\e$ is
  a value or there exists $\e'$ such that $\e \lred \e'$.
\end{lem}

Type soundness is such a simple corrollary of preservation and progress
that it is often not even stated, but since the remaining chapters
prove type soundness in a few different ways, we state the full result
to complete the example metatheoretical development.

\begin{thm}[Type Soundness for STLC]
  For any term $\e$, if $\judgeSTLC \emptyset \e T$ then either
  $\e$ is nonterminating or $\e$ evaluates to some value
  $\val$ such that $\judgeSTLC \emptyset \val T$
\end{thm}

\chapter{Executable Refinement Types}

\label{chapter:lambdah}


\newcommand{\arraypredty}[3]{
  \left\{
    \begin{array}{l@{~:~}l|l} 
      #1 & #2 & \begin{array}{ll} #3 ~ \end{array}
    \end{array}
  \right\} 
} 

\renewcommand{\fix}[1]{\t{fix}_{#1}}
\renewcommand{\inl}[1]{{\tt inl}~#1}
\renewcommand{\inr}[1]{{\tt inr}~#1}
\renewcommand{\fold}[2]{\t{fold}\,[#1]\,#2}
\renewcommand{\unfold}[2]{\t{unfold}\,[#1]\,#2}
\renewcommand{\top}{{\tt Top}}
\renewcommand{\Nat}{\t{Nat}}
\renewcommand{\Int}{\t{Int}}
\newcommand{\TrueNat}{\mathbb{N}}
\newcommand{\TrueInt}{\mathbb{Z}}
\newcommand{\IntList}{\t{IntList}}
\renewcommand{\Bool}{\t{Bool}}
\renewcommand{\Array}{\t{Array}}
\renewcommand{\Unit}{\t{Unit}}
\renewcommand{\Matrix}{\t{Matrix}}
\renewcommand{\fsucc}{\t{succ}}
\renewcommand{\pred}{\t{pred}}
\renewcommand{\true}{\t{true}}
\renewcommand{\false}{\t{false}}

\renewcommand{\dpairt}[3]{#2\!:\!#1\times#3}
\renewcommand{\sumtype}[2]{#1+#2}
\renewcommand{\rectype}[2]{\mu #1.\,#2}
\renewcommand{\castlapp}[3]{\app {\castl {#1} {#2}} {#3}}
\renewcommand{\castapp}[2]{\castlapp {} {#1} {#2}}
\renewcommand{\dpair}[3]{\langle #1=#2,#3\rangle}
\renewcommand{\letpl}[5]{\t{let}^{#1}~\langle #2,#3 \rangle = #4~\t{in}~#5}
\renewcommand{\letp}[4]{\letpl{}{#1}{#2}{#3}{#4}}
\renewcommand{\letexp}[3]{\t{let}~{#1}={#2}~\t{in}~{#3}}
\renewcommand{\ifexp}[3]{\t{if-XXXXXXXXXXXX}~ #1~#2~#3}
\renewcommand{\ifexpt}[4]{\t{if}_{#4}~#1~#2~#3}
\renewcommand{\arrowd}[3]{\lamt{#2}{#1}{#3}}
\renewcommand{\ascriptl}[3]{#2~\t{as}^{#1}~#3}

\newcommand{\lambdaexcoresyntax}{
  \[
  \begin{array}{llr}
    \mydefhead{d,e,f,g,p,q ::=\qquad\qquad\qquad\qquad\qquad} {Expressions:} 
    \mydefcase x                                              {variable} 
    \mydefcase \primfun                                       {primitive function}
    \mydefcase {\lam x S e}                                   {abstraction}
    \mydefcase {\app f e}                                     {application} 
    \mydefcase \constructor                                   {\bf \emph{constructor}}
    \mydefcase {\caseexp e {\listof {c \Mapsto f}}}           {\bf \emph{case discrimination}}
    \\
    \mydefhead{v ::=\qquad\qquad\qquad\qquad\qquad}{Values: ($\subset$ Terms)}
    \mydefcase \primfun                                  {primitive function}
    \mydefcase {\lam x S e}                              {abstraction}
    \mydefcase {\constructor~\listof e}                  {\bf \emph{constructor application}}
    \\
    \mydefhead{S,T,U::=}{Types:} 
    \mydefcase{\predty x B \predp}      {refinement type} 
    \mydefcase{\lamt x S T}             {dependent function type}
    \mydefcase{\existsbind x S T}       {\bf \emph{existential type}}
  \end{array}
  \]

  \begin{eqnarray*}
    c~\listof \e & \defeq & c~{e_1}~ \cdots~ {e_n} 
    \\
    \lamtmulti x S {T} 
    & \defeq &
    \lamt {x_1} {S_1} {} \cdots \arrowt {} {\lamt {x_n} {S_n} {T}}
    \\
    \existsmulti x S T
    & \defeq &
    \existsbind {x_1} {S_1} {} \cdots \existsbind {x_n} {S_n} {T}
    \\
    \listof{\patternexp \constructor \f}
    & \defeq &
    \patternexp {\constructor_1} {\f_1}, \cdots, \patternexp {\constructor_n} {f_n}
  \end{eqnarray*}
}

\newcommand{\synlabel}[1]{\mbox{\emph{#1}}}

\newcommand{\restrictedsyntax}{
  \[
  \begin{array}{llr}
    \mydefhead{p ::=}                     {Predicates:}
    \mydefcase l                          {atomic predicate}
    \mydefcase{p \wedge p}                {conjunction}
    \mydefcase{
      \caseexp w {\listof{\patternexp {\constructor~\listof{x}} \predp}}
    }{
      case discrimination
    }
    \\
    \mydefhead{d,e,f,g,p,q ::=\qquad\qquad\qquad\qquad\qquad} {Terms:} 
    \mydefcase x                                              {variable} 
    \mydefcase \primfun                                       {primitive function}
    \mydefcase {\lam x R e}                                   {abstraction}
    \mydefcase {\app f e}                                     {application} 
    \mydefcase \constructor                                   {constructor}
    \mydefcase {
      \caseexp w {\listof{\patternexp {\constructor~\listof{x}} \e}}
    }{
      case discrimination
    }
    \\
    \mydefhead{v ::=\qquad\qquad\qquad\qquad\qquad}{Values: ($\subset$ Terms)}
    \mydefcase \primfun                                  {primitive function}
    \mydefcase {\lam x S e}                              {abstraction}
    \mydefcase {\constructor~\listof e}                  {constructor application}
    \\
    \mydefhead {Q,R::=}                    {Restricted Types:}
    \mydefcase {\predty x B p}             {refinement type}
    \mydefcase {\lamt x Q R}               {restricted function type}
    \\
    \mydefhead {E,F::=}                    {Types with Covariant Existentials:}
    \mydefcase {\predty x B p}             {refinement type}
    \mydefcase {\lamt x R E}               {augmented function type}
    \mydefcase {\existsbind x E F}         {augmented existential type}
    \\
    \mydefhead {\existsenv ::=}            {Algorithmic Typing Environments:} 
    \mydefcase \emptyset                   {empty environment}
    \mydefcase {\existsenv,x:E}            {environment extension}
  \end{array}
  \]
}

\newcommand{\tycasedefinition}{
  \[
  \begin{array}{r@{~~}c@{~~}l}
    \caseexp w {\listof{\patternexp {\constructor~\listof{x}} \e}}
    & \defeq &
    \caseexp w {\listof{\patternexp \constructor {\lammulti x R {\lam y E \e}}}}
    \\
    & \mbox{ where } & ty(c) = \lamtmulti x R {\predty y B \predp}\\
    & \mbox{ and }   & \mbox{$E$ has the form $\existsmulti z F {\predty y B {\predp \wedge \predq}}$} \\
    & & \mbox{($\listof {z:F}$ and $\predq$ will be clear from context)} 
  \end{array}
  \]

  \[
  \begin{array}{r@{~~}c@{~~}l}
    \tycase w {\listof{\patternexp {\constructor~\listof{x}} {\existsmulti {z_c} {E_c} {F_c}}}}
    & \defeq &
    \existsmulti {x_c} {E_c} {\tycase w {\listof{\patternexp \constructor {F_c}}}}
    \\
    \multicolumn{3}{r}{
      \mbox{ 
        where $\listof{x_c:E_c}$ contains all of the bindings for each $c$,
        (possibly empty)
      }
    }
    \\
    \\
    \tycase w {\listof{\patternexp {\constructor~\listof{x}} {\predty z B {\predq}}}}
    & = &
    \predty y B {\caseexp w {\listof{
          \patternexp {\constructor~\listof{x}} \predq
        }
      }
    }
    \\
    && \mbox{ where $ty(c) = \lamtmulti {x_c} {R_c} {\predty z B \predp}$}
    \\
    \\
  \end{array}
  \]
}

\newcommand{\lambdaexsyntax}{
    \[
    \begin{array}{llr}
      \mydefhead{e ::=\qquad\qquad\qquad\qquad\qquad}{Expressions:} 
      \mydefcase x                              {variable} 
      \mydefcase c                              {constructor} 
      \mydefcase f                              {primitive function}
      \mydefcase {\lam x S e}                   {abstraction}
      \mydefcase {\app e e}                     {application} 
      \mydefcase {\caseexp e {c~\listof x} e e} {case}
      \\
      \mydefhead{S,T::=}{Types:} 
      \mydefcase{\predty x B e}      {refinement type} 
      \mydefcase{\lamt x S T}        {dependent function type}
      \\
      \mydefhead{A::=}{Augmented Types:}
      \mydefcase{\predty x B e}             {refinement type}
      \mydefcase{\lamt x S A}               {augmented function type}
      \mydefcase{\existsty \existsenv A}    {existential type}
      \\
      \mydefhead{B::=}{Base types:} 
      \mydefcase{\Int 			}{base type of integers}
      \mydefcase{\Bool			}{base type of booleans}
      \mydefcase{\IntList       }{base type of lists of integers}
      \\
      \mydefhead{\tyenv,\existsenv ::=}{Environments:} 
      \mydefcase{\emptyset			}{empty environment}
      \mydefcase{\tyenv,x:S			}{environment extension}
    \end{array}
    \]
}

\newcommand{\lambdahcsyntax}{
    \[
    \begin{array}{llr}
      \mydefhead{e ::=\qquad\qquad\qquad\qquad\qquad}{Expressions:} 
      \mydefcase x                              {variable} 
      \mydefcase c                              {constructor} 
      \mydefcase f                              {primitive function}
      \mydefcase {\lam x S e}                   {abstraction}
      \mydefcase {\app e e}                     {application} 
      \mydefcase {\caseexp e {c~\listof x} e e} {case}
      \\
      \mydefhead{S,T::=}{Types:} 
      \mydefcase{\lamt x S T}        {dependent function type}
      \mydefcase{\predty x B e}      {refinement type} 
      \mydefcase{\existsty {x:S} T}  {existential type}
      \\
      \\
      \mydefhead{B::=}{Base types:} 
      \mydefcase{\Int 			}{base type of integers}
      \mydefcase{\Bool			}{base type of booleans}
      \mydefcase{\IntList       }{base type of lists of integers}
      \\
      \mydefhead{\tyenv ::=}{Environments:} 
      \mydefcase{\emptyenv			}{empty environment}
      \mydefcase{\tyenv,x:S			}{environment extension}
    \end{array}
    \]
    \mynote{Since lambdas can only have $S$, note that I allowed only
      these in the environment.}
}

\newcommand{\lambdaexevalrules}{
  \[
  \begin{array}{@{}rclr}
    \headrule{${e_1}\red{e_2}$}{Redex \underline{E}valuation}
    \\
    {\app{(\lam{x}{T}{e_1})}{e_2}}
    &\red&
    {\subst{x}{e_2}{T}\, e_1}
    &\rel{E-$\beta$}
    \\
    \app \primfun \e
    &\red&
    \deltaf{\primfun}{e} 
    &\rel{E-$\delta$}
    \\
    \caseexp {\constructor~\listof e} {\patternexp \constructor \f, \cdots}
    &\red&
    \f ~ {\listof e} ~ (\constructor ~ \listof e)
    &\rel{E-Case}
    \\
    \\
    \\
    \headrule{$\tmd \lred \e$}{Contextual \underline{E}valuation}
    \\
    \qquad\qquad\qquad	\tmctx[\tmd] & \lred & \tmctx[\e] \quad \mbox{if $\tmd \red \e$}
    &\rel{E-Compat} \\
    \qquad\qquad\qquad	\tyctx[\tmd] & \lred & \tyctx[\e] \quad \mbox{if $\tmd \red \e$}
    &\rel{E-TyCompat} \\
    \\
    \headrule{\tmctx, \tyctx}{Evaluation Contexts}
    \\
    \multicolumn{4}{l}{
      \begin{array}{lll}
      \tmctx  & ::= &
      \hole
      ~|~ \app \tmctx \e
      ~|~ \app \e \tmctx
      ~|~ \lam x \tyS \tmctx
      ~|~ \lam x \tyctx \e
      ~|~ \caseexp \tmctx {\listof {\patternexp \constructor \f}}
      ~|~ \caseexp \e {\patternexp \constructor \tmctx, \cdots}
      \\
      \tyctx & ::= &
      \lamt x \tyctx \tyT
      ~|~ \lamt x \tyS \tyctx
      ~|~ \predty x B \tmctx
      ~|~ \existsbind x \tyctx T
      ~|~ \existsbind x S \tyctx
    \end{array}
    }
    \\
  \end{array}
  \]
}

\newcommand{\instanceof}{\mbox{\,\,$\lessdot$\,\,}}
\newcommand{\tycase}{\caseexp}

\newcommand{\compositionaltyperules}{
  \headtrule{$\judge \tyenv \e T$}{\underline{T}yping}
  \begin{trules}
    \trule{T-Const}{~
    }{
      \JudgeH {} \constructor {\fun{ty}(\constructor)}
    }

    \trule{T-Prim}{~
    }{
      \JudgeH {} \constant {\fun{ty}(\constant)}
    }

    \trule{T-Var}{
      (x:T)\in \tyenv
    }{
      \judgeH \tyenv x {\self(T,x)}
    }
    
    \trule{T-Fun}{
      \judgeT{\env}{S} 
      \and
      \judge{\env,x:S}{\e}{\tyT}
    }{
      \judge{\env}{(\lam{x}{\tyS}{\e})}{(\lamt{x}{\tyS}{\tyT})}
    }

    \trule{T-App}{
      \judgeH \tyenv \e S 
      \and
      \judgeH \tyenv \f {\arrowt S T}
    }{
      \judgeH \tyenv {\f ~ \e} T
    }

    \\\\

    \trule{T-Case}{
      \judge \tyenv e {\existsmulti z T {\predty y B \predq}}
      \\\\
      \mbox{For each $c \Mapsto f$:}
      \\\\
      ty(c) = \lamtmulti x S {\predty y B {\predp}}
      \and
      \judge \env \f {
        (
        \lamtmulti x S {}
        \lamt y {\existsmulti z T {\predquick B {\predp \wedge \predq}}} U
        )
      }
    }{
      \judge \tyenv {\caseexp e {\listof {c \Mapsto f}}} U
    }
  \end{trules}
}
  
\newcommand{\lambdaexwellformedtypes}{
  \headtrule{$\judgeT {} T$}{\underline{W}ell-formed \underline{T}ypes}
  \begin{trules}
    \trule{WT-Arrow}{
      \judgeT \env \tyS 
      \and 
      \judgeT {\env,x:S} \tyT
    }{
      \judgeT \env {\lamt x \tyS \tyT}
    }

    \trule{WT-Base}{
      \judge {\env,x:B} \predp \Bool	
    }{
      \judgeT \env {\predty x B \predp}
    }

    \trule{WT-Exists}{
      \JudgeHT {} {S} 
      \and
      \JudgeHT {,x\!:\!S} {T}
    }{
      \JudgeHT {} {\existsbind x S T}
    }
  \end{trules}
}

\newcommand{\selfdefinition}{
  \[
  \begin{array}{r@{~~=~~}l}
    \self(\predty x B \predp, e) & \predty x B {\predp \wedge (x =_B e)} \\
    \self(\lamt x S T, e) & \lamt x S {\self(T, e~x)} \\
    \self(\existsbind x S T, e) & \existsbind x S {\self(T, e)}
    \vspace{12pt}
  \end{array}
  \]
}

\newcommand{\compositionalsubtyping}{
  \headtrule{$\judgesub \env S T$}{\underline{S}ubtyping}
  \begin{trules}
    \trule{S-Arrow}{
      \judgesub \env {\tyT_1}{\tyS_1}
      \and
      \judgesub{\env,x:\tyT_1}{\tyS_2}{\tyT_2}
    }{
      \judgesub \env {(\lamt x {\tyS_1} {\tyS_2})} {(\lamt x {\tyT_1}{\tyT_2})}
    }
    
    \trule{S-Base}{
      \judgeimp {\env,x:B} \predp \predq
    }{
      \judgesub \env {\predty x B \predp} {\predty x B \predq}
    }

    \trule{S-Witness}{
      \JudgeH {} e T 
      \and
      \JudgeHsub {} S {\subst x e S\, U}
    }{
      \JudgeHsub {} S {\existsty {x\!:\!T} U}
    }
    
    \trule{S-Bind}{
      \JudgeHsub {,x\!:\!S} T U
      \and
      x\not\in FV(U)
    }{
      \JudgeHsub {} {\existsty {x:S} T} U
    }
  \end{trules}
}

\newcommand{\typingalgorithm}{
  \headtrule{$\JudgeAlg {} e E$}{\underline{A}lgorithmic Typing}
  \begin{trules}
    \trule{A-Const}{~
    }{
      \JudgeAlg {} \constructor {\fun{ty}(\constructor)}
    }

    \trule{A-Prim}{~
    }{
      \JudgeAlg {} \primfun {\fun{ty}(\primfun)}
    }

    \trule{A-Var-Base}{
      (x:A) \in \existsenv 
    }{
      \JudgeAlg {} x {\self(E,x)}
    }

    \trule{A-Fun}{
      \JudgeAlgT {} R
      \and
      \JudgeAlg {,x\!:\!R} e E
    }{
      \JudgeAlg {} {(\lam x R e)} {(\lamt x R E)}
    }

    \trule{A-App}{
      \JudgeAlg {} {e_1} {\existsmulti z {E_z} {\lamt x R {E_1}}}
      \and
      \JudgeAlg {} {e_2} {E_2}
      \and
      \JudgeAlgsub {,\listof {z\!:\!E_z}} {E_2} R
    }{
      \JudgeAlg {} 
      {\app {e_1} {e_2}} 
      {\existsmulti z {E_z} {\existsbind x {E_2} {E_1}}}
    }

    \\\\

    \trule{A-Case}{
      \judgeAlg \existsenv w {\existsmulti z E {\predty y B q}}
      \\\\
      \mbox{For each $\patternexp {\constructor ~ \listof x} \tmd$}:
      \\\\
      ty(c) = \lamtmulti x Q {\predty y B {p}}
      \and
      \JudgeAlg {,\listof{x:Q}, y:\existsmulti z E {\predquick B {q\wedge p}}} \tmd F
    }{
      \JudgeAlg {} {
        (\caseexp w {\listof{
            \patternexp {\constructor~\listof{x}} \tmd
        }})
      }{
        (\caseexp w {\listof{
            \patternexp {\constructor~\listof{x}} F
        }})
      }
    }
  \end{trules}
}

\newcommand{\lambdaextypewellformednessalgorithm}{
  \headtrule{$\JudgeAlgT {} R$}{\underline{A}lgorithmic Well-formed \underline{T}ypes}
  \begin{trules}
    \trule{AT-Arrow}{
      \JudgeAlgT {} {R_1} \qquad
      \JudgeAlgT {,x\!:\!R_1} {R_2}
    }{
      \JudgeAlgT {} {\lamt x {R_1} {R_2}}
    }

    \trule{AT-Base}{
      \JudgeAlg {,x:B} p \Bool	
    }{
      \JudgeAlgT {} {\predty x B p}
    }
  \end{trules}
}
 
\newcommand{\compositionalsubalg}{ 
  \headtrule{$\JudgeAlgsub{} E R$}{\underline{A}lgorithmic \underline{S}ubtyping}
  \begin{trules}
    \trule{AS-Arrow}{
      \JudgeAlgsub {} {R_2} {R_1} \qquad
      \JudgeAlgsub {,x:R_2} {E_1} {R_3}
    }{
      \JudgeAlgsub {} {(\lamt x {R_1} {E_1})} {(\lamt x {R_2} {R_3})}
    }

    \trule{AS-Base}{
      \judgeimp {\existsenv,x:B} {q} {p}
    }{
      \JudgeAlgsub {} {\predty x B {q}} {\predty x B {p}}
    }

    \trule{AS-Bind}{
      \JudgeAlgsub {,x\!:\!E_1} {E_2} {R_3} 
    }{
      \JudgeAlgsub {} {\existsbind x {E_1} {E_2}} {R_3}
    }
  \end{trules}
  
}

\newcommand{\instancerules}{
  \begin{mathpar}
    \trule{I-Refl}{
      
    }{
      \judgeinstance \tyenv T T
    }
    \\
    \trule{I-Arrow}{
      \judgeinstance {\tyenv,x:S} T A
    }{
      \judgeinstance \tyenv {\lamt x S T} {\lamt x S A}
    }
    \\
    \trule{I-Exists}{
      \judgeH \tyenv e S \qquad \judgeinstance \tyenv T {\subst x e S\, A}
    }{
      \judgeinstance \exists T {\existsty {x:S} A}
    }
  \end{mathpar}
}

\newcommand{\Trans}{\ensuremath{Trans}}
\newcommand{\Pred}{\ensuremath{Pred}}

\newcommand{\imptranslation}{
  \begin{eqnarray*}
    \Trans(\judgeimp {} {e_1} {e_2}) & = & 
    {e_1} \lred^* \true \Implies {e_2} \lred^* \true
    \\
    \Trans(\judgeimp {x:T,E} {e_1} {e_2}) & = &
    \forall x:\Base(T). \Pred(T,x) \\ && 
    \qquad \Implies \Trans(\judgeimp E {e_1} {e_2}) \\
    \\
    \Base(\predty x B p) & = & B \\
    \Base(\lamt x {T_1} {T_2}) & = & \arrowt {\Base(T_1)} {\Base(T_2)} \\
    \Base(\existsbind x {T_1} {T_2}) & = & \Base(T_2) \\
    \\
    \Pred(\predty x B e, e') & = & \subst x {e'} {}\, e \lred^* \true \\
    \Pred((\lamt x {T_1} {T_2}), e') & = &
    \forall x:\Base(T_1). \Pred(T_1,x) \\ && 
    \qquad\Implies \Pred(T_2,e'~x) \\
    \Pred((\existsbind x {T_1} {T_2}), e') & = & 
    \exists x:\Base(T_1). \Pred({T_1},x) \\ && 
    \qquad \wedge ~ \Pred({T_2},e')
  \end{eqnarray*}
}

\renewcommand{\bisim}{\ensuremath{\simeq}}
\newcommand{\judgeequiv}[3]{\ensuremath{#1\vdash #2\,\equiv\,#3}}
\renewcommand{\judgesimt}[4]{\ensuremath{#1\vdash #2\,\bisim\,#3:#4}}

\renewcommand{\judgesim}[3]{\ensuremath{#1\vdash #2\,\bisim\,#3}}
\renewcommand{\judgesimmt}[2]{\ensuremath{{#1} \bisim {#2}}}

\renewcommand{\bisimilarityrules}{
  \mynote{Let this slide until we have a good story on the basic theory}

  \emph{Definitional equivalence of types modulo redundant casts:}
  \[
  \begin{array}[t]{@{}c@{}r@{}}
    \trule{Eq-Arrow}{
      \judgeequiv {E,x:S_1} {S_2} {T_2}
    }{
      \judgeequiv E {\lamt x {S_1} {S_2}} {\lamt x {T_1} {T_2}}
    }
  \end{array}
  \qquad
  \begin{array}[t]{@{}c@{}r@{}}
    \trule{Eq-Base}{
      \judgesimt {E,x:B} s t \Bool
    }{
      \judgeequiv E {\predty x B s} {\predty x B t}
    }
  \end{array}
  \]

  \

  \emph{Reflexivity and congruence bisimilarity rules:}
  \[
  \begin{array}[t]{@{}c@{}r@{}}
    \trule{Up-Refl}{
      \judge E t T
	}{
	  \judgesimt E t t T
	}
    \trule{Up-Cast}{
	  \judgeequiv E S T  
    }{
	  \judgesimt E {\cast S} {\cast T} {\arrowt {Top(S)} S}
    }
  \end{array}
  \qquad
  \begin{array}[t]{@{}c@{}r@{}}
    \trule{Up-Fun}{
      \judgeequiv E S {S'}
      \qquad
      \judgesimt {E,x:S} s t T
    }{
	  \judgesimt E {(\lam x S s)} {(\lam x S' t)} {(\lamt x S T)}
    }
    \trule{Up-App}{
	  \judgesimt E s {s'} {\lamt x S T}
      \qquad
	  \judgesimt E t {t'} S
    }{
	  \judgesimt E {s~t} {s'~t'} {{\subst x {t'} S} \, T}
    }
    \trule{Up-Equiv}{
      \judgesimt E s t S
      \qquad
      \judgeequiv E S T
    }{
      \judgesimt E s t T
    }
  \end{array}
  \]
  \[
  \begin{array}[t]{@{}c@{}r@{}}
    \trule{Up-Checking}{
      \judgesimt {E,x:B} p q \Bool
      \qquad
      \subst x c B\, p \red^* s
      \qquad
      \subst x c B\, q \red^* t
    }{
      \judgesimt E {
        \checkingpred x B p s c
      } { 
        \checkingpred x B q t c
      } {
        \predty x B p
      }
    }
  \end{array}
  \]

  \

  \emph{Extensional bisimilarity modulo redundant casts (Function casts
    perform eta-expansion):}
  \[
  \begin{array}[t]{@{}c@{}r@{}}
    \trule{Up-Add}{
      \judgesimt E s t T
	}{
	  \judgesimt E s {\castapp T t} T
	}
    \trule{Up-Eta}{
      \judgesimt E s t {(\lamt x S T)}
	}{
	  \judgesimt E s {(\lam x S t~x)} {(\lamt x S T)}
	}
    \trule{Up-AddChecking}{
      \judge E s {\predty x B q}
      \qquad
      \judgesimt E s c {\predty x B q}
      \qquad
      {\subst x c B}\, q \red^* t
    }{
      \judgesimt E s {\checkingpred x B q t c} {\predty x B q}
    }
  \end{array}
  \]
}

\newcommand{\lambdaexextensionalequivalence}{
  \[
  \begin{array}{@{\qquad}lcl}  
    \multicolumn{3}{@{}l}{\mbox{ \fbox{$\judgesimt \env {\e_1} {\e_2 : T}$}~~ (defined by induction
        on $\Base {T}$) }} \\~\\
    \judgesimt \env {\e_1} {\e_2:\predty x B \predp} 
    & \defiff &
    \judgeIn \env {\e_1} {\predty x B \predp}
    \\ & &
    \judgeIn \env {\e_2} {\predty x B \predp}
    \\ & &
    \forall \closingsubst $ such that $ \judge {} \closingsubst \env, ~
    (\closingsubst(\e_1) \lred^* \constant) \Leftrightarrow (\closingsubst(\e_2) \lred^* \constant) 
    \\
    \\
    \judgesimt \env {\f_1} {\f_2 : {\plamt x S T}} 
    & \defiff &
    \forall \e_1, \e_2 $ such that $ \judgesimt \env {\e_1} {\e_2:S},
    \\ & & 
    \judgesimt \env {\f_1~\e_1} {\f_2~\e_2 : {\subst x {\e_1} {}}\, T \curlywedge \subst x {\e_2} {}\, T}
    \\
    \\
    \judgesimt \env {\e_1} {\e_2 : {\pexistsbind x S T}} 
    & \defiff &
    \exists \tmd_1, \tmd_2 $ such that $ \judgesimt \env {\tmd_1} {\tmd_2:S},
    \\ & & 
    \judgesimt \env {\e_1} {\e_2 : {\subst x {\tmd_1} {}}\, T \curlywedge \subst x {\tmd_2} {}\, T}
    \\
    
    \multicolumn{3}{@{}l}{\mbox{ \fbox{$S \curlywedge T$}}}
    \\~\\
    \predty x B \predp \curlywedge \predty x B \predq & \defeq & \predty x B {\predp \wedge \predq}
    \\
    \plamt x {S_1} {S_2} \curlywedge \plamt x {T_1} {T_2} & \defeq & \lamt x {(S_1 \curlywedge T_1)} {(S_2 \curlywedge T_2)}
    \\
    \pexistsbind x {S_1} {S_2} \curlywedge \pexistsbind x {T_1} {T_2} & \defeq & \existsbind x {(S_1 \curlywedge T_1)} {(S_2 \curlywedge T_2)}
    \\
  \end{array}
  \]
}

\newcommand{\StateBisimulation}{
  \rnlemma{bisim}{Bisimulation}{
    Suppose $\judgesimt E s t T$ 
    \begin{enumerate}
    \item
      If $s\lred^* s'$ 
      then  $\exists t'$ 
      such that $t\lred^* t'$ 
      and $\judgesimt E {s'} {t'} T$ 
    \item
      If $t\lred^* t'$ 
      then $\exists s'$ 
      such that $s\lred^* s'$ 
      and $\judgesimt E {s'} {t'} T$ 
    \end{enumerate}
  }
}

\newcommand{\StateCastInsertBisim}{
  \rnlemma{castinsertbisim}{(Cast insertion is Upcasting)}{
    Suppose $\judgeE{E}$.
    \begin{enumerate}
    \item
      If $\judge E s S$ and $\judgec E s t T$ 
      then $\judgesimty E S T$ and $\judgesimt E s t S$.
    \item
      If $\judge E s S$ and $\judgecc E s t {} T$ and $\judgesimty E S T$ 
      then $\judgesimt E s t T$.
    \item
      If $\judgeT E S$ and $\judgect E S T$  then $\judgesimty E S T$.
    \end{enumerate}
  }
}

\newcommand{\StateWtProgAccepted}{
  \rnlemma{wtprogaccepted}{(Compilation Completeness)}{
    Suppose $\judgeE{E}$.
    \begin{enumerate}
    \item
      If $\judge{E}{s}{S}$ then $\exists t, T$
      such that $\judgec{E}{s}{t}{T}$.
    \item
      If $\judge{E}{s}{S}$ and $\judgesub{E}{S}{T}$ then $\exists t$  
      such that $\judgecc{E}{s}{t}{}{T}$.
    \item
      If $\judgeT{E}{S}$ then $\exists T$
      such that $\judgect{E}{S}{T}$.
    \end{enumerate}
  }
}

This chapter describes the metatheoretical core of this dissertation:
\emph{executable refinement types}. This is a type system with the
expressivity of dynamic contracts. Basic built-in types such as $\Int$
and $\Bool$ may be refined by arbitrary predicates written in the
programming language at hand.  These refined base types are combined
via dependent function types.  Because these types share their design
with higher-order dependent contracts, they can be enforced at
run-time using the same techniques.

Before delving into formal definitions, consider these illustrative
examples. The type of integers greater than zero may be written as
\[
\predty x \Int {x > 0}
\]
and the type of binary search trees may be written as
\[
\predty x {\t{Tree}} {\t{isBinarySearchTree}(x)}
\]
where $\t{isBinarySearchTree}$ is a hand-coded integrity checking
routine. The operation to add an integer $x$ to a set $y$ may be
given the type
\[
\lamt x {\Int} {}
\arrowt {\t{Set}} {}
\predty z {\t{Set}} {x \in z}
\]

Any semi-decidable predicate over base types is expressible, including
undecidable predicates. One such undecidable predicate is ``integers
representing the binary encoding of a turing machine which halts on
empty input'' which might appear as $\predty x \Int
{\t{haltsOnEmptyInput}(x)}$. While such precise types make type
checking undecidable, the type system still enjoys the usual notion of
type soundness (``well typed programs don't go wrong'').  Thus, we can
use it as a reference point from which to explore pragmatic techniques
for working with expressive types. Rather than building a full
programming language suitable for general use, the type system is
presented as part of a small calculus, in order to focus on the
theoretical aspects of the type system. 

\subsection*{Chapter Outline}

\noindent
The metatheoretical contributions of this chapter are organized as
follows:

\begin{itemize}
\item[-] \hhref{Section}{section:lambdah:syntax} presents the
  constructs of the calculus $\lambdah$.

\item[-] \hhref{Section}{section:lambdah:opsem} gives formal meaning
  to the terms of $\lambdah$ via a small-step operational semantics.

\item[-] \hhref{Section}{section:lambdah:denotation} formalizes the
  intuition that refinement types express subsets of types of the
  simply-typed lambda calculus, providing a semantic foundation
  for the syntactic formalisms of the $\lambdah$ type system.

\item[-] \hhref{Section}{section:lambdah:typesystem} presents
  the type system of $\lambdah$ as a set of inference rules
  for assigning types to terms and proving subtyping between
  $\lambdah$ types. 

\item[-] \hhref{Section}{section:lambdah:typesoundness} establishes
  type soundness for $\lambdah$ (any well-typed term is either
  nonterminating or results in a value of the same type) via the
  syntactic approach of proving lemmas for \emph{progress} (any
  well-typed term is either a value or can evaluate further) and
  \emph{preservation} (a term's type remains unchanged by evaluation).

\item[-] \hhref{Section}{section:lambdah:logicalrelation} establishes
  that $\lambdah$ terms respect \emph{extensionality} (functions which
  map equal inputs to equal output are observably equivalent) via a
  new \emph{logical relation}~\citep{statman_logical_1985}. This
  provides a powerful technique for proving the observable equivalence
  of $\lambdah$ terms.

\item[-] \hhref{Section}{section:lambdah:relatedwork} discusses
  other systems with similar approaches to specifying or reasoning
  about programs.
\end{itemize}

\section{The Language $\lambdah$}
\label{section:lambdah:syntax}

This section defines our calculus $\lambdah$, an enhancement of STLC
with more precise types to model programming languages with types that
can express the same specifications as dynamic contracts.
\hhref{Figure}{figure:lambdah:syntax} summarizes the syntax of
$\lambdah$ for reference.  The grammar of terms is unchanged from
STLC; it is the type language of $\lambdah$ that forms the core of
this dissertation: Basic types refined by executable predicates and
dependent function types.

\begin{displayfigure}{Syntax of $\lambdah$}
  \ssp
  \label{figure:lambdah:syntax}
  \basicsyntax
\end{displayfigure}

\begin{itemize}
\item[-] A dependent function type $\lamt x S T$ classifies functions
  of domain $S$ and codomain $T$.\footnote{We use the notation from
    Cayenne~\citep{augustsson_cayenne_1998} $\lamt x S T$ in
    preference to the equivalent syntax $\Pi x\!:\!S.~T$ for dependent
    function types. The use of $\Pi$ indicates that $\Pi x\!:\!S.T$ may
    be considered to be the type of $S$-indexed products of types
    $T[x]$.} However, $x$ may appear in $T$, so the type of the result
  depends on the input. If it does not occur free in $T$, yielding the
  simple function type syntax $\arrowt S T$.

\item[-] The basic types $B$ of STLC are fairly coarse and cannot
  denotes types of interest that are common in contracts, such as
  integer subranges. To overcome this limitation, we
  introduce \emph{executable refinement types} of the form $\predty x
  B \predp$.  Here, the variable $x$ (of basic type $B$) can occur
  within the boolean term \predp.  Informally, this refinement type
  denotes the set of primitive constants $\constant$ of type $B$ that
  satisfy the predicate $p$. Thus, $\predty x B \predp $ denotes a
  refined subtype of $B$.  Occasionally, the predicate $\predp$ will
  not contain any free occurrences of $x$, and we will write
  $\predquick B \predp$.  We also write simply $B$ as an abbreviation
  for the trivial refinement type $\predquick B \true$. These
  refinement types are inspired by prior work on decidable refinement
  type systems discussed
  in \hhref{Section}{section:intro:refinementtypes}. However, our
  refinement predicates are arbitrary boolean expressions, so every
  recursively enumerable subset of the integers is actually a
  $\lambdah$ type.
\end{itemize}

To illustrate the precision of this type language, let us adjust the
$ty$ function for $\lambdah$ to give primitives more precise types, such
as
\[
\begin{array}[t]{r@{~:~}l}
	\true 	& \predty{b}{\Bool}{b} \\
	\false 	& \predty{b}{\Bool}{\t{not}~b} \\
	\Iff	& \lamt{b_1}{\Bool}{\lamt{b_2}{\Bool}{\predty{b}{\Bool}{b \Iff (b_1\Iff b_2)}}} \\
	\t{not}	& \lamt{b}{\Bool}{\predty{b'}{\Bool}{b \Iff \t{not}~b}} \\
	n	& \predty{m}{\Int}{m=n} \\
	+	& \lamt{n}{\Int}{\lamt{m}{\Int}{\predty{z}{\Int}{z=n+m}}}\\
	\t{if}_{T}& \arrowt{\Bool}\arrowt{T}\arrowt{T}T  \\
	\fix{T}	& \arrowt{(\arrowt{T}{T})}{T}
\end{array}
\]

\noindent
In particular, we assume that each primitive constant of boolean and
integer type is assigned a singleton type that denotes exactly that
constant. For example, the type of an integer $n$ denotes the
singleton set $\setz{n}$. This reflects the fact that a literal
constant itself, being present in the code, need not be abstracted to
any larger type.

The types for primitive functions may also be quite precise: The type
\[
+: \lamt{n}{\Int}{\lamt{m}{\Int}{\predty{z}{\Int}{z=n+m}}}
\]
exactly specifies that this function performs addition.  That is, the
term $n+m$ has the type $\predty{z}{\Int}{z=n+m}$ denoting the
singleton set $\setz{n+m}$.  Note that even though the type of ``$+$''
is defined in terms of ``$+$'' itself, this does not cause any
problems in our technical development, since the semantics of $+$ do
not depend on its type.\footnote{The definition of a primitive in
  terms of itself is, however, unsatisfying. In the case of functions,
  we may hope that the type is an abstraction of the value. In
  \hhref{Chapter}{chapter:compositionality} we explore some of the
  consequences of giving such precise types to functions.}

The constant $\fix{T}$ is the fixpoint constructor of type $T$, and
enables the definition of recursive functions.  For example, the
factorial function can be defined as:

\[
\ssp
\begin{array}{l}
	{\fix{\arrowt{\Int}{\Int}}}\\\quad
	    {\lam{f}{(\arrowt{\Int}{\Int})}{}}\\\qquad
		{~\lam{n}{\Int}
			{}}
	\\
	\qquad\qquad
	\t{if}_{\Int}~
	\begin{array}[t]{@{}l}
	{(n=0)}\\
	1\\
	{(n*(\app{f}{(n-1)}))}
	\end{array}

\end{array}
\]

The types of $\lambdah$ can express many other precise specifications,
such as the following (where we assume that $\Unit$, $\Array$, and
$\t{RefInt}$ are additional basic types, and the primitive function
$\t{sorted}:\arrowt{\Array}{\Bool}$ identifies sorted arrays.)
\begin{itemize}
\item[-] $\t{printDigit} ~:~\arrowt{\predty{x}{\Int}{0\leq x \wedge x\leq 9}}{\Unit}$.
\item[-] $\t{swap}~:~
	\lamt{x}{\t{RefInt}}
		{\arrowt{\predty{y}{\t{RefInt}}{x\neq y}}
			{\Bool}}$.
\item[-] $\t{binarySearch} ~:~ \arrowt{\predty{a}{\Array}{\app{\t{sorted}}{a}}}{\arrowt{\Int}{\Bool}}$.
\end{itemize}

\section{Operational Semantics of $\lambdah$}
\label{section:lambdah:opsem}

We next give formal meaning to $\lambdah$ terms via small-step
operational semantics, as we did for STLC in
\hhref{Section}{section:stlc:opsem}.  In fact, redex evaluation for
$\lambdah$ is identical to that of STLC. However, $\lambdah$ types may
contain terms, so we extend evaluation to types via additional
contexts in a natural way. \hhref{Figure}{figure:lambdah:opsem}
presents the operational semantics in full, for easy reference.

Since types contain terms, evaluation is also defined for types. When
a term $\tmd$ reduces to another term $\e$, then the two terms are
definitionally equivalent, so the corresponding types $\tyctx[\tmd]$
and $\tyctx[\e]$ should also be equivalent. This is proven directly
once the appropriate notion of type equivalence is
defined.~(\hhrefpref{Lemma}{lemma:TypeEval})

\begin{displayfigure}{Operational Semantics of $\lambdah$}
  \ssp
  \label{figure:lambdah:opsem}
  \basicopsem
\end{displayfigure}

\section{Denotation of $\lambdah$ types to sets of $STLC$ terms}
\label{section:lambdah:denotation}

Intuitively, $\lambdah$ types describe subsets of the basic types of
$STLC$ and functions upon those subsets. To validate this intuition
and provide a solid foundation for $\lambdah$, as well as demonstrate
how to do so for future systems, we ``bootstrap'' the type system by
using a denotational interpretation into STLC. This corresponds with
the intuition -- also supported by the identical redex evaluation
rules -- that a program with executable refinement types is
essentially an STLC program with a more precise specification.

Underlying every $\lambdah$ type $T$ and term $\e$ is an STLC type or
term, respectively, which we call its \emph{shape} and write $\Base T$
or $\Base \e$. \hhref{Figure}{figure:lambdah:shape} shows the formal
definition of the function $\Base - : \lambdah \rightarrow
STLC$.\footnote{This mapping is similar in spirit to \emph{erasure}
  which maps a term of the typed $\lambda$-calculus to the
  corresponding term of the untyped $\lambda$-calculus, but ``shape''
  is more concise than ``refinement erasure''.} 

\begin{displayfigure}{Shape of $\lambdah$ types and terms}
\label{figure:lambdah:shape}
\ssp
\vspace{-20pt}
\begin{eqnarray*}
  \Base {\predty x B \predp} & \defeq & B \\
  \Base {\lamt x S T} & \defeq & \arrowt {\Base S} {\Base T}\\
  \\
  \Base \constant & \defeq & \constant\\
  \Base x & \defeq & x\\
  \Base {\lam x S \e} & \defeq & \lam x {\Base S} {\Base \e} \\
  \Base {\tmf~\e} & \defeq & \Base {\tmf} ~ \Base {\e} \\
\end{eqnarray*}
\vspace{-35pt}
\end{displayfigure}

We give each type $T$ an interpretation $\meaningf{T}$ (defined
by induction on $\Base T$) that is the set of terms $\e$ such that
$\Base{\e}$ is of type $\Base{T}$ and $\e$ obeys the contractual
aspect of $T$. \hhref{Figure}{figure:lambdah:denotation} contains the
formal definition.

\begin{displayfigure}{Denotation from $\lambdah$ types to sets STLC terms}
\label{figure:lambdah:denotation}
\vspace{-40pt}
\begin{eqnarray*}
  \meaningf {\predty x B \predp}
  & \defeq & 
  \{ \e ~|~ \judgeSTLC {} {\Base \e} B
  ~\wedge~ 
  (\e \lred^* \constant \mbox{ implies } \subst x \constant B\, \predp \lred^* \true) \} 
  \\
  \meaningf {\lamt x S T} 
  & \defeq &
  \{
  \tmf ~|~ \judgeSTLC {} {\Base \tmf} {\arrowt {\Base S} {\Base T}}
  ~\wedge~ \forall \e \in \meaningf{S},~ 
  \tmf~\e \in \meaningf{\subst x \e S\, T}
  \}
\end{eqnarray*}
\vspace{-40pt}
\end{displayfigure}
 
\begin{itemize}
\item[-] A refined basic type $\predty x B \predp$ is interpreted as
  the set of closed terms $\e$ of type $B$ for which the predicate
  $\predp$ holds, in the sense that whenever $\e \lred^* \constant$
  for some constant $\constant$, then $\subst x \constant B\, \predp
  \lred^* \true$.  Note that we cannot insist on the simpler
  requirement that $\subst x \e B\, \predp \lred^* \true$ because that
  would forbid assigning types to divergent terms, but we intend that
  any type $T$ is populated by at least the divergent term
  $\t{fix}_T~(\lam x T x)$.  In other terminology, our types specify
  \emph{partial} correctness, not \emph{total} correctness.\footnote{
    An alternate fomulation used by \citet{xu_static_2009} also
    included divergent terms in each type but only required for
    convergent $\e$ that $\subst x \e B\, p \not\lred^* \false$.  In
    other words, the semantics of a predicate included all terms $\e$
    for which membership in $\predty x B \predp$ cannot be effectively
    refuted. The treatment of the case where $\subst x \constant B\,
    \predp$ diverges is subtle and is concerned primarily with blame
    for dynamic contracts. See \citet{dimoulas_correct_2011} for a
    detailed discussion.  }

\item[-] A dependent function type $\lamt x S T$ is interpreted as the
  set of closed terms of simple type $\arrowt {\Base S} {\Base T}$
  that give output in $\meaningf{\subst x \e S\, T}$ whenever their
  input $\e$ is in $\meaningf{S}$.
\end{itemize}

\noindent
With this denotation defined, we can proceed to define the $\lambdah$
type system.

\section{The  $\lambdah$ Type System }
\label{section:lambdah:typesystem}

\hhref{Figure}{figure:lambdah:typerules} and
\hhref{Figure}{figure:lambdah:implication} present the type system of
$\lambdah$ via rules defining the following collection of judgments,
where a typing environment $\env$ maps variables to types.

\begin{itemize}
\item[-] The typing judgment $\judge \env \e \tyT$ states that,
  assuming the bindings in environment $\env$, term $\e$ has type
  $\tyT$. (\hhref{Figure}{figure:lambdah:typerules})

\item[-] The type well-formedness judgment $\judgeT \env \tyT$ states
  that $T$ is a well formed type under the assumptions in environment
  $\env$. (\hhref{Figure}{figure:lambdah:typewellformedness})

\item[-] The environment well-formedness judgment $\judgeT {} \env$
  states that $\env$ is a well formed
  environment. (\hhref{Figure}{figure:lambdah:envwellformedness})

\item[-] The subtyping judgment $\judgesub \env S T$ states that $S$
  is a subtype of $\tyT$ under the assumptions in environment
  $\env$. (\hhref{Figure}{figure:lambdah:subtyping})

\item[-] The implication judgment $\judgeimp \env \predp \predq$
  states that the predicate $\predp$ implies $\predq$ under the assumptions
  in $\env$. (\hhref{Figure}{figure:lambdah:implication})

\item[-] The closing substitution judgment $\judge {} \sigma \env$
  states that the substitution $\closingsubst$ is consistent
  with environment $\env$.  Defining this appropriately is somewhat
  subtle and is the subject of
  \hhref{Section}{section:lambdah:closingsubstitutions}. (\hhref{Figure}{figure:lambdah:implication})
\end{itemize}

\subsection{Typing for $\lambdah$}

The inference rules in \hhref{Figure}{figure:lambdah:typerules} defining
the typing judgment
\[
\judge \env e T
\]
are analogous to those for STLC with the addition of a
\emph{subsumption} rule (\rel{T-Sub}) that allows subtyping to be
applied at any point in a typing derivation. As will always be the
case, we assume that variables are bound at most once in an
environment and implicitly utilize $\alpha$-renaming of bound
variables to maintain this assumption and to ensure substitutions are
capture-avoiding.

\begin{displayfigure}{Typing Rules for $\lambdah$}
  \label{figure:lambdah:typerules}
  \vspace{-35pt}
  \basictyperules       
  \vspace{-40pt}
\end{displayfigure}

\begin{rulelist}
\item[\rel{T-Var}:] A variable $x$ is assigned the type it is bound to in the
  environment $\env$.

\item[\rel{T-Prim}:] A constant $\constant$ is given a type by the
  function $ty$, as updated for $\lambdah$ in
  \hhref{Section}{section:lambdah:syntax}.

\item[\rel{T-Fun}:] For a $\lambda$-abstraction $\plam x S \e$, if the
  domain type $S$ is well-formed and the body $\e$ has type $T$ under
  the additional assumption $x:S$, then the function is given the
  dependent function type $\lamt x S T$. Note again that $x$ may
  appear free in $T$.

\item[\rel{T-App}:] For an application $(\f~\e)$, if the term $\f$ in
  function position has the dependent function type $\lamt x S T$,
  then the resulting type of the application is $\subst x \e {}\, T$,
  where free occurences of the formal argument $x$ in $T$ have been 
  replaced by the actual argument $\e$.

\item[\rel{T-Sub}:] Whenever a term $\e$ has type $S$ and $S$ is
  a subtype of $T$ the that term also has type $T$.
\end{rulelist}

\subsection{Type well-formedness for $\lambdah$}

In STLC any syntactically valid type is well-formed, but in $\lambdah$
types may contain terms (predicates) that may themselves be ill-typed.
The type well-formedness judgment
\[
\judgeT \env T
\]
ensures that each predicate contained anywhere within a type
has type $\Bool$.

\begin{displayfigure}{Type Well-formedness for $\lambdah$}
  \label{figure:lambdah:typewellformedness}
  \vspace{-20pt}
  \lambdahtypewellformedness
  \vspace{-35pt}
\end{displayfigure}

\begin{rulelist}
\item[\rel{WT-Arrow}:] A function type $\lamt x S T$ is well-formed
  if the domain type is well-formed and the range type is
  well-formed with the added assumption of $x:S$.

\item[\rel{WT-Base}:] A refined basic type $\predty x B \predp$
  is well-formed if the predicate $\predp$ is a well-typed
  boolean term under the additional assumption $x:B$.
\end{rulelist}

\subsection{Environment well-formedness for $\lambdah$}

As types may be ill-formed, so an arbitrary environment built from the
syntactic definition may contain ill-formed types. The judgment
\[
\judgeE \env
\]
ensures that an environment is well-formed by checking that each type
is well-formed using only variables bound prior to that types
introduction into the environment. The typing rules maintain
well-formedness of the environment, but when discussing an
arbitrary environment it is necessary to be able to distinguish
those that make sense from those that do not.

\begin{displayfigure}{Environment Well-formedness for $\lambdah$}
  \label{figure:lambdah:envwellformedness}
  \vspace{-30pt}
  \lambdahenvwellformedness
  \vspace{-35pt}
\end{displayfigure}

\subsection{Subtyping for $\lambdah$}
\label{section:lambdah:subtyping}

The subtyping judgment 
\[
\judgesub \env S T
\]
defines when terms of a type $S$ may be considered to have type
$T$. The judgment has no analogue in STLC because STLC does not
support subtyping but requires types to match
exactly.\footnote{Equivalently, one may consider STLC a system with a
  degenerate subtyping relation that relates each type only to itself,
  and thus the subsumption rule is simply elided.}  This judgment may
be invoked via occurrences of the typing rule $\rel{T-Sub}$ at any
point within a typing derivation.

\begin{displayfigure}{Subtyping for $\lambdah$}
  \label{figure:lambdah:subtyping}
  \vspace{-30pt}
  \lambdahsubtyping
  \vspace{-35pt}
\end{displayfigure}

\begin{rulelist}
\item[\rel{S-Arrow}:] Subtyping for function types is completely
  standard, adjusted for dependent types: A dependent function type
  $\lamt x {S_1} {S_2}$ is a subtype of $\lamt x {T_1} {T_2}$ if the
  $T_1 \sword S_1$ (contravariant, as always) and the result types are
  subtypes in a context with $x$ bound to $T_1$. Note that $T_1$ must
  be used in preference to the coarser argument type $S_1$ because
  $T_2$ may not be well-formed in an environment binding $x$ to
  $S_1$.

\item[\rel{S-Base}:] Since a refined basic type $\predty x B \predp$
  roughly denotes the set of constants $\constant$ of type $B$ for
  which ${\subst x \constant B}\, \predp$ is valid, subtyping between
  refinement types reduces to implication between their refinement
  predicates. As an example, the rule $\rel{S-Base}$ states that the
  subtyping judgment
  \[
  \judgesub \emptyset {\predty x \Int {x > 7}} {\predty x \Int {x > 0}}
  \]
  follows from the validity of the implication:
  \[
  \judgeimp {x:\Int} {(x> 7)} {(x > 0)}
  \] 
\end{rulelist}

Basic properties that should hold of any subtyping relationship can
now be stated, though their proofs depend on further formalisms from
\hhref{Section}{section:lambdah:typesoundness}. The
first property is that subtyping should be reflexive -- the narrative
reading of the subtyping judgment $\judgesub \env T T$ is the 
tautology ``any term of type $T$ has type $T$''.  Secondly subtyping
should be transitive, both to respect the intuition that it is an
enriched form of containment and also since two applications of the
rule \rel{T-Sub} express transitivity in another way.

\begin{lem}[Subtyping is a Preorder]\
  \label{lemma:SubtypingPreorder}

  \begin{enumerate}
  \item If $\judgeT \env T$ then $\judgesub \env T T$
  \item If $\judgesub \env S T$ and $\judgesub \env T U$ then $\judgesub \env S U$
  \end{enumerate}
\end{lem}

\noindent\proof\

\begin{enumerate}
\item By induction on the derivation of $\judgeT \env T$, since implication
  is a preorder as well (\hhrefpref{Lemma}{lemma:ImplicationPreorder}).
\item By induction on the derivation of $\judgesub \env S T$ (followed by
  inversion on $\judgesub \env T U$) using {\hhref{Lemma}{lemma:NarrowingOfSubtyping}} for the
  bindings in function types, and the transitivity of implication (\hhrefpref{Lemma}{lemma:ImplicationPreorder}).~\qed
\end{enumerate}

This establishes that subtyping for $\lambdah$ obeys the expected properties
of any subtyping system. But since terms appear within $\lambdah$ types,
the more novel condition of type invariant under reduction is particular
to dependent type systems.

\begin{lem}[Type Equivalence under Reduction]
  If $\judgeT \env S$ and $S \lred T$ then $\judgesub \env S T$ and $\judgesub \env T S$.
  \label{lemma:TypeEval}
\end{lem}

\noindent\proof\ By induction on the derivation of $S$. In
the base case of \rel{WT-Base} invoking the definition of implication.~\qed

Just as confluence of reduction ensures that reduction produces
well-defined equivalence classes of terms, type equivalence under
reduction ensures the same for types and that these equivalence
classes correspond to the equivalence closure of subtyping.

\subsection{Implication and Closing Substitutions}
\label{section:lambdah:closingsubstitutions}
\label{section:lambdah:implication}

Subtyping for $\lambdah$ is determined by the implication judgment
\[
\judgeimp \env p q
\]
defined by the single rule \rel{Imp} which informally reads ``for all
closing substitutions consistent with the types in the environment, if
the now-closed predicate $p$ holds, then so must $q$''. However,
defining closing substitutions in such a way as to yield a consistent
implication relation is new to this research, and merits discussion.

\begin{displayfigure}{Implication for $\lambdah$}
  \ssp
  \label{figure:lambdah:implication}
  \implicationrules
\end{displayfigure}

One approach taken by \citet{flanagan_hybrid_2006}\footnote{Corrected
  by \citet{knowles_hybrid_2010} to use the technique described here.}
is to define closing substitutions in terms of the typing judgment,
as follows:
\[
\inferrule{
	\forall x\in dom(\env),~
	 \judge {} {\sigma(x)} {\env(x)}
}{
  \judge {} \sigma {\env}
}
\]

This approach leads to mutual recursion between the typing, subtyping,
implication, and closing substitution judgments. Unfortunately, the
implication rule~\rel{Imp} from
\hhref{Figure}{figure:lambdah:typerules} refers to the closing
substitution relation in a contravariant position, so standard
monotonicity arguments are not sufficient to show that the resulting
collection of mutually recursive inference rules converge to a fixed
point, and it is not obvious if there are interesting type systems
that satisfy these rules.

An alternative approach to defining implication is to axiomatize its
logic~\citep{denney_refinement_1998,ou_dynamic_2004,gronski_sage:_2006},
but the underlying problem remains, in that it is still not obvious
if there are interesting implication relations, and hence type systems,
that satisfy the axioms.

A third approach, taken by \citet{belo_polymorphic_2011}, is to
restrict refinements by insisting they be well-typed predicates in an
empty environment, so that closing substitutions are not
necessary. Unfortunately, refinements then cannot express even simple
facts such as that the integer stored in $y$ is greater than the
integer stored in $x$! 

Instead, in rule \rel{Subst} we have leveraged the denotational interpretation of
$\lambdah$ types from \hhref{Section}{section:lambdah:denotation} to
define closing substitutions in a direct manner with no circularity
problems: A substitution $\sigma : Var \rightarrow Term$ is a
\emph{closing substition} for environment $\env$ if for all bindings
$(x:T) \in \env$ we have $\sigma(x) \in \meaningf{\sigma(T)}$.

\begin{lem}[Implication is a Preorder]\
  \label{lemma:ImplicationPreorder}

  \begin{enumerate}
  \item For any $\env$ and $\predp$ we have $\judgeimp \env \predp \predp$.
  \item If $\judgeimp \env {\predp_1} {\predp_2}$ and $\judgesub \env {\predp_2} {\predp_3}$ then $\judgesub \env {\predp_1} {\predp_2}$
  \end{enumerate}
\end{lem}

\noindent\proof\ Both of these statements reduce to tautologies by inverting
\rel{Imp}.

\section{Type Soundness for \lambdah}
\label{section:lambdah:typesoundness}

We prove soundness via the usual syntactic ``progress'' and
``preservation'' lemmas. However, due to the semantic nature of
implication, many of the steps along the way are new, thus routine
lemmas usually omitted from a formal development (because
practitioners have seen them so many times) are less obvious and are
included here. Throughout this section, environments are assumed to be
well formed, to elide many uninteresting antecedents.

As a preliminary measure, let us formally state additional
expectations on the relationship between a primitives's operational
semantics and the type assigned to it, analogous to
\hhref{Requirement}{req:stlc:prim} for STLC primitives.

\begin{requirement}[Types of $\lambdah$ Primitives]
  \label{assume:consts}~

  \noindent
  For each primitive $\constant$:
  \begin{enumerate}
  \item
    The type of $\constant$ is closed and well-formed, \ie\
    $\judgeT{\emptyset}{\fun{ty}(\constant)}$.
  \item If $\constant$ is a primitive function then it cannot get stuck and
    its behavior is compatible with its type, \ie\
    if $\judge \emptyset {\app \constant v} T$ then $\deltaf \constant v$ is defined  
    and $\judge \emptyset {\deltaf \constant v} T$
  \item If $\constant$ is a primitive constant then it is a member of its type,
    which is a singleton type, \ie
      if $\fun{ty}(\constant)={\predty x B \predp}$ then ${\subst x \constant B}\, \predp\lred^*\true$
      and $\forall \constant'\neq \constant.~{\subst x {\constant'} B}\, \predp\not\lred^*\true$. 
  \end{enumerate}
\end{requirement}

Because we define closing substitutions semantically, many of our
proofs connect the semantic relation with our syntactic type system.
First, we define the \emph{semantic subtyping} relation $\judgesubset
\env S T$ and the \emph{semantic typing} relation $\judgeIn \env \e T$
induced by our semantic notion of types, following
\citet{frisch_semantic_2008}:
\begin{eqnarray*}
  \judgeIn \env \e T & \quad\defeq\quad & \forall \closingsubst, ~\judge {} \closingsubst \env \mbox{ implies } \closingsubst(\e) \in \meaningf{\sigma(T)} \\
  \judgesubset \env S T  & \quad\defeq\quad & \forall \sigma,~ \judge {} \sigma \env \mbox{ implies } \meaningf{\sigma(S)} \subseteq \meaningf{\sigma(T)} 
\end{eqnarray*}
Our formal subtype system is sound with respect to this model:

\begin{lem}[Semantic typing soundness]
  \mbox{}
  \label{lemma:lambdah:subtypingsoundone}
  \label{lemma:lambdah:typingsoundtwo}
  \begin{enumerate}
  \item
    If $\judgesub \env S T$ then $\judgesubset \env S T$
  \item
    If $\judge \env \e T$ then $\judgeIn \env \e T$
\end{enumerate}
\end{lem}

\noindent\proof\

\begin{enumerate}
\item
  Proceed by induction on the derivation of $\judgesub \env S T$,
  considering the final rule applied. If $S$ and $T$ are refined
  basic types, then semantic and formal subtyping have identical
  definitions. If $S$ and $T$ are function types, then the result
  follows by induction.
\item
  Proceed by induction on the derivation of $\judge \env \e T$
  considering the final rule applied. If the final rule is \rel{T-Sub}
  then part 1 maps subtyping to semantic sutyping.~\qed
\end{enumerate}

In fact, semantic soundness is essentially a form of type soundness
that leverages soundness of STLC: Because of type soundness of STLC, a
well-typed type is either nonterminating or terminates in a value of
the proper shape, and the additional semantic conditions on the
denotational definition ensure that the value satisfies its
refinements. However, to illustrate other desirable properties of the
syntactic typing discipline, we proceed with syntactic proofs.

The semantic substitution properties for these relations mirror the
usual syntactic properties, but have a completely different
basis. Note, for example, the contravariance of the substitution lemma
for closing substitutions. In the following proofs, we use
juxtoposition of substitutions to indicate composition or application
in order to reduce the number of extraneous symbols; no ambiguity
arises.

\begin{lem}[Semantic Substitution]
  \label{lemma:semanticsubstitution}\

  \noindent
  Suppose $\judgeIn \env \d S$. Let $\theta = \subst x \d {}$, and $\sigma_\env$ represent
  $\sigma$ limited to the domain of $\env$.

  \begin{enumerate}
  \item 
    If 
    $\judge {} \closingsubst {\env, \theta\envtail}$ 
    then 
    $\judge {} {\closingsubst_\env \theta \closingsubst_\envtail} {\env,x:S,\envtail}$

  \item 
    If $\judgesubset {\env,x:S,\envtail} S T$ 
    then $\judgesubset {\env,\theta\envtail} {\theta S} {\theta T}$

  \item 
    If $\judgeIn {\env,x:S,\envtail} \e T$ 
    then $\judgeIn {\env, \theta \envtail} {\theta e} {\theta T}$
  \end{enumerate}
\end{lem}

\noindent\proof\

\begin{itemize}
  \item[]1. By definition of $\judgeIn \env \tmd S$ we know $\sigma_\env(\tmd) \in \meaningf{\sigma_\env(S)}$

  \item[]2 and 3. For each lemma, fix a closing substitution $\sigma$
    such that $\judge {} \sigma {\env,\theta\envtail}$. Then $\judge
    {} {\sigma_\env\theta\sigma_\envtail} {\env,x:S,\envtail}$ by part
    1. Since $\sigma\tyS = \sigma_\env\theta\sigma_\envtail\tyS$ and
    $\sigma\tyT = \sigma_\env\theta\sigma_\envtail\tyT$ and $\sigma\e
    = \sigma_\env\theta\sigma_\envtail\e$ each conclusion follows by
    set-theoretic calculation.
\end{itemize}

\begin{corollary}
  All the conclusions of \hhref{Lemma}{lemma:semanticsubstitution} hold if $\judge \env \tmd S$.
\end{corollary}

A key notion of health for any inference system is that additional
assumptions should never lessen the conclusions that may be drawn.
The act of adding additional assumptions to the environment is
known as \emph{weakening}.

\begin{lem}[Weakening]
  \label{lemma:lambdah:weakening}\

  \noindent
  Let $\env  = \env_1,\env_2$  and $\env' = \env_1,x:U,\env_2$.
  
  \begin{enumerate}
  \item
    If $\judge \env p \Bool$ and $\judge \env q \Bool$ and $\judgeimp \env p q$
    then $\judgeimp {\env'} p q$
  \item
    If $\judgeT \env T$ and $\judgeT \env S$ and
    $\judgesub \env S T$ then $\judgesub {\env'} S T$.
  \item 
    If $\judge \env \e T$ then $\judge {\env'}\e T$.
  \item 
    If $\judgeT \env T$ then $\judgeT {\env'} T$.
  \end{enumerate}
\end{lem}

\noindent\proof\

\begin{proofcases}
\item 1. The quantification over closed terms after weakening is the
  same as the quantification before weakening, as the new binding is
  irrelevant.
  
\item 2. By induction on the derivation of
  $\judgesub \env S T$, using part 1 in the case for rule \rel{S-Base}.
  
\item 3 and 4. By mutual induction on the 
  derivations of $\judge \env \e T$ and $\judgeT \env T$, using part 2 in
  the case of rule \rel{T-Sub}
  $\square$
\end{proofcases}

Likewise, shrinking the domain of a quantification should only
increase the available conclusions, as there are fewer potential
counterexamples to any conclusion. For a subtyping system, this
operation of \emph{narrowing} indicates that subtyping behaves
according to this intuition.

\begin{lem}[Narrowing]
  \label{lemma:NarrowingOfImplication}
  \label{lemma:NarrowingOfSubtyping}

  \noindent
  Suppose $\judgesub \env S T$. Then
  
  \begin{enumerate} 
    \item If $\judgeimp {\env,x:T,\envtail} \predp \predq$ then $\judgeimp {\env,x:S,\envtail} \predp \predq$.
    \item If $\judgesub {\env,x:T,\envtail} {U_1} {U_2}$ then $\judgesub {\env,x:S,\envtail} {U_1} {U_2}$.
    \item If $\judge {\env,x:T,\envtail} \e U$ then $\judge {\env,x:S,\envtail} \e U$.
  \end{enumerate}
\end{lem}

\begin{lamproof}
  \pf\ 
  
  \begin{enumerate}
  \item By \hhref{Lemma}{lemma:lambdah:typingsoundtwo} (Soundness of Semantic Typing)
    we know $\judgesubset \env S T$ so the quantification over closed terms in
    the conclusion is contained in that of the premise.~\qed
    
  \item By induction over the derivation of $\judgesub
    {\env,x:T,\envtail} {U_1} {U_1}$ using part 1.

  \item By induction over the derivation of $\judge
    {\env,x:T,\envtail} \e U$, using part 2 as needed and applying
    subsumption immediately when $x$ is pulled from the
    environment.~\qed
  \end{enumerate}
\end{lamproof}

Via \hhref{Lemma}{lemma:lambdah:weakening} and
\hhref{Lemma}{lemma:NarrowingOfSubtyping} the reflexivity and
transitivity of subtyping may be
established. The reader interested in formalism may now wish to
revisit \hhref{Section}{section:lambdah:subtyping} and review the
these properties and their proofs.

The central lemma for proving preservation is the so-called
\emph{substitution lemma}, which states that a variable of a certain
type may be soundly replaced by any term of the same type. Each of
the judgments of the $\lambdah$ enjoys a substitution property.

\begin{lem}[Substitution]
  \label{lemma:substitution}
  Suppose $\judge \env \e S$,

  \begin{enumerate}
  \item If $\judgeimp {\env,x:S,\envtail} \predp \predq$ then $\judgeimp {\env,\subst x \e S\, \envtail} {\subst x \e S\, \predp} {\subst x \e S\, \predq}$
  \item If $\judgesub {\env,x:S,\envtail} T U$ then $\judgesub {\env,\subst x \e S\, \envtail} {\subst x \e S\, T} {\subst x \e S\, U}$
  \item If $\judge {\env,x:S,\envtail} \tmd T$ then $\judge {\env,\subst x \e S\, \envtail} {\subst x \e S\, \tmd} {\subst x \e S\, T}$
  \item If $\judgeT {\env,x:S,\envtail} T$ then $\judgeT {\env,\subst x \e S\, \envtail} {\subst x \e S\, T}$
  \end{enumerate}
\end{lem}

\noindent\proof\

\begin{enumerate}
\item Immediately follows from \hhref{Lemma}{lemma:semanticsubstitution} (Semantic
  Substitution).
  
\item By induction on the derivation of $\judgesub
  {\env,x:S,\envtail} T U$.  In the case of \rel{S-Base} invoking
  part 1.
  
\item and 4. By mutual induction on the derivations of $\judge
  {\env,x:S,\envtail} \tmd T$ and $\judgeT {\env,x:\tyS,\envtail}
  \tyT$, invoking part 2 in the case of \rel{T-Sub}.\qed
\end{enumerate}

Leveraging these lemmas, progress and preservation are routine and
ensure type soundness for $\lambdah$ in the usual way.

\begin{lem}[Preservation]
  \label{thm:preserve} 
  If $\judgeE{\env}$ and $\judge{\env}{\d}{T}$ and $\d \lred \e$ then
  $\judge{\env}{\e}{T}$
\end{lem}

\noindent\pf\ By induction on the typing derivation
$\judge \env \tmd T$, invoking \hhref{Lemma}{lemma:substitution} when
evaluation proceeds via \rel{E-$\beta$} and
\hhref{Requirement}{assume:consts} when evaluation proceeds via
\rel{E-$\delta$}.~\qed

\begin{lem}[Progress]
  \label{thm:progress}
  If $\judge \emptyset \e T$ then either $\e$ is a value,
  or there exists some $\e'$ such that $\e \lred \e'$.
\end{lem}

\begin{lamproof}
  \pf\ Proceed by induction on the derivation of $\judge {} \e T$,
  considering the final rule applied. The entire proof is standard and routine~\qed
\end{lamproof}

\begin{thm}[Type Soundness for $\lambdah$]
  For any term $\e$, if $\judge \emptyset \e T$ then either
  $\e$ is nonterminating or $\e$ evaluates to some value
  $\val$ such that $\judge \emptyset \val T$
\end{thm}

\noindent\pf\ This is a corollary of progress and preservation.

\section{Extensional Equivalence for $\lambdah$}
\label{section:lambdah:logicalrelation}

Type soundness assures that the $\lambdah$ type system relates to the
semantics in the intended manner, but is not particularly useful for
reasoning about meaning of terms. The most fundamental relationship
between terms is that of equality, and the defining trait of equality
is substitutability: If a term $\e$ is equal to $\tmd$, then replacing
$\e$ with $\tmd$ never affects any conclusion. Relative to the
semantics of a calculus, the authoritative definition of equivalence
between terms is \emph{contextual equivalence} (sometimes called
\emph{observable equivalence}) which states that no program context
can distinguish the two terms.

\begin{definition}[Contextual Equivalence]\

  \noindent
  Two terms $\e_1$ and $\e_2$ are \emph{contextually equivalent}, written $\e_1 \ctxequiv \e_2$,
  if and only if:

  Given any context $\tmctx$ such that $\judge \emptyset {\tmctx[\e_1]} B$ and $\judge \emptyset {\tmctx[\e_2]} B$,
  \[
  \tmctx[\e_1] \lred^* \constant ~ \Leftrightarrow ~ \tmctx[e_2] \lred^* \constant
  \]
\end{definition}
  
Reasoning about contextual equivalence between terms is notoriously
difficult.  The equivalence closure of reduction is much weaker than
contextual equivalence, so establishing equality between terms that
are not obviously syntactically related often involves complex
syntactic techniques such as bisimulation (of which many varities
exist) which are extremely sensitive to the form (not content) of a
calculus.

In contrast, when reasoning about mathematical functions, extensional
equivalence -- that equal functions map equal arguments to equal
results -- is the definition of equality. For a computational
calculus, this sort of equality is known as a \emph{logical
  relation}~\citep{statman_logical_1985}. That such a form of
reasoning is applicable confirms that what we call a ``function'' in a
calculus does, indeed, define a function relative to contextual
equivalence.

The logical relation we establish is extensional equivalence of terms
of a particular type under reduction, written $\judgesimt \env {\e_1}
{\e_2:T}$ and defined in \hhref{Figure}{figure:lambdah:equivalence}.
Two terms $\e_1$ and $\e_2$ of basic type are (extensionally)
equivalent if, for any closing substitution $\closingsubst$, whenever
$\closingsubst(\e_1)$ evaluates to a constant $\constant$ so does
$\closingsubst(\e_2)$, and vice versa.  Two functions $\tmf_1$ and
$\tmf_2$ of type $\lamt x S T$ are equivalent if they yield equivalent
output when given equivalent arguments $\e_1$ and $\e_2$.

\begin{displayfigure}{Extensional Equivalence Under Reduction for $\lambdah$} 
  \label{figure:lambdah:equivalence}
  \label{figure:lambdah:extensionalequivalence}
  \extensionalequivalence
\end{displayfigure}

A question that immediately arises is whether these two applications
$\tmf_1~\e_1$ and $\tmf_2~\e_2$ should have type $\subst x {\e_1} {}\,
{T}$ or $\subst x {\e_2} {}\, {T}$. In some sense it does not matter,
since $\e_1$ and $\e_2$ are extensionally equivalent by assumption.
But this assumes the premise: that $\lambdah$ operational semantics
respects extensional equivalence. So, since this is not yet
demonstrated, the definition requires that $\tmf_1~\e_1$ and
$\tmf_2~\e_2$ are extensionally equivalent at type
\[
\subst x {\e_1} {}\, {T} \curlywedge \subst x {\e_2} {}\, {T}
\]
where we {combine} both types via the \emph{wedge product}
($\curlywedge$), also defined in
\hhref{Figure}{figure:lambdah:equivalence}.  For basic types, wedge
product is type intersection, \emph{i.e.}, conjunction of refinement
predicates.  For function types, the wedge product is covariant in
both the function domain and the range. This covariance is an
illusion: we we only ever combine equivalent types with the wedge
product, so co- or contravariance is a formality: If the wedge product
were made more ``intuitively'' contravariant the \rel{T-App} case of
\hhref{Theorem}{thm:ExtensionalSoundness} would not be possible to
prove as stated. Since extensional equivalence is uncommon in type
systems research exposition, we present the following proofs in great
detail.

To work fluidly with this wedge product,  we note that it is a commutative
and associative operator with respect to the equivalence closure of subtyping, 
and it distributes through subtyping.

\begin{lem}
  \label{lemma:WedgeSubtyping}
  Assuming the underlying shapes of all mentioned types are equal,
  and each type is well-formed in the environment,
  \begin{enumerate}
  \item
    $\judgesub \env {S \curlywedge T} {T \curlywedge S}$
  \item
    $\judgesub \env {S\curlywedge (T \curlywedge U)} {(S \curlywedge T)\curlywedge U}$
  \item
    If $\judgesub \env {S_1} {T_1}$ and $\judgesub \env {S_2} {T_2}$
    then $\judgesub \env {S_1 \curlywedge S_2} {T_1 \curlywedge T_2}$
  \end{enumerate}
\end{lem}

\begin{lamproof}
  \pf\ By induction on the underlying shape of the types.~\qed
\end{lamproof}

We require that primitive functions do not violate extensionality, and
so each primitive function $\constant$ is extensionally equivalent to itself.

\begin{requirement}
  \label{requirement:primequiv}\

  For any primitive $\constant$, 
  \[ \judgesimt \emptyset {\constant} {\constant:ty(\constant)} \]
\end{requirement}

Equivalence is preserved under subtyping, allowing free manipulation
of $\curlywedge$ in the type assigned to an equivalence.

\begin{lem}[Extensional equivalence under subtyping]
  \label{lemma:EquivSubty}\

  If $\judgesimt \env {\e_1} {\e_2:S} $ and $\judgesub \env S T$ 
  then
  $\judgesimt \env {\e_1} {\e_2:T} $
\end{lem}

\begin{lamproof}
  \pf\ By induction on $\Base S$ (which equals $\Base T$), with proof
  cases elaborated by inversion of subtyping rules.

  \begin{proofcases}
  \item[\proofcase S = \predty x B \predp]: Then we have a derivation
    \[
    \inferrule*[right=\rel{S-Base}]{
      \judgeimp \env p q
    }{
      \judgesub \env {\predty x B p} {\predty x B q}
    }
    \]
    By definition of $\judgesimt \env {\e_1} {\e_2:{\predty x B p}}$,
    both $\e_1$ and $\e_2$ are semantically members of $\predty x B p$
    hence also semantic members of $\predty x B q$, which 
    yields $\judgesimt \env {\e_1} {\e_2:{\predty x B q}}$.
    
  \item[\proofcase S = \lamt x {S_1} {S_2}]: Then we have a derivation
    \[
    \inferrule*[right=\rel{S-Arrow}]{
      \judgesub \env {T_1} {S_1}
      \and
      \judgesub {\env,x:T_1} {S_2} {T_2}
    }{
      \judgesub \env {\lamt x {S_1} {S_2}} {\lamt x {T_1} {T_2}}
    }
    \]

    Fix $d_1$ and $d_2$ such that \(\judgesimt \env {d_1} {d_2 : T_1}\).
    By induction, $\judgesimt \env {d_1} {d_2 : S_1}$.
    
    By definition of $\e_1$ and $\e_2$ being extensionally equivalent,
    \[ 
    \judgesimt \env {\e_1~d_1} {\e_2~d_2 : \subst x {\d_1} {}\, {S_2} \curlywedge \subst x {d_2} {}\, {S_2}} 
    \]
    
    By \hhref{Lemma}{lemma:substitution} (Substitution),
    \[
    \judgesub \env {\subst x {d_1} {}\, {S_2}} {\subst x {d_1} {}\, {T_2}}
    \textnormal{~~~and~~~}
    \judgesub \env {\subst x {d_2} {}\, {S_2}} {\subst x {d_2} {}\, {T_2}}
    \]
    
    By \hhref{Lemma}{lemma:WedgeSubtyping}(3) the above two subtyping relationship combine with the wedge product to yield
    \[ 
    \judgesub \env {\subst x {d_1} {}\, {S_2} \curlywedge \subst x {d_2} {}\, {S_2}} {\subst x {d_1} {}\, {T_2} \curlywedge \subst x {d_2} {}\, {T_2}}
    \]

    By applying induction to the above unfolding of the definition of extensional equivalence,
    \[ 
    \judgesimt \env {\e_1~d_1} {\e_2~d_2 : \subst x {d_1} {}\, {T_2} \curlywedge \subst x {d_2} {}\, {T_2}}
    \]

    Hence $\judgesimt \env {\e_1} {\e_2 : \plamt x {T_1} {T_2}}$.~\qed
  \end{proofcases}
\end{lamproof}

Next, we show that extensional equivalence of reducts implies extensional
equivalence of the original terms.

\begin{lem}[Extensional equivalence under reduction]
  \label{lemma:simreduce}\

  If $\judgesimt \env {e_1'} {e_2' : U}$ and $e_1 \lred^* e_1'$ and $e_2 \lred^* e_2'$ then $\judgesimt \env {e_1} {e_2 : U} $
\end{lem}

\begin{lamproof}
  \pf\ By induction on $\Base U$.

  \begin{proofcases}
    \item[\proofcase $\predty x B p$:] If $e_1' \lred^* \constant$ then $e_1 \lred^* e_1' \lred^* \constant$; likewise for $e_2$ and $e_2'$.

  \item[\proofcase $\lamt x S T$:]
  Fix $d_1, d_2$ such that 
  \[\judgesimt \env {d_1} {d_2 : S}.\]

  By definition
  \[
  \judgesimt \env {e_1'~d_1} {e_2'~d_2 : \subst x {d_1} {}\, T \curlywedge \subst x {d_2} {}\, T}
  \]

  By applying an evaluation context, 
  \[ 
  e_1~d_1 \lred^* e_1'~d_1
  \textnormal{~~~and~~~}
  e_2~d_2 \lred^* e_2'~d_2
  \]
  
  By induction,
  \[
  \judgesimt \env {e_1~d_1} {e_2~d_2 : \subst x {d_1} {}\, T \curlywedge \subst x {d_2} {}\, T}
  \]

  Hence, \judgesimt \env {e_1} {e_2 : \plamt x S T}.~\qed
\end{proofcases}
\end{lamproof}

To prove the fundamental soundness theorem for extensional
equivalence, we extend the notion of extensional equivalence to
closing substitutions.
\begin{eqnarray*}
  \judgesimt {} \sigma {\gamma : \env}
  & ~~\defeq~~ &
  \forall x \in dom(\env), ~
  \judgesimt \emptyset {\sigma(x)} {\gamma(x) : (\sigma\curlywedge\gamma)(\env(x))}
\end{eqnarray*}
For convenience, we overload the wedge product operator for closing
substitutions, so $(\sigma \curlywedge \gamma)$ maps a type $T$ to the
wedge product of all possible choices of which variables to use from
$\sigma$ and which from $\gamma$. Note that we are only interested in
closing substitutions for the same environment, so they always have the
same domain.

\begin{eqnarray*}
  (\sigma \curlywedge \gamma)(T) 
  & \defeq & 
  \bigcurlywedge_{A \in \mathcal{P}(dom(\sigma))}
  (\sigma|_A \circ \gamma|_{dom(\sigma)\setminus A})(T)
\end{eqnarray*}

\begin{lem}[Extensional equivalence under substitution]
  \label{thm:ExtensionalSoundness}\

  If $\judge \env \e T$ then for any closing substitutions $\sigma$ and $\gamma$
  such that $\judgesimt {} \sigma {\gamma : \env}$,
  we have
  \[
  \judgesimt \emptyset {\sigma(\e)} {\gamma(\e): (\sigma\curlywedge\gamma)(T)}
  \]
\end{lem}

\begin{lamproof}
  \pf\ By induction on the height of the derivation of $\judge \env \e T$,
  rebinding metavariables for each case.

  \begin{proofcases}
    
  \item[\proofcase\ \rel{T-Var}:] Then $(x:T) \in \env$, so by
    definition of equivalent substitutions $\judgesimt \emptyset
    {\sigma(x)} {\gamma(x) : (\sigma\curlywedge\gamma)(T)}$
    
  \item[\proofcase\ \rel{T-Prim}:]
  Given by \hhref{Requirement}{requirement:primequiv}.

  \item[\proofcase\ \rel{T-App}:] Then the derivation concludes with
  \[                
  \inferrule*{
    \judge \env \f {\lamt x S T}
    \and
    \judge \env \e S
  }{
    \judge \env {\f~\e} {\subst x \e S\, T}
  }
  \]

  Fix $\sigma$ and $\gamma$.
  By induction, 
  \[
  \begin{array}{l@{~~~~~}l}
    \judgesimt \emptyset {\sigma(\f)} {\gamma(\f) 
      : (\sigma\curlywedge\gamma)(\lamt x {S} {T})
    }
    \\
    \judgesimt \emptyset {\sigma(\e)} {\gamma(\e) : (\sigma\curlywedge\gamma)(S)}
  \end{array}
  \]

  Applying the definition of extensional equivalence,
  \[
  \begin{array}{rcl}
    \judgesimt \emptyset {\sigma(\f) ~ \sigma(\e)} {\gamma(\f) ~ \gamma(\e)} 
    & : & \subst x {\sigma(\e)} {}\, (\sigma\curlywedge\gamma)(T)
    \\
    & \curlywedge & \subst x {\gamma(\e)} {}\, (\sigma\curlywedge\gamma)(T)
  \end{array}
  \]
  
  which simplifies to the conclusion for this case:
  \[
  \begin{array}{l@{~~~~~}l}
    \judgesimt \emptyset {\sigma( \f ~ \e )} {\gamma( \f~\e ): (\sigma\curlywedge\gamma)(\subst x {\e} {}\, T)}
  \end{array}
  \]

  \item[\proofcase\ \rel{T-Lam}:]
    The the derivation concludes with
    \[
    \inferrule*{
      \judge {\env,x:S} \e T
    }{
      \judge \env {\lam x S \e} {\lamt x S T}
    }
    \]

    Fix $\sigma$ and $\gamma$, then fix $d_1$ and $d_2$ such that $\judgesimt \emptyset {d_1} {d_2 : (\sigma\curlywedge\gamma)(S)}$.
    The definition of extensional equivalence of substitution yields
    \[ 
    \judgesimt {} {(\sigma \circ \subst x {d_1} {})} {(\gamma \circ \subst x {d_2} {}) : (\env,x:S)} 
    \]
    
    By induction,
    \[
    \begin{array}{l}
      \judgesimt \emptyset {(\sigma \circ \subst x {d_1} {})~\e} { (\gamma \circ \subst x {d_2} {}) ~ \e
        : ((\sigma \circ \subst x {d_1} {})\curlywedge(\gamma \circ \subst x {d_2} {})) (T)
      }
    \end{array}
    \]
  
    By \hhref{Lemma}{lemma:simreduce}, we perform $\beta$-expansion and rearrange some substitutions to yield 
    \[
    \begin{array}{rcl}
      \judgesimt \emptyset {\sigma(\lam x {S} \e)~{d_1}} {\gamma(\lam x {S} \e)~d_2}
      & : & \subst x {d_1} {}\, (\sigma\curlywedge\gamma) (T) \\
      & \curlywedge & \subst x {d_2} {}\, (\sigma\curlywedge\gamma) (T)
    \end{array}
  \]

  which simplifies to the conclusion of this case,
  \[
  \begin{array}{l}
    \judgesimt \emptyset {\sigma(\lam x {S} \e)} {\gamma(\lam x S \e) : (\sigma\curlywedge\gamma) (\lamt x S T)}
  \end{array}
  \]

  \item[\proofcase\ \rel{T-Sub}:]
    The derivation concludes with
    \[
    \inferrule*{
      \judge \env \e S
      \and
      \judgesub \env S T
    }{
      \judge \env \e T
    }
    \]

    Fix $\sigma$ and $\gamma$. By induction,
    \[
    \judgesimt \emptyset {\sigma(\e)} {\gamma(\e) : (\sigma\curlywedge\gamma)(S)}
    \]
    
    Applying the substitution lemma and
    distributing $\curlywedge$ through the subtyping,
    \[
    \judgesub \emptyset {(\sigma\curlywedge\gamma)(S)} {(\sigma\curlywedge\gamma)(T)}
    \]

    Thus by \hhref{Lemma}{lemma:EquivSubty} we conclude with
    \[
    \judgesimt \emptyset {\sigma (\e)} {\gamma (\e) : (\sigma\curlywedge\gamma)(T)}
    \]
  \end{proofcases}
  This concludes the case-by-case analysis.~\qed
\end{lamproof}

As a corollary, we have contextual equivalence of related terms. 
\begin{thm}[Soundness of extensional equivalence]
  If $\judgesimt \env \d {\e : T}$ then $\d \ctxequiv \e$. 
\end{thm}

\begin{proof}
  From the definition of $d \ctxequiv e$, fix any $\tmctx$ such that
  $\judge \emptyset {\tmctx[\tmd]} B$ and $\judge \emptyset {\tmctx[e]}
  B$. Apply \hhref{Lemma}{thm:ExtensionalSoundness} to the typing
  $\judge {x:T} {\tmctx[x]} B$ with the related substitutions
  $\judgesimt {} {\subst x d {}} {\subst x \e {} : (x:T)}$ to yield
  $\judgesimt \emptyset {\tmctx[\tmd]} {\tmctx[\e] : B}$, which
  is exactly the definition of $\tmd \equiv \e$.
\end{proof}

\section{Related Work}
\label{section:lambdah:relatedwork}

Research on refinement type systems discussed in
\hhref{Section}{section:intro:refinementtypes} has influenced our
choice of how to express program invariants, with the major difference
that none of the referenced systems include arbitrary executable
refinement predicates.

Liquid types \citep{rondon_liquid_2008} also refines basic types with
predicates, but the research agenda is more aligned with the question
``how can we develop usable and effective static verification of
modern higher-order programs?'' rather than ours of ``how shall we
proceed into the realm of undecidable type systems?''. Thus the
metatheory of Liquid Types is a model of an implementation for OCaml
(subsequently for C~\citep{rondon_csolve:_2012},
Javascript~\citep{chugh_dependent_2012}, and
Haskell~\citep{vazou_abstract_2013}) rather than a foundational
calculus, resulting in different choices for formalization: The
meaning of validity is given via a direct mapping to the The
refinement predicates of Liquid Types, while written in the
programming language at hand, are given meaning by a mapping to the
logic of Satisfiability Modulo Theories (SMT). Evaluation is that of
the host language (call-by-value for most implementations, and
call-by-need for Haskell) rather than a general theory of
$\beta$-equivalence, thus proof techniques for program equivalence are
largely out of scope. 

ESC/Haskell \citep{xu_extended_2006,xu_static_2009} features a similar
language for expressing specifications, but is concerned with the
practicalities of statically checking contracts for Haskell (hence the
name analogous to its predecessor ESC/Java) and takes a different
formal approach not primarily concerned with providing a foundational
calculus. A program meeting its specification is \emph{defined} by
symbolic execution, versus our presentation as a type system later
demonstrated sound by relating it to the semantics of a
program. However, ESC/Haskell gives a much more thorough treatment to
the notion of blame, which is integral to a complete contract system.

Manifest contracts \citep{greenberg_contracts_2012} is a term
synonymous with executable refinement types, chosen to emphasize that
predicates from contracts are made ``manifest'' in the type system,
versus being only ``latent'' properties of the program as it
executes. The work shares with this chapter a foundation built upon
from \citet{knowles_hybrid_2010}, hence uses a nearly identical
approach to defining subtyping between contracts. Subsequent to this
research, \citet{belo_polymorphic_2011} extended manifest to include
polymorphism and apply refinements to any type, not just basic
types. However as a foundation it is somewhat lacking due to the
restriction that refinements may only refer to the value being refined
and may not have other free variables to relate different portions of
a program.  It also does not include subtyping, but simply ensures
types are compatible and then defers contract enforcement to runtime,
much like gradual typing. Subtyping between function types is the
principle remaining challenge for allowing refinements of arbitrary
types.

None of the above systems provide a robust method for reasoning about
extensional equivalence of terms.

\chapter{Hybrid Type Checking}

\label{chapter:htc}

\newcommand{\castinsertionrules}{
  \headtrule{$\judgec \env \tmd \e \tyT$}{\underline{C}ast insertion and synthesis}
  \begin{trules}
    \trule{C-Var}{
      (x:\tyT)\in \env
    }{
      \judgec \env x x \tyT
    }
    
    \trule{C-Const}{
    }{
      \judgec \env \constant \constant {\fun{ty}(\constant)}
    }
    
    \trule{C-Fun}{
      \judgect \env {S_1} {T_1} 
      \and
      \judgec {\env,x:T_1,} \d {\e} {T_2}
    }{
      \judgec \env 
              {(\lam x {S_1} \d)} 
              {(\lam x {T_1} {\e})}
              {(\lamt x {T_1} {T_2})}
    }

    \trule{C-App}{
      \judgec \env {\f} {\g} {(\lamt x {T_1} {T_2})} \qquad
      \judgecc \env {\d} {\e} {T_1}
    }{
      \judgec \env 
      {\app {\f} {\d}}
      {\app {\g} {\e}}
      {\subst x \e {T_1}\, T_2}
    }
  \end{trules}
}

\newcommand{\castinsertionandchecking}{
  \headtrule{$\judgecc \env \d \e T$}{\underline{C}ast insertion and \underline{c}hecking}
  \begin{trules}
    \trule{CC-Ok}{
      \judgec \env \d \e \tyS
      \and
      \judgesubalg \pyes \env \tyS \tyT
    }{
      \judgecc \env \d \e \tyT
    }

    \trule{CC-Chk}{
      \judgec \env \d \e S
      \and
      \judgesubalg \pmaybe \env S T
    }{
      \judgecc \env \d {\casttwoapp S T \e} T
    }
  \end{trules}
}

\newcommand{\algorithmicsubtyping}{
  \headtrule{$\judgesubalg{a}{\env}{S}{T}$}{\underline{S}ubtyping \underline{A}lgorithm}
  \begin{trules}
    \trule{SA-Arrow}{
      \judgesubalg{a}{\env}{T_1}{S_1}
      \and
      \judgesubalg{b}{\env,x:T_1}{S_2}{T_2}
    }{
      \judgesubalg {a\sqcap b} \env {(\lamt x {S_1}{S_2})} {(\lamt{x}{T_1}{T_2})}
    }
    \\

    \trule{SA-Base}{
      \judgeimpalg a {\env,x:B} \predp \predq
    }{
      \judgesubalg a \env {\predty x B \predp} {\predty x B \predq}
    }
  \end{trules}
}

\newcommand{\typecastinsertionrules}{ 
  \headtrule{$\judgect{\env}{S}{T}$}{\underline{C}ast insertion on types}
  \begin{trules}
    \trule{C-Arrow}{
      \judgect{\env}{S_1}{T_1} \qquad
      \judgect{\env,x:T_1}{S_2}{T_2} 
    }{
      \judgect{\env}{(\lamt{x}{S_1}{S_2})}{(\lamt{x}{T_1}{T_2})}
    }
    \\

    \trule{C-Base}{
      \judgecc {\env,x:B} \predp \predq \Bool
    }{
      \judgect \env {\predty x B \predp} {\predty x B \predq}
    }
  \end{trules}
}


With $\lambdah$ as a sound foundation for the study of executable
refinement type systems, this chapter presents \emph{hybrid type checking},
a new technique for combining static type checking with dynamic contract checking.

To illustrate the key idea of hybrid type checking in general, before
developing a system specialized to $\lambdah$, consider again this
possible type rule for function application:
\[
\begin{array}{cc}
  \trulenoname{
	\judge \env \tmf {\arrowt{\tyT}{\tyU}}
    \qquad
	\judge \env \e \tyS \qquad
	\judgesub \env \tyS \tyT
  }{
	\judge{\env}{\tmf ~ \e}{\tyU}
  }
\end{array}
\]

The situation we focus on is when the antecedent
$\judgesub \env \tyS \tyT$ can be neither definitively proven nor refuted.

\begin{itemize}\item[-]
\emph{Statically rejecting} such programs would cause the compiler to
reject some programs that, on deeper analysis, could be shown to be
well-typed. \citet{flanagan_extended_2002} illustrates that this
approach is brittle in practice because there are many such programs
and it is difficult to predict which programs the compiler would
accept.

\item[-]

\emph{Statically accepting} such programs (based on the optimistic
assumption that the unproven subtype relations actually hold) may
result in specifications being violated at run time.
\end{itemize}
%
%
Instead, hybrid type checking is the general approach of accepting
such programs on a provisional basis, but inserting sufficient dynamic
checks to ensure that specification violations never occur at run
time.  Of course, checking that $\judgesub \env \tyS \tyT$ at run time
is still a difficult problem and would violate the principle of
\emph{phase distinction}~\citep{cardelli_phase_1988}.  Instead, our
hybrid type checking approach transforms the above application into
the code
\[ 
\tmf ~ (\casttwoapp \tyS \tyT \e)
\]
where the additional \emph{cast} (or \emph{coercion}~\citep{breazu-tannen_inheritance_1991})
$\casttwoapp \tyS \tyT$ dynamically enforces that $\e$ adheres to
$\tyT$. 

\begin{displayfigure}{Behavior of hybrid type checking for various categories of programs}
  \label{figure:htc:relate}
  \begin{center}
    \begin{tabular}{|c|c|c|c|}
      \hline
      \multicolumn{2}{|c|}{\bf Ill-typed programs} &
      \multicolumn{2}{|c|}{\bf Well-typed programs} 
      \\[.03in]
      \hline
      Clearly ill-typed &
      \multicolumn{2}{|c|}{Subtle programs} &
      Clearly well-typed 
      \\
      \hline
      ~~~Rejected by type checker~~~	& \multicolumn{2}{|c|}{Accepted}	& Accepted without casts \\
	  & \multicolumn{2}{|c|}{with casts}	& \\
      \hline
      \multicolumn{1}{c|}{} 	& Casts 	& Casts		 \\
      \multicolumn{1}{c|}{} 	& may		& never 	 \\
      \multicolumn{1}{c|}{} 	& fail		& fail 		 \\
      \cline{2-3}
    \end{tabular}
  \end{center}~ 
\end{displayfigure}

\hhref{Figure}{figure:htc:relate} illustrates the behavior of hybrid
type checking on various kinds of programs.  Every program is either
ill-typed or well-typed, but it is not always possible to make this
classification statically.  However, a particular algorithm may still
identify some (hopefully many) clearly ill-typed programs, which are
rejected, and similarly can identify some clearly well-typed programs,
which are accepted unchanged.

For the remaining programs, dynamic type casts are inserted to check
any unverified correctness properties at run time. We refer to this
category of programs as \emph{subtle}.

\begin{definition}
  A program is called \emph{subtle} relative to a hybrid type checker
  if the checker cannot statically prove or refute its well-typedness.
\end{definition}

If the original program is actually well-typed, these casts are
redundant and will never fail. Conversely, if the original program is
ill-typed in a manner that is not demonstrable at compile time, the
inserted casts may fail.  As static analysis technology improves, the
category of subtle programs in \hhref{Figure}{figure:htc:relate} will
shrink, as more ill-typed programs are rejected and more well-typed
programs are fully verified at compile time.

Hybrid type checking may facilitate the evolution and adoption of
advanced static analyses, by allowing software engineers to experiment
with sophisticated specification strategies that cannot (yet) be
verified statically or efficiently.  Such experiments can then
motivate and direct static analysis research.

For example, if a hybrid type checker fails to verify or refute a
subtyping query, it could send that query back to the compiler
writer. Similarly, if a cast $\casttwoapp S T v$ fails, the value $v$
is a witness that refutes an undecided subtyping query $S\sword T$,
and such witnesses could also be sent back to the compiler writer.
This information would provide concrete and quantifiable motivation
for subsequent improvements in the type checker's analysis. In
\hhref{Chapter}{chapter:sage} we explore the design of such an
implementation.

Just as different compilers (or different configurations of a single
compiler) for the same language may yield object code of varying
quality, we might imagine a variety of hybrid type checkers with
different trade-offs between static and dynamic checks. Fast
interactive hybrid compilers might perform only limited static
analysis to detect obvious type errors, while production compilers
could perform deeper analyses to detect more defects statically and to
generate improved code with fewer dynamic checks.

Hybrid type checking is inspired by prior work on soft
typing~\citep{fagan_soft_1991, wright_practical_1994, aiken_soft_1994,
  flanagan_catching_1996}, but it extends soft typing by rejecting
many ill-typed programs, in the spirit of static type checkers.  The
interaction between static typing and dynamic checks has also been
studied in the context of type systems with the type
\t{Dynamic}~\citep{thatte_quasi-static_1990, abadi_dynamic_1991}, and
in systems that combine dynamic checks with dependent
types~\citep{ou_dynamic_2004}. Hybrid type checking extends these
ideas to support more precise specifications.

The general approach of hybrid type checking appears applicable to a
variety of programming languages and specification languages.  This
chapter develops the design and correctness criteria for such a
system using $\lambdah$ as a representative core calculus.

The combination of $\lambdah$ and HTC enjoys the following
benefits:
\begin{enumerate}
\item
  It supports precise interface specifications, which facilitate
  modular development of reliable software.
\item
  As many defects as {possible and practical} are detected at
  compile time (and we expect this set will increase as static analysis
  technology evolves).
\item
  All well-typed programs are accepted by the checker, and their
  semantics are unchanged.
\item
  Due to decidability limitations, the hybrid type checker may
  statically accept some \emph{subtly ill-typed} programs, but it will
  insert sufficient dynamic casts to guarantee that specification violations 
  are always detected, either statically or dynamically.
\item
  The output of the hybrid type checker is always a well-typed program
  (and so, for example, type-directed optimizations are applicable).
\item
  If the source program is well-typed, then the inserted casts are guaranteed
  to succeed, and so the source and output programs are behaviorally
  equivalent.
\item
  If the subtyping algorithm can decide all reflexive queries, then
  cast insertion is idempotent: All necessary casts are inserted in a
  single pass.
\end{enumerate}

\subsection*{Chapter Outline}

The remainder of this chapter is organized as follows

\begin{itemize}
\item[-] \hhref{Section}{section:htc:syntax} introduces
  the extensions to $\lambdah$ to support hybrid type checking.
  
\item[-] \hhref{Section}{section:htc:opsem} describes the
  operational semantics of dynamic casts.

\item[-] \hhref{Section}{section:htc:typesystem}
  presents the new typing rules for casts.

\item[-] \hhref{Section}{section:htc:htc} formally
  defines hybrid type checking.

\item[-] \hhref{Section}{section:htc:matrix} walks
  through a sizable example in depth.

\item[-] \hhref{Section}{section:htc:correctness} 
  introduces the correctness properties of HTC
  beyond those of traditional type systems,
  and proves them for our context.

\item[-] \hhref{Section}{section:htc:relatedwork} surveys
  work most closely related to HTC.

\end{itemize}

\section{The Language $\lambdahtc$}
\label{section:htc:syntax}

In order to perform hybrid type checking augment $\lambdah$ with terms
for casts, yielding $\lambdahtc$. The extension includes two new forms,
one for casts and one for refinement checks in progress:
\[
\e ~ ::= ~ \cdots ~~|~~ {\casttwo S T} ~~|~~ {\checking {\predty x B \predp} \predq \constant}
\]
For easy reference, \hhref{Figure}{figure:htc:syntax} includes the
full syntax of the augmented language with new forms highlighted.

\begin{displayfigure}{Syntax for $\lambdahtc$}
  \label{figure:htc:syntax}
  \ssp
  \htcsyntax
\end{displayfigure}

\begin{itemize}

\item[-] The term $\casttwo S T$ denotes a cast from $S$ to $T$ as described
  in the introduction to this chapter.

\item[-] The term $\checking {\predty x B \predp} \predq \constant$
  denotes a cast-in-progress where a determination of whether $\constant$
  has type $\predty x B \predp$ is underway. If $\predq$ eventually
  evaluates to $\true$, then the cast will succeed.

\end{itemize}

\section{Operational Semantics of $\lambdahtc$}
\label{section:htc:opsem}

\hhref{Figure}{figure:htc:opsem} recapitulates the redex reduction rules and
evaluation contexts for $\lambdahtc$ with additions to include casts
used in hybrid type checking.  The new evaluation rules governing the
meaning of casts are:

\begin{displayfigure}{Operation Semantics of $\lambdahtc$}
  \label{figure:htc:opsem}
  {\def\baselinestretch{1.5}\large\normalsize \htcopsem}
\end{displayfigure}

\begin{rulelist}

\item[\rel{E-Cast-F}] Casting a function $\f$ of type $\lamt x {S_1} {S_2}$
to the type $\lamt{x}{T_1}{T_2}$ yields a new function
\[
\funcasttwo x {S_1} {S_2} \f {T_1} {T_2}
\]
Though large, this term of type $\lamt x {T_1} {T_2}$ is actually
simple: Consider applying it to an argument $\e$ of type $T_1$,
and $\beta$-reduce to the following:
\[
\casttwo {\subst x {\casttwo {T_1} {S_1}\, \e} {}\, S_2} {\subst x \e {}\, T_2} 
~~
(f ~~(\casttwo {T_1} {S_1} ~ \e))
\]
First the argument $\e$ is cast to type $S_1$, then it is passed to
the original function $\f$, yielding a result of type \mbox{$\subst x
  {\casttwo {T_1}{S_1}\, \e} {}\, S_2$}; call it $\e'$. The cast
inserted into the return type is necessary as $\e$ may not actually
have type $S_1$, so substituting it directly would make the resulting
type ill-formed and could cause actual errors not caught by run-time
checks. Then the result $\e'$ is cast to type \mbox{$\subst x \e {}\, T_2$}
as required.  Thus, higher-order casts are performed a lazy fashion --
the new casts $\casttwo {S_2} {T_2}$ and $\casttwo {T_1} {T_2}$ are
performed at every application of the resulting function, in a manner
reminiscent higher-order coercions~\citep{thatte_quasi-static_1990} or
higher-order contracts~\citep{findler_contracts_2002}.\footnote{We
  ignore the issue of blame assignment in the event of a run-time cast
  failure -- see \citep{gronski_unifying_2007} for a detailed
  analysis.}

\item[\rel{E-Cast-Begin} and \rel{E-Cast-End}] A basic constant $\constant$
is cast to a base refinement type $\predty x B \predp$ via multiple
evaluation steps. Via rule \rel{E-Cast-Begin} a cast application
$\casttwoapp {\predty x B \predq} {\predty x B \predp} \constant$
evaluates to a cast-in-progress $\checking {\predty x B \predp}
{\subst x \constant B\, \predp} \constant$.  The instantiated
predicate $\subst x \constant B\, t$ then evaluates via the closure
rule \rel{E-Ctx}, either diverging or terminating in a value $\val$.
If $\val$ is $\true$ then the cast-in-progress evaluates to
$\constant$, as it has been dynamically verified that $\constant$ has
type $\predty x B \predp$, otherwise the cast-in-progress is
``stuck'' and we call it a failed cast.

\end{rulelist}

\begin{definition}[Failed Cast]
  A \emph{failed cast} is a term of the form $\checking {\predty x B \predp} \predq \constant$
  where $\predq$ cannot evaluate further, yet $\predq \neq \true$.
\end{definition}

Note that these casts involve only familiar dynamic operations: tag
checks, predicate checks, and creating checking wrappers for
functions.  Thus, our approach adheres to the principle of phase
separation~\citep{cardelli_phase_1988}, in that there is no type checking
of actual program syntax at run time.

\section{The Type System of $\lambdahtc$}
\label{section:htc:typesystem}

The typing rules for casts are shown in
\hhref{Figure}{figure:htc:typerules}. Together with the collective typing
rules of $\lambdah$ (\hhref{Figure}{figure:lambdah:typerules},
\hhref{Figure}{figure:lambdah:typewellformedness},
\hhref{Figure}{figure:lambdah:subtyping},
\hhref{Figure}{figure:lambdah:envwellformedness},
\hhref{Figure}{figure:lambdah:implication}, pages
\pageref{figure:lambdah:typerules} -
\pageref{figure:lambdah:implication}), these define the type system of
$\lambdahtc$.

\begin{displayfigure}{Type rules for $\lambdahtc$}
  \label{figure:htc:typerules}
  \htctyperules
\end{displayfigure}

\begin{rulelist}

\item[{\rel{T-Cast}}] A cast $\casttwo S T$ is simply assigned the
  function type $\arrowt S T$ provided both $S$ and $T$ are
  well-formed types of the same shape. 

\item[{\rel{T-Checking}}] A cast-in-progress $\checking {\predty x B
  \predp} \predq \constant$ requires that the refinement type $\predty
  x B \predp$ is well-formed and that the predicate $\predq$ is a
  boolean term.  Moreover, since this cast succeeds whenever $\predq$
  evaluates to $\true$, we require that $\predq$ actually implies the
  property being checked, ${\subst x \constant B}\, \predp$.  In
  practice, this always holds because the evaluation rules ensure that
  $\predq$ is arrived at by evaluating $\subst x \constant B\, p$.

\end{rulelist}

A cast in a program may be considered unnecessary in two different
ways: via the operational semantics or via the type system.
Operationally, a cast is not useful if it can never fail. The type
system indicates a cast $\casttwo S T$ is redundant if $S \sword
T$. We take the term \emph{redundant} to formally indicate this
latter.

\begin{definition}[Redundant Cast]
  A \emph{redundant cast}, or \emph{upcast}, is a cast $\casttwo S T$ in a term $\tmctx[\casttwo S T]$ such
  that $\judge \env {\tmctx[\casttwo S T]} U$ by a derivation
  \[
  \begin{array}{c}
    \inferrule*{\vdots}{\judge \envtail {\casttwo S T} {\arrowt S T}} \\
    \inferrule*{\vdots}{\judge \env {\tmctx[\casttwo S T]} U}
  \end{array}
  \]
  and ${\judgesub \envtail S T}$.
\end{definition}

The fundamental correctness property for casts is that any redundant
cast can never fail. In a higher-order setting, where casts are
evaluated lazily, defining ``can never fail'' is an interesting
problem with multiple approaches.

\citet{wadler_well-typed_2009} present a syntactic approach to this
definition for a simpler calculus without dependent type or refinement
types, but with careful tracking of blame for failed casts.
Through the blame-based analogues of ``progress'' (if a term $\e$
cannot blame a principal $A$ then it is not stuck blaming $A$) and
``preservation'' (if a term $\e$ cannot blame $A$ and it evaluates one
step to $\e'$ then neither can $\e'$ blame $A$).  Together these prove
that an upcast (in their case, a contract with blame labels mediating
between a simply-typed language and a dynamically-typed language) can
never blame a principal that provides well-typed input.

Instead of attempting a translation of their results to our
dependently-typed setting, we prove a stronger result using the
extensional definition of equivalence: A redundant cast is always
contextually equivalent to the identity function.\footnote{This result
  was actually proved prior to the publication of their work.}

The definition of extensional equivalence for $\lambdah$ in
\hhrefpref{Figure}{figure:lambdah:extensionalequivalence} applies
unmodified to $\lambdah$ with hybrid type checking, and all the
theorems surrounding it extend without difficulty. Thus correctness of
the operational semantics of casts reduces to the following theorem:

\begin{thm}
  \label{theorem:equiv}
  If $\judgesub \env S T$ then $\judgesimt \env {\casttwo S T} {(\lam x S x) : (\arrowt S T)} $
\end{thm}

\begin{lamproof}
  \pf\ Proceed by induction on the shape of $S$ (which is necessarily also the shape of $T$) 
  proceeding by considering the outermost syntactic form.

  \begin{proofcases}
  \item[\proofcase $\tyS = \predty x B \predp$ and $T = \predty x B \predq$] \

    To test equivalence of $\casttwo {\predty x B \predp} {\predty x B
      \predq}$ and $\plam x {\predquick B \predq} x$, we apply them
    to two terms $\tmd$ and $\e$ such that $\judgesimt \env \tmd {\e :
      \predty x B \predq} $.  For any closing substitution $\closingsubst$, we are
    guaranteed that $\sigma(\tmd) \lred^* \constant$ if and only if
    $\closingsubst(\e) \lred^* \constant$, so it suffices to consider
    applying the cast to such a $\constant$.  By the semantic definition
    of implication and the assumption that constants are assigned
    appropriate types, $\closingsubst(\subst x \constant B\, \predq)
    \lred^* \true$ so the cast succeeds, behaving as the identity
    function on $\constant$.

  \item[\proofcase $\tyS = \lamt x {S_1} {S_2}$ and $T = \lamt x {T_1} {T_2}$] \

    To show that this cast is equivalent to the identity function,
    fix a term $\f$ and apply the cast, yielding
    \[
    \funcasttwo x {S_1} {S_2} \f {T_1} {T_2} 
    \]
    which must be shown equivalent to $\f$. So now fix an argument
    an argument $\e$ of type $T_1$ and $\beta$-reduce to
    \[
    \casttwo {\subst x {\casttwo {T_1} {S_1}\, \e} {}\, S_2} {\subst x \e {}\, T_2} 
    ~~
    (f ~~(\casttwo {T_1} {S_1} ~ \e))
    \]
    By induction, $\casttwo {T_1} {S_1} ~ \e$ is equivalent to $\e$, so
    the whole term is equivalent to
    \[
    \casttwo {\subst x \e {}\, S_2} {\subst x \e {}\, T_2} 
    ~~
    (\f ~ \e)
    \]
    Again by induction (and the substitution property of subtyping)
    this is equivalent to $(f ~ e)$.~\qed
  \end{proofcases}
\end{lamproof}

\begin{corollary}
  For a derivation
  \[
  \inferrule*{
    \inferrule*[right=\rel{T-Cast}]{
      \vdots
    }{
      \judge \env {\casttwo S T} {\arrowt S T}
    }
    \\\\
    \vdots
  }{
    \judge \emptyset {\tmctx[\casttwo S T\, e]} U
  }
  \]
  if the cast leads to a failure (a formal definition involving labels
  is beyond the scope of this chapter) then $\njudgesub \env S T$ and
  $\e$ is a \emph{counterexample}, a term such that $\judge \env \e S$
  but $\njudge \env \e T$.
\end{corollary}

To conclude this section, before treating the hybrid type checking
algorithm, we note that all the lemmas of the preceding chapter extend
in routine manner to casts, and we review just the key lemmas of
progress and preservation.  The preservation lemma holds as stated,
with four new cases to consider.

\begin{lem}[Preservation]
  If $\judge \env \tmd \tyT$ and $\tmd \lred \e$ then $\judge \env \e \tyT$.
\end{lem}

\begin{lamproof}
  \pf\ The proof is by straightforward induction as before; we discuss only
  the cases involving casts. Note the free rebinding of metavariables,
  which is immaterial to the argument.

  \begin{proofcases}
  \item[\proofcase \rel{T-Cast}:] Evaluation within types does
    not affect typing.

  \item[\proofcase \rel{T-App} for a function cast:] We have
    \[ {\casttwo {\lamt x {S_1} {S_2}} {\lamt x {T_1} {T_2}}} ~ \f\]
    evaluating to
    \[ \funcasttwo x {S_1} {S_2} \f {T_1} {T_2} \]
    and we invite the reader to verify that preservation holds in
    this case.
    
  \item[\proofcase \rel{T-App} for a cast to basic type:] We have
    \[ {\casttwo {\predty x B \predp} {\predty x B \predq}} ~ \constant \]
    evaluating to
    \[ \checking {\predty x B \predq} {\subst x \constant B \, \predq} \constant \]
    All the antecedents of \rel{T-Checking} are satisfied obviously
    since the implication judgment is reflexive.

  \item[\proofcase \rel{T-Checking} still in progress:] Evaluation of the predicate does
    not affect the implication judgment.

  \item[\proofcase \rel{T-Checking} completed:] The implication antecedent
    of \rel{T-Checking} ensures that $\constant$ can be assigned
    the necessary type by subsumption.~\qed
  \end{proofcases}
\end{lamproof}

In this setting, the Progress theorem holds only modulo failed casts;
a term may be ``justifiably'' stuck because a cast has failed.

\begin{lem}[Progress]
  If $\judge \emptyset \e T$ then either $\e$ is a value, $\e$ contains a failed cast,
  or there is some $\e'$ such that $\e \lred \e'$.
\end{lem}
  
\begin{lamproof}
  \pf\ By induction on the derivation of $\judge \emptyset \e T$ as
  before.~\qed
\end{lamproof}


\myclearpage
\section{Hybrid Type Checking for $\lambdahtc$}
\label{section:htc:htc}

We now describe how to perform hybrid type checking. We work in the
specific context of the language $\lambdahtc$, but have also
demonstrated that the general approach extends to other languages with
similarly expressive type systems, such as \sage{} of
\hhref{Chapter}{chapter:sage}.

The key judgments constituting our hybrid type checking system utilize
a three-element lattice of certainty, where ``$\pyes$'' means a
judgement certainly holds, ``$\pno$'' means a judgment has been refuted,
and ``$\pmaybe$'' means that a judgment can be neither proven nor
refuted. In the following, $a \in \{\pyes, \pno, \pmaybe\}$
and $\pno < \pmaybe < \pyes$.

\begin{itemize}
\item[-] The implication algorithm $\judgeimpalg a \env \predp \predq$
  soundly approximates implication, yield a certainty $a$ about whether
  the implication holds. (This is a parameter to the
  system, hence not in any figure)

\item[-] The subtyping algorithm $\judgesubalg a \env S T$
  soundly approximates subtyping. (\hhref{Figure}{figure:htc:subtypingalgorithm})
  
\item[-] Cast insertion and checking $\judgecc \env \d \e T$ inserts casts
  as into $\d$ as necessary to enforce type $T$, producing a new term $\e$.
  (\hhref{Figure}{figure:htc:castinsertionandchecking})

\item[-] Cast insertion and synthesis $\judgec \env \d \e T$ traverses $\d$
  to produce a new term $\e$ and its type $T$. 
  (\hhref{Figure}{figure:htc:castinsertionandsynthesis})

\item[-] Cast insertion on types $\judgect \env S T$ traverses the type $S$
  and inserts necessary casts to ensure that $T$ is a well-formed type.
  (\hhref{Figure}{figure:htc:typecastinsertion})
\end{itemize}

The difference between the two cast insertion judgments of one of
``polarity'', \ie\ which arguments of the judgment are considered
inputs and which are considered outputs. The defining rules for these
judgments make the distinction clear and rigorous. Each judgment
is discussed in depth below.

\subsection{Implication Algorithm}

The implication algorithm is a parameter of the hybrid type checking
system. Since implication is undecidable, it is intended that our theories
adapt to embrace the latest technology in best-effort automated
proving, so it is most effective to provide the specification to
which any such implementation must adhere. For any implication
$\judgeT{\env}{\predp\Implies \predq}$, the algorithmic judgment
\[
\judgeimpalg a \env \predp \predq
\]
always must hold with $a \in \setz{\pyes, \pno, \pmaybe}$:

\begin{itemize}
\item[-] The judgment $\judgeimpalg \pyes \env \predp \predq$ means
  the algorithm finds a proof that $\judgeimp \env \predp \predq$.

\item[-] The judgment $\judgeimpalg \pno \env \predp \predq$ means the
  algorithm finds a proof that $\njudgeimp \env \predp \predq$.

\item[-] The judgment $\judgeimpalg \pmaybe \env \predp \predq$ means
  the algorithm terminates without either discovering a proof of
  either $\judgeimp \env \predp \predq$ or $\njudgeimp \env \predp
  \predq$.
\end{itemize}

The requirement placed upon this result is the subject of \hhref{Requirement}{ass:imp}:

\begin{requirement}[Soundness of $\judgeimpalg a \env \predp \predq$]
  \label{ass:imp}\

  \begin{enumerate}
  \item
    If $\judgeimpalg \pyes  \env \predp \predq$ then $\judgeimp \env \predp \predq$.
  \item
    If $\judgeimpalg \pno \env \predp \predq$ then $\njudgeimp \env \predp \predq$.
  \end{enumerate}
\end{requirement}

This algorithm cannot ever be complete. An extremely naive algorithm
could simply return $\judgeimpalg \pmaybe \env \predp \predq$ in all
cases.

\subsection{Subtyping Algorithm}

The implication-checking algorithm $\judgeimpalg a \env \predp \predq$
naturally induces a 3-valued algorithmic subtyping judgment:
\[
	\judgesubalg a \env \tyS \tyT
\]
as shown in \hhref{Figure}{figure:htc:algsub}.  

\begin{displayfigure}{Algorithmic Subtyping for $\lambdahtc$}
  \label{figure:htc:algsub}
  \label{figure:htc:subtypingalgorithm}
  \vspace{-25pt}
  \algorithmicsubtyping
  \vspace{-35pt}
\end{displayfigure}

\begin{rulelist}
\item[\rel{SA-Base}:] Algorithmic subtyping between refined basic
  types $\predty x B p$ and $\predty x B q$ reduces to a corresponding
  three-valued implication judgment between $p$ and $q$.

\item[\rel{SA-Arrow}:] Subtyping between function types reduces to
  subtyping between corresponding contravariant domain and covariant
  range types.  This rule uses the standard meet (greater lower bound)
  operation $\sqcap$ on the three-element lattice of possible
  implication results:
  \[
  \begin{array}{|c||c|c|c|}
    \hline
    \sqcap		& \pyes 	& \pmaybe	& \pno \\
    \hline
    \hline
    \pyes		& \pyes		& \pmaybe	& \pno \\
    \pmaybe		& \pmaybe	& \pmaybe	& \pno \\
    \pno		& \pno		& \pno		& \pno \\
    \hline
  \end{array}
  \]
  If the subtyping relation can be proven between the domain and range
  components then the subtyping relation holds between the function
  types.  If the appropriate subtyping relation does not hold between
  either the domain or range components, then the subtyping relation
  does not hold between the function types.  Otherwise, in the uncertain
  case, subtyping \emph{may} hold between the function types.
\end{rulelist}

A consequence of the required soundness of the implication algorithm
is that the algorithmic subtyping judgment $\judgesubalg a \env S T$
is also sound.

\begin{lemma}[Soundness of $\judgesubalg{a}{\env}{S}{T}$]
  \label{lemma:htc:AlgorithmicSubtypingSoundness}
  Suppose $\judgeE{\env}$.
  \begin{enumerate}
  \item
    If $\judgesubalg{\pyes}{\env}{S}{T}$ then $\judgesub{\env}{S}{T}$.
  \item
    If $\judgesubalg{\pno}{\env}{S}{T}$ then $\njudgesub{\env}{S}{T}$.
  \end{enumerate}
\end{lemma}
\begin{lamproof}
  \pf\ By induction on the derivation of $\judgesubalg a \env S T$
  using \hhref{Requirement}{ass:imp}.~\qed
\end{lamproof}

\hhref{Lemma}{lemma:htc:AlgorithmicSubtypingSoundness} should also be
read as a requirement for future hybrid systems: It may not always be
that algorithmic subtyping statically reduces to algorithmic
implication.  From a broader perspective, the subtyping algorithm
itself is a parameter to hybrid type checking, but this fact should
still hold of the algorithmic approximation to subtyping.

\subsection{Cast Insertion}

Hybrid type checking uses this subtyping algorithm to type check the
source program, and to simultaneously insert dynamic casts to
compensate for any indefinite answers returned by the subtyping
algorithm. The judgment which translates subtyping results to
appropriate casts is the \emph{cast insertion and checking judgment}
\[
\judgecc \env \d \e T
\]
which intuitively takes $\env$, $\d$ and $T$ as inputs and provides
$\e$ either with casts or without. There are only two defining rules,
presented in \hhref{Figure}{figure:htc:castinsertionandchecking},
which are mutually recursive with the \emph{cast insertion and
  synthesis} judgment described next:

\begin{displayfigure}{Cast insertion and Checking for $\lambdahtc$}
  \label{figure:htc:castinsertionandchecking}
  \vspace{-25pt}
  \castinsertionandchecking
  \vspace{-35pt}
\end{displayfigure}

\begin{rulelist}
\item[\rel{CC-Ok}:] If the subtyping algorithm succeeds in proving that $S$
  is a subtype of $T$ (\ie, $\judgesubalg \pyes \env S T$), then $\e$ is
  clearly of the desired type $T$, and so is returned unchanged.
  
\item[\rel{CC-Chk}:] In the uncertain case where $\judgesubalg \pmaybe \env S
  T$, the cast $\casttwo S T$ is inserted to dynamically ensure that
  $\e$ is actually of the desired type $T$.
\end{rulelist}

Note that if $\judgesubalg \pno \env S T$ then no rule applies with
inputs $\env$, $\tmd$, and $T$, so the term $\tmd$ is rejected.

The cast insertion and synthesis judgment,
\[
\judgec \env \d \e T
\]
take $\env$ and $\d$ as inputs and produces $\e$ and $T$ as outputs,
where $T$ is the type of the resulting term $\e$. Since types contain
terms, we extend this cast insertion process to types via the judgment
\[
\judgect \env \tyS \tyT
\]
which serves only to traverse types and insert casts to the terms
contained therein. The rules defining each of these this judgments are
presented in \hhref{Figure}{figure:htc:castinsertion}; Most of the rules
are straightforward enhancements of the typing rules:

\begin{displayfigure}{Cast insertion and synthesis for $\lambdahtc$ terms}
  \label{figure:htc:castinsertion}
  \label{figure:htc:castinsertionandsynthesis}
  \vspace{-25pt}
  \castinsertionrules
  \vspace{-35pt}
\end{displayfigure}

\begin{displayfigure}{Cast insertion for $\lambdahtc$ types}
  \label{figure:htc:typecastinsertion}
  \vspace{-25pt}
  \typecastinsertionrules
  \vspace{-35pt}
\end{displayfigure}

\begin{rulelist}
\item[\rel{C-Var} and \rel{C-Const}:] Variable references and constants do not require additional casts. 

\item[\rel{C-Fun}:] An abstraction $\plam x {S_1} \tmd$ by processed by
  first inserting casts into the type $S_1$ to yield $T_1$ and then
  processing $\tmd$ to yield a term $\e$ of type $T_2$; the resulting
  abstraction $\lam x {T_1} \e$ has type $\lamt x {T_1} {T_2}$.

\item[\rel{C-Cast}:] If a cast $\casttwo S T$ occurs in a program,
  then either of $S$ or $T$ may contain terms, so casts are inserted
  into each of these types. In an implementation of hybrid type
  checking, one might simply forbid casts in input programs, but we
  include this rule for completeness and to discuss idempotence of
  cast insertion.

\item[\rel{C-App}:] For an application $(\f~\tmd)$, first casts are
  inserted into $\f$, yielding $\g$ of type $\lamt x {T_1}
  {T_2}$. Then invokes the cast insertion \emph{and checking}
  judgment to convert the argument $\tmd$ into a term $\e$ of the
  appropriate argument type $T_1$.

\item[\rel{C-Arrow}:] Casts are inserted into the domain and codomain
  of a function type $\lamt x \tyS \tyT$ by induction.

\item[\rel{C-Base}:] The refinement predicate $\predp$ of a
  refined base type is checked against type $\Bool$.

\end{rulelist}

\subsection{Correctness of Cast Insertion}
\label{section:htc:correctness}

Since hybrid type checking relies on necessarily incomplete algorithms
for subtyping and implication, this section contributes a clear
specification for hybrid type checking in general and proves that our
description of hybrid type checking for $\lambdahtc$ meets that
specification.

A modern typed compiler relies on type safety for a variety of
factors, including memory layout and optimizations. So it is important
that during hybrid type checking while the input program may be subtly
ill-typed, the output program is well-typed.  Thus any terms that
would violate guarantees a compiler relied on are safely wrapped.

\begin{thm}[Cast insertion produces well-typed terms]
\label{lemma:htc:CompilationSoundness}
Suppose $\judgeE{\env}$.
\begin{enumerate}
\item
If  $\judgec \env \e {\e'} \tyT$ then $\judge \env {\e'} \tyT$.  
\item
If  $\judgecc \env \e {\e'}{\tyT}$ and $\judgeT \env \tyT$ then $\judge \env {\e'} \tyT$.  
\item
If $\judgect \env T {T'}$ then $\judgeT \env {\tyT'}$.  
\end{enumerate}
\end{thm}
\begin{lamproof}
  \pf\ Proceeds directly by mutual induction on cast insertion derivations.~\qed
\end{lamproof}

This fact ensures that the cast insertion rules actually do their job
of inserting casts everywhere that they are needed. Furthermore, the
cast insertion process is idempotent provided algorithmic subtyping
meets the most basic requirement of reflexivity.

\begin{thm}[Idempotence of Cast Insertion]
  \label{thm:htc:idempotence}\

  Suppose that for any $\env$ and $T$ such that $\judgeT \env T$, the subtyping
  algorithm also determines $\judgesubalg \pyes \env T T$.
  
  \begin{enumerate}
    \item If $\judgecc \env \d \e T$ then $\judgecc \env \e \e T$.
    \item If $\judgec \env \d \e T$ then $\judgec \env \e \e T$
  \end{enumerate}
\end{thm}

\begin{lamproof}
\pf\ By mutual induction on the derivations.~\qed
\end{lamproof}

Converse to the fact that hybrid type checking always produces a
well-typed program, it also accepts all well-typed programs. The
program may be subtle and thus have some casts added, but these casts
are all redundant.

\begin{thm}[Cast insertion is the identity on well-typed terms]
  \label{proved:htc:castequiv}\

  \begin{enumerate}
  \item If $\judge \env \d T$ then there exists a term $\e$ and type
    $S$ such that $\judgec \env \d \e S$ and $\judgesub \env S T$ and
    $\judgesimt \env \d {\e:T}$.

  \item If $\judge \env \d S$ then for any $T$ such that $\judgesub
    \env S T$ there exists a term $\e$ such that $\judgecc \env \d \e
    T$ and $\judgesimt \env \d {\e:T}$.

  \item If $\judgeT \env S$ then there exists $T$ such that $\judgect \env S T$ and
    $\judgesub \env S T$ and $\judgesub \env S T$.

  \end{enumerate}
\end{thm}

\begin{lamproof}
  \pf\ By mutual induction on the derivations of $\judge \env \e T$
  and $\judgeT \env T$. Because algorithmic subtyping can only return
  $\pmaybe$ or $\pyes$, the output of cast insertion is simply the
  input with redundant casts inserted.  By \hhref{Theorem}{theorem:equiv}
  these are all equivalent to identity functions.~\qed
\end{lamproof}

\label{section:htc:behavioralcorrectness}


\myclearpage
\section{An Example of HTC}
\label{section:htc:matrix}

To illustrate the behavior of the cast insertion algorithm, consider a
function \t{serializeMatrix} that serializes an $n$ by $m$ matrix into
an array of size $n\times m$.  We extend the language $\lambdahtc$
with two additional base types:
\begin{itemize}
\item[-]
\Array, the type of one dimensional arrays containing integers.
\item[-]
\Matrix, the type of  two dimensional matrices, again containing integers.
\end{itemize}
The following primitive functions return the size of an array; create
a new array of the given size; and return the width and height of a
matrix, respectively:
\[
\begin{array}{rcl}
  \t{asize} &:& \lamt{a}{\Array}{\Int}\\
  \t{newArray} &:& \arrowd{\Int}{n}{\predty a \Array {\t{asize}~a}=n}\\
  \t{matrixWidth} &:& \lamt{a}{\Matrix}{\Int}\\
  \t{matrixHeight} &:& \lamt{a}{\Matrix}{\Int}\\
\end{array}
\]
We introduce the following type abbreviations to denote arrays of size $n$ and matrices of size $n$ by $m$:
\[
\begin{array}{rcl}
\Array_n	&\defeq& 
	\predty{a}{\Array}{(\app{\t{asize}}{a}=n)}
\\
\Matrix_{n,m} &\defeq& 
	\predty a \Matrix {\left(
	\begin{array}{@{~}l@{~}l@{~}}
		& \app{\t{matrixWidth}~~}{a}=n \\
	\wedge 	& \app{\t{matrixHeight}}{a}=m
	\end{array}
	\right)
	}
\end{array}
\]
The shorthand $\ascriptl l \e T$ ensures that the term $\e$ has
type $T$ by passing $\e$ as an argument to the identity function of
type $\arrowt T T$:
\[
	\ascriptl l \e T	
	~\defeq~
	(\lam{x}{T}{x}) ~ \e
\]
\medskip
We now define the function
\t{serializeMatrix} as:
\[
	\left(\begin{array}{@{}l@{}}
		\lam{n}{\Int}{} ~
		\lam{m}{\Int}{} ~
		\lam{a}{\Matrix_{n,m}}{}  \\
		\quad \t{let}~r~=~\t{newArray}~e~\t{in} ~\dots ~;~ r\\
	\end{array}\right) 
~\t{as}~T
\]
The elided term $\dots$ initializes the
new array $r$ with the contents of the matrix $a$,
and
we will consider several possibilities
for the size expression $e$.  
The type $T$ is the specification of \t{serializeMatrix}:
\[
T~~\defeq~~(\lamt{n}{\Int}{} ~
\lamt{m}{\Int}{} ~
\arrowt{\Matrix_{n,m}}{} ~
\Array_{n \times m})
\]

For this declaration to type check, the inferred type $\Array_e$ of
the function's body must be a subtype of the declared return type:
\[ 
\judgesub{n:\Int,~m:\Int~}{\Array_e}{\Array_{n\times m}} 
\]
Checking this subtype relation reduces to checking the implication:
\[
  {n:\Int,~m:\Int,~r:\Array}~\vdash
  \begin{array}[t]{ll}
	&	{ (\app{\t{asize}}{r}=e)} \\
	\Implies &
		     { (\app{\t{asize}}{r}=(n\times m)) }
  \end{array}
  \]
  which in turn reduces to checking the equality:
  \[ \forall n,m\in\Int.~~e=n\times m \]
  The implication checking algorithm might use an automatic theorem
  prover to verify or refute such conjectured equalities.
 
We now consider three possibilities for the expression $e$. 
\begin{enumerate}
\item 
If $e$ is the expression $n\times m$, the equality is trivially true,
and no additional casts are inserted (even when using a rather weak theorem prover).
\item 
If $e$ is $m\times n$ (\ie, the order of the multiplicands is
reversed), and the underlying theorem prover can verify
\[
	\forall n,m\in\Int.~~m\times n=n\times m
\]
then again no casts are necessary.  Note that a theorem prover
that is not complete for arbitrary
multiplication might still have a specific axiom about the
commutativity of multiplication.

If the theorem prover is too limited to verify   this
equality, the hybrid type checker will still accept this
program. However, to compensate for the limitations of the theorem
prover, the hybrid type checker will insert a redundant cast,
yielding the function (where we have elided the source type of the cast):
\[
\left(
\casttwoapp {\dots} T {
  \left(
  \begin{array}{@{}l@{}}
	\lam{n}{\Int}{} ~
	\lam{m}{\Int}{} ~
	\lam{a}{\Matrix_{n,m}}{}  \\
	\quad \t{let}~r~=~\t{newArray}~e~\t{in} ~\dots ~;~ r\\
  \end{array}
  \right)
}
\right)
~\t{as}~T
\]
This term can be optimized, via~\rel{E-Beta}
and~\rel{E-Cast-Fn} steps and via removal of clearly redundant $\casttwo{\Int}{\Int}{}$
casts, to:  
\[
\begin{array}{@{}l@{}}
  \lam{n}{\Int}{} ~
  \lam{m}{\Int}{} ~
  \lam{a}{\Matrix_{n,m}}{}  \\
  \quad \t{let}~r~=~\t{newArray}~(m\times n)~\t{in} \\
  \qquad\quad ~\dots ~;~\\
  \qquad\quad \casttwoapp {\Array_{m \times n}}{\Array_{n \times m}} r
\end{array}
\]
The remaining cast checks that the result value $r$ is of the declared
return type $\Array_{n \times m}$, which reduces to dynamically
checking that the predicate
\[
\t{asize}~r~~=~~n \times m
\]
evaluates to $\true$, which it does.
\item 
  Finally, if $\e$ is erroneously $m\times m$, the function is ill-typed. 
  By performing random or directed~\citep{godefroid_dart:_2005} testing of
  several values for $n$ and $m$ until it finds a counterexample,  the theorem prover might reasonably refute the conjectured
  equality:
  \[
  \forall n,m\in\Int.~~m\times m=n\times m
  \]
  In this case,
  the hybrid type checker reports a static type error.
  
  Conversely, if the theorem prover is too limited to refute the
  conjectured equality, then the hybrid type checker will produce (after optimization) the
  program:
  \[
  \begin{array}{@{}l@{}}
	\lam{n}{\Int}{} ~
	\lam{m}{\Int}{} ~
	\lam{a}{\Matrix_{n,m}}{}  \\
	\quad \t{let}~r~=~\t{newArray}~(m\times m)~\t{in} \\
	\qquad\quad ~\dots ~;~\\
	\qquad\quad \casttwoapp {\Array_{m \times m}}{\Array_{n \times m}} r\\
  \end{array}
  \]
  If this function is ever called with arguments for which $m\times
  m\not=n\times m$, then the cast will detect the type error. 
\end{enumerate}

Note that prior work on practical dependent
types~\citep{xi_dependent_1999} could not handle these cases, since
the type $T$ uses non-linear arithmetic expressions. In contrast, case
2 of this example demonstrates that even fairly partial techniques for
reasoning about complex specifications (\eg, commutativity of
multiplication, random testing of equalities) can facilitate static
detection of defects.  Furthermore, even though catching errors at
compile time is ideal, catching errors at run time (as in case 3) is
still an improvement over not detecting these errors at all, and
getting subsequent crashes or incorrect results.


\myclearpage
\section{Static Checking vs. Hybrid Checking}
\label{section:htc:comparison}

Given the proven benefits of traditional, purely-static type systems,
an important question that arises is how hybrid type checkers relate
to static type checkers.  To study this question theoretically,
suppose we are given a static type checker that targets a restricted
subset of $\lambdahtc$ for which type checking is statically decidable.
Specifically, suppose $\dterm$ is a subset of $\fun{Term}$ such that
for all $\predp,\predq\in\dterm$ and for all singleton environments
$x:B$, the judgment $\judgeimp {x:B} \predp \predq$ is decidable.  We
introduce a statically-decidable language $\lambdas$ that is obtained
from $\lambdahtc$ by only permitting $\dterm$ predicates in refinement
types.  We also assume all types in $\lambdas$ are closed, to avoid
the complications of substituting arbitrary terms into refinement
predicates via the rule~\rel{T-App} (and hence the above environment
$x:B$ suffices).  It then follows that subtyping and type checking for
$\lambdas$ are decidable, and we denote this type checking judgment as
$\judgestatic{\env}{\e}{T}$.
 
As an extreme, we could take $\dterm=\setz{\true}$, in which case the
$\lambdas$ type language is essentially that of STLC.  However, to
yield a more general argument, we assume only that $\dterm$ is a
subset of $\fun{Term}$ for which implication is decidable.

Clearly, the hybrid implication algorithm can give precise answers on
(decidable) $\dterm$-terms, and so we assume that for all
$\predp,\predq\in\dterm$ and for all environments $x:B$, the judgment
$\judgeimpalg{a}{x:B}{\predp}{\predq}$ holds for some
$a\in\setz{\pyes,\pno}$.  Under this assumption, hybrid type checking
behaves identically to static type checking on (well-typed or
ill-typed) $\lambdas$ programs, formalized as:

\begin{thm}
  For all $\lambdas$ terms $\e$, $\lambdas$ environments $\env$, and
  $\lambdas$ types $T$, the following three statements are equivalent:
  \begin {enumerate}
  \item
    $\judgestatic \env \e T$
  \item
    $\judgehybrid \env \e T$
  \item
    $\judgehybridc \env \e \e T$
  \end {enumerate}
\end {thm}
\begin{proof}
  The hybrid implication algorithm is complete on $\dterm$-terms, and
  hence the hybrid subtyping algorithm is complete for $\lambdas$
  types. The proof then follows by induction on typing derivations.~\qed
\end{proof}

\noindent
Thus, to a $\lambdas$ programmer, a hybrid type checker behaves
exactly like a traditional static type checker.

\medskip
\myclearpage

We now compare static and hybrid type checking from the perspective of
a $\lambdahtc$ programmer.  To enable this comparison, we need to map
expressive $\lambdahtc$ types into the more restrictive $\lambdas$
types, and in particular to map arbitrary boolean predicates into
$\dterm$ predicates.  We assume the computable function
\[
	\approxfun: \fun{Term} \rightarrow \dterm
\]
performs this mapping.  The function $\fun{erase}$ then maps
$\lambdahtc$ refinement types to $\lambdas$ refinement types by using
$\approxfun$ to abstract boolean terms:
\[
	\fun{erase}(\predty x B \predp) ~\defeq~ \predty x B {\approxfun(\predp)}
\]
We extend $\fun{erase}$ in a compatible manner to map $\lambdahtc$
types, terms, and environments to corresponding $\lambdas$ types,
terms, and environments.  Thus, for any $\lambdahtc$ program $\e$, this
function yields the corresponding $\lambdas$ program $\fun{erase}(\e)$.

As might be expected, the $\fun{erase}$ function must lose
information, and we now explore the consequences of this information
loss for programs with complex specifications. As an extreme example,
let $\fun{Halt}_{n,m}$ denote a closed formula that encodes ``Turing
machine $m$ eventually halts on input $n$''. Suppose that $\approxfun$
maps $\fun{Halt}_{n,m}$ to $\false$ for all $n$ and $m$, and consider
the collection of programs $P_{n,m}$, which includes both well-typed
and ill-typed programs:
\[
	P_{n,m} ~~\defeq~~ 
			(\lam x {\predty x \Int {\fun{Halt}_{n,m}}}{x})~1
\]
Decidability arguments show that the hybrid type checker will accept
some ill-typed program $P_{n,m}$ based on its inability to statically
prove that $\fun{Halt}_{n,m}=\false$.  In contrast, the static type
checker will reject (the erased version of) $P_{n,m}$.  Hence:
\begin{quote}
\emph{The static type checker statically rejects (the erased version
  of) an ill-typed program that the hybrid type checker accepts.
}\end{quote} Thus, the static type checker performs better in this
situation.

However, this particular static type checker will also reject (the
erased version of) many well-typed programs $P_{n,m}$ that the hybrid
type checker accepts. The following theorem generalizes this argument,
and shows that for any computable mapping $\approxfun$ there exists some
program $P$ such that hybrid type checking of $P$ performs better than
static type checking of $\fun{erase}(P)$.

\begin{thm}
\label{theorem:hybridbetter}
For any computable mapping $\approxfun$ either:
\begin{enumerate}
\item
the static type checker rejects the erased version of some well-typed
$\lambdahtc$ program, or
\item
the static type checker accepts the erased version of some ill-typed
$\lambdahtc$ program for which the hybrid type checker would statically
detect the error.
\end{enumerate}
\end{thm}

\noindent
\begin{proof}
Let $\env$ be the environment $x:\Int$.

By reduction from the halting problem, the judgment $
\judgeimp{\env}{\predp}{\false} $ for arbitrary boolean terms $\predp$
is undecidable. However, the implication judgment
$\judgeimp{\env}{\approxfun(\predp)}{\approxfun(\false)}$ is
decidable. Hence these two judgments are not equivalent, \ie:
\[
	\setc{t}{(\judgeimp{\env}{\predp}{\false})}
	~~\not=~~
	\setc{t}{(\judgeimp{\env}{\approxfun(\predp)}{\approxfun(\false)})}
\]
It follows that there must exists some \emph{witness} $w$ that is in one of these sets but not the other, and so one of the following two cases must hold.
\begin{enumerate}
\item
Suppose:
\[
\begin{array}{l}
	\judgeimp{\env}{w}{\false}\\
	\njudgeimp{\env}{\approxfun(w)}{\approxfun(\false)}
\end{array}
\]
We  construct as a counter-example the program $P_1$:
\[
	P_1  ~~\defeq~~
		\lam{x}{\predty{x}{\Int}{w}}
			{(\ascriptl{}{x}{\predty{x}{\Int}{\false}})}\\
\]
From the assumption $\judgeimp{\env}{w}{\false}$ the subtyping judgment
\[
\begin{array}{l}
	\judgesub{\emptyset}
		{\predty{x}{\Int}{w}}
		{\predty{x}{\Int}{\false}} \\
\end{array}
\]
holds. Hence, $P_1$ is well-typed, and (by
\hhref{Theorem}{proved:htc:castequiv} is accepted by the hybrid type
checker.
%
However, from the assumption
$\njudgeimp{\env}{\approxfun(w)}{\approxfun(\false)}$ the erased
version of the subtyping judgment does not hold:
\[
\begin{array}{l}
	\njudgesub{\emptyset}
		{\fun{erase}(\predty{x}{\Int}{w})}
		{\fun{erase}(\predty{x}{\Int}{\false})} \\
\end{array}
\]
Hence $\fun{erase}(P_1)$ is ill-typed and rejected by the static type checker.

\item
Conversely, suppose:
\[
\begin{array}{l}
	\njudgeimp{\env}{w}{\false}\\
	\judgeimp{\env}{\approxfun(w)}{\approxfun(\false)}
\end{array}
\]
From the first supposition and by the definition of the implication judgment, there exists integers $n$ and $m$ 
such that  
\[
	{\subst x n \Int}\, w\lred^m \true
\]
We now construct as a counter-example the program $P_2$:
\[
\begin{array}{@{}l}
	P_2 ~~\defeq~~ 		\lam{x}{\predty{x}{\Int}{w}}
			{(\ascriptl{}{x}{\predty{x}{\Int}{\false\wedge(n=m)})}} \\
\end{array}
\]
In the program $P_2$, the term $n=m$ has no semantic meaning since it
is conjoined with $\false$. The purpose of this term is to serve only
as a ``hint'' to the following rule for refuting implications (which
we assume is included in the reasoning performed by the implication
algorithm). In this rule, the integers $a$ and $b$ serve as hints, and
take the place of randomly generated values for testing if $\predp$ ever
evaluates to $\true$.
\[
\trulenoname{
		{\subst x a \Int}\, \predp \lred^b \true
	}{
		\judgeimpalg{\pno}{\env}{\predp}{(\false\wedge a=b)}
	}
\]
This rule enables the implication algorithm to conclude that:
\[
\begin{array}{l}
	\judgeimpalg{\pno}{\env}{w}{\false\wedge(n=m)}\\
\end{array}
\]
Hence, the subtyping algorithm can conclude:
\[
\begin{array}{l}
	\judgesubalg{\pno}
		{}
		{\predty{x}{\Int}{w}}
		{\predty{x}{\Int}{\false\wedge(n=m)}} \\
\end{array}
\]
Therefore, the hybrid type checker correctly rejects $P_2$,
which by \hhref{Theorem}{proved:htc:castequiv} is therefore ill-typed.

We next consider how the static type checker behaves on the program $\fun{erase}(P_2)$.
We consider two cases, depending on whether the following implication judgment holds:
\[
	\judgeimp{\env}{\approxfun(\false)}{\approxfun(\false\wedge(n=m))}
\]
\begin{enumerate}
\item
If this judgment holds then by the transitivity of implication and the assumption $\judgeimp{\env}{\approxfun(w)}{\approxfun(\false)}$ we have that:
\[
\begin{array}{l}
	\judgeimp{\env}
		{\approxfun(w)}
		{\approxfun(\false\wedge(n=m))} \\
\end{array}
\]
Hence the subtyping judgment
\[
\begin{array}{l}
	\judgesub{\emptyset}
		{\predty{x}{\Int}{\approxfun(w)}}
		{\predty{x}{\Int}{\approxfun(\false\wedge(n=m))}} \\
\end{array}
\]
holds and the program $\fun{erase}(P_2)$ is incorrectly accepted by
the static type checker:
\[
\begin{array}{l}
	\judgehybrid{\emptyset}{\fun{erase}(P_2)}{
					\arrowt{\predty{x}{\Int}{\approxfun(w)}}
					{\predty{x}{\Int}{\approxfun(\false\wedge(n=m))}}
			}
\end{array}
\]

\item
If the above judgment does not hold then consider as a counter-example the program $P_3$:
\[
\begin{array}{l}
	P_3 \defeq 		\lam{x}{\predty{x}{\Int}{\!\false}}
			{(\ascriptl{}{x}{\predty{x}{\Int}{\false\wedge(n\!=\!m)})}} \\
\end{array}
\]
This program is well-typed, from the subtype judgment:
\[
\begin{array}{@{}l}
	\judgesub{\emptyset}
		{\predty{x}{\Int}{\false}}
		{\predty{x}{\Int}{\false\wedge(n=m)}} \\
\end{array}
\]

However, the erased version of this subtype judgment does not hold:
\[
	\njudgesub{\emptyset}
		{\fun{erase}(\predty{x}{\Int}{\!\false})}
		{\fun{erase}(\predty{x}{\Int}{\false\wedge\!(n\!=\!m)})} 
\]
Hence, $\fun{erase}(P_3)$ is rejected by the static type checker.~\qed
\end{enumerate}
\end{enumerate}
\end{proof}


\myclearpage
\section{Related Work}
\label{section:htc:relatedwork}

Many systems feature a combination of static and dynamic checking of
specifications. The details of what languages the specifications are
written in (if any), which checks are performed during which phase of
a program's lifecycle, and how this decision is made, differ
considerably.

\subsection{Static structural types, dynamic contracts}

Widespread languages such as Java and C\# enforce structural types
statically, while enforcing the implicit contracts on array bounds
dynamically. The programming language
Eiffel~\citep{meyer_object-oriented_1988} also includes a structural
type system while making dynamic contracts explicit; the languages of
statically checked specifications and dynamically checked
specifications are completely different.

All implementations of ``contracts as a library'' such as
\citet{hinze_typed_2006}, \citet{chitil_practical_2012} fall into the
same classification as Eiffel, with types forming the language of
statically checked specifications and boolean terms of the language
at hand expressing dynamically checked specifications.

Having multiple specification languages is somewhat awkward since it
requires the programmer to factor each specification into its static
and dynamic components. Furthermore, the factoring is too rigid, since
the specification needs to be manually refactored to exploit
improvements in static checking technology. Contracts implemented
as a library have the additional drawback that there is not possible
to take advantage of any formal correspondence between the static
and dynamic specifications languages.

\subsection{Optimizing implied dynamic checks}

The limitations of purely-static and purely-dynamic approaches have
also motivated other work on hybrid analyses. For example,
CCured~\citep{necula_ccured:_2005} is a sophisticated hybrid analysis
for preventing the ubiquitous array bounds violations in the C
programming language. Unlike our proposed approach, it does not detect
errors statically - instead, the static analysis is used to optimize
the run-time analysis.

Soft typing~\citet{fagan_soft_1991} performs best-effort static type
  checking in order to eliminate casts which are implicit in the
  semantics of any dynamically typed programming language.
\citet{henglein_dynamic_1994} characterized the \emph{completion
  process} of inserting the necessary coercions, and presented a
rewriting system for generating minimal completions~
\citet{thatte_quasi-static_1990} developed a similar system in which
the necessary casts are implicit. 

Any of these analyses could be viewed as hybrid type systems where the
subtyping algorithm never rejects a query. Rather, they start from
dynamically typed languages which are equivalent to a subtyping
algorithm that always returns ``$\pmaybe$'' and enhance it to
sometimes return ``$\pyes$''. Optionally, they may be enhanced to
determine that ``$\pno$'' is the correct answer, but will only issue
warnings since rejection of a program is not within their purview.

\subsection{Dynamic typing in a statically-typed language}

Many authors have proposed systems that combine static structural type
systems and dynamic typing, using casts to mediate between the
dynamically typed portions of a program and the statically typed
portions.  \citet{abadi_dynamic_1991} extended a static type system
with a type {\tt Dynamic} that could be explicitly cast to and from
any other type (with appropriate run-time
checks). \citet{siek_gradual_2006} independently developed a system
dubbed \emph{gradual typing} that combines dynamic and static typing
via casts similar to ours. However, they do not address dependent
types or refinement predicates, so it remains to be investigated
whether the approaches from gradual typing can shed additional light
on our more advanced setting.  A longer discussion of subsequent work
on gradual typing is deferred to \hhref{Chapter}{chapter:relatedwork}.

\citet{ou_dynamic_2004} developed a type system similar to
$\lambdahtc$, adding an interface between untyped code and code with
refinement and dependent types. Unlike the hybrid type checking
approach, their type system restricts refinement predicates to ensure
decidability and it supports mutable data.  In addition, whereas
hybrid type checking uses dynamic checks to circumvent decidability
limitations, their system uses dynamic checks to permit
interoperability between precisely typed code fragments (that use
refinement types) and more coarsely typed code fragments (that do
not), and so it permits the introduction of refinement predicates into
a large program in an incremental manner.

These systems are intended to support looser type specifications,
blending classic simple type systems with a dynamically typed
programming style. In contrast, our work uses similar,
automatically-inserted casts to support more precise type
specifications. \hhref{Chapter}{chapter:sage} combines both
approaches to support a large range of specifications, from
\t{Dynamic} at one end to precise hybrid-checked specifications at the
other.

\chapter{Type Reconstruction}

\newcommand{\judgecon}[4]{\ensuremath{#1\vdash #2\,:\,#3 ~\&~ #4}}
\newcommand{\judgeTcon}[3]{\ensuremath{#1\vdash #2 ~\&~ #3}}
\newcommand{\judgeBcon}[5]{\ensuremath{#1\vdash {#3} ~ @ ~ {#2} : {#4} ~\&~ {#5} }}

\newcommand{\compositionalconstraintgenrules}{
  \begin{trules}
    \headtrule{$\judge{\tyenv}{e}{T}$}{\underline{C}onstraint \underline{G}eneration rules}
    \trule{CG-Var}{
      (x:T)\in \tyenv
    }{
      \judgecon \tyenv x T \emptyset
    }
    \trule{CG-Prim}{
    }{
      \judgecon \tyenv f {\fun{ty}(f)} \emptyset
    }
    \trule{CG-Fun}{
      \judgeTcon \tyenv S {C_S} \qquad
      \judgecon  {\tyenv,x:S} e T {C_e}
    }{
      \judgecon \tyenv {(\lam x S e)} {(\lamt x S T)} {C_S, C_e}
    }
    \trule{CG-Case}{
      \judgeBcon \tyenv e {ms} T C
    }{
      \judgecon \tyenv {\caseexp e {ms}} T C
    }
    \\
    \trule{CG-App}{
      \judgecon \tyenv {e_1} {\existsty {\bar y} {\bar T} {\lamt x S T}} {C_1}\\
      \judgecon \tyenv {e_2} {\existsty {\bar z} {\bar {T'}} S'} {C_2} \\\
      C_3 = \judgesub {\tyenv,\bar y:\bar T, \bar z:\bar {T'}}{S'}{S}
    }{
      \judgecon \tyenv 
      {\app {e_1} {e_2}} 
      {\existsty {\bar y, \bar z, \bar x} {\bar T, \bar {T'}, S} T}
      {C_1, C_2, C_3}
    }
    \\
    \headtrule{$\judgeT{\tyenv}{T}$}{\underline{B}ranch \underline{C}onstraint}
    \trule{BC-Empty}{
    }{
      \judgeBcon \tyenv e \cdot T \emptyset
    }
    \trule{BC-Extend}{
      \judgeBcon \tyenv e {ms} T {C_{ms}} \qquad
      ty(c) = \lamt {\listof x} {\listof S} T \\
      \judgecon {\tyenv, \listof x : \listof S, e = c ~ \listof x} {e'} T C
    }{
      \judgeBcon \tyenv e {(\branch {c~x} {e'} ~|~ ms)} T {C,C_{ms}}
    }
    \headtrule{$\judgeT{\tyenv}{T}$}{\underline{W}ell-formed \underline{T}ypes \underline{C}onstraints}
    \trule{WTC-Arrow}{
      \judgeTcon \tyenv S {C_S} \qquad
      \judgeTcon {\tyenv,x:S} T {C_T}
    }{
      \judgeTcon \tyenv {\lamt x S T} {C_S, C_T}
    }
    \trule{WTC-Base}{
      \judgecon {\tyenv,x:B} e \Bool C
    }{
      \judgeTcon \tyenv {\predty x B e} C
    }
    \trule{WT-Ex}{
      \judgeTcon {\tyenv,x:S} T C
    }{
      \judgeTcon \tyenv {\existsty x S T} C
    }
  \end{trules}
}

\label{chapter:inference}

We have addressed the problem of type-checking for a language like
\lambdah{}, given complete type annotation by the programmer. But, even
for small examples, writing explicitly typed terms can be tedious and
would become truly onerous for larger programs.  To reduce the
annotation burden, many typed languages -- such as ML, Haskell, and
their variants -- perform type reconstruction,\footnote{Type
reconstruction is sometimes referred to as ``type inference'',
especially in mainstream programming practice where it is often an
inextricable part of a type checker. But type checking itself is an
inference process, and has also been called ``type inference'', so for
clarity and precision we carefully distinguish the terminology.}
often stated as: \emph{Given a program containing type variables, find
a replacement for those variables such that the resulting program is
well typed.}  If there exists such a replacement, the program is said
to be \emph{typeable}.  Under this definition, type reconstruction
subsumes type checking.  Hence, for expressive and undecidable type
systems, such as that of \lambdah{}, type reconstruction is clearly
undecidable.

Instead of surrendering to undecidability, we separate type
reconstruction from type checking, and define the type reconstruction
problem as: \emph{Given a program containing type variables, find a
  replacement for those variables such that typeability is preserved.}
In a decidable type system, this definition coincides with the
previous one, since the type checker can decide if the resulting
explicitly typed program is well typed.  Our generalized definition
also extends to undecidable type systems, since alternative
techniques, such as hybrid type checking of \hhref{Chapter}{chapter:htc},
can be applied to the resulting program. In particular, type
reconstruction for \lambdah{} is decidable!

Our approach to inferring refinement predicates is inspired by
techniques from axiomatic semantics, most notably the strongest
postcondition (SP) transformation~\citep{back_calculus_1988}.  This
transformation supports arbitrary predicates in some specification
logic, and computes the most precise correctness predicate for each
program point.  It is essentially syntactic in nature, deferring all
semantic reasoning to a subsequent theorem-proving phase.  For
example, looping constructs in the program are expressed simply as
fixed point operations in the specification logic.

In the richer setting of $\lambdah{}$, which includes higher order
functions with dependent types, we must infer both the structural
shape of types and also any refinement predicates they contain.  We
solve the former using traditional type reconstruction techniques and
the latter using a syntactic, SP-like, transformation.  Like SP, our
algorithm infers the most precise predicates possible. The calculated
higher-order strongest postconditions may contain complex terms that
cannot be feasibly checked dynamically due to fixed-point operations,
existential quantification, and parallel boolean connectives, that are
only partially executable, but such execution will never be necessary
as the predicates are all correct by construction.

Though the most obvious novelty in type reconstruction for $\lambdah$
lies in higher-order strongest postcondition calculation, we present
an end-to-end solution for the full type reconstruction problem,
including pieces that resemble traditional type reconstruction. There
are two reasons for this:

\begin{enumerate}
\item This full presentation demonstrates exactly how ordinary type
  reconstruction is separable from the higher-order postcondition
  calculation.
\item The metatheoretical approach is of key importance for guiding
  future work in how to provide generalized type reconstruction in
  other contexts.
\end{enumerate}

\subsection*{Chapter Outline}

The organization of this chapter is as follows

\begin{itemize}
\item[-] \hhref{Section}{section:inference:problemstatement} formally
  defines our generalized notion of type reconstruction.

\item[-] \hhref{Section}{section:inference:constraintgeneration} describes
  the first phase of type reconstruction, \emph{constraint generation},
  which results in a provisional typing derivation and a set of subtyping
  constraints of the form $\judgesub{\env}{S}{T}$ (the same as the subtyping judgment)
  that must be satisfied for this typing derivation to be valid.

\item[-] \hhref{Section}{section:inference:shapereconstruction} describes the
  second phase of type reconstruction, \emph{shape reconstruction},
  which converts the subtyping constraints to implication constraints,
  each of the form \judgeImp{\env}{\predp}{\predq} (the same as the
  implication judgment).

\item[-] \hhref{Section}{section:inference:satisfiability} describes the final phase
  of type reconstruction, the calculation of \emph{strongest
    refinements} (SR), which solves all remaining implication
  constraints.

\item[-] \hhref{Section}{section:inference:correctness} establishes
  the end-to-end correctness of the phases of type reconstruction.

\item[-] \hhref{Section}{section:inference:relatedwork} surveys
  closely related work.
\end{itemize}

\section{Type Reconstruction for $\lambdahr$}
\label{section:inference:problemstatement}

To explore type reconstruction for $\lambdah$ we work with an extended
calculus $\lambdahr$, which the sections of this chapter will extend
as needed to present further aspects of type reconstructions. The
input to type reconstruction is described by the syntax of
\hhref{Figure}{figure:inference:reconsyntax}, which extends $\lambdah$
with \emph{type variables} $\tyvar$, representing types omitted in an
input program.
\[
T ~ ::= ~ \cdots ~~|~~  \tyvar
\]
\begin{displayfigure}{Syntax for $\lambdahr$ with type variables}
  \label{figure:inference:reconsyntax}
  \ssp
  \syntaxtostaterecon
\end{displayfigure}
A type variable may appear anywhere a type is expected.  A candidate
solution to type reconstruction is a \emph{type replacement} $\tysol$
that maps type variables to types. Application of a type replacement
is lifted compatibly to all syntactic sorts, and is not capture
avoiding.

\begin{definition}[Typeability]
  A closed term $\e$ containing type variables in place of some or all
  type annotations is \emph{typeable} if there is some type $T$ and
  type replacement $\tysol$ from type variables to types such that
  $\judge \emptyset {\tysol(\e)} {\tysol(T)}$. 
\end{definition}

\begin{definition}[The Type Reconstruction Problem]
  Given a term $\e$ containing type variables in place of some or all
  type annotations, find a type replacement $\tysol$ such that $\e$ is
  typeable if and only if $\tysol(\e)$ is typeable, and $\tysol(\e)$
  contains no type variables.  \footnote{Note that the condition
  requiring the replacement of \emph{all} type variables could be
  relaxed to allow partial type reconstruction, in the event that even
  this generalized notion of type reconstruction were not decidable or
  feasible.}
\end{definition}

Note that for the type reconstruction problem, we consider only the
basic language of $\lambdah$, omitting casts from
\hhref{Chapter}{chapter:htc} since they add little interest to type
reconstruction. 

\section{Constraint Generation}
\label{section:inference:constraintgeneration}

The first phase of type reconstruction is constraint generation, which
essentially mirrors type checking but collects subtyping constraints
instead of enforcing
them. \hhref{Figure}{figure:inference:constraintgensyntax} presents the
syntax of constraints and extensions to $\lambdahr$ necessary.

\begin{itemize}
\item[-] A \emph{subtyping constraint} has the same form as the
  subtyping judgment, $\judgesub \env S T$. Constraint generation
  results in a set of such constraints that essentially record the flow
  of data through the program.

\item[-] A \emph{type well-formedness constraint} has the same form as
  the type well-formedness judgment, $\judgeT \env T$. Constraint
  generation results in type well-formedness constraints that record
  the environment in which a type variable must be well-formed,
  constraining the variables that may appear free in $T$.
\end{itemize}

\begin{definition}[Constraint Satisfaction]
  \label{defn:subconsat}
  A type replacement $\tysol$ \emph{satisfies} a subtyping constraint
  $\judgesub \env S T$ (respectively, type well-formedness constraint
  $\judgeT \env T$) if $\judgesub {\tysol\env} {\tysol S} {\tysol T}$
  holds as a subtyping judgment (respectively $\judgeT {\tysol\env}
  {\tysol S}$ holds as a type well-formedness judgment).  A type
  replacement $\tysol$ satisfies a constraint set $C$ if it satisfies
  all the constraints in $C$.
\end{definition}

To facilitate our development, we require that the type language be
closed under substitution. But a substitution cannot immediately be
applied to a type variable, so each type variable $\tyvar$ is equipped
with a \emph{delayed substitution} $\theta$.
\[
T ~ ::= ~ \cdots ~~|~~  \dsub\dapp\tyvar
\]
If the substitution $\dsub$ is empty, then $\dsub\dapp\tyvar$ may
simply be written as $\tyvar$.  The usual definition of
capture-avoiding substitution is then extended to type variables,
which simply delay that substitution:
\[
\begin{array}{r@{~~}c@{~~}l}
  {\substT x \e T} ~ (\theta\dapp\tyvar)
  & = & 
  ({\substT x \e T} \circ \theta)\dapp\tyvar
\end{array}
\]
Here we use the notation $\substT x \e T$ for a delayed substitution,
adding a type annotation that we use for later syntactic processing --
it does not affect the semantics of substitution. When a type
replacement is applied to a type variable $\alpha$ with a delayed
substitution $\theta$, the substitution $\tysol(\theta)$ is
immediately applied to $\tysol(\alpha)$:
\[
\tysol(\dsub\dapp\tyvar) ~ = ~ \tysol(\dsub)(\tysol(\tyvar))
\]

\begin{displayfigure}{Syntax for $\lambdahr$ constraint generation}
  \label{figure:inference:constraintgensyntax}
  \ssp
  \constraintgensyntax\
\end{displayfigure}

Constraint generation is defined by the judgments in
\hhref{Figure}{figure:inference:constraintrules}. Each rule is derived
from the corresponding type rule in
\hhrefppref{Figure}{figure:lambdah:typerules} with subsumption
distributed throughout the derivation to make the rules
syntax-directed.

\begin{displayfigure}{Constraint Generation Rules for $\lambdahr$ terms}
  \label{figure:inference:constraintrules}
  \constraintgenrules
\end{displayfigure}

\begin{itemize}
\item[-] The judgment $\judgeC \env \e T C$ states that under environment $\env$
  the term $\e$ has type $T$, subject to constraint set $C$. (\hhref{Figure}{figure:inference:constraintrules})

\item[-] The judgment $\judgeTC \env T C$ states that under environment $\env$
  the type $T$ is well-formed, subject to constraint set $C$. (\hhref{Figure}{figure:inference:typeconstraintrules})
\end{itemize}

\subsection{Constraint Generation for $\lambdahr$ terms}

The constraint generation judgment for $\lambdahr$ terms,
\[
\judgeC \env e T C
\]
is derived from the typing judgment but, like cast insertion, is
made syntax-directed by removing the subsumption rule. The use
of subtyping is replaced by the emission of subtyping constraints
which, when satisfied, ensure that the term is well typed.

\begin{rulelist}
\item[\rel{CG-Var}:] A variable $x$ is assigned the type to which it is
  bound in the environment $\env$ with no constraints.

\item[\rel{CG-Const}:] A primitive $\constant$ is assigned a type
  by $ty$ with no constraints.
  
\item[\rel{CG-Fun}:] For a $\lambda$-abstraction $\lam x S \e$ we
  first generate constraint sets $C_1$ and $C_2$ from its type
  annotation $S$ and body $\e$, respectively, and assign it the usual
  type under the union of their constraints, $C_1 \cup C_2$.

\item[\rel{CG-App}:] Most of the interest lies in this rule for
  function applications $\f~\e$. Given that $f$ has type $T$, it may
  be that $T$ is a type variable, so we cannot extract its domain
  type. Instead we constrain the lower bound of its domain to be $S$,
  the type of $\e$, and its return type to be a fresh unknown type
  variable $\tyvar$. Thus the return type of the application include a
  delayed substitution ${\subst x \e S} \dapp \tyvar$. Since any
  solution for $\tyvar$ must be well-formed in the environment
  $\env,x:S$, this type-wellformedness constraint is emitted as well.
\end{rulelist}

\subsection{Constraint Generation for $\lambdahr$ types}

The constraint generation judgment
\[
\judgeTC \env T C
\]
emits in $C$ any constraints which must be satisfied in order for $T$
to be a well formed type in environment $\env$.

\begin{displayfigure}{Constraint Generation Rules for $\lambdahr$ types}
  \label{figure:inference:typeconstraintrules}
  \typeconstraintgenrules
\end{displayfigure}

\begin{rulelist}
\item[\rel{WTC-Arrow} and \rel{WTC-Base}:] Well-formedness of types
  is subject to those constraints that come from checking the
  terms which they contain.

\item[\rel{WTC-Var}:] Type variables are well-formed subject to
  a new type well-formedness constraint defining the variables that
  may appear free in any solution. Furthermore, only
  type variables without a delayed substitution are allowed
  in the source program. This constraint could probably be relaxed
  but this reflects the reality that delayed substitutions are
  a formality not present in the statement of the type reconstruction
  problem itself.
\end{rulelist}

The constraint set $C$ establishes a firm interface between this phase
and the remaining phases; remaining phases need not inspect the
original program at all, but merely satisfy the constraints: When
applied to a typeable \lambdahr{} program, the constraint generation
rules emit a satisfiable constraint set. Conversely, if the constraint
set derived from a program is satisfiable, then that program is
typeable.

\begin{lem}[Constraint Generation is Sound and Complete]
  \label{lemma:constraint-gen}\

  \begin{enumerate}
    \item
      For any environment $\env$ and term $\e$:
      \[
      \begin{array}{r@{\qquad}c@{\qquad}l}
        \exists \tysol, T.~ \judge {\tysol \env} {\tysol \e} {\tysol T}
        & \Longleftrightarrow &
        \exists \tysol', S, C. \left\{
        \begin{array}{l}
          \judgeC \env \e S C \\
          \tysol' $ satisfies $ C \\
        \end{array}
        \right.
      \end{array}
      \]

    \item
      For any environment $\env$ and type $T$,
      \[
      \begin{array}{r@{\qquad}c@{\qquad}l}
        \exists \tysol.~ \judgeT {\tysol \env} {\tysol T}
        & \Longleftrightarrow &
        \exists \tysol', C. \left\{
        \begin{array}{l}
          \judgeTC \env T C \\
          \tysol' $ satisfies $ C \\
        \end{array}
        \right.
      \end{array}
      \]
  \end{enumerate}
\end{lem}

\begin{lamproof}

\pf\ We examine the details of the proof for terms only.

\begin{proofcases}
\item[($\Leftarrow$)] Soundness of constraint generation is
  simple: let $\tysol = \tysol'$ and let $T = S$.

\item[($\Rightarrow$)] To demonstrate completeness of constraint
  generation -- that every typeable term generates a satisfiable
  constraint set -- we will prove a stronger statement in order to
  strengthen our induction hypothesis to deal with new type variables
  arising during constraint generation:
  \[
  \begin{array}{r@{\qquad}c@{\qquad}l}
    \exists \tysol, T.~ \judge {\tysol \env} {\tysol \e} {\tysol T}
    & \Longrightarrow &
    \exists \tysol', S, C. \left\{
    \begin{array}{l}
      \judgeC \env \e S C \\
      \tysol\tysol' \mbox{ satisfies } C \\
      \judgesub{\tysol \env}{\tysol\tysol' S}{\tysol T} \\
      \tysol' \mbox{ and } \tysol \mbox{ have disjoint domains}
    \end{array}
    \right.
  \end{array}
  \]

  Proceeding by (mutual) induction, each case requires calculations with
  substitutions to produce the required $\tysol'$. All the interest
  lies in a single case:
  
  \begin{proofcases}
  \item[\proofcase \rel{T-App}:] We have a derivation (to minimize subscripts,
    all metavariables used here are fresh -- the ``\e'' and ``$T$''
    below are not that of the theorem but new for this proof case)

    \[
    \inferrule*{
      \inferrule*{ \vdots }{ \judge{\tysol \env}{\tysol \f}{\tysol(\lamt x T U)} } \\
      \inferrule*{ \vdots }{ \judge{\tysol \env}{\tysol \e}{\tysol T} }
    }{
      \judge{\tysol \env}{ \tysol \f ~ \tysol \e }{ \substT{x}{\tysol \e}{\tysol T} \, (\tysol U) }
    }
    \]

    By induction we have corresponding subderivations that we can put together with $\rel{CG-App}$

    \[
    \inferrule*{
      \inferrule*{ \vdots }{ \judgeC \env \f {S_\f} {C_\f} } \\
      \inferrule*{ \vdots }{ \judgeC \env \e {S_\e} {C_\e} } \\
      \tyvar \mbox{ fresh }
    }{
      \judgeC \env {f ~ e} {\substT x e {S_\e} \dapp \tyvar} {C_\f \cup C_\e \cup \{\judgeT {\env,x\!:\!S_e} \tyvar,~ \judgesub \env {S_\f} {\lamt x {S_\e} \tyvar}\}}
    }
    \]

    Our strengthened hypothesis provides us with $\tysol_\e$ and
    $\tysol_\f$ with disjoint domains that combine with $\tysol$ to
    satisfy $C_\e$ and $C_\f$, respectively, along with the facts
    \begin{enumerate}
    \item
      \label{arg}
      \judgesub {\tysol \env} {\tysol\tysol_\e S_\e} {\tysol T}

    \item
      \label{fun}
      \judgesub {\tysol \env} {\tysol\tysol_\f S_\f} {\tysol (\lamt x T U)}
    \end{enumerate}

    We need to account for the fresh $\alpha$ and the natural choice is
    the corresponding type from the premise. Namely, we use this solution:
    \[
    \tysol' = [\tyvar \mapsto \tysol U] \circ \tysol_\f\tysol_\e
    \]
    
    Since the domains of all substitutions are disjoint, this also
    satisfies $C_\e$ and $C_\f$ immediately, and the remaining
    constraint simplifies to
    \[
    \judgesub {\tysol\tysol' \env} {\tysol' \tysol S_\f} {\tysol \tysol' (\lamt x {S_\e} {\tysol T})}
    \]
    and we can combine \ref{arg} and \ref{fun} along
    with the fact that subtyping is a preorder (Lemma~\refpref{lemma:SubtypingPreorder})
    to calculate that this judgment holds.

    Then to satisfy the strengthened induction hypothesis we also
    need
    \[
    \judgesub {
      \tysol \env
    }{
      \tysol' \tysol ({\substT x \e {S_\e}} \dapp\tyvar)
    }{
      \substT x {\tysol e} {\tysol T}\, \tysol U
    }
    \]
    which simplifies to
    \[
    \judgesub {
      \tysol \env
    }{
      \substT x {\tysol\tysol' \e} {\tysol\tysol' S_\e} \, {\tysol U}
    }{
      \substT x {\tysol e} {\tysol T}\, \tysol U
    }
    \]
    in which both sides are syntactically identical since
    $dom(\tysol')$ contains no type variables appearing in $\e$ and
    the type annotation on the substitution is irrelevant.~\qed

  \end{proofcases}
  
\end{proofcases}
\end{lamproof}

Consider the following \lambdahr{} term $\e$ (the expression
\( \lethead{x}{T}{\d}~\e \) is syntactic sugar for \( (\lam{x}{T}{\e})~\d \)).

\[
\begin{array}{l}
  \lethead{id}{(\lamt{x}{\tyvar_1}{\tyvar_2})}{\lam{x}{\tyvar_3}{x}} \\
  \lethead{w}{\predty{n}{\Int}{n = 0}}{0} \\
  \lethead{y}{\predty{n}{\Int}{n > w}}{3} \\
  id~(id~ y)
\end{array}
\]

Eliding some generated type variables and type well-formedness
constraints for clarity, the corresponding constraint generation
judgement is
\[ 
\judgeC \emptyenv \e {{\substT x {(id~y)} {\tyvar_1}} \dapp \tyvar_2} C
\]
where $C$ contains the following constraints, in which $T_{id} \equiv
(\lamt{x}{\tyvar_1}{\tyvar_2}$) and the declaration of some variables
($w$ and $y$ in this case) is left implicit at their leftmost
occurence.
\[
\begin{array}{r@{~\vdash~}r@{~\sword~}l}
  \emptyenv
  & \lamt x {\tyvar_3} {\tyvar_3} 
  & \lamt x {\tyvar_1} {\tyvar_2}
  \\
  id:T_{id}
  & \predty n \Int {n = 0}
  & \predty n \Int {n = 0}
  \\
  id:T_{id},~
  w=0
  & \predty n \Int {n = 3}
  & \predty n \Int {n > w}
  \\
  id:T_{id},~
  w=0,
  y>w
  & \predty{n}{\Int}{n > w}
  & \tyvar_1
  \\
  id:T_{id},~
  w=0,
  y>w
  & \substT{x}{y}{\tyvar_1}\dapp\tyvar_2
  & \tyvar_1
  \\
  \emptyenv & 
  \multicolumn{2}{l}{\lamt x {\tyvar_1} {\tyvar_2}}
  \\
  \emptyenv & 
  \multicolumn{2}{l}{\tyvar_3}
\end{array}
\]

\section{Shape Reconstruction}
\label{section:inference:shapereconstruction}

The second phase of reconstruction is to infer a type's basic shape,
deferring resolution of refinement predicates. To defer reconstruction
of refinements, we introduce placeholders $\ph \in Placeholder$ to
represent unknown refinement predicates (in the same way that type
variables represent unknown types). The net effect of shape
reconstruction is to reduce subtyping contraints to implication
constraints and type well-formdness constraints to refinement
well-typedness constraints.
\hhref{Figure}{figure:inference:shapereconsyntax} shows the full
syntax for the resulting language after shape reconstruction. During
reconstruction, there may be terms from the union of the two
languages, but when it terminates successfully all type variables are
eliminated and all constraints are reduced.

\begin{displayfigure}{Syntax for $\lambdahr$ shape reconstruction}
  \label{figure:inference:shapereconsyntax}
  \ssp
  \shapereconsyntax\
\end{displayfigure}

\begin{itemize}
\item[-] An implication constraint $\judgeimp \env \predp \predq$
  indicates that for the program to be typeable, the corresponding
  implication must hold.
\item[-] A refinement well-typedness constraint $\judge \env \predp \Bool$
  constrains the variables that may appear free in $\predp$.
\end{itemize}

Like type variables, each
placeholder has an associated delayed substitution.
\[
\begin{array}{l}
  \begin{array}{r@{~~~}c@{~~~}l}
    \predp & ::= & \cdots  ~|~  \dsub\dapp\ph 
  \end{array}
\end{array}
\]
A placeholder replacement is a function $\rho : Placeholder
\rightarrow Term$ and is lifted compatibly to all syntactic
structures.  As with type replacements, applying placeholder
replacement allows any delayed substitutions also to be applied.
\[
\begin{array}{rcl}
  \substT{x}{\e}{T}(\dsub\dapp\ph) & = & (\substT{x}{\e}{T},\dsub)\dapp\ph \\
  \tmsol(\theta\dapp\ph) & = & \tmsol(\theta)(\tmsol(\ph)) \\
\end{array}
\]
For a placeholder replacement $\tmsol$, the definition of constraint
satisfaction is analogous to \hhref{Definition}{defn:subconsat} for
type replacements.

\begin{definition}
  \label{defn:impconsat}
  A placeholder replacement $\tmsol$ \emph{satisfies} a constraint set
  $P$ if and only if $\tmsol(P)$ contains only valid implications
  and typing judgments.
\end{definition}

\begin{displayfigure}{Shape Reconstruction Algorithm}
  \label{figure:inference:shape}\

  \ssp
  Input: $C$ \\
  Output: $\tysol, P$ \\
  Initially: $P = \emptyset$ and $\tysol = \emptysubst$ \\
  match some constraint in $C$, removing it, until quiescent: \\
  \[
  \begin{array}{@{~~~}r@{~~}lcl}
    & ~~~~~~~~~~ \\
    1.
    &
    \begin{array}[t]{ll}
      &          \judgesub{\env}{\theta \dapp \alpha}{\lamt{x}{T_1}{T_2}} \\
      $ or $   & \judgesub{\env}{\lamt{x}{T_1}{T_2}}{\theta \dapp \alpha} \\
    \end{array}
    & \Longrightarrow &
    \begin{array}[t]{l}
      $if $ occurs(\alpha, \lamt{x}{T_1}{T_2}) \\
      $then$ fail \\
      $otherwise for fresh $ \alpha_1, \alpha_2 \\
      \tysol := [\alpha \mapsto (\lamt x {\alpha_1} {\alpha_2})] \circ \tysol \\
      C := \tysol(C) \\
      P := \tysol(P)
    \end{array}
    \vspace{5pt}
    \\
    \\
    2. 
    &
    \begin{array}[t]{ll}
      & \judgesub{\env}{\theta\dapp\alpha}{\predty{x}{B}{\e}} \\
      $ or $ & \judgesub{\env}{\predty{x}{B}{\e}}{\theta\dapp\alpha}
    \end{array}
    & \Longrightarrow &
    \begin{array}[t]{l}
      $for fresh $ \ph \\
      \tysol := [\alpha \mapsto \predty{x}{B}{\ph}] \circ \tysol \\
      C := \tysol(C) \\
      P := \tysol(P)
    \end{array}
    \vspace{5pt}
    \\
    \\
    3. & 
    \judgesub{\env}{(\lamt{x}{S_1}{S_2})}{(\lamt{x}{T_1}{T_2})}
    & \Longrightarrow &
    C := C \cup 
    \left\{ 
    \begin{array}{l}
      \judgesub{\env}{T_1}{S_1} \\
      \judgesub{\env,x:T_1}{S_2}{T_2}
    \end{array}
    \right\}
    \vspace{5pt}
    \\
    \\
    4. & \judgesub{\env}{\predty{x}{B}{p}}{\predty{x}{B}{q}}
    & \Longrightarrow &
    P := P \cup \{ \judgeT{\env}{p \Implies q} \}
    \vspace{5pt}
    \\
    \\
    5. & \judgeT \env {(\lamt x S T)}
    & \Longrightarrow &
    C := C \cup 
    \left\{ 
    \begin{array}{l}
      \judgeT \env S \\
      \judgeT {\env,x:S} T 
    \end{array} 
    \right\}
    \vspace{5pt}
    \\
    \\
    6. & 
    \judgeT \env {\predty x B \predp}
    & \Longrightarrow &
    P := P \cup \{ \judge {\env,x:B} \predp \Bool \}
    \vspace{5pt}
    \\
    \\
    7. 
    &
    \begin{array}{ll}
      & \judgesub \env {\predty x B p} {\lamt x S T} \\
      $ or $ 
      & \judgesub  \env {\lamt x S T} {\predty x B p} \\
      $ or $ 
      & \judgesub  \env {\predty x B \predp} {\predty x {B'} \predq} \\
      & (B \neq B')
    \end{array}
    & \Longrightarrow &
    \mbox{\emph{fail}}
  \end{array}
  \]
\end{displayfigure}
The shape reconstruction algorithm, detailed
in \hhref{Figure}{figure:inference:shape} takes as input a subtyping
constraint set $\subconset$ and processes the constraints in
$\subconset$ nondeterministically according to the rules
in \hhref{Figure}{figure:inference:shape}.  When the conditions on the
left-hand side of a rule are satisfied, the updates described on the
right-hand side are performed.  The set $\impconset$ of implication
constraints, each of the form $\judgeT{\env}{p \Implies q}$, and the
type replacement $\tysol$ are outputs of the algorithm.  Each rule in
\hhref{Figure}{figure:inference:shape} resembles a step of traditional
type reconstruction; this algorithm is presented as formal evidence
that this step of reconstruction can, in fact, be separated from any
reasoning about refinement predicates.

\begin{enumerate}
  \item When a type variable $\tyvar$ must have the shape of a
    function type, it is replaced by \lamt{x}{\tyvar_1}{\tyvar_2},
    where $\tyvar_1$ and $\tyvar_2$ are fresh type variables.  The
    standard function $occurs$ checks that $\tyvar$ has a finite
    solution, since \lambdahr{} does not have recursive types.
    Occurences of $\alpha$ which appear in refinement predicates or in
    the range of a delayed substitution are ignored -- these
    occurences do not require a solution involving recursive types.
    \begin{eqnarray*}
      occurs(\tyvar, \predty{x}{B}{\e}) & \defeq & false \\
      occurs(\tyvar, \dsub\dapp\tyvar') & \defeq & false ~~~ (\tyvar \neq \tyvar')\\
      occurs(\tyvar, \dsub\dapp\tyvar) & \defeq & true \\
      occurs(\tyvar, \lamt{x}{S}{T}) & \defeq & occurs(\tyvar, S) \vee occurs(\tyvar, T)
    \end{eqnarray*}

\item When a type variable must be a refinement of a base type $B$,
  the type variable is replaced by $\predty{x}{B}{\ph}$ where $\ph$ is
  a fresh placeholder.  

\item A subtyping constraint between two function types induces
  additional constraints between the domains and codomains of the
  function types.

\item A subtyping constraint between two refined base types is
  simplified to an implication constraint between their refinements.

\item A type well-formedness constraint on a function type
  is decomposed into type well-formedness constraints on both
  of its components.

\item A type well-formedness constraint on a refined base type
  is replaced by the equivalent refinement well-typedness constraint.

\item Whenever two types have incompatible shapes, the algorithm
  fails immediately.

\end{enumerate}

The algorithm terminates once no more progress can be made.  At this
stage, any type variables remaining in $\tysol(\subconset)$ are not
constrained to be subtypes of any concrete type but may be subtypes of
each other.  We set these type variables equal to an arbitrary
concrete type to eliminate them (the resulting subtyping judgements
are trivial by reflexivity).

For a set of subtyping constraints $C$, one of the following occurs:
\begin{enumerate}
\item Shape reconstruction fails, in which case $C$ is unsatisfiable, or
\item Shape reconstruction succeeds, yielding $\tysol$ and $P$.
  Then $P$ is satisfiable if and only if $C$ is satisfiable.
  Furthermore, if $\tmsol$ satisfies $P$ then $\tmsol\circ\tysol$
  satisfies $\subconset$.
\end{enumerate}

More formally,

\begin{lem}[Shape Reconstruction is Sound and Complete]
  \label{lemma:shape-correctness}\

  For any set of subtyping constraints $C$,
  \[
  \begin{array}{r@{~~}c@{~~}l}
    \exists \tysol.~ \tysol \mbox { satisfies } C
    & \Longleftrightarrow &
    \exists \tysol', P, \tmsol.~
    \left\{
    \begin{array}{l}
      \tysol', P \mbox{ result from shape reconstruction } \\
      \tmsol \mbox{ satisfies } P
    \end{array}
    \right.
  \end{array}
  \]

\end{lem}

\begin{lamproof}
  \pf\
  Each step of the algorithm maintains the invariant
  that $C$ and $P$ are equisatisfiable with their
  prior state, so consider only the case where shape
  reconstruction fails. The algorithm can only fail when
  two types are utterly incompatible, and by the invariant
  this constraint originates from a constraint in the
  original set of subtyping constraints.~\qed
\end{lamproof}

Returning to our example, shape reconstruction returns the type
replacement
\[
\tysol = [~\tyvar_1 \mapsto \predty{n}{\Int}{\ph_1},~
  \tyvar_2 \mapsto \predty{n}{\Int}{\ph_2},~
  \tyvar_3 \mapsto \predty{n}{\Int}{\ph_3}~]
\]
and the following implication constraint set $P$, where now
$T_{id} = \lamt{x}{\predty{n}{\Int}{\ph_2}}{\predty{n}{\Int}{\ph_3}}$,
noting that the type variables have been elaborated into refinement
types containing placeholders.
\[
\begin{array}{r@{~\vdash~}r@{~}c@{~}l}
  n:\Int & \ph_1 
  & \Implies
  & \ph_3 
  \\
  x:\predty n \Int {\ph_1},~
  n:\Int 
  & \ph_3 
  & \Implies
  & \ph_2
  \\
  id: T_{id},~
  n:\Int
  & (n = 0) 
  & \Implies
  & (n = 0)
  \\
  id: T_{id},~
  w=0,~
  n:\Int
  & (n = 3) 
  & \Implies
& (n > w)
  \\
  id: T_{id},
  w=0,~
  y>w,~
  n:\Int
  & (n > w) 
  & \Implies
  & \ph_1
  \\
  id: T_{id},
  w=0,~ 
  y>w,~
  n:\Int
  &
  \multicolumn{3}{c}{\substT{x}{y}{\predty{n}{\Int}{\ph_1}}\dapp\ph_2
  \Implies \ph_1 }
  \\
  n:\Int & \ph_1 & : & \Bool
  \\
  n:\Int, x:\predty n \Int {\ph_1} & \ph_2 & : & \Bool
  \\
  n:\Int & \ph_3 & : & \Bool
\end{array}
\]

\section{Strongest Refinements}
\label{section:inference:satisfiability}

The final phase of type reconstruction solves the residual implication
constraint set $P$ by finding a placeholder replacement that preserves
satisfiability.  

Our approach is based on the intuition -- temporarily ignoring the
complication of dependent types -- that implications are essentially
dataflow paths that carry the specifications of data sources
(constants and function postconditions) to the requirements of data
sinks (function preconditions), with placeholders functioning as
intermediate nodes in the dataflow graph. Thus, if a placeholder \ph{}
appears on the right-hand side of two implication constraints
\judgeImp{\env}{p}{\ph} and \judgeImp{\env}{q}{\ph}, then our
replacement for \ph{} is simply the disjunction $p\vee q$ (the
strongest consequence) of these two lower bounds.  Our algorithm
repeatedly applies this transformation until no placeholders remain,
but several difficulties arise that are not present in the simpler
first-order setting of SP:

\begin{enumerate}
\item $p$ or $q$ may contain variables that cannot
  appear in a solution for \ph{}
\item \ph{} may have a delayed substitution
\item \ph{} may appear in $p$ or $q$
\end{enumerate}
To help resolve these issues, we extend the language of terms with
parallel boolean connectives and existential quantification and will
utilize the fixed point operator assumed to be included in the
primitives of the language.
\[
\begin{array}[t]{l@{~~~}c@{~~~}ccccccc}
  \e & ::= & \cdots 
  & ~|~ & \e \vee \e
  & ~|~ & \e \wedge \e
  & ~|~ & \existstm{x}{T}{\e}
\end{array}
\]

\hhref{Figure}{figure:inference:finalsyntax} summarizes the syntax of
the final language in which the output of type reconstruction lies,
and \hhref{Figure}{figure:inference:more-eval} presents the
operational semantics of the parallel boolean connectives.

\begin{displayfigure}{Syntax for $\lambdahr$ with strongest refinements}
\label{figure:inference:finalsyntax}
\ssp
\finalinferencesyntax\
\end{displayfigure}

\begin{itemize}
\item[-]
  The parallel disjunction $\e_1 \vee \e_2$ (respectively conjunction
  $\e_1 \wedge \e_2$) evaluates $\e_1$ and $\e_2$ nondeterministically,
  reducing to \true{} (resp. \false{}) if either of them reduces to
  \true{} (resp. \false{}). 
\item[-] The existential term $\existstm{x}{T}{\e}$
  binds $x$ in $\e$, and evaluates by nondeterministically replacing $x$
  with a closed term of type $T$.  The evaluation rules are summarized
  in \hhref{Figure}{figure:inference:more-eval}.  Note that neither
  parallel boolean connective (nor, obviously, existential
  quantification) can be encoded via $\beta$ or $\delta$
  reduction.\footnote{In synthetic
    topology~\citep{escardo_synthetic_2004}, where open sets correspond
    to semidecidable predicates very much like executable refinement
    types, the ``parallel or'' connective is required in order for the
    usual defining property of a topology to apply. Synthetic topology
    uses a single-valued boolean, denoting falsity by
    nontermination. Because we denote falsity by termination at $\false$
    we also require ``parallel and''. Our inputs are essentially open
    sets, though our syntactic approach obscures this to the point where
    formally establishing the connection must be left for future
    research, and our output must also be an open set, which we
    construct by union and intersection.  Further, any \emph{overt} data
    type -- dual to \emph{compact} -- is equipped with a semidecidable
    existential quantifier, which would become relevant if existential
    quantification were ever to be supported in runtime checks.}
\end{itemize}

\begin{displayfigure}{Additional Redex Reduction for Parallel Logical Connectives}
\label{figure:inference:more-eval}

\begin{tabular}{c@{~~~~~~~~~~}c}
  \(
  \begin{array}{r@{~~}c@{~~}l@{~~}r}
    \true \vee \e & \red & \true & \rel{E-Or-L} \\\\
    \e \vee \true & \red & \true & \rel{E-Or-R} \\\\
    \false \vee \false & \red & \false & \rel{E-Or-F} \\
  \end{array}
  \)
  &
  \(
  \begin{array}{r@{~~}c@{~~}l@{~~}r}
    \false \wedge \e & \red & \false & \rel{E-And-L} \\\\
    \e \wedge \false & \red & \false & \rel{E-And-R} \\\\
    \true \wedge \true & \red & \true & \rel{E-And-T} \\
  \end{array}
  \)
  \\
  \\
  \multicolumn{2}{c}{
    \(
    \existstm{x}{T}{\predp} ~\red~ \substT{x}{\e}{T}\, \predp
    ~~~~~ \mbox{if \judge{\emptyenv}{\e}{T}}
    ~~~~~\rel{E-Exists}
    \)
  }
  \\
  \\
  \multicolumn{2}{c}{
    \(
    \tmctx ~::=~ \cdots ~|~  \e \vee \tmctx  ~|~  \tmctx \vee \e
       ~|~  \tmctx \wedge \e  ~|~  \e \wedge \tmctx
       ~|~ \existstm x \tyctx \predp ~|~ \existstm x T \tmctx
    \)
  }
  \\
  \\
\end{tabular}
\end{displayfigure}


\subsection{Free Variable Elimination}

The immediate lower bounds for a placeholder may contain free
variables that may not appear free in any solution. In our example
program, the type variable $\tyvar_1$ appeared in the empty
environment and $\tysol(\tyvar_1) = \predty{n}{\Int}{\ph_1}$, yielding
a refinement well-typedness constraint $\judge {n\!:\!\Int} {\ph_1}
\Bool$.  The only variable that can appear in a solution for $\ph_1$
is therefore $n$.  But consider the following lower bound for $\ph_1$:
\[
\judgeImp{
  id:T_{id},~ w = 0,~ y>w,~ n:\Int 
}{
  (n > w)
}{
  \ph_1
}
\]
Since $id$, $w$, and $y$ cannot appear in a solution for $\ph_1$,
we need a way to identify and remove them from the lower bound without
changing its logical interpretation.

In general, each placeholder \ph{} introduced by shape reconstruction
has an associated environment $\envfor{\ph}$ in which it must have
type \Bool, recorded in the refinement well-typedness constraints.
This gives us a reasonable definition for the free variables of a
placeholder (with its associated delayed subtitution):
\[
fv(\theta\dapp\ph) = 
( dom(\envfor{\ph}) \setminus dom(\theta) )
\cup fv(rng(\theta))
\]
We then rewrite each implication constraint \judgeImp{\env,y:T}{p}{q}
where $y \not\in fv(q)$ into the constraint
\judgeImp{\env}{(\existstm{y}{T}{p})}{q}. This corresponds to the
following tautology\footnote{A ``De Morgan law'' for quantification.} $(\forall
x.\, p \Rightarrow q) \Leftrightarrow ((\exists x. p) \Rightarrow q)$
when $x \not\in fv(q)$; the executable existential is carefully
defined so that the tautology holds.

Thus we rewrite the constraint 
\[
\judgeImp{
  id:T_{id},~ w = 0,~ y>w,~ n:\Int 
}{
  (n > w)
}{
  \ph_1
}
\]
to the new constraint 
\[
\judgeImp{
  n:\Int
}{
  (\existstm{id}{T_{id}}{
    \existstm{w}{w=0}{
      \existstm{y}{y>w}{n > w}
    }
  })
}{
  \ph_1
}
\]

This transformation is semantics-preserving:

\begin{lem}[Correctness of Free Variable Elimination]
  \label{lemma:existential}
  For $y \notin fv(q)$, $\judgeimp {\env,y:T} p q$ if and only if
  $\judgeimp \env {(\existstm y T p)} q$
\end{lem}

\begin{lamproof}
\pf\ Consider each direction of entailment separately.

\begin{proofcases}
\item[$(\Rightarrow)$] Suppose \judgeimp{\env,y:T}{p}{q}
  and \( y \not\in fv(q) \) and consider any $\csub$ such that $\judge
  {} \csub \env$ and $\csub(\existstm{y}{T}{p}) \lred^* \true$.  Then
  there is some $\e$ such that \judge \emptyenv \e T and
  $\csub(\existstm{y}{T}{p}) \lred \csub(\substT{x}{\e}{T} p) \lred^*
  \true$.

  Let $\csub' = \csub \circ \subst y \e {}$ so that $\judge {} {\csub'}
  {\env,y:T}$.  Then by assumption, $\csub(p) \lred^* \true$ implies
  $\csub(q) \lred^* \true$.  Since $y \notin fv(q)$, $\csub'(q) =
  \csub(q)$, and we have proved $\judgeimp \env {\existstm y T p} q$.
  
\item[$(\Leftarrow)$] Conversely, suppose $\judgeimp \env
  {\existstm y T p} q$ where $y \not\in fv(q)$ and consider $\csub$
  such that $\judge {} \csub {\env,y:T}$ and $\csub(p) \lred^* \true$.

  Since $y \not\in dom(\env)$ we can ignore it, so $\judge {} \csub
  \env$, and the nondeterministic existential can replace $y$ with
  $\csub(y)$, so $\csub(\existstm y T p) \lred^* \csub(p) \lred^*
  \true$.  By assumption, this implies that $\csub(q) \lred^* \true$,
  and we have proved $\judgeimp {\env,y:T} p q$.~\qed
\end{proofcases}
\end{lamproof}

Repeatedly applying this transformation, we rewrite each implication
constraint until the domain of the environment (and hence the free
variables of the left-hand side) is a subset of the free variables of
the right-hand side.

\subsection{Delayed Substitution Elimination}

At this stage, all constraints are of the form
\[
\judgeImp{\env}{p}{\dsub\dapp\ph}
\]
where $dom(\env) \subseteq fv(\dsub\dapp\ph)$ The next issue is the
presence of delayed substitutions in constraints. To eliminate the
delayed substitution $\dsub$ we add the variables in its domain to the
environment and encode its semantic content as an additional
predicate. Formally,
\[
\begin{array}{c@{\qquad}c}
  \begin{array}[t]{rcl}
    env(\emptysubst) & = & \emptyenv \\
    env(\substT{x}{\e}{T}, \dsub) & = & x:T, env(\dsub)
  \end{array}
  &
  \begin{array}[t]{rcl}
    \dsubpred{~\emptysubst~} & = & \true \\
    \dsubpred{~\substT{x}{\e}{T}, \dsub~} & = & (x = \e) \wedge \dsubpred{\dsub}
  \end{array}
\end{array}
\]
The environment $env(\dsub)$ binds all the variables in
$dom(\dsub)$ while the term $\dsubpred{\dsub}$ represents
the semantic content of $\dsub$.

We then transform the constraint
\[
\judgeImp{\env}{p}{\dsub\dapp\ph} 
\]
into the equivalent constraint
\[
\judgeImp{\env, env(\dsub)}{\dsubpred{\dsub}\wedge p}{\ph}.
\]
But we can rewrite the constraint even more cleanly: $\env$ must be
some prefix of $\envfor{\ph}$ since by the previous transformation
$dom(\env) \subseteq fv(\dsub\dapp\ph) \subseteq dom(\envfor{\ph})$.
Any $x\in dom(\dsub)$ such that $x \not\in dom(\envfor{\ph})$ can be
dropped from $\dsub$ and we see that $\env, env(\dsub)$ is then
exactly $\envfor{\ph}$.  So our constraint is
\[ \judgeImp{\envfor{\ph}}{\dsubpred{\dsub} \wedge p}{\ph} \]

To prove this transformation correct, we use the following
well formedness judgement \judgewfsub{\env}{\dsub} which distinguishes
those delayed substitutions that may actually occur in context $\env$.
\begin{mathpar}
  \trule{WF-Empty}{
    \\
  }{
    \judgewfsub{\env}{\emptysubst}
  }

  \trule{WF-Ext}{
    \judge \env \e T 
    \and
    \judgewfsub {\env,x:T} {\dsub'}
  }{
    \judgewfsub \env {{\substT x \e T} \circ \dsub'}
  }
\end{mathpar}

\begin{lem}[Correctness of Substitution Elimination]
  \label{lemma:subst-elimination}
  Suppose $\tmsol$ is a placeholder replacement such that
  $\judgewfsub{\tmsol(\env)}{\tmsol(\dsub)}$. Then $\tmsol$ satisfies
  $\judgeimp{\env}{p}{\dsub\dapp\ph}$ if and only if $\tmsol$ satisfies
  $\judgeimp{\env,env(\dsub)}{\dsubpred{\dsub}\wedge p}{\ph}$.
\end{lem}

\noindent\pf\ Assume throughout that
\judgewfsub{\tmsol \env}{\tmsol\dsub}

\begin{proofcases}

\item[$(\Rightarrow)$] Suppose
  $\tmsol$ satisfies \judgeimp{\env}{p}{\dsub\dapp\ph}, i.e.
  \judgeimp{\tmsol \env}{\tmsol p}{\tmsol\dsub(\tmsol\ph)}
  and
  consider any $\csub$ s.t. $\judge {} \csub {\tmsol \env, env(\tmsol\dsub)}$
  and $\csub(\tmsol(\dsubpred{\dsub}\wedge p)) \lred^* \true$.
  Then by definition of the conjunction we know that
  $\csub(\tmsol p) \lred^* \true$, hence
  $\csub(\tmsol (\dsub\dapp\ph)) \lred^* \true$ by assumption,
  which is $\csub(\tmsol\dsub(\tmsol\ph)) \lred^* \true$.
  
  Thus since $\csub$ is a closing substitution for $\tmsol(\env,env(\dsub))$,
  we know for any $x \in dom(\dsub)$ that $\csub(x) = (\tmsol\dsub)(x)$
  hence $\csub(\tmsol\ph) = \csub(\tmsol\dsub(\tmsol\ph))$, which
  we already know evaluates to \true.  Hence $\tmsol$ satisfies
  \judgeimp{\env, env(\dsub)}{\dsubpred{\dsub}\wedge p}{\ph}.
  
\item[$(\Leftarrow)$] 
  Suppose $\tmsol$ satisfies
  \judgeimp{\env, env(\dsub)}{\dsubpred{\dsub}\wedge p}{\ph} i.e.
  \judgeimp{\tmsol \env, env(\tmsol\dsub)}{
    \tmsol(\dsubpred{\dsub} \wedge p)
  }{
    \tmsol\ph
  }
  and consider
  any $\csub$ s.t. $\judge {} \csub {\tmsol \env}$
  and $\csub(\tmsol p) \lred^* \true$.
  Let $\csub' = \csub \circ (\tmsol\dsub)$; note that
  $\judge {} {\csub'} {\tmsol \env, env(\tmsol\dsub)}$.
  
  Obviously, $\csub'(\tmsol p) \lred^* \true$ since $dom(\dsub) \cap
  fv(p) = \emptyset$.  Furthermore, by intensional equality, we
  see that $\csub'(\dsubpred \dsub) \lred^* \true$.  So 
  the conjunction in our assumption evaluates to \true,
  and we infer that $\csub'(\tmsol\ph) \lred^* \true$.
  
  And $\csub'(\tmsol\ph) = \csub(\tmsol\dsub(\tmsol\ph))$ which is
  exactly what we need to conclude that
  $\tmsol$ satisfies
  \judgeimp{\env}{p}{\dsub\dapp\ph}~\qed
\end{proofcases}

\subsection{Placeholder Solution}

After the previous transformations, all lower bounds of a placeholder
$\ph$ appear in constraints of the form
\[ \judgeImp{\envfor{\ph}}{p_i}{\ph} \]
for $i \in \{1..n\}$, assuming $\ph$ has $n$ lower bounds.  We want to
set $\ph$ equal to the parallel disjunction $p_1 \vee p_2 \vee \cdots
\vee p_n$ of all its lower bounds (the disjunction must be parallel
because some subterms may be nonterminating).  However, $\ph$ may
appear in some $p_i$ due to recursion or self-composition of a
function.  In this case we use a least fixed point operator,
conveniently already available in our language, to find a solution to
the equation $\ph = p_1 \vee \cdots \vee p_n$.

More formally, suppose $\envfor{\ph} = x_1:T_1, \cdots, x_k:T_k$.
Then $\ph$ is a predicate over $x_1 \cdots x_k$ and we can interpret
it as a function $\funcfor{\gamma} : T_1 \rightarrow \cdots
\rightarrow T_k \rightarrow \Bool$ and give a recursive definition for
the function. Using the shorthands of
\hhref{Figure}{figure:inference:shorthands} the function
$\funcfor{\ph}$ can be defined as the following least fixed point
computation:
\begin{displayfigure}{Shorthands for $\lambdahr$}
\label{figure:inference:shorthands}
\vspace{-35pt}
\begin{eqnarray*}
  \listof{T} \rightarrow \Bool & \defeq & 
  T_1 \rightarrow T_2 \rightarrow \cdots \rightarrow T_k \rightarrow \Bool \\
  \lammulti x T \e & \defeq &
  \lam{x_1}{T_1}{\lam{x_2}{T_2}{\cdots\lam{x_k}{T_k}{\e}}}
  \\
  f~\listof{x} & \defeq & f~x_1~x_2~\cdots~x_k 
\end{eqnarray*}
\vspace{-40pt}
\end{displayfigure}
\[
\funcfor{\ph} ~=~ 
\fix{\listof{T} \rightarrow \Bool} ~ (
\lam{f}{\listof{T} \rightarrow \Bool}{
  \lammulti x T {
    [\ph \mapsto f~\listof x](p_1 \vee \cdots \vee p_n)
  }
})
\]

Our solution for \ph{} is \( SR(\ph) = \funcfor{\ph}~\listof{x} \).
This is the \emph{strongest refinement} (SR) that is implied by all
lower bounds of $\ph$. This fixed-point calculation will terminate
immediately in the event that $\ph$ does not appear in any of its
lower bounds. When the program is recursive, the presence of a
fixed-point calculation is expected. Applying existing and new
techniques to this carefully isolated subproblem remains an open and
fruitful area for future research, but has already borne fruit:
Subsequent to this work, \citet{rondon_liquid_2008} applied predicate
abstraction to calculate enough of these fixed points that the program
annotations to ensure fully static type-checking were reduced from
31\% of the text to less than 1\%.
\begin{lem}[SR is the Strongest Refinement]
  \label{lemma:strongest-consequence}\

  If $\tmsol$ satisfies $P$, then $\tmsol$
  satisfies \( \judgeImp{\envfor{\ph}}{SR(\ph)}{\ph} \).
\end{lem}

\begin{lamproof}
  \pf\ Consider any $\tmsol$ satisfying $P$ and $\sigma$ such that
  $\judge {} \csub {\tmsol {\envfor \ph}}$ and $\csub\tmsol(SR(\ph)) \lred^*
  \true$ minimal $k$.  Since the possible nondeterministic
  evaluations of $\csub\tmsol SR(\ph)$ correspond to
  simultaneously evaluating all lower bounds for $\ph$ simultaneously,
  one of them must eventually evaluate to $\true$.  Then
  by the assumption that $\tmsol$ satisfies $P$, we know
  that $\csub\tmsol\ph \lred^* \true$.~\qed
\end{lamproof}

The result of equisatisfiability follows from the fact that we
have chosen the strongest possible solution for \ph{}.

\begin{lem}[Correctness of Strongest Refinement Calculation]
  \label{lemma:satisfiability}\

  $P$ is satisfiable if and only if $P[\ph := SR(\ph)]$ is satisfiable.
\end{lem}

\begin{lamproof}
  \pf\
  
  \begin{proofcases}
  \item[$(\Leftarrow)$] For any $\tmsol : PlaceHolders
    \rightarrow Terms$ that satisfies $P[\ph := SR(\ph)]$,
    we have $\tmsol \circ [\ph := SR(\ph)]$ that
    satisfies $P$.

  \item[$(\Rightarrow)$] Consider any $\tmsol :
    PlaceHolders \rightarrow Terms$ that satisfies $P$.
    By \hhref{Lemma}{lemma:strongest-consequence} if
    $\tmsol(\ph) \Implies p$ occurs in $P$, then
    $SR(\ph) \Implies \tmsol(\ph) \Implies p$; covariant occurences of
    $\ph$ in environments are analogous.  If $p \Implies \tmsol(\ph)$
    occurs in $P$, then $p \Implies SR(\ph)$ by construction of
    $SR(\ph)$; contravariant occurences of types in environments do
    not affect satisfiability.~\qed \end{proofcases}
\end{lamproof}

    In our example, the only lower bound of $\ph_3$ is $\ph_1$ and the
    only lower bound of $\ph_2$ is $\ph_3$, so let us set $\ph_3 := \ph_1$
    and $\ph_2 := \ph_3$ in order to discuss the more interesting solution
    for $\ph_1$.  The resulting unsatisfied constraints (simplified for
    clarity) are:
\[
\begin{array}{r@{~\vdash~}r@{~\Implies~}l}
  n:\Int
  & \existstm{w}{\predty{n}{\Int}{n = 0}}{ (n > w)}
  & \ph_1
  \\
  n:\Int
  &
  \multicolumn{2}{c}{
    \existstm{w}{w = 0}{
      \existstm{y}{y > w}{
        \subst x y {}\dapp\ph_1
      }
    }
    \Implies 
    \ph_1 
  }
\end{array}
\]
The exact text of $SR(\ph_1)$ is too large to print here, but it is
equivalent to \\ \( \existstm w {w= 0}{n > w} \)
and thus equivalent to $(n > 0)$.  The resulting explicitly typed
program (simplified accordingly) is:
\[
\begin{array}{l}
  \lethead{id}{(
    \lamt{x}{
      \predty{n}{\Int}{n>0}
    }{
      \predty{n}{\Int}{n>0}
    })
  }{
    \lam{x}{\predty{n}{\Int}{n>0}}{x}
  } \\
  \lethead{w}{\predty{n}{\Int}{n = 0}}{0} \\
  \lethead{y}{\predty{n}{\Int}{n > w}}{3} \\
  id~(id~ y)
\end{array}
\]

\section{Type Reconstruction is Typability-Preserving}
\label{section:inference:correctness}

The output of our algorithm is the composition of the type replacement
returned by shape reconstruction and the placeholder replacement
returned by the satisfiability routine.  Application of this composed
replacement is a typeability-preserving transformation.
Moreover, for any typeable program, the algorithm succeeds in producing
such a replacement.

\begin{thm}
  For any $\lambdahr$ program $\e$, one of the following occurs:
  \begin{enumerate}
  \item Type reconstruction fails, in which case $\e$ is untypeable, or
  \item Type reconstruction returns a type replacement $\tysol$ such that
    $\e$ is typeable if and only if $\tysol(e)$ is well typed.
  \end{enumerate}
\end{thm}

\begin{lamproof}\ 
  \pf\

  \begin{enumerate}
  \item Only shape reconstruction can fail. If it does, then by
    \hhref{Lemma}{lemma:shape-correctness} the subtyping constraints are
    unsatisfiable.  Then by \hhref{Lemma}{lemma:constraint-gen}, $e$ is
    not typeable.
    
  \item Type reconstruction solved constraints that were faithful,
    by \hhref{Lemma}{lemma:constraint-gen}. Thus
    by \hhref{Lemma}{lemma:shape-correctness} we have $\tysol$ and by
    \hhref{Lemma}{lemma:satisfiability} we have $\tmsol$ such that
    $\tmsol\tysol\e$ is typeable (well typed) if and only if $\e$ is
    typeable.~\qed \end{enumerate}
\end{lamproof}

\section{Related Work}
\label{section:inference:relatedwork}

Constraint-based type reconstruction for systems with subtyping is a
tremendously broad topic that we cannot fully review here.  The
problem is studied in some generality
by \citet{mitchell_coercion_1984}, \citet{fuh_type_1988},
\citet{lincoln_algorithmic_1992}, \citet{aiken_type_1993}, and
\citet{hoang_lower_1995}.  Type inference systems parameterized by a
subtyping constraint system are developed by
\citet{pottier_simplifying_1996} and \citet{odersky_type_1999}.  This
work is complementary to generalized systems in that it focuses on the
solution of our particular instantiation of subtyping constraints; we
also do not investigate parametric polymorphism, which is included in
the mentioned frameworks.  Set-based analysis presents many similar
ideas, and this chapter also draws inspiration from the works of
\citet{heintze_set_1992}, \citet{cousot_formal_1995},
\citet{fahndrich_making_1996}, and \citet{flanagan_componential_1997}.

Many of the refinement type systems discussed in
\hhref{Section}{section:intro:refinementtypes} perform some manner
of local type
reconstruction~\citep{pierce_local_1998,dunfield_tridirectional_2004},
but require type annotations to guide the type checking process, and
none address the general problem of the satisfaction of a set of
subtyping constraints between refinement types.  Dependent
ML~\citet{xi_dependent_1999} solves systems of linear inequalities to
infer a restricted class of type indices.

Liquid types~\citep{rondon_liquid_2008}, discussed at length in
\hhref{Section}{section:relatedwork:liquidtypes} take the solution
presented here a step further by using abstract interpretation to
solve for closed forms of fixed points such as those produced by our
algorithm. The particular abstract interpretation used is
\emph{predicate abstraction} where a finite set of predicate templates
are instantiated with the variables of the program, so common
predicates for a given domain can be outlined and then the system
performs full type reconstruction. If a solution cannot be found by
the predicate abstraction process, then the program is rejected.
A great benefit of Liquid Types is that the predicates inferred
can be very readable as they come from a small set of templates
and do not include fixed points or existential quantification.
Our deliberate separation of the \emph{generation} of the fixed point
constraints from the \emph{solving} of them for a closed form
leaves other avenues open. For example, it is not generally
necessary to find a closed form solution except for nice error
messages: our reconstructed predicates are never used
as preconditions, but only to satisfy other preconditions, so
strategic unrolling and approximation may be sufficient to
perform (possibly hybrid) type checking of the resultant program.

\citet{rastogi_ins_2012} developes a type inference system
for the much more complex context of Actionscript and evaluate
the performance asymptotically and through an implementation
with benchmarks time trials.

\chapter{Compositional and Decidable Checking}

\label{chapter:compositional}
\label{chapter:compositionality}

Having established that executable refinement types can be effectively
reconstructed from a program with no type annotations and subsequently
enforced using hybrid type checking, we return to the question of when
-- if ever -- we can predict that the type checking question \emph{for
  a particular program} will be decidable. Note that this is distinct
from, though related to, the design of a decidable restriction of the
type system.

Unfortunately, the essential difference between dependent types
and simple structural type systems prevents a satisfactory answer.
Simple type systems such as that of $STLC$ perform \emph{compositional
  reasoning} in that the type of a term depends only on the types of
its subterms, not on their semantics.  Dependent types offer more
expressive abstractions, but violate those abstractions and base their
reasoning directly upon the semantics of terms.\footnote{This is, in
  fact, the origin of the term ``dependent type'': the types
  \emph{depend} on terms.} Pragmatically, noncompositionality makes
the decidability of static checking unpredictable.

This chapter shows how compositional reasoning may be restored using
standard type-theoretic techniques, namely a form of existential types
and subtyping.  Despite its compositional nature, the type system is
\emph{exact}, in that the type of a term can completely capture its
semantics, thus demonstrating that precision and compositionality are
compatible.  We then address predictability of static checking for
executable refinement types by giving a type-checking algorithm for an
important class of programs with refinement predicates drawn from a
decidable theory.  Our algorithm relies crucially on the fact that the
type of a term depends only the types of its subterms (which fall into
the decidable theory) and not their semantics (which will not, in
general).

\subsection*{Chapter Outline}

\noindent
This chapter is organized as follows:

\begin{itemize}
\item[-] \hhref{Section}{section:compositional:intro} clarifies the
  non-compositionality of dependent type systems and its consequences
  for executable refinement types.

\item[-] \hhref{Section}{section:compositional:lambdaex} Describes
  $\lambdaex$, a variant of $\lambdah$ augmented to support
  compositional reasoning.

\item[-] \hhref{Section}{section:compositional:opsem} provides an
  operational semantics to $\lambdaex$.

\item[-] \hhref{Section}{section:compositional:typesystem} presents the
  $\lambdaex$ type system and proves its compositionality.

\item[-] \hhref{Section}{section:compositional:exactness} proves that
  the $\lambdaex$ type system is \emph{exact}: Whenever primitives
  have suitably precise types, every term may be assigned a type that
  precisely describes its semantics.

\item[-] \hhref{Section}{section:compositional:safety} proves type
  soundness for $\lambdaex$, relying crucially on exactness.

\item[-] \hhref{Section}{section:compositional:logicalrelation}
  extends the definition of extensional equivalence for $\lambdah$ to
  $\lambdaex$

\item[-] \hhref{Section}{section:compositional:algorithmics} provides a
  type checking algorithm for the important special case of programs
  whose refinement predicates fall in a decidable theory.

\item[-] \hhref{Section}{section:compositional:bst} demonstrates the
  verification in $\lambdaex$ of the classic example of binary search
  trees.

\item[-] \hhref{Section}{section:compositional:relatedwork} discusses work with related goals and
surveys other applications of similar type-theoretic techniques.
\end{itemize}

\section{Compositional Reasoning}
\label{section:compositional:intro}

\newcommand{\lete}[4]{\t{let}~#1:#2=#3~\t{in}~#4}

During program development, compositional reasoning separates a
program into cognitively manageable pieces.  During analysis and
verification, compositional reasoning limits the amount of information
that must be stored and processed. Before we embark upon a formal
definition, consider the following illustration of how structural type
systems reason compositionally. In a well-typed program such as
\[
\lete x T e d
\]
the type $T$ provides an abstract specification of $e$ that is concise
and yet sufficiently precise to verify that $e$ and $d$ interact
properly. This verification process is compositional in that the
verification of $d$ cannot rely on any properties of $e$ other than
those exposed via its type $T$; the analysis of the superterm relies
only on the results of analysis of the subterm(s).  Consequently,
replacing $e$ with any equivalently-typed term does not affect the
typeability of the overall program. Formally, we can define
compositionality of any type system via its type rules.\footnote{This
  assumes that a type rule does not inspect the derivation of its
  antecedents, but only their content. This is, in fact,
  compositionality (proof irrelevance) of the metatheory.}

\begin{definition}[Compositional Reasoning]
  A type rule $\rel{A}$ performs \emph{compositional reasoning} if for 
  any derivation $\judge \env \e S$ used in the premise, as in
  \[
  \begin{array}{c}
    \inferrule*[right=\rel{A}]{
      \inferrule*{ \vdots }{ \judge \env e S } \\
      \mbox{possibly other premises}
    }{ 
      \judge {\env'} {\tmctx[\e]} T
    }
  \end{array}
  \]
  any derivation of $\judge \env \d S$ may be used in place,
  as in
  \[
  \begin{array}{c}
    \\
    \inferrule*[right=\rel{A}]{
      \inferrule*{ \vdots }{ \judge \env \tmd S } \\
      \mbox{possibly other premises}
    }{ 
      \judge {\env'} {\tmctx[\d]} T
    }
  \end{array}
  \]
  Such a rule is said to be compositional. A type system containing
  only compositional rules is also called compositional.
\end{definition}

Seen through the lens of the Curry-Howard-Lambek correspondence
\citep{curry_functionality_1934, curry_combinatory_1958,
  howard_formulae-as-types_1980, lambek_introduction_1986} this
corresponds to a sort of proof irrelevance: The particular proof of a
proposition cannot affect how that property is invoked in proofs of
other propositions.

\subsection{Dependent Types Perform Non-compositional Reasoning}
\label{section:compositional:dependenttypes}

Executable refinement types provide a natural means to abstract a
term's behavior to a greater or lesser degree (or, in the extreme, not
abstract its behavior at all).  For example, the constant $1$ may be
given an infinite number of types, including $\Int$, $\t{Pos}$ and the
exact type $\predty{y}{\Int}{y=1}$. These types are related via
subtyping in the expected manner:
\[
	\predty{y}{\Int}{y=1}
	~~~\sword~~~
	\t{Pos}
	~~~\sword~~~
	\Int
\]

\noindent
As demonstrated throughout this dissertation, executable refinement
types cooperate in a clean and expressive manner with dependent
function types.  For example, the function $\t{add1}$ may also be
given many types, including the simple type $\arrowt \Int \Int$, the
an exact type $\lamt x \Int {\predty y \Int {y = x + 1}}$, or the
intermediate specification $\lamt x \Int {\predty y \Int {y>x}}$
describing only that $\t{add1}$ is increasing.  Again, subtyping
appropriately relates these various specifications for \t{add1}:
\[
\begin{array}{ll}
  & \lamt{x}{\Int}{\predty{y}{\Int}{y=x+1}} \\
  \sword & \lamt{x}{\Int}{\predty{y}{\Int}{y>x}} \\
  \sword & \Int\rightarrow\Int \\
\end{array}
\]

At first, it appears that a programmer has precise control, via type
annotations, over the specifications that must be verified during
type-checking. Unfortunately, this is not the case at all!  Standard
dependent type systems, such as those of Cayenne
\citep{augustsson_cayenne_1998}, Epigram \citep{mcbride_view_2004},
Agda \citep{norell_dependently_2009} and Coq
\citep{the_coq_development_team_coq_2012}, when given a program
$\tmctx[e]$, reason about the interaction between the term $\e$ and
its context $\tmctx$ using both:
\begin{itemize}
\item[-] compositional reasoning based on the \emph{type} of $e$; and
\item[-] non-compositional reasoning based on the \emph{syntax} of
  $e$ (hence its behavior).
\end{itemize}

\noindent
The non-compositional nature of dependent type systems originates from 
the following standard type rule for dependent function
application. (For simplicity, we elide the typing environment).
\[
\begin{array}{cc}
\trulenoname{
  \plainjudge \f {(\lamt x S T)} \qquad
  \plainjudge e S
}{
  \plainjudge {f~e} {{\subst x e S}\, T}
}
\end{array}
\]
The inferred type ${\subst x e S\, T}$ (denoting $T$ with $x$ replaced
by $e$) of $f~e$ includes the term $e$ itself, and not just its type $S$.

To highlight the consequences of the non-compositional nature of this
rule, suppose that $e$ is an arbitrary term of type $\t{Pos}$, and
that $f$ is the exactly-typed identity function on integers:
\[
\begin{array}{rcl}
	e &:& \t{Pos} \\
	f &:&(\lamt x \Int {\predty y \Int {y=x}})
\end{array}
\]
Using compositional reasoning, we should only be able to infer that
$f$ returns its argument, which is of type $\t{Pos}$, and so $f~e$ has
type $\t{Pos}$.

Under the above type rule, however, the application $f~e$ has the more
precise (indeed, exact) type stating that $f~e$ returns exactly the
value of $e$.
\[
	f~e~~:~~\predty{y}{\Int}{y=e}
\]
Thus, the abstraction $\t{Pos}$ of $e$ has now been circumvented. To explicitly
see the counterexample to compositionality, consider the corresponding derivation
\[
\inferrule*{
  \plainjudge e {\t{Pos}} \\
  \plainjudge f {\plamt x \Int {\predty y \Int {y=x}}}
}{
  \plainjudge {f~e} {\predty y \Int {y=e}}
}
\]
and consider replacing the derivation of $e:\t{Pos}$ with another
for $d:\t{Pos}$, yielding the erroneous derivation
\[
\inferrule*[right=Invalid]{
  \plainjudge d {\t{Pos}} \\
  \plainjudge f {\plamt x \Int {\predty y \Int {y=x}}}
}{
  \plainjudge {f~e} {\predty y \Int {y=e}}
}
\]

Non-compositional reasoning causes multiple difficulties.  First, an
arbitrary program term $e$ has leaked into a refinement predicate.
Since $e$ could be a large term, any type-checking procedure now needs
to deal with extremely large types, probably much larger than any of
the types present in the source program. Since the types in source
programs are abstract specifications carefully chosen by the
programmer, it is not clear that these new, larger, types generated
during type checking are at all the right abstractions for performing
lightweight verification.

Second, even if the refinement predicates in source programs are
carefully chosen from a decidable theory, the fact that program terms
leak into refinement predicates during type checking means that static
type checking still requires deciding implications between arbitrary
program terms, which is of course undecidable.

\subsection{Compositional Reasoning for Executable Refinement Types}
\label{section:compositional:compositionalreasoning}

Given the universal and precise abstractions provided by dependent and
refinement types, and the benefits of compositional reasoning, this
chapter explores an alternative strategy for static checking of
executable refinement types with dependent function types. The key
idea underlying our approach is illustrated by the following rule for
function application:
\[
\inferrule*{
  f : (\lamt x T U)
  \and
  e : S
  \and
  S \sword T
}{
  f~e : \pexistsbind x S U
}
\]
The application $f~e$ must return a value of type $U$, where $x$ is
bound appropriately, but what is the appropriate binding for $x$?  

Compositionality dictates that all we are permitted to know about $e$
(and hence about $x$) is that it has type $S$.  Thus, the inferred
type of the application $f~e$ is the existential type
\[
\existsbind x S U
\]
Roughly speaking, this type denotes the union of all types of the form
$\subst x \tmd S\, U$, where $\tmd$ ranges over terms of type
$S$.\footnote{Existential types are also written as dependent pairs
  $\Sigma x\!:\!S.U$, with two equally valid informal readings. The
  first is that the type is the (disjoint) union of $\subst x \tmd {}
  \, U$ for each $\tmd$ of type $S$. The second is as the type of
  pairs of a term $\tmd$ of type $S$ along with a corresponding type
  $\subst x \tmd {}\, U$. The latter is more prevalent in constructive
  logic, as it emphasizes the reification of the existential witness.}

For the application $f~e$ considered earlier, this rule infers the
existential type 
\[
	f~e ~~:~~ (\existsty{x:\t{Pos}}{\predty{y}{\Int}{y=x}})
\]
that is (informally) equivalent to the type $\t{Pos}$, and indeed is a
subtype of $\t{Pos}$.
\[
	(\existsty{x:\t{Pos}}{\predty{y}{\Int}{y=x}})
	~~\sword~~
	\t{Pos}
\]
Thus, the type rule achieves the desired goal
of compositional reasoning: $f~e$ has type $\t{Pos}$ simply because
the argument expression has type $\t{Pos}$. All we can know about the
argument is its type.  Furthermore, note that the expression $e$
itself no longer leaks into the refinement predicate, which provides
benefits in terms of smaller inferred types and facilitates decidable
type checking.
 
\subsection{Expressiveness and Exactness}
\label{section:compositional:exactnessintro}

Given that this type system is strictly limited to compositional
reasoning, and cannot, for example, include the traditional rule for
dependent function application, a key question that arises is to what
degree the requirement for compositional reasoning limits the
expressiveness of the type system. Certainly, the refinement type
language itself is sufficiently expressive to describe exact types,
such as the type
\[
	\predty x \Int {x=1}
\]
for the constant $1$. More interestingly, we show that, if every
constant in the language is assigned an exact type, then for any
well-typed program $e$, our type system will infer an {exact} type
that exactly captures the run-time semantics of $e$.  This result
holds even if $e$ itself includes coarse type specifications such as
$\Int$. This curious result occurs because all annotations in a
program are preconditions to functions, which will be made as precise
as allowed by the type of the argument to the function. Thus it is the
types of opaque constants that drives the precision of type
checking. This is discussed at length in
\hhref{Section}{section:compositional:exactness}.

Assigning exact types to primitives is straightforward, via
\emph{selfification}.  For example, the fixed point primitive $\t{fix}_T$
is typically given the inexact type $(T\rightarrow T)\rightarrow T$.
For the case where $T = \lamt x S B$, an exact type for $\t{fix}_T$
is:
\[
\lamt f {(\arrowt T T)} {\lamt x S {\predty y B {y = \t{fix}_T~f~x}}}
\]

In practice, of course, primitives are typically assigned somewhat
coarser types.  Nevertheless, this completeness result indicates that
the precision of the type system is entirely parameterized by the
types of these primitives; \emph{the type system itself neither
  requires nor performs any additional abstraction.}

Put differently, the precision of any compositional analysis is
naturally driven by the precision of the abstractions that it
composes, and careful choice of abstractions is crucial for achieving
precise, scalable analyses.  The type system is entirely configurable
in this regard; it does not perform any abstraction itself, and
instead simply propagates the abstractions inherent in the types of
constants and recursive functions, with the result that the precision
of the analysis is largely under the control of the programmer.

\subsection{Decidable Type Checking}

A key benefit of compositional type checking is that program terms
never leak into refinement predicates, so the language of refinement
predicates can be cleanly separated from that of program terms.  In
particular, if refinement predicates in source programs are drawn from
the decidable theory of linear inequalities, then all proof
obligations generated during type checking also fall within a
decidable theory, and so type checking is decidable.  We use this
result to develop a decidable type checking algorithm for an important
and easily-identifiable subset of our target language. 


\section{The Language \lambdaex} 
\label{section:compositional:lambdaex}

We present our ideas via the language $\lambdaex$, a variant of
$\lambdah$ augmented with existential types to support
compositionality and also new case discrimination constructs (also
called ``pattern matching'') to enable more interesting and
challenging examples. \hhref{Figure}{figure:compositional:syntax}
presents the syntax of $\lambdaex$, highlighting forms that are not
present in $\lambdah$.

\begin{displayfigure}{Syntax of \lambdaex}
  \label{figure:compositional:syntax}
  \ssp
  \lambdaexcoresyntax
\end{displayfigure}

As before, primitives are presumed to include basic operations of the
language, such as boolean and arithmetic operations ($+$, $\times$,
$\t{not}$, $\Leftrightarrow$, etc.), and a fixed point operator
$\t{fix}_T$ for each type $T$. However, in $\lambdaex$ primitives are
separated into syntactic classes of primitive functions $\primfun$ and
data constructors $\constructor$. Constructors include the
boolean constants ($\true$, $\false$), integer constants ($1$, $2$,
$6$, $-17$, etc), and any constructors for algebraic data types
such as the list constructors $\t{nil}$ and $\t{cons}$.

A constructor is eliminated by using the case discrimination
expression
\[ 
\caseexp \e {\listof {\patternexp \constructor \f}}
\] 
where $\listof{\patternexp \constructor \f}$ is a sequence of cases to
whose constructor $\constructor$ to which $\e$ is compared, and the
appropriate one chosen. For simplicity, we assume all case discrimination
is disjoint and complete, introducing nonterminating alternatives as
necessary.  Thus the ordering of the cases is irrelevant and they
can be freely reordered.

The $\lambdaex$ type language includes existential types of the form
\[
\existsbind x S T
\]
which classify terms of type $T$ where $x$ represents some unknown
term of type $S$.  Since $\lambdaex$ lacks polymorphism, in addition
to $\Bool$ and $\Int$ we also assume a base type of lists of integers
\IntList\ to develop our examples; adding other monomorphic data types
is straightforward.

\section{Operational Semantics of $\lambdaex$}
\label{section:compositional:opsem}

\hhref{Figure}{figure:compositional:evalrules} presents the
operational semantics of $\lambdaex$. The notable enhancement is case
discrimination, but the complete set of redex reduction rules,
contextual evaluation rules, and evaluation contexts is presented for
ease of reference.

\begin{rulelist}
\item[\rel{E-case}:] The case discrimination expression $\caseexp
  {\constructor~\listof{\e}} {\listof {\patternexp \constructor \f}}$
  reduces by selecting the appropriate case $\patternexp \constructor
  \f$ and applying $\f$ to all the arguments of $\constructor$.  After
  all of $\listof e$, the full term $\constructor~\listof e$ is also
  passed to $\f$; this is a formality (an important one) that
  enables path sensitivity, discussed later in
  \hhref{Section}{section:compositional:typesystem}.
\end{rulelist}

\begin{displayfigure}{Redex Evaluation for \lambdaex}
  \label{figure:compositional:evalrules}
  \ssp
  \lambdaexevalrules
\end{displayfigure}

\section{Denotation of \lambdaex\ types to sets of $STLC$ terms}

The definition of implication for $\lambdaex$ is unchanged from that
of $\lambdah$ (\hhrefpref{Figure}{figure:lambdah:implication}) but the
quantification over closing substitutions depends on a denotational
interpretation of all $\lambdaex$ types as sets of terms. The
denotational semantics of \hhref{Section}{section:lambdah:denotation}
extends naturally to the existential types of $\lambdaex$, mirroring
the denotation of function types.

The extended definition of the shape of a $\lambdaex$ types is given in
\hhref{Figure}{figure:compositional:shape}. The shape of an
existential type $\existsbind x S T$ is simply the shape of the
underlying type $T$, since bound variables are irrelevant in
$STLC$. The shape of a term is still simply the same term where all
types have been replaced by their shape.

\begin{displayfigure}{Shape of $\lambdaex$ types and terms}
\label{figure:compositional:shape}
\ssp
\vspace{-20pt}
\begin{eqnarray*}
  \Base {\predty x B \predp} & \defeq & B \\
  \Base {\lamt x S T} & \defeq & \arrowt {\Base S} {\Base T}\\
  \Base {\existsbind x S T} & \defeq & \Base T\\
\end{eqnarray*}
\vspace{-35pt}
\end{displayfigure}

\hhref{Figure}{figure:compositional:denotation} shows the
interpretation of $\lambdaex$ types as sets of terms. The
interpretations of refined basic types and function types are
unchanged from $\lambdah$. An existential types $\existsbind x S T$ is
interpreted as the set of $STLC$ terms $\e$ of simple type $\Base T$
for which there exists some existential witness $\d$ in $\meaningf{S}$
such that $\e \in \meaningf{\subst x \e S\, T}$.

\begin{displayfigure}{Denotation from $\lambdaex$ types to sets STLC terms}
\label{figure:compositional:denotation}
\vspace{-40pt}
\begin{eqnarray*}
  \meaningf {\predty x B \predp}
  & \defeq & 
  \{ \e ~|~ \judgeSTLC {} {\Base \e} B
  ~\wedge~ 
  (\e \lred^* \constant \mbox{ implies } \subst x \constant B\, \predp \lred^* \true) \} 
  \\
  \meaningf {\lamt x S T} 
  & \defeq &
  \{
  \tmf ~|~ \judgeSTLC {} {\Base \tmf} {\arrowt {\Base S} {\Base T}}
  ~\wedge~ \forall \e \in \meaningf{S},~ 
  \tmf~\e \in \meaningf{\subst x e S\, T}
  \}
  \\
  \meaningf {\existsbind x S T}
  & \defeq &
  \{
  \e ~|~ \judgeSTLC {} {\Base \e}
  ~\wedge~ \exists \tmd \in \meaningf{S},~
  \e \in \meaningf{\subst x \tmd S\, T}
  \}
\end{eqnarray*}
\vspace{-40pt}
\end{displayfigure}

\section{The \lambdaex\ Type System}
\label{section:compositional:types}
\label{section:compositional:typesystem}

The judgments of the $\lambdaex$ type system are those
of $\lambdah$ (\hhrefpref{Section}{section:lambdah:typesystem})
but their definitions are considerably more involved due to the
enriched type and term languages.

\begin{itemize}
\item[-] The typing judgment $\judge \env e T$ states that term $\e$ may be assigned
  type $T$ under the assumptions in environment $\env$. (\hhref{Figure}{figure:compositional:typerules})
\item[-] The type well-formedness judgment $\judgeT \env T$ ensures that $T$ is a well-formed type under
  the environment $\env$. (\hhref{Figure}{figure:compositional:wellformedtyperules})
\item[-] The subtyping judgment $\judgesub \env S T$ retains the standard reading: Any term of type $S$ may also be
  assigned type $T$. (\hhref{Figure}{figure:compositional:subtyping})
\end{itemize}

\subsection{Typing for $\lambdaex$}

The typing judgment 
\[
\judge \env e T
\]
for $\lambdaex$ is enhanced from that of $\lambdah$ to support
existentially quantified variables occurring in types, and case
discrimination. These two constructs interact in important ways to
provide path-sensitivity, without which precise types for constructors
do not provide information during case discrimination.

\begin{displayfigure}{Type Rules for $\lambdaex$}
  \label{figure:compositional:typerules}
  \vspace{-20pt}
  \compositionaltyperules
  \vspace{-35pt}
\end{displayfigure}

\begin{rulelist}
\item[\rel{T-Const} and \rel{T-Prim}:] Constructors $c$ and primitive
  functions $f$ are assigned types according to $ty$.

\item[\rel{T-Var}:] Variables are assigned types yielded by $\self$,
  defined in \hhref{Figure}{figure:compositional:selfdefinition}, a
  key feature discussed in \hhref{Section}{section:compositional:self}.

\item[\rel{T-Fun}:] This standard rule for $\lambda$-abstractions
  is unchanged.

\item[\rel{T-App}:] Typing of function a application $(\f~\e)$ is
  interesting, since it appears not to support dependent function types
  at all. We discuss this in detail in \hhref{Section}{section:compositional:nondependent}.

\item[\rel{T-Case}:] This rule for a case discrimination expression
  $\caseexp \e {\listof {\patternexp \constructor \f}}$ is rather
  complex, because it uses only the information contained in the type
  of $\constructor$ to provide a measure of path-sensitivity, as the
  type of $\t{if}$ did in $\lambdah$. 

  Each constructor $c$ must necessarily return some refinement
  $\predty x B \predp$ of the same base type $B$ (for example,
  $\t{nil}$ and $\t{cons}$ both return refinements of $\t{IntList}$).
  The type of the discriminand $\e$ must also be some refinement of
  $B$, but may have existentially quantified variables $\listof {z :
    T}$, thus the type of $\e$ must have the form $\existsmulti z T
  {\predty y B q}$.
  
  For each branch $\patternexp \constructor \f$, where the particular
  constructor $\constructor$ has a type of the form $\lamtmulti x S
  {\predty y B p}$, if this branch is chosen, then we have ``learned''
  that $\e$ was constructed with constructor $\constructor$, so the
  predicate $\predp$ holds for $\e$. So we conjoin $\predp$ to the
  type of $\e$, yielding $\existsmulti z T {\predty y B {\predp \wedge
      \predq}}$. The handling function $\f$ must then take parameters
  $\listof {x:S}$ corresponding to the arguments to the constructor as
  well as the constructed term itself.  Though $\listof x$ may not
  occur in $U$, their occurrences in $\predp \wedge \predq$ capture
  relationships between the variables so that the body of $\f$ may
  assume anything already known about $\e$ as well as anything learned
  by the fact that the constructor used was $\constructor$. In this
  way, the precision of the path-sensitivity of \rel{T-Case} is
  determined by the types of constructors.
  
  For example, \t{cons} may be assigned a range of types,
  including: \[ \begin{array}{rcl} \t{cons} & : & \arrowt \Int {}
    \arrowt \IntList \IntList \\ \t{cons} & : & \arrowt \Int {} \lamt
    x \IntList {} \\ && \qquad \predty y \IntList {\t{length}(y) =
      \t{length}(x) + 1} \\ \t{cons} & : & \lamt n \Int {} \lamt x
    \IntList {} \\ && \qquad \predty y \IntList {y =
      \t{cons}~n~x} \end{array} \] In particular, if the constructor
  $\constructor$ has the precise return type \mbox{$\predty y B {y =
      c~\listof x}$}, then the new binding implies $\subst y
  {c~\listof x} {}\, e'$.  However, if $c$ has a more coarse type,
  then the na\"ive approach of adding an assumption such as $e =
  c~\listof x$ is reasoning noncompositionally using the exact
  semantics of $c$, ignoring its type.
\end{rulelist}

\begin{displayfigure}{Definition of $\self$}
  \label{figure:comp:selfdefinition}
  \label{figure:compositional:selfdefinition}
  \vspace{-20pt}
  \selfdefinition
  \vspace{-30pt}
\end{displayfigure}

\subsection{Type well-formedness for $\lambdaex$}

The type well-formedness judgment
\[
\judgeT \env T
\]
for $\lambdaex$ is that of $\lambdah$ with the addition of
well-formedness for existential types. The rule $\rel{WT-Exists}$
for an existential type $\existsbind x S T$ ensures that $T$ is
well formed in the extended environment $\env,x:S$.

\begin{displayfigure}{Type well-formedness for for $\lambdaex$}
  \label{figure:compositional:wellformedtyperules}
  \vspace{-30pt}
  \lambdaexwellformedtypes
  \vspace{-35pt}
\end{displayfigure}

\subsection{Subtyping for $\lambdaex$}

Subtyping between existential types
(\hhref{Figure}{figure:compositional:subtyping}) is derived from the
logical interpretation of subtyping as ``implication'' between
propositions.\footnote{A similar inspiration may underlie
  \citet{dreyer_type_2003} where subtyping rules like our own are
  briefly mentioned, but ultimately not used, as their work had
  different goals.}

\begin{displayfigure}{Subtyping for $\lambdaex$}
  \label{figure:compositional:subtyping}
  \vspace{-30pt}
  \compositionalsubtyping
  \vspace{-35pt}
\end{displayfigure}

\begin{rulelist}
\item[\rel{S-Arrow} and \rel{S-Base}:] Subtyping between function
  types and base types is standard.

\item[\rel{S-Bind}:] This rule to introduce an existential on the left
  corresponds to the tautology of first-order logic:
  \[ 
  (\forall x \in
  A.\, P(x) \Rightarrow Q) \Longrightarrow (\exists x \in A.\, P(x))
  \Rightarrow Q \mbox{ where } x \notin fv(C)
  \]

\item[\rel{S-Witness}:] To introduce an existential on the right
  requires a witness term $e$ of type $T$ for the variable $x$.  If
  $S$ is a subtype of $\subst x e T\, U$, then we can disregard
  the identity of the witness and retain only its specification to
  conclude that $T$ is a subtype of $\existsbind x T U$.
\end{rulelist}

\noindent
Corresponding to our intuition, a combination of \rel{S-Witness} and
\rel{S-Bind} shows that the following reassuringly covariant rule is
admissible:\footnote{ The full derivation is not too complex, eliding
  some the environment prefix and some weakening:
  \[
  \inferrule*[right=\rel{T-Bind}]{
    \inferrule*[right=\rel{T-Witness}]{
      \inferrule*[right=\rel{T-Sub}]{
        \judge {x:S_1} x {S_1}
        \and
        \judgesub {} {S_1} {T_1}
      }{
        \judge {x:S_1} x {T_1}
      }
      \and
      \judgesub {x:S_1} {S_2} {T_2}
    }{
      \judgesub {x:S_1} {S_2} {\existsbind x {T_1} {T_2}}
    }
  }{
    \judgesub {} {\existsbind x {S_1} {S_2}} {\existsbind x {T_1} {T_2}}
  }
  \]
}
\[
\inferrule*{
  \judgesub \tyenv {S_1} {T_1}
  \and
  \judgesub {\tyenv,x\!:\!S_1} {S_2} {T_2}
}{
  \judgesub \tyenv {\existsbind x {S_1} {S_2}} {\existsbind x {T_1} {T_2}}
}
\]

%
%
%

\subsection{Non-dependent Function Application}
\label{section:compositional:nondependent}

As mentioned above, the typing rule for a function application
$(\f~e)$ is extremely simple, and does not refer to dependent types at
all.  Instead, the combination of existential types and subtyping
provides enough power in other areas of the type system that this
straightforward rule is sufficiently expressive.  Suppose $\f$ has a
dependent function type $\lamt x T U$ and the argument $\e$ has type
$S$ where $S \sword T$.  Then, we use subtyping to specialize the
function type as follows:
\begin{eqnarray*}
(\lamt x T U)
& \sword & (\lamt x S U)  \\
& \sword & (\lamt x S {\existsbind x S U}) \\
& \equiv & (\arrowt S {\existsbind x S U}) 
\mbox{~~~ since $x \notin fv(S)$}
\end{eqnarray*}
The new function type $\arrowt S {\existsbind x S U}$ is only
applicable to terms of type $S$.  More interestingly, its existential
return type now internalizes the assumption that $x$ will be of type
$S$, so we have a non-dependent function type that precisely
characterizes the behavior of the function $\f$ on arguments of type
$S$, and which does not refer to the exact semantics of the argument
$\e$.\footnote{ Non-dependent type rules for function application in a
  dependent type system are also used by
  \citet{harper_type-theoretic_1994} and \citet{dreyer_type_2003} in
  ML module systems where computational effects make the standard
  substitution-based rule unsound.  In the latter, a specialization of
  the technique we present allows the power of dependent function
  application when the argument is effect-free.}

Based on the above discussion, the following rule (mentioned in the
introduction) is admissible:
\[
\inferrule*{
  \judge {\tyenv} \f {\lamt x T U}
  \and
  \judge {\tyenv} \e S
  \and
  \judgesub {\tyenv} S T
}{
  \judge {\tyenv} {\f ~ \e} {\pexistsbind x S U}
}
\]

\subsection{Self Types}
\label{section:compositional:self}

To motivate our non-standard rule for variable references, consider
the expression $x-x$, where $x$ has type $\Int$ and
subtraction has an exact type:
\[
	- ~~:~~ (
\lamt w \Int {} \lamt y \Int {\predty z \Int {z = w - y}} )
\]

Under the standard rule, whereby a reference to $x$ simply has type
$\Int$, the type of $x-x$ is given by the judgment
\[
\judgeH {x\!:\!\Int} {(x - x)} 
{(\existsbind w \Int {\existsbind y \Int {\predty z \Int {z = w - y}}}})
\]
which could be any integer. This is needlessly coarse, since the fact
that $x - x = 0$ does not depend on the particular value of
$x$. Informally, it would appear that even via compositional reasoning
we should be able to give an exact type $\predty z \Int {z=0}$. What
happened?  The type system has lost a key notion of identity, that the
two variable references in $x-x$ refer to the same value.

To maintain the necessary information, \emph{selfification}
is used to assign to the variable reference $x$ the type
$\self(Int, x)$, or $\predty y \Int {y=x}$, which captures the
identity of $x$, and yields the conclusion:
\[
\judgeH {x\!:\!\Int} {(x - x)} 
{
  \begin{array}[t]{l}
    \existsbind w {\predquick \Int {w = x}} {} \\ \qquad 
    \existsbind y {\predquick \Int {y = x}} {} \\ \qquad\qquad 
    \predty z \Int {z = w - y}
  \end{array}
}
\]
Since $(w=x) \wedge (y=x) \wedge (z=w-y)$ implies $z = 0$, this is a
subtype of $\predty z \Int {z = 0}$, as intended.  In addition to this
example, $\self$ types are also crucial for achieving path-sensitivity
in case discrimination, which requires a notion of identity for the
matched expression.

Selfification interacts in delicate ways with our goal of
compositional reasoning. For example, \citet{ou_dynamic_2004}, in
their declarative system, give a general selfification rule
\[
\trule{T-Self}{
  \JudgeH {} e T
}{
  \JudgeH {} e {\self(T, e)}
}
\]
But this rule is non-compositional as they cannot replace the
subderivation $\JudgeH {} e T$ with another $\JudgeH {} {e'} T$.  In
their algorithmic presentation, they give self types only to constants
and variables. A key point not highlighted, however, is that giving
self types to variables \emph{is} compositional: Since there are no
subderivations, compositionality is vacuous. This is not simply a
syntactic trick, but the syntactic manifestation of the essence of
compositional reasoning: We may know nothing about a variable but its
type \emph{and} we may remember from which variable we ``fetch'' a
value. This insight is instrumental to reconciling compositional
reasoning with the precision of selfification.  The restriction of
selfification to variables enables us to prove the following key
compositionality theorem for our system:

\begin{thm}[Compositionality]
  The $\lambdaex$ type system is compositional.
\end{thm}

\pf\ This is directly observable by inspection of the typing
rules.~\qed\smallskip

\section{Extensional equivalence for $\lambdaex$}
\label{section:compositional:logicalrelation}

\hhref{Figure}{figure:compositional:equivalence} presents the full
definition of extensional equivalence for $\lambdaex$. It is that of
$\lambdah$ from \hhrefpref{Figure}{figure:lambdah:equivalence} augmented
with a definition for equivalence at existential type: Two terms
$\e_1$ and $\e_2$ of existential type $\existsbind x S T$ are
equivalent if there exist equivalent witness terms $\tmd_1$ and
$\tmd_2$ of type $S$ under which $\e_1$ and $\e_2$ are equivalent at
type $\subst x {\tmd_1} S\, T \curlywedge \subst x {\tmd_2} S\, T$.

\begin{displayfigure}{Extensional Equivalence Under Reduction for $\lambdaex$} 
  \label{figure:compositional:equivalence}
  \label{figure:compositional:extensionalequivalence}
  \lambdaexextensionalequivalence
\end{displayfigure}

As for $\lambdah$, extensional equivalence in $\lambdaex$ implies
contextual equivalence; the required proofs are straightforward
extensions of those in
\hhref{Section}{section:lambdah:logicalrelation}.

\section{The $\lambdaex$ Type System is Exact}
\label{section:compositional:exactness}

The requirement for compositional reasoning restricts the kinds of
inference rules the $\lambdaex$ type system can include.  For example,
we have seen that it forbids both the standard rule for dependent
function application and the selfification of arbitrary terms.  The
type system described above carefully adds the most important
capabilities while maintaining compositional reasoning.  We now
address the important question of how much precision or expressive
power our type system has lost because of its restriction to
compositional type rules.

In fact, under moderate assumptions, our type system is \emph{exact}:
it can assign to each program term a type that completely captures the
semantics of that term. Informally, an exact type is any type that
captures a terms semantics and an exact type system is one that can
capture \emph{any} term's semantics. The intent of the function
$\self$ is to assign exact types (to variables, in this case) so that
quantifications over self types are equivalent to quantifications over
a single value. By defining exactness clearly we can prove this is the
case.

\begin{definition}[Exactness]
  If $\judge \env \e T$ then the type $T$ is \emph{exact} for term
  $\e$ in environment $\env$ if for any other term $\tmd$ the judgment
  $\judge \env \tmd T$ implies $\tmd \ctxequiv \e$.  A type system is
  exact when it can assign an exact type to every well-typed term.
\end{definition}

\begin{lem}[Exactness of self type denotation]
  If $\tmd \in \meaningf{\self(T, e)}$ then $\tmd \ctxequiv \e$.
\end{lem}

\begin{lamproof}
  \pf\ This is equivalent to showing $\judgesimt \emptyset \tmd {\e : \self(T, e)}$,
  which is proved by induction on $\Base T$.~\qed
\end{lamproof}

Since self types have the desired denotation, the following chain of
corollaries establishes that all of our judgments interact with
\self{} types as expected.

\begin{corollary}
  \label{lemma:selfquant}\
  
  \begin{enumerate}
  \item If $\judge \env \tmd {\self(T, \e)}$ then $\tmd \ctxequiv \e$.
  \item If $\judge {} \sigma {\env,x:\self(T, \e),\envtail}$ then $\sigma(x) \ctxequiv \sigma(e)$.
  \item $\judgeimp {\env,x:\self(T,\e),\envtail} p {\subst x \e {}\, p}$
  \item $\judgesub {\env,x:\self(T,\e),\envtail} T {\subst x \e {}\, T}$
  \item $\judgesub {\env,\envtail} {\existsbind x {\self(S,\e)} T} {\subst x \e {}\, T}$
  \end{enumerate}
\end{corollary}

Establishing exactness for $\lambdaex$ requires that each primitive
function $\primfun$ and constructor $\constructor$) be given an exact
type. For the duration of this section, we assume that $ty$ has this
property:

\begin{requirement}[Exact Types for Primitives]
\label{assume:comp:primexact}
In the exact variant of $\lambdaex$ each primitive $f$ and constructor
$c$ is given an exact type by $ty$.
\end{requirement}

The exact information provided by types of constants and the self
types of variables is propagated losslessly by the $\lambdaex$ type
system.

\begin{thm}[The $\lambdaex$ type system is exact]
  \label{theorem:selfinductive}
  Supposing \hhref{Requirement}{assume:comp:primexact} holds,
  if $\judge \env e T$ then also $\judge \env e {T'}$
  where $T'$ is an exact type for $\e$ and $\env$.
\end{thm}

\pf\ This is proved by demonstrating that $\judge \env e {\self(T,e)}$
using \hhref{Corollary}{lemma:selfquant}, by induction on the typing
derivation $\judge \env e T$.~\qed
\smallskip

\hhref{Theorem}{theorem:selfinductive} seems very strong: for an
arbitrary term $e$ of type $T$ (where $e$'s subterms may include
coarse types), we can assign $e$ the type $\self(T,e)$ that
\emph{exactly} captures the semantics of $e$, all via compositional
reasoning.  In exploring the conflict upon which this paper is
premised, we have developed a compositional type system as powerful as
one based on substitution, but without the violation of
abstraction.\footnote{We do not claim to have improved upon dependent
  types in their general use as a foundation for mathematics, but only
  for the particular use of verifying a program specified by
  executable refinement types.} Rather, our type system insists that
whatever precision is desired must be explicitly expressed via types.

As an illustration of \hhref{Theorem}{theorem:selfinductive}, consider
the type of the fixed point primitive $\t{fix}_T$ used to express
recursion, which is typically given the inexact type $(T\rightarrow
T)\rightarrow T$.  We can make this type exact via selfification. For
example, if $T = \lamt x U B$, then an exact type for $\t{fix}_T$ is
$\self(~(T\rightarrow T)\rightarrow T, ~\t{fix}_T)$ which is
\[
\lamt f {(\arrowt T T)} {\lamt x U {\predty y B {y = \t{fix}_T~f~x}}}
\]

For a more interesting example, consider Euclid's algorithm for
computing greatest common divisors, where we use \t{if} expressions to
abbreviate case discrimination on type $\Bool$:
\[
\begin{array}{rcl}
\t{gcd} &~~=~~&
		\t{fix}_T~(\lam g T {\lam a \Int {\lam b \Int e}})
\\
\mbox{where~~~~~~} T	&=& \Int \rightarrow \Int \rightarrow \Int 
\\
\mbox{and~~~~~~} e	&=&
	\begin{array}[t]{@{}l}
		\t{if}~b=0~\t{then}~a \\
		\t{else}~ \t{if}~a>b ~\t{then}~g~(a-b)~b \\
        \t{else}~g~a~(b-a)
	\end{array}
\end{array}
\]
If $\t{fix}_T$ has the exact type $\self(~(T\rightarrow T)\rightarrow
T, ~\t{fix}_T)$, then the type systems infers  the following exact type for \t{gcd}:
\[
\begin{array}{l}
  \existsty {h\!:\! (\lamt g T {\lamt a \Int {\lamt b \Int {\predty r \Int {r=e}}}})} 
  {\\~~~~~~
    \lamt a \Int {\lamt b \Int {\predty r \Int {r=\t{fix}_T~h~a~b}}}}
\end{array}
\]		
Note that this type includes within refinement predicates both fixed
point computations and also the body $e$ of \t{gcd}.  Thus, the type
system is essentially translating the entire computation into the
predicate language. While this illustrates the
expressiveness of the type system, reasoning about fixed point
computations is notoriously difficult for automated theorem provers.

However, primitive functions in this calculus represent terms whose
semantics are unavailable; the predicates that occur in inferred types
are determined by the specifications of primitives. The primitives of
$\lambdaex$ model built-in constructs of a language implementation but
also names from external modules\footnote{In a well-designed language,
  built-in primitives and imported symbols are indistinguishable},
which will typically be assigned approximate types. The import of the
theorem is not that all abstractions can be broken even without
dependent types, but that \emph{the type system neither requires nor
  performs any abstraction itself}.

Returning to the example, a more practical approach is to use the
simple type $(T\rightarrow T)\rightarrow T$ for $\t{fix}_T$, and for
the programmer to provide an appropriately-precise specification $T$
for the recursive function \t{gcd}, corresponding to the imperative
idea of a loop invariant.\footnote{The wide variety of methods for
  inferring loop invariants may be applied to avoid this annotation
  burden, as mentioned in \hhref{Chapter}{chapter:inference}.  The
  most closely related progress on this variant of the problem is
  discussed in \hhref{Chapter}{chapter:relatedwork}.} The structural
specification $T=\Int \rightarrow \Int \rightarrow \Int$ is rather
coarse; a more precise (but not exact) specification is:
\[
T~~=~~
\lamt a \Int {\lamt b \Int {\predty r \Int {a\t{~mod~}r=0~\wedge b\t{~mod~}r=0}}}
\]
The type system can then verify this specification for \t{gcd}, with
simple proof obligations such as ``{if $b\t{~mod~}r=0$ and
  $(a-b)\t{~mod~}r=0$ then $a\t{~mod~}r=0$.}''

\section{Type Soundness for \lambdaex}
\label{section:compositional:implication}
\label{section:compositional:safety}

This chapter proves type soundness for $\lambdaex$ 
not by the usual means, but by proving type soundness for
the exact variant -- where all primitives have exact types --
which ensures that any more approximate variant is also
safe.

Given exactness, an existentially quantified variable with a singleton
type is equivalent to having simply performed a substitution (it is
essentially an explicit substitution \citep{abadi_explicit_1990}). This
intuition is formalized in \hhref{Corollary}{lemma:selfquant}
and the following complementary lemma:
\begin{lem}
  \label{lemma:self:subst}
  For any expression $e$ and type $T$, if $\JudgeH {} e S$ and
  $\JudgeHT {,x\!:\!S} T$ then $\JudgeHsub {} {\subst x e S\, T}
  {\existsbind x {\self(S, e)} T}$
\end{lem}

\pf\ By induction on the size of $T$, utilizing
\hhref{Theorem}{theorem:selfinductive} to provide the exact witness
$\e$ via \rel{S-Witness}.~\qed
\smallskip

From exactness, the standard substitution, preservation, and progress
lemmas apply unchanged. The proofs follow the same structure as those
for $\lambdah$, so we omit the details and simply state type safety.  If a
term is well-typed in any variant of our system, then it is certainly
well-typed in the exact variant.  Hence, a well-typed term cannot
``get stuck'' even in those imprecise variants where the substitution
lemma cannot be proved.

\begin{thm}
  If $\judge \emptyset e T$, then $e$ either reduces to a value or is
  nonterminating.
\end{thm}

\pf\ Map the derivation of $\judgeH \tyenv e T$ into
the exact variant by induction over the derivation, then apply
progress and preservation.~$\square$\smallskip

There may be another syntactic approach to type soundness that applies
in these systems, 
but it is informative to consider such systems as coarsenings of the
exact variant.


\section{A Type-checking Algorithm for \lambdaex}
\label{section:compositional:algorithmics}

We now investigate how to provide decidable compositional type
checking for an interesting subset of $\lambdaex$ programs.  Type
checking for $\lambdaex$ is undecidable, despite the recovery of
compositional reasoning, because implication remains undecidable.
However, arbitrary terms are no longer injected into types. So by
restricting refinement predicates to a decidable theory we can now
provide a decidable, compositional, dependent type system.  Note that
this decidability result crucially relies on compositional reasoning:
in a traditional dependent type system, even when all annotations fall
within a decidable theory, substitutions made during type checking may
result in proof obligations outside of that theory.

\hhref{Figure}{figure:compositional:restrictedsyntax} defines the type
language for which we provide a type-checking algorithm. We leave the
exact language of atomic decidable predicates $l$ abstract, so the
approach is applicable to any decidable predicate language. During
type-checking, atomic predicates will be combined with conjunction and
case-splitting on variables, so those are also included in the
predicate sublanguage of terms, denoted by metavariables $p$ or $q$,
which are now no longer used for arbitrary terms.

\begin{displayfigure}{Syntax for Algorithmic Typing of \lambdaex}
  \label{figure:compositional:restrictedsyntax}
  \ssp
  \restrictedsyntax
\end{displayfigure}

\begin{displayfigure}{Shorthand for pushing case expressions inside types}
  \label{figure:compositional:tycasedef}
  \tycasedefinition
\end{displayfigure}

We distinguish two sublanguages of types.  The first, ranged over by
metavariable $Q$ and $R$, does not include existential quantifiers and
may appear in source programs.  Augmented types ranged over by $E$ and
$F$ generated during type checking may contain existential
quantification, but only in positive positions.  Thus $Q \subseteq E
\subseteq T$. Only augmented types $E$ are bound in the environment
during type checking, so we use a new metavariable $\existsenv$ to
range over these $E$-environments.

The type-checking algorithm is presented as a set of syntax-directed
inference rules.  The judgments are analogous to those of the type
system.
\begin{itemize}
\item[-] The algorithmic typing judgment $\JudgeAlg {} e E$ reads that
  $\e$ is assigned type $E$, which may contain covariant existentials.

\item[-] The algorithmic type well-formedness judgment $\JudgeAlgT {}
  R$ is only invoked for programmer-provided restricted types $R$, and
  ensures that $R$ is well-formed in environment $\env$.

\item[-] The algorithmic subtyping judgment $\judgeAlgsub \env E R$
  checks that the type $E$ is a subtype of the restricted type
  $R$. Though $E$ may contain covariant existentials, they are not
  problematic for algorithmic checking. The upper bound $R$ will never
  contain existentials.
\end{itemize}

\subsection{Algorithmic typing for $\lambdaex$}
Our algorithmic type checking judgment
\[
\JudgeAlg {} e E
\]
is shown in \hhref{Figure}{figure:compositional:typingalgorithm}.  It is
essentially derived from the typing judgment $\judge \env e T$.  As
usual for syntax-directed type checking, we remove the subsumption
rule and instead inline subtyping where needed in function
applications to ensure the argument is of the appropriate type, and in
case discrimination to merge the types of all the branches.

\begin{rulelist}
\item[\rel{A-Const} and \rel{A-Prim}:] Constructors and primitive functions
  are given types by $ty$.

\item[\rel{A-Var}:] Variables are still given types via $\self$.
  Adding uninterpreted function symbols to the decidable predicate
  language retains decidability. Other approximations may be
  considered for efficiency; because of our nonstandard approach to
  soundness, any approximation is sound.

\item[\rel{A-App}:] For an application $\f~\e$, this rule introspects
  the type of $\f$ to discover a functional type within existential
  quantifiers, of the form $\existsmulti z {E_z} {\lamt x R {E_1}}$
  (performing the work of as many applications of \rel{S-Bind} as
  necessary). Then the type $E_2$ of the argument $\e$
  is checked for compatibility with $R$. The result of the application
  is assigned type $\existsmulti z {E_z} {\existsbind x {E_2} {E_1}}$

\item[\rel{A-Case}] In typing a case discrimination expression, we
  lift the \t{case} syntax to types to concisely express the resulting
  type. Intuitively, $\caseexp e {\listof {\patternexp
      {\constructor~\listof x} E}}$ is equivalent to whichever case
  succeeds.  \hhref{Figure}{figure:compositional:typingalgorithm}
  includes the exact definition. To avoid arbitrary expressions
  appearing in refinement predicates, we require that the inspected
  subexpression in a \t{case} construct be a variable. This
  requirement can be satisfied by rewriting the more general form
  $\caseexp e {\listof {\patternexp {\constructor~\listof x} \f}}$
  into the semantically equivalent $\plam y S {\caseexp y {\listof
      {\patternexp {\constructor~\listof x} \f}}} ~ \e$.  This
  rewriting step, which is a burden imposed on the input program,
  introduces a type $R$ for the variable $y$, and so ensures that any
  expression used in case discrimination will have a specification
  that falls in the restricted type language.\footnote{A similar
    restriction is present in \citet{rondon_liquid_2008}, where only
    variables may be inserted into predicate templates, so introducing
    bindings of intermediate results to variables is crucial to
    achieving precision while maintaining decidability.}
\end{rulelist}

\begin{displayfigure}{Algorithmic Typing for \lambdaex}
  \label{figure:compositional:typingalgorithm}
  \typingalgorithm
\end{displayfigure}

\subsection{Algorithmic type-wellformedness for $\lambdaex$}

\hhref{Figure}{figure:compositional:algsub} presents the rules defining the algorithmic type well-formedness
judgment
\[
\judgeAlgT \env R
\]
which checks if programmer-supplied types $R$ are well-formed. It is
easy to check well-formedness of types with existential
quantification, but for illustrative purposes the rules are
deliberately limited to those cases that actually may occur during
type checking.

\begin{displayfigure}{Algorithmic Type Well-formedness for \lambdaex}
  \label{figure:compositional:typewellformednessalgorithm}
  \lambdaextypewellformednessalgorithm
\end{displayfigure}

\subsection{Algorithmic subtyping for $\lambdaex$}
Like typing, algorithmic type checking relies on subtyping, but
interestingly only for the asymmetric form
\[
\judgeAlgsub \existsenv E R
\]
which checks subtyping between an inferred type $E$, with covariant
existential quantifications, and a programmer-specified, restricted
type $R$.

\begin{displayfigure}{Algorithmic Subtyping for \lambdaex}
  \label{figure:compositional:algsub}
  \vspace{-20pt}
  \compositionalsubalg\
  \vspace{-35pt}
\end{displayfigure}

\begin{rulelist}
\item[\rel{AS-Bind}:] Since existential types never appear on the
  right hand side, we (fortunately!) never need to ``guess'' witnesses
  for existentials.  Existentials on the left are unproblematic,
  because the existential quantification in a negative context
  transforms into a universal quantification in a positive context,
  resulting in just another binding in the typing environment.  We
  could add the covariant rule from the end of
  \hhref{Section}{section:compositional:intro} to allow some
  existential quantification on the right side of subtyping, but our
  syntactic conditions make that unnecessary.
\end{rulelist}

Algorithmic typeability is not closed under evaluation; instead
soundness is guaranteed by the correspondence between algorithmic
typing and the full type system.

\begin{thm}[Soundness of Algorithmic Typing]
  \label{thm:algsoundness}~\\
  If $\judgeAlg \existsenv e E$ then $\judgeH \existsenv e E$
\end{thm}

\begin{lamproof}
  \pf\ Proceed by induction on the derivation of $\judgeAlg \existsenv e
  E$.~\qed
\end{lamproof}

While the algorithm is not complete with respect to the full type
system, it is complete for the class of sublanguages we have
described.

\begin{thm}[Relative Completeness]~\\
  If $\judgeH \existsenv e T$ and $e$ is annotated with restricted
  types, and case discrimination inspects only variables, then there
  exists a type $E$ such that $\judgeAlg \existsenv e E$ and $\judgeHsub
  \existsenv E T$
\end{thm}

\begin{lamproof}
  \pf\ By induction on the derivation of $\judgeH \existsenv e
  T$.~\qed
\end{lamproof}

These algorithmic rules characterize a class of programs for which
type checking is decidable: If all program annotations are restricted
types whose refinements are expressions in some decidable theory
(where this means that queries of the form \mbox{\judgeAlgimp
  \existsenv q p} are decidable, such as linear inequalities with
conjunction and case splits), then all implication queries also fall
in this theory, and are decidable.  Interestingly, because
existentials only appear as assumptions in proof obligations, the
theory need only support a restricted form of existential
quantification that is ``productive'' in that each existential
corresponds to positive existence of some term, and there is never any
existential proof obligation.

If a program includes some predicates that do not fall in the
predetermined language, this algorithm may still succeed in verifying
the program if the theorem prover used is able to dispatch
the generated proof obligations. However, in this case a rejection
by this algorithm does not indicate that the program is ill-typed.
Hybrid type checking may be used as a fallback for such programs. 
The following informal architecture yields an algorithm always
verifies a well-defined subset of programs, but never rejects
a well-typed program:

\begin{enumerate}
  \item Given a program $\e$, does the algorithm yield $\judgeAlg \emptyset \e E$?
  \item If so, then accept the program.
  \item If not, and the program's annotations fall in the language of restricted types,
    then reject the program.
  \item Otherwise, perform hybrid type checking on the program.
\end{enumerate}



\section{Example: Binary Search Trees in \lambdaex}
\label{section:compositional:bst}
\newcommand{\BST}{\t{BST}}
\newcommand{\fmax}{\t{upper}}
\newcommand{\fmin}{\t{lower}}
\newcommand{\cmaxInt}{\t{maxInt}}
\newcommand{\cminInt}{\t{minInt}}
\newcommand{\cempty}{\t{empty}}
\newcommand{\cnode}{\t{node}}
\newcommand{\vhi}{hi}
\newcommand{\vlo}{lo}

As an illustration of our decidable type system, we now revisit the
example of binary search trees.  We assume an additional base type
\BST{} to represents binary search trees, with auxiliary functions:
\[
	\begin{array}{rcl}
	\fmin &:& \BST \rightarrow \Int\\
	\fmax &:& \BST \rightarrow \Int\\
	\end{array}
\]
that return the lower and upper bounds of integers in a BST,
and return $\cmaxInt$ and $\cminInt$, respectively, on empty BSTs.
We also assume some additional constructors and primitives:
\[
	\begin{array}{rcl}
	\cminInt &:&  \Int\\
	\cmaxInt &:&  \Int\\
	\t{min}	 &:&  \lamt x \Int {} \lamt y\Int {} \predty z \Int {z\leq x \wedge z\leq y} \\
	\t{max}	 &:&  \lamt x \Int {} \lamt y\Int {} \predty z \Int {z\geq x \wedge z\geq y} \\
	\end{array}
\]
The type of a binary search tree with integers in the range $[\vlo,\vhi)$ is defined by the refinement type:
\[
\BST_{\vlo,\vhi} ~ = ~ 
\predty x \BST { 
  ~\vlo \leq \fmin(x) 
  \wedge  \fmax(x) < \vhi~
}
\]

Binary search tree are created using the constructors \cempty{} and
\cnode{}, which are assigned the following precise types:
\begin{eqnarray*}
  \cempty & : & \lamt \vlo \Int {} \lamt \vhi \Int {} \BST_{\vlo,\vhi} \\
  \cnode & : & 
  \begin{array}[t]{l}
    \lamt \vlo \Int {} \lamt \vhi \Int {} \\
    \lamt v {\predty v \Int {\vlo \leq v < \vhi}} {} \\
    \lamt x {\BST_{\vlo,v}} {} \lamt y {\BST_{v,\vhi}} {} 
    \BST_{\vlo,\vhi}
  \end{array}
\end{eqnarray*}
Here, these constructors take additional ``index'' arguments
$\vlo$ and $\vhi$, and so we are using an index-type-like
implementation of binary search trees.  Using these constructors, we
can define the \t{insert} operation on \BST{}s:
\[
\begin{array}[t]{l}
  \t{fix}_T (
  \begin{array}[t]{@{}l}
  \lam f T {} \\
  \lam \vlo \Int {} \lam \vhi \Int {}  \\
  \lam v {\predty y \Int {\vlo \leq y < \vhi}} {} \\
  \lam x {\BST_{\vlo,\vhi}} \\
  \quad \t{case}~x~\t{of} \\
  \qquad\qquad (\cempty~\vlo~\vhi)      ~ \triangleright (\cnode~lo~hi~v~(\cempty~\vlo~v)~(\cempty~v~\vhi)) \\
  \qquad\qquad (\cnode~\vlo~\vhi~n~l~r) ~ \triangleright 
  \\
  \qquad\qquad\qquad (\t{case}~{v<n}~\t{of} \\
  \qquad\qquad\qquad\qquad \true ~ \triangleright (\cnode~\vlo~\vhi~n~(\t{insert}~\vlo~n~x~l)~r) \\
  \qquad\qquad\qquad\qquad \false ~ \triangleright ~ (\cnode~\vlo~\vhi~n~l~(\t{insert}~n~\vhi~x~r))
  \end{array}
\end{array}
\]
where \t{insert} has type 
\begin{eqnarray*}
  T & = & 
  \begin{array}[t]{ll}
    \lamt \vlo \Int {} \lamt \vhi \Int {}  \\
    \lamt v {\predty v \Int {\vlo \leq v < \vhi}} {} \\
    \arrowt {\BST_{\vlo,\vhi}} {\BST_{\vlo,\vhi}}
  \end{array}
\end{eqnarray*}

\noindent
All the proof obligations generated for this program are formulae over
linear integer inequalities, and hence decidable.

In this example, the variables $\vlo$ and $\vhi$ are somewhat
awkward, and provide an indirect specification of the behavior of
\t{insert}.  
We can express this program more naturally by removing these
parameters instead expressing the key data invariants directly in
terms of the underlying data structure.  In this formulation, the
\BST\ constructors have the more natural types:
\begin{eqnarray*}
  \cempty & : & \BST_{\cmaxInt,\cminInt} \\
  \cnode & : & 
  \begin{array}[t]{l}
    \lamt v \Int {} \\
    \lamt x {\predty x \BST {\fmax(x) < v}} {} \\
    \lamt y {\predty y \BST {v \leq \fmin(y)}} {} \\
    \BST_{\fmin(x),\fmax(y)}
  \end{array}
\end{eqnarray*}

\noindent
The revised \t{insert} implementation is identical to the one shown above but elides index variables:
\[
\begin{array}[t]{l}
  \t{fix}_{T'} (
  \begin{array}[t]{@{}l}
  \lam f {T'} {} ~~
  \lam v \Int  {} ~~
  \lam x \BST {} \\
  \qquad \t{case} ~ x ~ \t{of} \\
  \qquad\qquad \cempty ~ \triangleright ~ (\cnode~v~\cempty~\cempty) \\
  \qquad\qquad (\cnode~n~l~r) ~ \triangleright\\
  \qquad\qquad\qquad (\t{case}~{v<n}~\t{of} ~ \\
  \qquad\qquad\qquad\qquad \true ~ \triangleright ~ (\cnode~n~(\t{insert}~x~l)~r) \\
  \qquad\qquad\qquad\qquad \false ~ \triangleright ~ (\cnode~n~l~(\t{insert}~x~r)) \\
  \end{array}
\end{array}
\]
and has the following type $T'$:
\begin{eqnarray*}
  T' & = & 
  \begin{array}[t]{ll}
    \lamt v \Int {} \lamt x \BST {} \\
    \BST_{\t{min}(\fmin(x),v), \t{max}(\fmax(x),v)}
  \end{array}
\end{eqnarray*}


\section{Related Work}
\label{section:compositional:relatedwork}

First, we present a high-level comparison of refinement types
with indexed types, an alternative approach towards similar goals.
We next survey other applications of existential
quantification in type systems.  Finally, we briefly outline
the development of refinement and refinement types which this
work directly builds upon.

\subsection{Indexed Types}
\label{section:compositional:append}
\label{section:compositional:indexed}

Indexed types are discussed in
\hhref{Section}{section:intro:refinementtypes} in general, so here we
focus on their relationship to the concept of
compositionality. Formally, the type systems of Dependent ML
\citep{xi_dependent_1999}, ATS \citep{cui_ats:_2005}, and $\Omega$mega
\citep{sheard_putting_2005}, distinguish compile-time from run-time
data, and allow types to depend only on compile-time data. This
approach makes compositionality a formally vacuous proposition, and
indeed can sometimes be encoded in existing polymorphic type systems
without use of dependent types
\citep{zenger_indexed_1997,mcbride_faking_2002}.

As an illustrative example, consider the family of types $\IntList_n$,
where $n$ indicates the length of lists inhabiting the type.  The
types of list constructors and the \t{append} function are as follows.
We use the type $\TrueNat$ to distinguish the compile-time type of
natural numbers.
\[
\begin{array}[t]{rcl}
  \t{nil} & : & \IntList_0 \\
  \t{cons} 
  & : &
  \lamt n \TrueNat {}
  \arrowt \Int {}
  \arrowt {\IntList_n} {\IntList_{n+1}}
  \\
  \t{append} 
  & : &
  \lamt m \TrueNat {} 
  \lamt n \TrueNat {}   \\
  && \qquad \arrowt {\IntList_m} {}
  \arrowt {\IntList_n} {}
  {\IntList_{m+n}}
\end{array}
\]
The connection between run-time invariants and compile-time data is
based essentially on injecting an abstraction of run-time data into
compile-time data; for example lists indexed by their length use a
compile-time copy of the natural numbers.  The expressions that are
reflected into indices can be carefully controlled to ensure that type
compatibility remains decidable.

We can naturally express the indexed type $\IntList_n$ as the
refinement type $\predty x \IntList {\t{length}(x) = n}$, and the
above types and all proof obligations remain equivalent.  In this way,
by reifying the abstraction function from a value to its associated
index, arbitrary indexed types can be embedded into executable
refinement types.  The abstraction function ($\t{length}$ in this
case) may be treated as an uninterpreted symbol -- its definition
cannot be necessary in proof obligation since it is not even available
to indexed types.

The above type for \t{append} is somewhat awkward, though, as it
requires the additional index parameters $n$ and $m$.\footnote{Note
  that these index parameters can be inferred in many cases.}
In contrast, executable refinement types allow a more natural
expression of the same specification for \t{append}, that eliminates
these index parameters:
\[
\begin{array}{l}
\lamt x \IntList {}
  \lamt y \IntList {} \\
  \qquad
  \predty z \IntList {\t{length}(z) = \t{length}(x) + \t{length}(y)}
\end{array}
\]

More critical than the aesthetic issue of index parameters, the index
of a data structure, decided by its implementor, determines which
properties may be reasoned about, inhibiting reuse and composition of
data structures. For example, if one is interested in verifying the
ordering of a list, rather that its length, then a different index is
required, specifically the minimum element (at the head of the list)
to specify \t{cons}, and also the maximum element (at the tail) to
specify \t{append}.  We also need integers indexed by their exact
values $\Int_i$, where $i$ is drawn from compile-time integers
$\TrueInt$ (with $\pm \infty$ for corner cases).  We define the type
$\IntList_{i,j}$ of lists in ascending order with minimum element $i$
and maximum element $j$ using the following constructors.  Note that
predicate subtypes are used on indices of compile-time type $\TrueInt$
but not on runtime terms.
\[
\begin{array}[t]{rcl}
  \t{nil} & : & \IntList_{\infty,-\infty} 
  \\
  \t{cons}
  & : &
  \lamt i \TrueInt {}
  \lamt j {\predty j \TrueInt {j \geq i}} {}
  \lamt k \TrueInt {}
  \\
  && \qquad
  \arrowt {\Int_i} {}
  \arrowt {\IntList_{j,k}} {\IntList_{i,\t{max}(i,k)}}
  \\
  \t{append}
  & : &
  \lamt i \TrueInt {}
  \lamt j \TrueInt {}
  \lamt k {\predty k \TrueInt {k \geq j}} {}
  \lamt l \TrueInt {} \\
  && ~~
  \arrowt {\IntList_{i,j}} {}
  \arrowt {\IntList_{k,l}} {}
  \IntList_{\t{min}(i,k),l}
\end{array}
\]
In general, since the index of a data type determines the properties
that may be reasoned about, more complex properties require embedding
of more data as indices.  In the limit, giving the precise type $\lamt
x \IntList {\predty y \IntList {y=x}}$ to the identity function on
lists (or any type) requires embedding the entire type of lists (or
any type) into the index language, which reduces the utility of the
syntactic distinction of compile-time data.

In contrast, executable refinement types can naturally express a wide
variety of refinements over the same \IntList\ type, such as
\[
\predty x \IntList {\t{minElem}(x) = i \wedge \t{maxElem}(x) = j}
\]
or the most natural $\predty x \IntList {\t{isSorted}(x)}$.  The proof
obligations for the latter may be more problematic, and the current
work is progress towards characterizing those situations where they
are not difficult.

Reasoning about how the constructors of a type relate to the global
predicate upon the type is an open question.
\citet{kawaguchi_type-based_2009} defines a class of ``measures''
which are global predicates calculated by catamorphism over the
defining equations of a recursive datatype, and automatically
generates the appropriate refinements for the constructors.
\citet{atkey_when_2011} generalizes this approach to a
category-theoretic analysis of when a type defined as the initial
algebra of a functor is actually a refinement of a type defined by
another functor. 

\subsection{Existential Types}
\label{section:compositional:existentials}

Formal existential quantification in type-theory has found a variety
of uses, including closure conversion
\citep{morrisett_system_1999,grossman_existential_2002}, module
systems
\citep{mitchell_abstract_1988,harper_type-theoretic_1994,dreyer_type_2003},
and semantics of object orientation \citep{bruce_comparing_1999}.
Most similar to the present work are ML module systems, which also
combine existentials with subtyping and a different (formal) form of
singleton types.

The standard substitution-based typing rule for function application
is unsound in the presence of effects, so dependencies have to be
``forgotten'' \citep{harper_type-theoretic_1994}.
\citet{dreyer_type_2003} ameliorate this restriction with a
sophisticated effect system that restores the power of dependent
application in the event that the argument to a function is
effect-free.

Also similar is the need to express sharing constraints.  Just as we
give self types to variables, \citet{stone_deciding_2000} and
subsequently \citet{dreyer_type_2003} assign singleton kinds (a
special syntactic form) to modules in order to preserve information
when applying a dependently-kinded functor.  In contrast to module
languages, we examine the consequences and form of existentials and
selfification in the context of executable refinements with case discrimination and automatic theorem proving.  Our singletons are expressed
in the existing type language rather than adding a special form, and
in our setting a type of a variable $x$ may constrain other bound
variables so we maintain this information while adding the identity
information to the known type of $x$.

Our different approaches and goals led us to emphasize compositional
reasoning as a key aspect of our system, which we present in a more
minimal calculus than the large and pragmatism-oriented ML module
type systems.  Compositionality is obviously important in their work
as well, since it affects separate (re-)compilation.

Our existential types may be considered a limited expression of
$\Sigma$ types for dependent pairs in powerful type theories such as
those underlying Coq \citep{the_coq_development_team_coq_2012}, Hoare
Type Theory \citep{nanevski_polymorphism_2006}, and Epigram
\citep{mcbride_view_2004}.  Dependent ML, in fact, makes use of types
such as $\Sigma x\!:\!T_1.T_2$ with explicit introduction and
elimination forms (i.e. without subtyping) to allow functions
operating over arguments of unknown index.

With respect to $\Sigma$ types, the work presented in this chapter can
be interpreted as a way of providing automation for a simple and
common case where the full power of these type theories is not
necessary.

\subsection{Liquid Types}

Liquid Types \citep{rondon_liquid_2008} present a similar system, with
predicates drawn from a carefully designed language of predicate
templates. To avoid inserting arbitrary terms into predicates, only
variables are used to instantiate these templates, and variables are
required to have types within the selected predicate language. For
programs in \emph{a-normal form}, where every subexpression is bound
to a variable via a $\t{let}$ expression, every subexpression is given
a type within the language of predicate templates. This provides some
of the benefits of compositionality, as no predicate will need to be
proved that falls outside of the selected predicate language. Instead,
as with our approximate variant, the approximation occurs because a
precise type for each expression may not be expressible within the
decidable predicate language.

\chapter{\sage{}: An Implementation of Executable Refinement Types}

\label{chapter:sage}
\label{section:sage:intro}

\newcommand{\sagesubtyperules}{
  \headtrule{$\judgesub{\env}{S}{T}$}{\underline{S}ubtype rules}
  \begin{trules}
    \trule{S-Refl}{
    }{
      \judgesub \env T T
    }

    \trule{S-Dyn}{
    }{
      \judgesub \env T \dynamic
    }
    
    \trule{S-Fun}{
      \judgesub{\env}{T_1}{S_1}  \qquad
      \judgesub{\env,x:T_1}{S_2}{T_2}  
    }{
      \judgesub{\env}
               {(\funtyd{x}{S_1}{S_2})}
               {(\funtyd{x}{T_1}{T_2})}
    }

    \trule{S-Var}{
      \judgesub{\env,\subst x v {}\, \envtail}{\subst x v {}\, S}{\subst x v {}\, T} 
    }{
      \judgesub{\env,x=v:U,\envtail}{S}{T}
    }

    \trule{S-Eval-L}{
      \d \red \e \qquad
      \judgesub{\env}{\redctx[\d]}{T} 
    }{
      \judgesub{\env}{\redctx[\e]}{T}
    }
    
    \trule{S-Eval-R}{
      \d \red \e \qquad
      \judgesub{\env}{S}{\redctx[\d]} 
    }{
      \judgesub{\env}{S}{\redctx[\e]}
    }
    
    \trule{S-Ref-L}{
      \judgesub{\env}{S}{T}
    }{
      \judgesub{\env}{(\t{Refine}~S~f)}{T}
    }

    \trule{S-Ref-R}{
      \judgesub{\env}{S}{T}  \qquad
      \judgeTP{\env,x:S}{f~x}
    }{
      \judgesub{\env}{S}{(\t{Refine}~T~f)} 
    }
  \end{trules}
}

\newcommand{\sagesubtypingalg}{
  \headtrule{$\judgesubalg{a}{\env}{S}{T}$}{Algorithmic subtyping rules}
  \begin{trules}
    \trule{AS-Db}{
      \judgesubdb \pno \env S T
    }{
      \judgesubalg \pno \env S T
    }
    
    \trule{AS-Refl}{
    }{
      \judgesubalg \pyes \env T T
    }

    \trule{AS-Fun}{
      \judgesubalg a \env {T_1} {S_1} \\
      \judgesubalg b {\env,x:T_1} {S_2} {T_2}
    }{
      \judgesubalg {a\sqcap b} \env {(\funtyd x {S_1} {S_2})} {(\funtyd x {T_1} {T_2})}
    }

    \trule{AS-Dyn-L}{
    }{
      \judgesubalg{\pmaybe}{\env}{\dynamic}{T} 
    }

    \trule{AS-Dyn-R}{
    }{
      \judgesubalg{\pyes}{\env}{S}{\dynamic} 
    }

    \trule{AS-Ref-L}{
      \judgesubalg{a}{\env}{S}{T} ~~~~ a \in \{\pyes, \pmaybe\}
    }{
      \judgesubalg{a}{\env}{(\t{Refine}~S~\f)}{T}
    }
    
    \trule{AS-Ref-R}{
      \judgesubalg{a}{\env}{S}{T}  \\
      \judgeTPalg{b}{\env,x:S}{\f~x}
    }{
      \judgesubalg {a \sqcap b}{\env}{S}{(\t{Refine}~T~\f)} 
    }

    \trule{AS-Var}{
      \judgesubalg{a}{\env,\subst x v {}\, F}{\subst x v {}\, S}{\subst x v {}\, T} 
    }{
      \judgesubalg{a}{\env,x=v:u,F}{S}{T}
    }

    \trule{AS-Eval-L}{
      \e \red \e' \quad
      \judgesubalg{a}{\env}{D[\e']}{T} 
    }{
      \judgesubalg{a}{\env}{D[\e]}{T}
    }

    \trule{AS-Eval-R}{
      \e \red \e' \quad
      \judgesubalg{a}{\env}{S}{D_2[\e']} 
    }{
      \judgesubalg{a}{\env}{S}{D_2[\e]}
    }
  \end{trules}
  \vspace{-2.5em}
  \[
  D~::=~\bullet~|~N~D \mbox{\qquad where $N$ is a normal form}
  \]
}

This chapter presents \sage{}, an exploratory implementation of
executable refinement types and hybrid type checking.  In creating a
new language with experimental features, we hope not for mass
adoption, but to ascertain the technical feasibility of the
eventual incorporation of our ideas into mainstream languages.  Of
interest are question such as:

\begin{itemize}
\item[-]
  How long does type checking take?
\item[-]
  How effective are state-of-the-art theorem provers at dispatching
  the obligations that arise?
\item[-]
  Can realistic programs be easily written and specified? 
\end{itemize}

\noindent
To place these questions in the introductory context of this
dissertation, consider again this variety of specifications for a
function that inverts a matrix, reproduced here verbatim for
reference:
\begin{enumerate}
\item
  The argument can be any (dynamically-typed) value.
\item
  The argument must be an array of arrays of  numbers. 
\item
  The argument must be a \emph{matrix}, that is, a rectangular (non-ragged)
  array of arrays of  numbers. 
\item
  The argument must be a square matrix. 
\item
  The argument must be a square matrix with nonzero determinant.
\item        
  The argument must be a square matrix satisfying a custom
  invertibility test procedure \t{isInvertible}.
\end{enumerate}

\noindent
\sage{} supports \emph{all} of these specifications in a unified
type system.

\subsection*{Chapter Outline}

\begin{itemize}
\item[-]
  \hhref{Section}{section:sage:overview} gives a brief
  overview of the key features of \sage{}.

\item[-]
  \hhref{Section}{section:sage:examples} illustrates the \sage{}
  language through a series of examples written in
  \sage{}'s concrete syntax.

\item[-]
  \hhref{Section}{section:sage:syntax} presents the abstract
  syntax for \sage{} Core, a minimal calculus for discussing
  \sage{}'s operational semantics and type system.

\item[-]
  \hhref{Section}{section:sage:opsem} defines a small-step
  operational semantics for \sage{} Core.

\item[-]
  \hhref{Section}{section:sage:typesystem} presents the type system of
  \sage{} Core.

\item[-]
  \hhref{Section}{section:sage:htc} presents a hybrid
  type checking / cast insertion algorithm for the language.

\item[-] \hhref{Section}{section:sage:architecture}
  and~\hhref{Section}{section:sage:experiments} describe our
  implementation and experimental results.

\end{itemize}

\section{Overview of \sage{} features}
\label{section:sage:overview}

On a technical level, the \sage{} type system can be viewed as a
synthesis of three concepts: the type \dynamic{} of terms that ``may
be'' any value expressible in the language, executable refinement
types, and first-class types.  These features add expressive power in
three orthogonal directions, yet they all cooperate neatly within
\sage{}'s hybrid static/dynamic checking framework.

For this section and the next, \sage{} concrete syntax will be
used. \sage{}'s concrete syntactic style is borrowed from OCaml and
Coq so that the transliterations of mathematical symbols into ASCII
are standard and familiar. Here are some brief examples:
\begin{itemize}
\item[-] The dependent function type $\lamt x \Int \Int$ is written \t{x:Int -> Int}.
\item[-] The function $\lam x S {x + 1}$ is written \t{fn (x:S) => x + 1}.
\item[-] Top-level definitions are introduced with \t{let} and terminated
with a semicolon. For example, this defines the function \t{add1}.
\begin{verbatim}
  let add1 = fn (x:Int) => x + 1;
\end{verbatim}
The function arguments may also be placed to the left of the equals sign.
The definition of \t{add1} is equivalently written as
\begin{verbatim}
  let add1 (x:Int) = x + 1;
\end{verbatim}
\end{itemize}

Any further details of the concrete syntax should be, hopefully,
intuitive, but the full syntactic sugar provided by the \sage{}
implementation is technically uninteresting. 

\subsection{Type \dynamic} 

\sage{} supports dynamically-typed programming with the type
\dynamic{}~\citep{thatte_quasi-static_1990, henglein_dynamic_1994}.
\dynamic{} is a supertype of all types; any value can be upcast to
type \dynamic{}. A value of type \dynamic{} can be implicitly downcast
(via a run-time check) to a more precise type.  Downcasts are
implicitly inserted when necessary, such as when applying the
primitive addition operator $+$, which requires arguments of type
\t{Int}, to arguments of type \dynamic{}.  Thus, declaring variables
to have type \dynamic{} (which is the default if type annotations are
omitted) yields the usual programming style and runtime
characteristics of dynamically-typed programming languages such as
Scheme, Python, Perl, Ruby, etc.

These dynamically-typed programs can later be annotated with
traditional type specifications like \t{Int} and \t{Bool}.  The
programmer need not fully annotate the program with types in order to
reap some benefit.  Types enable \sage{} to check more properties
statically, but it is still able to fall back to dynamic checking
whenever the type \t{Dynamic} is encountered.

\subsection{Executable Refinement Types} 

\sage{} supports high precision specifications via executable
refinement types. For example, the following code snippet defines the
type of integers in the range from \t{lo} (inclusive) to \t{hi}
(exclusive):

\begin{verbatim} 
  { x:Int | lo <= x && x < hi } 
\end{verbatim}

\sage{} extends prior chapters by allowing refinement of any type, not
only base types. This results in additional challenges to
decidability.  For example, whether an arbitrary function may be
assigned the type

\begin{verbatim}
  { f : Int -> Int | f(0) == f(3) } 
\end{verbatim}
is undecidable. It may even be difficult to translate proof
obligations to the input language of an off-the-shelf theorem
prover. Nonetheless, hybrid type checking techniques still apply to
this language of types, with difficult cases resulting in simply more
run-time checks.

\subsection{First-Class Types} 
\label{section:sage:firstclasstypes}

Finally, \sage{} elevates types to be first-class values, following
the Type Theory of \citet{martin-lof_intuitionistic_1975} and Pure
Type Systems of \citet{barendregt_introduction_1991}\footnote{ Type
  Theory and Pure Type System have inspired other languages, including
  Cayenne, Agda, and Epigram.}.  To the usual functions from
\emph{terms} to \emph{terms} that exist in all languages, first-class
types add:

\begin{itemize}
\item[-] Functions from \emph{terms} to \emph{types}.  The following
  function \t{Range} takes two integers and returns the type of
  integers within that range:
\begin{verbatim}

  let Range (lo:Int) (hi:Int) : * = { x:Int | lo <= x && x < hi };
\end{verbatim}
  Here, \t{*} is the type of types and indicates that \t{Range} returns
  a type.  

\item[-] Functions from \emph{types} to \emph{types}. The following
  function \t{endofunction} accepts a type \t{T} as a parameter and
  returns the type of of endofunctions over \t{T}:

\begin{verbatim}
  let endofunction (T:*) : * = T -> T;
\end{verbatim}\

\item[-] Functions from \emph{types} to \emph{terms}. The following
  function \t{id} accepts a type \t{T} as its first parameter
  and returns the identity function on \t{T}:

\begin{verbatim}
  let id (T:*) (x:T) : T = x;
\end{verbatim}
\end{itemize}

The traditional limitation of both first-class types (in the presence
of nonterminating programs) and executable refinement types is type
checking is not statically decidable. \sage{} applies hybrid type
checking to both problems. The same mechanism supports dynamic
typing. In practice, most errors are detected statically, while only
more complicated violations escape static checking and are detected at
run time.

\section{Examples of \sage{} programs}
\label{section:sage:examples}

This section examines some code snippets and then several longer
example programs to illustrate the features of \sage{} and give a
sense of the language.  The following examples have fairly complete
specifications to highlight the specification features, rather than
leveraging the type \t{Dynamic}.

\subsection{Binary Search Trees}

We will first demonstrate \sage{} programming with the oft-studied
example of binary search trees, whose \sage{} implementation is shown
in Figure~\ref{figure:sage:bst}.  For completeness we include the
entirety of the code here, with line numbers for reference, beginning
with the definition of $\t{Range}$ from
Section~\ref{section:sage:firstclasstypes}.

\begin{verbatim}
     1: let Range (lo:Int) (hi:Int) : * =
     2:   {x:Int | lo <= x && x < hi};
\end{verbatim}

A binary search tree $\t{(BST~lo~hi)}$ is an ordered tree containing
integers in the range $[\t{lo},\t{hi})$.  A tree may either be
$\t{Empty}$, or a $\t{Node}$ containing an integer
$\t{v}\in[\t{lo},\t{hi})$ and two subtrees containing integers in the
ranges $[\t{lo},\t{v})$ and $[\t{v},\t{hi})$, respectively.  

\begin{verbatim}
     4: datatype BST (lo:Int) (hi:Int) =
     5:     Empty
     6:   | Node of (v:Range lo hi) * (BST lo v) * (BST v hi);
\end{verbatim}

This definition is desugared into the core language of
Section~\ref{section:sage:syntax}. The exact encoding 
is not important, but the types of the constructors
are, using \sage{}'s concrete syntax for types,

\begin{verbatim}
   Empty : (lo:Int) -> (hi:Int) -> BST lo hi
   Node  : (lo:Int) -> (hi:Int) -> 
           (v:Range lo hi) -> (BST lo v) -> (BST v hi) -> 
           BST lo hi
\end{verbatim}

Thus, the type of binary search trees explicates the requirement that
these trees must be ordered. This is a blend of the style of indexed
types and direct style: The \t{BST} itself is indexed, but the values
on the nodes are constrained by direct refinement.

The function \t{search} takes as arguments two integers \t{lo} and
\t{hi}, a binary search tree of type $\t{(BST~lo~hi)}$, and an integer
\t{x} in the range contained in the tree. It returns \t{true} if
\t{x} is contained within the tree.

\begin{verbatim}
     8: let rec search (lo:Int) (hi:Int) (t:BST lo hi)
     9:                (x:Range lo hi) : Bool =
    10:   case t of
    11:       Empty _ _ -> false
    12:     | Node v l r ->
    13:         if x = v then true
    14:         else if x < v
    15:              then search lo v l x
    16:              else search v hi r x;
\end{verbatim}

Note the use of dependent function types, where the types
of the parameters \t{t} and \t{x} to \t{search}
depend on the values of the parameters \t{lo} and \t{hi}.

The function \t{insert} takes similar arguments and extends the given
tree with the integer \t{x}. It is a useful demonstration of
the construction of the tree.

\begin{verbatim}
    18: let rec insert (lo:Int) (hi:Int) (t:BST lo hi)
    19:                (x:Range lo hi) : (BST lo hi) =
    20:   case t of
    21:       Empty -> Node lo hi x (Empty lo x) (Empty x hi)
    22:     | Node v l r ->
    23:         if x < v 
    24:         then Node lo hi v (insert lo v l x) r
    25:         else Node lo hi v l (insert v hi r x);
\end{verbatim}

The \sage{} system uses an automatic theorem prover to statically
verify that the specified ordering invariants on binary search trees
are satisfied by these two functions. No run-time checks are required
for this example.

\begin{displayfigure}{Binary Search Trees in \sage{}}
\noindent
\begin{verbatim}
     1: let Range (lo:Int) (hi:Int) : * =
     2:   {x:Int | lo <= x && x < hi};
     3:
     4: datatype BST (lo:Int) (hi:Int) =
     5:     Empty
     6:   | Node of (v:Range lo hi) * (BST lo v) * (BST v hi);
     7:
     8: let rec search (lo:Int) (hi:Int) (t:BST lo hi)
     9:                (x:Range lo hi) : Bool =
    10:   case t of
    11:       Empty -> false
    12:     | Node v l r ->
    13:         if x = v then true
    14:         else if x < v
    15:              then search lo v l x
    16:              else search v hi r x;
    17:
    18: let rec insert (lo:Int) (hi:Int) (t:BST lo hi)
    19:                (x:Range lo hi) : (BST lo hi) =
    20:   case t of
    21:       Empty -> Node lo hi x (Empty lo x) (Empty x hi)
    22:     | Node v l r ->
    23:         if x < v 
    24:         then Node lo hi v (insert lo v l x) r
    25:         else Node lo hi v l (insert v hi r x);
\end{verbatim}
\label{figure:sage:bst}
~
\end{displayfigure}


The precise type specifications enable \sage{} to detect various
common programming errors.  For example, suppose we inadvertently used
the wrong conditional test:
\begin{code}
~~~~23:~~~~~~if \underline{x <= v} 
\end{code}
For this (incorrect and ill-typed) program, \sage{} will
report that the specification for \t{insert} is violated by the first
recursive call:
\begin{code}
line 25: x does not have type (Range lo v)\\
\end{code}
Similarly, if one of the arguments to the constructor \t{Node} is incorrect,
\eg: 
\begin{code}
~~~~26:~~~~~~else Node lo hi v \underline{r} (insert v hi r x); 
\end{code}
\sage{} will report the  type error:
\begin{code}
line 26: r does not have type (BST lo v)
\end{code}

Using this \t{BST} implementation, constructing trees with specific
constraints is straightforward and verifiable.  For example, the
following code constructs a tree containing only positive numbers:

\begin{code}
let PosBST : * = BST 1 MAXINT;\\
\\
let nil : PosBST = Empty 1 MAXINT;\\
let add (t:PosBST) (x:Range 1 MAXINT) : PosBST = \\
~~~~~~~~insert 1 MAXINT t x;\\
let find (t:PosBST) (x:Range 1 MAXINT) : Bool = \\
~~~~~~~~search 1 MAXINT t x;\\~\\
let t : PosBST = add (add (add nil 1) 3) 5;
\end{code}

Note that this fully-typed BST implementation inter-operates with
dynamically-typed client code:
\begin{code}
let t : Dynamic = (add nil 1) in find t 5;
\end{code}

\subsection{Regular Expressions}
\label{section:sage:regex}

\begin{displayfigure}{Regular Expressions in \sage{}}
\label{figure:sage:regexp}
\vspace{-20pt}
\ttfamily
$\!\!\!\!\!$
\begin{tabular}{l}
datatype Regexp =\\
~~Alpha\\
| AlphaNum\\
| Kleene of Regexp\\
| Concat of Regexp * Regexp\\
| Or of Regexp * Regexp\\
| Empty;\\
~\\
let match (r:Regexp) (s:String) : Bool = ... \\
~\\
let Credential = \{s:String | match (Kleene AlphaNum) s\}; \\
\end{tabular}
\vspace{5pt}
\end{displayfigure}

Regular expressions are a common means of validating string data, and
serve as a good example of a specification beyond the common
constraints using linear arithmetic.  Figure~\ref{figure:sage:regexp}
declares the \t{Regexp} data type and the function \t{match}, which
determines if a string matches a regular expression.  To avoid
irrelevant complexity, \t{Regexp} only expresses regular expressions
over the two character classes of alphabetic characters (\t{Alpha})
and alphanumeric characters (\t{AlphaNum}).  In addition to these
primitive character classes, we include the usual the Kleene closure,
concatenation, and choice operators.  As an example, the regular
expression ``\t{[a-zA-Z0-9]*}'' would be represented in our datatype
as \t{(Kleene AlphaNum)}.\footnote{Assuming ASCII, since a more
  robust character set is not important for this demonstration.}

The code then uses \t{match} to define the type \t{Credential}, which
refines the type \t{String} to allow only alphanumeric strings.  We
use the type \t{Credential} to enforce an important, security-related
interface specification for the following function \t{authenticate}.
This function performs authentication by querying a SQL database
(where `\verb|^|' denotes string concatenation):
\begin{code}
let authenticate (user:Credential) (pass:Credential) : Bool =\\
~~  let query : String = \\
~~~~   ("SELECT count(*) FROM client WHERE name =" \verb@^@ \\
~~~~    user \verb@^@ " and pwd=" \verb@^@ pass) in\\
~~  executeSQLquery(query) > 0;\\
\end{code}
This code is prone to SQL injection attacks if given specially-crafted
non-alphanumeric strings.  For example, calling
\begin{code}
\t{authenticate ~"\t{admin --}" ~""} 
\end{code}
breaks the authentication mechanism because ``\t{--}'' starts a comment in SQL
and consequently ``comments out'' the password part of the query.  To prohibit
this vulnerability, the type:
\begin{code}
\t{authenticate} : \t{Credential}~$\rightarrow$~\t{Credential}~$\rightarrow$~\Bool
\end{code}
specifies that  \t{authenticate} should be applied only to alphanumeric
strings.

Next, consider the following user-interface code:
\begin{code}
let username : String = readString() in\\
let password : String = readString() in\\
authenticate username password;
\end{code}

\emph{This code is ill-typed}, since it passes arbitrary user input of type
\t{String} to \t{authenticate}.  However, proving that this code is
ill-typed is quite difficult, since it depends on complex reasoning
showing that the user-defined function \t{match} is not a tautology,
\ie\, not all \t{String}s are \t{Credential}s.

In fact, the current implementation of \sage{} cannot statically
verify or refute this code.  Instead, it inserts the following casts
at the call site to enforce the specification for \t{authenticate}
dynamically:
\begin{code}
authenticate ($\cast{\t{Credential}}{\t{username}}$) ($\cast{\t{Credential}}{\t{password}}$);
\end{code}
At run time, these casts check that \t{username} and \t{password} are
alphanumeric strings satisfying the predicate \t{match (Kleene
  AlphaNum)}.  If the \t{username} ``\t{admin --}'' is ever entered,
the cast \t{($\cast{\t{Credential}}{\t{username}}$)} will fail and halt
program execution.

As a bonus, a dynamic cast failure actually strength\-ens \sage{}'s
ability to detect type errors statically. In particular, the string
``\t{admin --}'' is a witness proving that not all \t{String}s are
\t{Credential}s, \ie, $\njudgesub{}{\t{String}}{\t{Credential}}$.
Rather than discarding this information, and potentially observing the
same error on later runs or in different programs, such refuted
subtype relationships are stored in a database.  If the above code is
later revisited, the \sage{} type checker will discover upon
consulting this database that \t{String} is not a subtype of
\t{Credential}, and it will statically reject the call to
\t{authenticate} as ill-typed.  Additionally, the database stores a
list of other programs previously type-checked under the assumption
that \t{String} may be a subtype of \t{Credential}.  These programs
may also fail at run time and \sage{} will also report that they
should be revisited using the more-informed type checker.  This
database is discussed further after the formal developments, in
Section~\ref{section:sage:db}.


\subsection{Printf}
\label{section:sage:printf}

Our final example is the classic \t{printf} function.  The number and
type of the expected arguments to \t{printf} depends in subtle ways on
the format string (the first argument).  In \sage{}, we can assign to
\t{printf} the precise type:
\begin{code}
~~printf : (format:String) -> (Printf\_Args format)
\end{code}
where the user-defined function
\begin{code}
~~\t{Printf\_Args} : String -> \type
\end{code}
returns the \t{printf} argument types for the given format
string.\footnote{The implementation of \t{Printf\_Args} is more
  complex than warrants textual inclusion.}  For example,
\begin{code}
~~\t{(Printf\_Args~"\%d\%d")} 
\end{code}
  evaluates to the type 
\begin{code}
~~Int -> Int -> Unit
\end{code}

Thus, the \sage{} language is sufficiently expressive to need no
special support for accommodating \t{printf} and catching errors in
\t{printf} clients statically. In contrast, other language
implementations require special rules in the compiler or run time to
ensure the type safety of calls to \t{printf}. For example,
Scheme~\citep{sperber_revised_2007} and
GHC~\citep{marlow_haskell_2010} leave all type checking of arguments
to the run time. OCaml~\citep{leroy_ocaml_2013}, on the other hand, performs
static checking, but it requires the format string to be constant.

\sage{} can statically check many uses of \t{printf} with non-constant
format strings, as illustrated by the following example:

\begin{verbatim}
  let repeat (s:String) (n:Int) : String =
    if (n = 0) then "" else (s \verb@^@ (repeat s (n-1)));

    // checked statically:
  printf (repeat "\%d" 2) 1 2;  
\end{verbatim}

The \sage{} type checker infers that \t{printf (repeat "\%d" 2)} has
type \\ \t{Printf\_Args~(repeat~"\%d"~2)}, which evaluates (at cast
insertion time) to $\Funty{\Int}\Funty{\Int}\Unit$, and hence this
call is well-typed. Conversely, the type checker would statically
reject the following ill-typed call:
\begin{code}
~~// compile-time error:\\
~~printf (repeat "\%d" 2) 1 false;
\end{code}
For efficiency, and to avoid non-termination, the type checker
performs only a bounded number of evaluation steps before resorting to
dynamic checking.  Thus, the following call requires a run-time check:
\begin{code}
~~// run-time error:\\
~~printf (repeat "\%d" 200) 1 2 ... 199 false;
\end{code}
As expected, the inserted dynamic cast catches the error.

The current \sage{} implementation is not yet able to statically
verify that the implementation of \t{printf} matches its specification,
\begin{verbatim}
  (format:String) -> (Printf_Args format)
\end{verbatim}
As a result, the type checker inserts a single dynamic type cast into
the \t{printf} implementation.  This example illustrates the
flexibility of hybrid checking --- the \t{printf} specification is
enforced dynamically on the \t{printf} implementation, but also
enforced (primarily) statically on client code.  We revisit this
example in Section~\ref{section:sage:printftwo} to illustrate
\sage{}'s cast insertion algorithm.


\section{The \sage{} Core Language}
\label{section:sage:lang}
\label{section:sage:syntax}

\sage{} programs are desugared into a small core language shown in
Figure~\ref{figure:sage:syntax}.  We use the same metavariable
conventions from the rest of this dissertation with few differences
that will be clear from context.  Most importantly, the syntactic sort
of types and terms is merged, so the metavariables used for terms
($\e$, $\f$, $\predp$, etc) and those used for types ($S$, $T$, etc)
now refer to the same syntactic sort. The choice of which metavariable
to use in a particular situation is only to provide some intuition.

\begin{displayfigure}{Syntax of \sage{} Core}
\label{figure:sage:syntax}
\sagesyntax
\end{displayfigure}

First-class types also allow us to express all type constructors other
than dependent function types as primitives, similar to the treatment
of conditionals (via \t{if}) and recursion (via \t{fix}) in
$\lambdah$.  This is the fruition of Type Theory: The core constructs
need not change, but form a meta-theoretical framework; we need only
discuss the semantics of the primitives we include in the framework.
The following primitives are crucial to the definition of \sage:

\begin{itemize}

\item[-] The primitive $\t{Refine}$ is used to construct executable
  refinement types. Suppose \mbox{$f:\Funty{T}{\Bool}$} is some
  arbitrary predicate over type $T$.  Then the type $\Refine~T~f$
  denotes the executable refinement type classifying all values of
  type $T$ that satisfy the predicate $f$.  Following our notation
  established in previous chapters, we use the shorthand
  \mbox{$\predty x T \predp$} to abbreviate \mbox{$\Refine{}~T~(\lam x
    T \predp)$}.

\item[-] The primitive \t{fix} enables the definition of recursive functions and
  recursive types.  For example, the type of integer lists is defined via the
  fixed point calculation:
  \[
  \t{fix}~~\type~~(\lam{L}{\type}{\t{Sum}~\Unit~(\t{Pair}~\Int~L)})
  \]
  which (roughly speaking) returns a type \t{L} satisfying the equation:
  \[
  L~~=~~\t{Sum}~\Unit~(\t{Pair}~\Int~L)
  \]
  (Here, \t{Sum} and \t{Pair} are the usual type constructors for sums and
  pairs, respectively.)

  The subtyping algorithm treats recursion in types differently than
  the operational semantics treats recursion in terms, as most recursive
  types are infinite and result in infinite reduction sequences.

\item[-] The primitive 
  \dynamic{}~\citep{thatte_quasi-static_1990,abadi_dynamic_1991} can
  be thought of as the most general type. Every value may be assigned type
  \dynamic{}, and casts can be used to convert values from type
  \dynamic{} to other types (and of course such downcasts may fail if
  applied to inappropriate values).
    
\item[-] The primitive \t{cast} performs dynamic coercions between
  types.  It takes as arguments a type $T$ and a value (of type
  $\dynamic$), and it attempts to cast that value to type $T$.  We use
  the shorthand $\cast T \, \e$ to abbreviate $\t{cast}~T~\e$. Thus,
  for example, the expression
  \[
  \cast {\predty x \Int {x\geq 0}} ~ y
  \]
  casts the integer $y$ to the refinement type of natural numbers, and fails if
  $y$ is negative. 

\item[-] The primitive $\t{Checking}$ is used to construct
  casts-in-progress while a refinement predicate is being checked.
  Following the notation established in previous chapters, we will
  write $\checking {\predty x T \predp} \predq \val$ to abbreviate
  \mbox{$\t{Checking}~T~(\lam x T \predp)~\val~\predq$}. 

\end{itemize}

\noindent
Shorthands for selected primitives are summarized in
Figure~\ref{figure:sage:shorthands}.

\begin{displayfigure}{\sage{} Shorthands}
\label{figure:sage:shorthands}
\vspace{-25pt}
\[
\begin{array}[t]{rcl}
  \cast T {}                                  &=& \t{cast}~T \\
  \predty x T \predp                         &=& \Refine~T~(\lam{x}{T}{\predp})\\
  \checking {\predty x T \predp} \predq \val &=& \t{Checking} ~ T ~ (\lam x T \predp) ~ \predq ~ \val \\
  \t{if}_T~\predp~\t{then}~\d~\t{else}~\e     &=& \t{if}~T~\predp~(\lam x {\predquick \Unit \predp} \d)~(\lam x {\predquick \Unit {\t{not}~\predp}} \e)
\end{array}
\]
\vspace{-10pt}
\end{displayfigure}

\section{Operational Semantics of \sage{} Core}
\label{section:sage:opsem}

\hhref{Figure}{figure:sage:semantics} and
\hhref{Figure}{figure:sage:castevalrules} describe the meaning of
\sage{} programs with a small-step operational semantics. Because much
of the interest lies in the meaning of primitives, we do not embed all
of this meaning in a $\delta$ function, but write out in full the
meaning of those primitives whose semantics bear discussion. We do,
however, elide standard semantics such as the basics of boolean and
arithmetic connectives. The semantics the primitive \t{if} are
presented as an example of a primitive with standard semantics.  The
semantics of \t{cast} are presented in full.

\begin{displayfigure}{Operational Semantics for \sage{} Core}
  \label{figure:sage:semantics}
  \vspace{-25pt}
  \sageevalrules
  \vspace{-20pt}
\end{displayfigure}

Evaluation is performed inside call-by-value evaluation contexts
$\cbvctx$, to model the actual \sage{} execution strategy. Arbitrary
contexts are still written $\tmctx$.

\begin{rulelist}
\item[\rel{E-App} and \rel{E-Let}] Both $\lambda$-abstractions
  and \t{let}-bindings evaluate according to $\beta$-reduction,
  replacing their bound variable with the actual value in their
  body.

\item[\rel{E-Fix}:] The operation \t{fix} is used to define recursive
  functions and types, which are considered values, and hence
  $\t{fix}~U~v$ is also a value.  However, when this construct
  $\t{fix}~U~v$ appears in a strict position (\ie, in a function
  position or in a cast), the rule~\rel{E-fix} performs one step of
  unrolling to yield $v~(\t{fix}~U~v)$.

\item[\rel{E-If1} and \rel{E-If2}:] These rules encode the
  standard definition for a Church-encoded conditional. Each
  of the branches of the conditional is a constant function 
  taking a $\unit$ parameter and returning the value of that
  branch.
\end{rulelist}

\hhref{Figure}{figure:sage:castevalrules} displays all redex evaluation
rules pertaining to casts in \sage{}. The expanded type language
and first-class types result in many more possibilities than
in $\lambdah$.

\begin{displayfigure}{Operational Semantics for \sage{} Casts}
  \label{figure:sage:castevalrules}
  \sagecastevalrules
\end{displayfigure}

\begin{rulelist}
\item[\rel{E-Cast-Dyn}] Casts to type $\dynamic$ always succeed.

\item[\rel{E-Cast-Basic}] Casts to primitive types such as \Int, \Bool,
and $\Unit$ succeed when applied to appropriate values.

\item[\rel{E-Cast-Fn}] Casts to function types are more involved.  The
  casts of \sage{} only retain their target type (an experimental
  design decision) so we require the following partial function
  \fun{domain} returns the domain of a function value.  The rule casts
  a function $\val$ to type $\funtyd x S T$ by creating a new
  function:
  \[
  \lam x S {\cast T ~ \val ~(\cast D ~ x)}
  \]
  where $D=\fun{domain}(\val)$ is the domain type of the function
  $\val$. Just as in $\lambdah$, this new function takes a value $x$
  of type $S$, casts it to $D$, applies the given function $v$, and
  casts the result to the desired result type $T$.

\item[\rel{E-Cast-Begin} and \rel{E-Cast-End}:] These two
  rules are analogous to those for $\lambdah$. To cast
  a term $\val$ to a refined type 
  $\t{Refine}~T~\f$, first $\val$ is cast to type $T$
  and the predicate $f~(\cast T \, v)$ is evaluated
  to determine its truth or falsity. If this check
  succeeds, then the result of the cast is
  $\cast T \, \val$.\footnote{There is clearly room for optimized evaluation
  strategies that duplicate less work when casting to a refinement type,
  but the current presentation has been chosen for clarity.}
  
\item[\rel{E-Cast-Type}:] Casts to type $\type$ succeed only for
  special values of type $\type$. With first-class types, the
  programmer may write nonsensical function ``types'' such as $\arrowt
  3 6$, since this is syntactically valid.  Suppose this type occurs
  in a cast $\cast {\arrowt 3 6}$. Though this term is ill-typed, the
  hybrid type checker may not be able to determine that this is so
  (for example if $3$ and $6$ were the result of a long-running
  computation) and may output the cast $\cast {\arrowt {\cast \type \,
      3} {\cast \type \, 6}}$. Later evaluation via $\rel{E-Cast-Fn}$
  will expand the cast on the domain type to $\cast {\cast \type\, 3}$
  which is not a value, and will result in a failed cast.

\end{rulelist}


\section{The \sage{} Core Type System}
\label{section:sage:types}
\label{section:sage:typesystem}

The \sage{} type system is defined via the following judgments:

\begin{itemize}
\item[-] The typing judgment $\judge \env \e T$ denotes that expression $\e$ has
  type $T$ under the assumptions in environment $\env$. 

\item[-] The subtyping judgment $\judgesub \env S T$ denotes that type $S$ is a subtype
  of $T$ under the assumptions in environment $\env$.

\item[-] The validity judgment $\judgeTP \env \predp$ denotes that $\predp$
  is valid in environment $\env$. This replaces the use
  of $\judgeimp \env \predp \predq$ in prior chapters.
\end{itemize}

All of the above judgments utilitize an environment $\env$ that binds
variables to types and, in some cases, to values.  We assume that
variables are bound at most once in an environment and, as usual, we
apply implicit $\alpha$-renaming of bound variables to maintain this
assumption and to ensure that substitutions are capture-avoiding.  In
addition, we assume that the typing environment is well-formed
according to the judgment $\judgeE{\env}$.

\subsection{Typing for \sage{} Core}

The main typing judgment 
\[
\judge \env \e T
\] 
assigns type $T$ to term $\e$ in the environment $\env$, and is defined
by the rules in Figure~\ref{figure:sage:typerules}. Note that the type
well-formedness judgment $\judgeT \env T$ from prior chapters is
merged with this judgment: Since types are terms, they are validated
by being assigned the type $\type$ via the typing judgment $\judge
\env T \type$. Thus, \Int, \Bool, and \Unit{} all have type $\type$.
Also, $\type$ itself has type $\type$.\footnote{The relationship
  $\type:\type$ results in a paradox known as Girard's
  Paradox~\citep{girard_interpretation_1972, coquand_analysis_1986,
    hurkens_simplification_1995} which allows one to construct a
  non-terminating term that is a proof of $\perp$ (falsity, or the
  empty type). When type theory is proposed as a foundation for
  mathematics, the solution is to introduce a stratification of
  $\type$ into $\type_n$ indexed by a natural number, and insist that
  $\type_n:\type_m$ only if $n < m$. However this complexity is not
  needed since the paradox (not the only one in \sage) does
  not affect the use of a language for programming.}

\begin{displayfigure}{Type Rules for \sage{} Core}
\ssp
\label{figure:sage:typerules}
\sagetyperules
\end{displayfigure}

\begin{displayfigure}{Example types of \sage{} constants}
\label{figure:sage:primtypes}
\ssp
\[
\begin{array}[t]{rcl}
  \type        &:& \type \\
  \Unit        &:& \type \\
  \Bool        &:& \type \\
  \Int         &:& \type \\
  \dynamic     &:& \type \\
  \Refine      &:& \funtyd X \type {\Funty {(\Funty X \Bool)} \type} \\
  \\
  \unit        &:& \Unit \\
  \true        &:& \predty b \Bool b \\
  \false       &:& \predty b \Bool {\t{not}~b} \\
  \t{not}      &:& \funtyd b \Bool {\predty{b'}{\Bool}{b' = \t{not}~b}} \\
  n            &:& \predty m \Int {m=n} \\
  +            &:& \funtyd n \Int {\funtyd{m}{\Int}{\predty{z}{\Int}{z=n+m}}}\\
  \t{=}        &:& \funtyd x \dynamic {\funtyd y \dynamic {\predty b \Bool {b = (x = y)}}} \\
  \\
  \t{if}       &:& \funtyd X \type {\funtyd p \Bool {}}\\
  \            & &                 \Funty   {(\Funty {\predty d \Unit p          } X)} {}\\
  \            & &                 \Funty   {(\Funty {\predty d \Unit {\t{not}~p}} X)} {}\\
  \            & &                 X\\
  \\
  \sagefix     &:& \lamt X \type {\Funty {(\Funty X X)} X} \\
  \t{cast}     &:& \lamt X \type {\Funty \dynamic X} \\
  \t{Checking} &:& \lamt X \type {} 
                   \lamt f {(\arrowt X \Bool)} {} 
                   \lamt x X {}
                   \arrowt {\predty \predp \Bool {\predp \Rightarrow f~x}} X 
\end{array}
\]
\label{figure:sage:constants}
\end{displayfigure}

\begin{rulelist}
\item[\rel{T-Prim}:] The auxiliary function $ty$ returns the type of
  the primitive $\constant$, as defined in
  Figure~\ref{figure:sage:primtypes}.  In \sage{} these types can be
  quite rich; note the heavy use of first-class types.

  Many types are precise as in $\lambdah$; an integer $n$ has the
  precise type $\predty{m}{\Int}{m=n}$ denoting the singleton set
  $\setz{n}$.

  In particular, note the type of \t{if}. The ``then'' parameter to
  \t{if} is a thunk of type $(\Funty{\predty{d}{\t{Unit}}{p}}{X})$.
  That thunk can be invoked only if the domain
  $\predty{d}{\t{Unit}}{p}$ is inhabited, \ie, only if the test
  expression $p$ evaluates to $\true$.  Thus the type of \t{if}
  precisely specifies its behavior; \sage{} has a
  \emph{path-sensitive} type system.

\item[\rel{T-Var}:] For a variable $x$ we extract the type $T$ of $x$
  from the environment, and assign to $x$ the singleton refinement
  type\footnote{The necessity of this ``selfification'' was discovered
    prior to the work of Chapter~\ref{chapter:compositionality} and
    was influenced more by \citet{ou_dynamic_2004} but seems prescient
    in retrospect.} \mbox{$\predty{y}{T}{y = x}$}.

\item[\rel{T-Fun}:] For a function
  \mbox{$\lam x S \e$} we infer the type $T$ of
  $\e$ in an extended environment and return the dependent function
  type $\funtyd x S T$, where $x$ may occur free in $T$.

\item[\rel{T-Arrow}:] The type \mbox{$\funtyd x S T$} is itself a
  term, which is assigned type $\type$, provided that both $S$ and $T$
  have type $\type$ in appropriate environments.

\item[\rel{T-App}:] This rule is completely standard for function
  application in dependent types.\footnote{The
    compositionality-restoring techniques of
    Chapter~\ref{chapter:compositional} were developed after \sage{}
    and have not been attempted in this more challenging context.} For
  an application $(\f ~ \e)$ we first check that $\f$ has a function
  type $\lamt x S T$ and that $\e$ is in the domain of $\f$.  The
  result type is $T$ with all occurrences of the formal parameter $x$
  replaced by the actual parameter $\e$.

\item[\rel{T-Let}:] For the term $\letexpt x \val S \e$ we first check
  that the type of the bound value $\val$ is $S$.  Then $\e$ is typed
  in an environment that contains both the type \emph{and} the value
  of $x$.\footnote{In the surface syntax of \sage{} a non-value may
    still be bound to an expression as in \mbox{\t{let x:S = d in e}}
    but this is desugared into $(\lam x S \e)~\d$.}  The precise
  bindings from $\t{let}$ expressions are used in the subtype
  judgment, as described below.

\item[\rel{T-Sub}:] Subtyping is allowed at any
  point in a typing derivation via this general subsumptoin rule.
\end{rulelist}

\subsection{Subtyping for \sage{} Core}

The subtype judgment 
\[
    \judgesub \env S T
\]
states that $S$ is a subtype of $T$ in the environment $\env$, and it
is defined as the greatest solution (to permit infinite types) to the
collection of subtype rules in Figure~\ref{figure:sage:subtypes}.

\begin{displayfigure}{Subtyping Rules for \sage{} Core}
\label{figure:sage:subtypes}
\sagesubtyperules
\end{displayfigure}

\begin{rulelist}
\item[\rel{S-Refl}:] Every type is a subtype of itself, axiomatically.

\item[\rel{S-Dyn}:] Every type is a subtype of the type \dynamic,
  which functions essentially as a ``top'' type in the static type
  system, though its algorithmic implementation during hybrid type
  checking differs.

\item[\rel{S-Fun}:] Subtyping between function types is entirely standard.

\item[\rel{S-Var}:] A variable may be hygienically replaced 
  with the value to which it is bound during subtyping.

\item[\rel{S-Eval-L} and~\rel{S-Eval-R}:] The subtype relation is
  closed under evaluation of terms in arbitrary positions.

\item[\rel{S-Ref-L}:] If $S$ is a subtype of $T$, then any
  refinement of $S$ is also a subtype of $T$.

\item[\rel{S-Ref-R}:] When $S$ is a subtype of $T$ we invoke the
  validity judgment $\judgeTP{\env}{f~x}$, discussed below, to
  determine if $f~x$ is valid for all values $x$ of type $S$.  If so,
  then $S$ is a subtype of $\Refine~T~f$.
\end{rulelist}

\subsection{Validity for \sage{} Core}

Our type system is parameterized with respect to the validity judgment
\[
    \judgeTP \env \predp
\]
which defines the validity of term $\predp$ in an environment
$\env$. Chapter~\ref{chapter:lambdah} demonstrates one way to
construct such a judgment using denotational semantics. Since this
chapter is concerned with the implementation scenario where validity
checking will be implemented by an off-the-shelf theorem prover, it is
more appropriate and useful to discuss the axioms to which an
implementation must adhere.  In the following, all environments and
terms are assumed to be well-formed.

\begin{requirement}[Validity Axioms]
\label{requirement:sage:axioms}
A validity judgment for \sage{} must obey the following axioms.

\begin{enumerate}
  \item Faithfulness: $\judgeTP \emptyset \true$ and $\judgenotTP
    \emptyset \false$.\footnote{Early presentations of this material
      erroneously included an environment in this axiom. We extend
      thanks to Michael Greenberg for pointing this out.}
  \item Hypothesis: If $(x : {\predty y S \predp}) \in \env$ then $\judgeTP{\env}{\subst y x {}\, \predp}$.
  \item Weakening: 
    If $\judgeTP {\env,\envextra} \predp$ then $\judgeTP {\env,\envtail,\envextra} \predp$.
  \item Substitution: 
    If $\judgeTP {\env, (x:S), \envtail} \predp$ and $\judge \env \e S$
    then $\judgeTP {\env, \subst x \e {}\, \envtail} {\subst x \e {}\, \predp}$.
  \item Definition:  
    $\judgeTP {\env, (x = \val:S), \envtail} \predp$ if and only if $\judgeTP {\env, \subst x \val {}\, \envtail}{\subst x \val {}\, \predp}$.
  \item Preservation: 
    If $\e \lred^* \e'$, then $\judgeTP{\env}{\redctx[\e]}$ if and only if $\judgeTP{\env}{\redctx[\e']}$.
  \item Narrowing:
    If $\judgeTP{\env,(x:T),\envtail} \predp$ and $\judgesub \env S T$ then $\judgeTP {\env,(x:S),\envtail} \predp$.
\end{enumerate}
\end{requirement}

In addition to the undecidability induced by theorem proving, the
subtype judgment may require an unbounded amount of evaluation during
type checking. These decidability limitations motivate the
development of the hybrid type checking techniques of the following
section.


The \sage{} type system guarantees type soundness through the standard
lemmas progress (well-typed programs can only get stuck due to failed
casts) and preservation (evaluation of a term preserves its type). The
proofs are lengthy but standard; for detail consult
\citet{gronski_sage:_2006}.

\section{Hybrid Type Checking for \sage{}}
\label{section:sage:htc}

Hybrid type checking for \sage{}, as for $\lambdah$, relies on the
cooperation between a subtyping algorithm and a cast insertion
algorithm, as well as an algorithmic theorem proving judgment. The key judgments are:

\begin{itemize}

\item[-]
  The cast insertion judgment $\judgec \env \d \e T$ elaborates the
  source term $\d$, in environment $\env$, to a safe term $\e$ adding
  casts where necessary and synthesizing $\tyT$.

\item[-]
  The cast insertion and checking judgment $\judgecc \env \d \e T$
  takes as an input the desired type $T$ and ensures that $\e$ has
  type $T$ by adding necessary casts as determined by the
  conclusions from the subtyping algorithm.

\item[-] The algorithmic subtyping judgment $\judgesubalg a \env S T$
  checks for a subtyping relationship between $S$ and $T$ and has a
  three-valued result $a$, as usual for HTC.

\item[-] The algorithmic validity judgment, also called ``theorem proving'', $\judgeTPalg a \env \predp$ attempts to prove or
  refute $\predp$, and has a three-valued result $a$ as well.

\item[-] The database of counterexamples $\judgesubdb \pno \env S T$
  refutes subtyping judgments for which a witness has been gathered
  via a runtime failure.

\end{itemize}

The cast insertion rules guarantee that the resulting program is
well-typed, and thus it can only go wrong due to failed casts.  In
addition, this property permits type-directed optimizations on
compiled code.

\subsection{Cast Insertion and Checking}

\noindent
The only judgment that directly inserts casts,
\[
\judgecc \env \d \e T
\]
is defined identically for \sage{} and for $\lambdah$.

\begin{rulelist}
  \item[\rel{CC-OK}:] When checking a whether term $\d$ of type $S$ may be considered to have
    type $T$, if the $S$ can be proven a subtype of $T$ then the term is unchanged.

  \item[\rel{CC-Chk}:] If, however, the subtype relationship can be neither proven nor refuted,
    then the check is deferred until run time by inserting the cast $\cast{T}$.
\end{rulelist}

If subtyping is refuted, the lack of an inference rule forces a static rejection of the program.

\begin{displayfigure}{Cast Insertion and Checking for \sage{} Core}
  \vspace{-2.7em}
  \label{figure:sage:algsubtyping}
  \sagecheckingrules
  \vspace{-3.5em}
\end{displayfigure}

\subsection{Cast Insertion and Synthesis}

The cast insertion and checking judgment is applied
for a full program by the cast insertion and synthesis
judgment
\[
\judgec \env \d \e T
\]
defined by rules in Figure~\ref{figure:sage:castinsert}. Many are
similar to the corresponding type rules or analogous to the rules
from Chapter~\ref{chapter:htc}, but a few have interesting new
consequences due to \sage{}'s additional features. 

\begin{displayfigure}{Cast Insertion for \sage{} Core}
  \label{figure:sage:castinsert}
  \sagecastinsertionrules
\end{displayfigure}

\begin{rulelist}
  \item[\rel{C-Var}:] This rule provides selfification of variables
    analogously to the type rule \rel{T-Var}.

  \item[\rel{C-Prim}:] Primitives are assumed to be
    well-typed, so no casts are needed.

  \item[\rel{C-Fun}:] For a function $\plam x S \e$ the input
    annotation $S$ is, itself, a term, and is checked against type
    \type{}.

  \item[\rel{C-Arrow}:] For a function type $\lamt x S T$, the two
    component types $S$ and $T$ are checked against type $\type$.

  \item[\rel{C-Let}:] The term $\letexpt x \val S \e$ is processed by
    recursively processing $\val$, $S$ and $\e$ in appropriate
    environments.

  \item[\rel{C-App1} and \rel{C-App2}:] 
    The rules for function application are more interesting.  The rule
    \rel{C-App1} processes an application $\f~\e$ by processing the
    function $\f$ to some term $\f'$ of some type $\tyU$. The type $U$ may
    be a function type embedded inside refinements as in
    \[
    \predty x {(\arrowt \Int \Int)} : x(0) > 0
    \]
    In order to extract the domain and codomain of the function, we
    use \fun{unrefine} to remove any outer refinements of $U$ before
    checking the type of the argument $\e$ against the expected type.
    Formally, \fun{unrefine} is defined as follows:
    \[
    \begin{array}{rcl} 
      \fun{unrefine}: \fun{Term} &\rightarrow& \fun{Term} \\
      \fun{unrefine}(\funtyd{x}{S}{T}) &=& \funtyd{x}{S}{T} \\
      \fun{unrefine}(\t{Refine}~T~f) &=& \fun{unrefine}(T) \\ \fun{unrefine}(S) &=&
      \fun{unrefine}(S') \qquad \mbox{if $S\lred S'$} 
    \end{array} 
    \]
    
    The last clause permits $S$ to be simplified via evaluation while
    removing outer refinements.  Given the expressiveness of the type
    system, this evaluation may not converge within a given time bound.
    Hence, to ensure that our compiler accepts all (arbitrarily
    complicated) well-typed programs, the rule~\rel{C-App2} provides a
    backup compilation strategy for applications that requires less static
    analysis, but performs more dynamic checking.  This rule checks that
    the function expression has the most general function type
    $\Funty{\dynamic}{\dynamic}$, and correspondingly coerces $\e$ to type
    \dynamic, resulting in an application with type \dynamic.
    
\end{rulelist}    

\subsection{Algorithmic Subtyping}

\noindent
For any subtype query $\judgesub{\env}{S}{T}$, the algorithmic subtyping judgment
\[ 
\judgesubalg{a}{\env}{S}{T}
\]
returns a result $a\in\setz{\pyes,\pno,\pmaybe}$ depending on whether
the algorithm succeeds in proving ($\pyes$) or refuting ($\pno$) the
subtype query, or whether it cannot decide the query ($\pmaybe$). 

Our algorithm conservatively approximates the subtype specification in
Figure~\ref{figure:sage:typerules}.  However, special care must be
taken in the treatment of \dynamic.  Since we would like values of
type \dynamic{} to be implicitly cast to other types, such as \t{Int},
the subtype algorithm should conclude
$\judgesubalg{\pmaybe}{\env}{\dynamic}{\Int}$ (forcing a cast from
$\dynamic$ to $\Int$), even though clearly
$\njudgesub{\env}{\dynamic}{\Int}$.

A na\"{\i}ve subtype algorithm could always return the result
``$\pmaybe$'' and thus trivially satisfy these requirements, but more
precise results enable \sage{} to verify more properties and to detect
more errors statically.

The rules in Figure~\ref{figure:sage:subtypingalgorithm} illustrate
the multiple sorts of reasoning that our algorithm uses. However,
these rules are not syntax directed so they do not define an algorithm
without some more description. The algorithm attempts to apply the
rules in the order in which they are presented in the figure, falling
into three main categories:

\begin{enumerate}
\item The counterexample database is checked first, so that subtyping
  can be immediately rejected if a counterexample is known.

\item Then the structural rules are tried in order to attempt to
  decompose the subtyping problem into smaller queries. If types have
  incompatible shapes, then the algorithm returns ``$\pno$''.

\item Finally, if none of the above apply, the type is evaluated for
  some finite number of steps to try to reduce it to a form where
  other rules apply. The form of evaluation is specialized to only
  perform useful reductions: In particular, we do not expand fixpoints
  unless they appear in function position.

\item If no rule applies, the algorithm returns ``\pmaybe''
\end{enumerate}

\begin{displayfigure}{Algorithmic Subtyping for \sage{} Core}
  \vspace{-2.7em}
  \label{figure:sage:subtypingalgorithm}
  \sagesubtypingalg
  \vspace{-3.5em}
\end{displayfigure}

\begin{rulelist}

\item[\rel{AS-Db}:] Before applying any other rules, the algorithm attempts to
refute that $\judgesub{\env}{S}{T}$ by querying the database of previously
refuted subtype relationships.  The judgment $\judgesubdb{\pno}{\env}{S}{T}$
indicates that the database includes an entry stating that $S$ is not a
subtype of $T$ in an environment $\env'$, where $\env$ and $\env'$ are compatible in
the sense that they include the same bindings for the free variables in $S$
and $T$.  This compatibility requirement ensures that we only re-use a
refutation in a typing context in which it is meaningful.

\item[\rel{AS-Eval-L} and \rel{AS-Eval-R}:] These two rules evaluate the terms
representing types.  The algorithm only applies these two rules a bounded
number of times before timing out and forcing the algorithm to use a different
rule or return ``?''.  This prevents non-terminating computation as well as
infinite  unrolling of recursive types.

\item[\rel{AS-Dyn-L} and~\rel{AS-Dyn-R}:] These rules ensure that any type can
be considered a subtype of $\dynamic$ and that converting from $\dynamic$ to
any type requires an explicit coercion.

\item[\rel{AS-Ref-R}:] This rule for checking whether $S$ is a subtype
  of a specific refinement type relies on a theorem-proving algorithm,
  $\judgeTPalg a \env \predp$, for checking validity. This algorithm
  is an approximation of some validity judgment $\judgeTP \env \predp$
  satisfying the axioms in
  \hhref{Requirement}{requirement:sage:axioms}. As with subtyping, the
  result $a\in\setz{\pyes, \pmaybe, \pno}$ indicates whether or not
  the theorem prover could prove or refute the validity of $\predp$.
  The algorithmic theorem proving judgment must be conservative with
  respect to the logic it is approximating, as captured in the
  following requirement:
  
  \begin{requirement}[Algorithmic Theorem Proving]
    \label{req:prover}\mbox{}
    \begin{enumerate}
      
    \item 
      If $\judgeTPalg \pyes \env \predp$ then $\judgeTP \env \predp$.

    \item
      If $\judgeTPalg \pno \env \predp$ then $\forall \env', \predp'$
      obtained from $\env$ and $\predp$ by replacing each occurrence
      of the type \dynamic{} by any type, we have that $\judgenotTP
      {\env'} {\predp'}$.
      
    \end{enumerate}
  \end{requirement}

  Our current implementation of this theorem-proving algorithm translates the
  query $\judgeTPalg a \env \predp$ into input for the Simplify theorem
  prover~\citep{detlefs_simplify:_2005}. For example, the query
  \[
  \judgeTPalg{a}{x:\predty{x}{\Int}{x \geq 0}}{x+x\geq 0}
  \]
  is translated into the Simplify query:
  \begin{center}\tt
    (IMPLIES (>= x 0) (>= (+ x x) 0))
  \end{center}
  for which Simplify returns  \t{Valid}.  Given the incompleteness of Simplify
  (and other theorem provers), care must be taken in how the Simplify
  results are interpreted. For example, for the translated version of the
  query
  \[
  \judgeTPalg{a}{x:\Int}{x*x\geq 0}
  \]
  Simplify returns \t{Invalid}, because it is incomplete for arbitrary
  multiplication.  In this case, the \sage{} theorem prover returns the result
  ``$\pmaybe$'' to indicate that the validity of the query is unknown. We
  currently assume that the theorem prover is complete for linear integer
  arithmetic.  Simplify has very effective heuristics for integer arithmetic,
  but does not fully satisfy this specification. More recent SMT provers
  are explicit about when they mean ``maybe'' instead of erroneously
  saying ``no''.

  An effective and mature hybrid type checking implementation will
  rely on compile-time evaluation of terms, a counterexample database,
  and a theorem prover that returns three-valued results.
\end{rulelist}

\noindent
Assuming that $\judgeTPalg{a}{\env}{\predp}$ satisfies Requirement
\ref{req:prover} and that $\judgesubdb{\pno}{\env}{S}{T}$ only if
$\njudgesub{\env}{S}{T}$ (\ie the database has not been corrupted), it
is straight-forward to show that the subtype algorithm
$\judgesubalg{a}{\env}{S}{T}$ satisfies Lemma~\ref{lemma:sage:algsub}.

\begin{lem}[Algorithmic Subtyping]
\label{lemma:sage:algsub}\

\begin{enumerate}

\item If $\judgesubalg{\pyes}{\env}{S}{T}$ then
  $\judgesub{\env}{S}{T}$.

\item If $\judgesubalg{\pno}{\env}{T_1}{T_2}$ then $\forall \env',S_1,S_2$
  that are obtained from $\env,T_1,T_2$ by replacing the type
  \dynamic{} by any type, we have that $\njudgesub{\env'}{S_1}{S_2}$.

\end{enumerate}
\end{lem}

\subsection{Walkthrough of hybrid type checking in \sage{}} 
\label{section:sage:printftwo}

To illustrate in detail the process of hybrid type checking for \sage{},
this section walks through an example, revisiting \t{printf}. Consider the
following term
\[
\e ~~\defeq~~ \verb|printf "%d" 4|
\]
For this term, the rule~\rel{C-App1} will first compile the
subexpression {\mbox{\t{(printf "\%d")}}} via the following
cast insertion judgment (based on the type of \t{printf} from
Section~\ref{section:sage:printf}):
\[
\judgec{\emptyset}{\mbox{\t{(printf "\%d")}}}{\mbox{\t{(printf "\%d")}}}{\mbox{\t{(Printf\_Args "\%d")}}} 
\]
The rule~\rel{C-App1} then calls the function \fun{unrefine} to
evaluate \mbox{\t{(Printf\_Args "\%d")}} to the normal form
$\Funty{\Int}{\Unit}$.  Since $4$ has type $\Int$, the term $\e$ is
therefore accepted as is; no casts are needed.

However, the computation for \mbox{\t{(Printf\_Args "\%d")}} may not
terminate within a preset time limit.  In this case, 
rule~\rel{C-App2} applies to $\e$:
\[
    (\cast{\Funty{\dynamic}{\dynamic}}{\mbox{\t{(printf "\%d")}}})~~4
\]
At run time, \mbox{\t{(printf "\%d")}} will evaluate to some function
$(\lam{x}{\Int}{\e'})$ that expects an $\Int$, yielding the
application:
\[
    (\cast{\Funty{\dynamic}{\dynamic}}{(\lam{x}{\Int}{\e'})})~~4
\]
The rule~\rel{E-Cast-Fn} then reduces this term to:
\[
    (\lam{x}{\dynamic}{\cast{\dynamic}{((\lam{x}{\Int}{\e'})~(\cast{\Int}{x}))}})~~4
\]
where the nested cast $\cast{\Int}{x}$ dynamically ensures that the
next argument to \t{printf} must be an integer.
 

\section{The \sage{} Architecture}
\label{section:sage:architecture}

\begin{displayfigure}{\sage{} Architecture}
\label{figure:sage:architecture}
\vspace{10pt}
\qquad\qquad\qquad\qquad\includegraphics{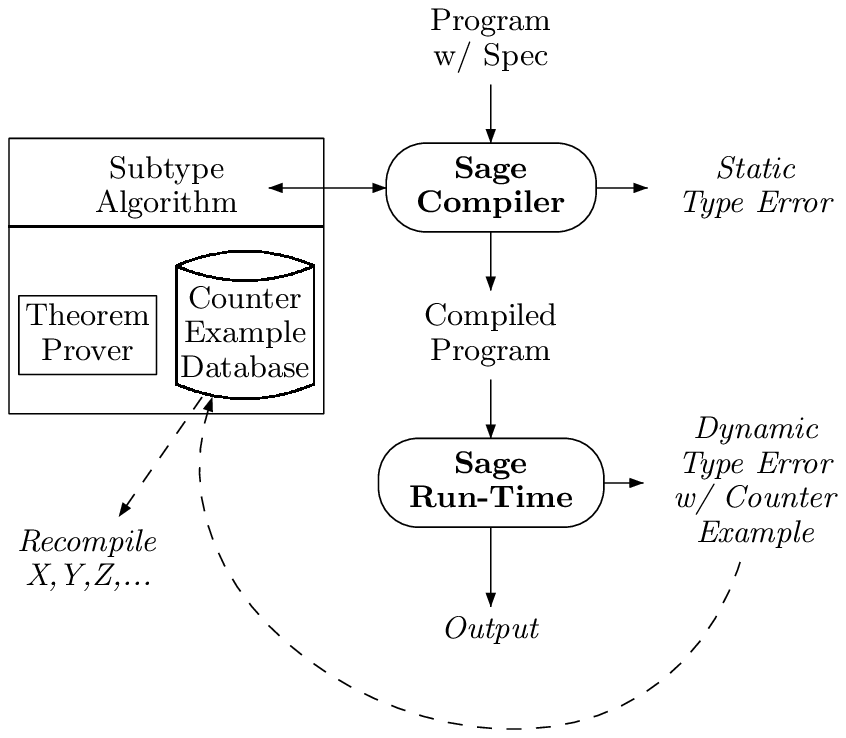}
\end{displayfigure}

The overall architecture of the \sage{} system is shown in
Figure~\ref{figure:sage:architecture}. The subtyping algorithm
described formally earlier is where the novelty of the implementation
architecture lies, specifically in the interplay between a
counterexample database and an automated theorem prover. 

\subsection{Theorem Prover} Testing subtyping between refinement types
reduces to testing implication between refinement predicates.  In
order to reason about these predicates, the subtype algorithm
translates each implication between \sage{} predicates into a validity
test of a logical formula, which can be passed to an automated theorem
prover. 
Some \sage{} terms may not readily translate to the language of the
prover, in which case we may either return $\pmaybe$ or approximate
the formula by one that is stronger (then a $\pyes$ answer
implies soundness, while $\pno$ from the prover must be translated
to $\pmaybe$) or one that is weaker (in this case it is $\pyes$ that
must be translated to $\pmaybe$).

\subsection{Counter-Example Database} 

Suppose \sage{} inserts the cast $(\cast T ~ \e)$ because it cannot
prove or refute some subtype test $\judgesub{\env}{S}{T}$.  If that
cast fails, the run time inserts an entry into the database asserting
that $\njudgesub{\env}{S}{T}$ (the syntactic representation of which
is is $\judgesubdb \pno \env S T$) The subtype algorithm consults this
database during type checking and will subsequently reject any program
that relies on $S$ being a subtype of $T$.  Thus, dynamic type errors
actually improve the ability of the \sage{} type checker to detect
type errors statically.

Moreover, when such a cast fails, \sage{} will report a list of known
programs that contain a cast that this counterexample refutes, since
these programs may also fail at run time.  Thus, the counter-example
database functions somewhat like a regression test suite, in that it
can detect errors in previously type-checked programs.

Function casts must be treated with care to ensure blame is assigned
appropriately upon failure~\citep{findler_contracts_2002}.  In
particular, if a cast inserted during the lazy evaluation of a
function cast fails, an entry for the original, top-level function
cast is inserted into the database in addition to the ``smaller'' cast
on the argument or return value. The ``larger'' counterexample more
directly corresponds to the error in the original source code, while
the ``smaller'' counterexample refutes potentially more subtyping
queries. The details of this process have not been formally modeled.

Currently we prototype via a local database, but in practice this
would take the form of a ``crash report'' sent back to a program's
maintainer. Such reports have become ubiquitous subsequent to this
research; this confluence of our experiments and social currents has
opened up an exciting avenue by which to integrate our techniques into
common practice.  Over time, we predict that the database will grow to
be a valuable repository of common but invalid subtype relationships,
leading to further improvements in the checker's precision and less
reliance on inserted casts.


\label{section:sage:db}

\section{Experimental Results}
\label{section:sage:experiments}
\label{section:sage:implementation}

Our prototype \sage{} implementation consists of roughly 5,000 lines
of OCaml code, including parsing, desugaring, primitive definitions,
operational semantics, and cast insertion. The theorem prover used is
Simplify~\citep{detlefs_simplify:_2005} and the counterexample
database is prototyped using a simple local file and serialized OCaml
data structures.  The complete implementation is available at
\url{http://sage.soe.ucsc.edu/}.

We evaluated the \sage{} language and implementation using the
benchmarks listed in Figure~\ref{tab:decisions}.  The program
\t{arith.sage} defines and uses a number of mathematical functions,
such as \t{min}, \t{abs}, and \t{mod}, where refinement types provide
precise specifications.  The programs \t{bst.sage} and \t{heap.sage}
implement and use binary search trees and heaps, and the program
\t{polylist.sage} defines and manipulates polymorphic lists.  The
types of these data structures ensure that every operation preserves
key invariants.  The program \t{stlc.sage} implements a type checker
and evaluator for the simply-typed lambda calculus (STLC), where
\sage{} types specify that evaluating an STLC-term preserves its
STLC-type.  We also include the sorting algorithm \t{mergesort.sage},
as well as the \t{regexp.sage} and \t{printf.sage} examples discussed
earlier.

Figure~\ref{tab:decisions} characterizes the performance of the
subtype algorithm on these benchmarks.  We consider two configurations
of this algorithm, both with and without the theorem prover.  For each
configuration, the figure shows the number of subtyping judgments
proved (denoted by $\pyes$), refuted (denoted by $\pno$), and left
undecided (denoted by $\pmaybe$).  The benchmarks are all well-typed,
so no subtype queries are refuted.  Note that the theorem prover
enables \sage{} to decide many more subtype queries.  In particular,
many of the benchmarks include complex refinement types that use
integer arithmetic to specify ordering and structure invariants;
theorem proving is particularly helpful in verifying these benchmarks.

Our subtyping algorithm verifies a large majority of subtype tests.
Only a small number of undecided queries result in casts.  For
example, in \t{regexp.sage}, \sage{} cannot statically verify
subtyping relations involving regular expressions (they are checked
dynamically) but it statically verifies all other subtype judgments.
Some complicated tests in \t{stlc.sage} and \t{printf.sage} must also
be checked dynamically.

Despite the use of a theorem prover, type-checking times for these
benchmarks is quite manageable.  On a 3GHz Pentium 4 Xeon processor
running Linux 2.6.14, type-checking required fewer than 10 seconds for
each of the benchmarks, except for \t{polylist.sage} which took
approximately 18 seconds.  We also measured the number of evaluation
steps required during each subtype test.  We found that 83\% of the
subtype tests required no evaluation, 91\% required five or fewer
steps, and only a handful of the the tests in our benchmarks required
more than 50 evaluation steps. Of course, evaluation steps are not
good measurements for run-time performance, as a mature implementation
may have a highly optimized method of normalizing types, but this
serves as a qualitative indication that the amount of evaluation is
extremely small or none.

\begin{table}
  \begin{tabular}{l|r|r@{~~}r@{~~}r|r@{~~}r@{~~}r}
    & Lines    &  \multicolumn{3}{c|}{Without Prover} &
    \multicolumn{3}{c}{With Prover} \\
    \multicolumn{1}{c|}{Benchmark} & \multicolumn{1}{@{}c@{}|}{of code} 
    & $ ~\surd ~ $ & $ ~~~~? $ & $ ~~\times $ 
    & $ ~\surd ~ $ & $ ~~~? $ & $ ~~\times $ 
    \\
    \hline   
    \texttt{arith.sage} & 45 & 132 & 13 & 0 & 145 & 0 & 0 \\
    \texttt{bst.sage} & 62 & 344 & 28 & 0 & 372 & 0 & 0 \\
    \texttt{heap.sage} & 69 & 322 & 34 & 0 & 356 & 0 & 0 \\
    \texttt{mergesort.sage} & 80 & 437 & 31 & 0 & 468 & 0 & 0 \\
    \texttt{polylist.sage} & 397 & 2338 & 5 & 0 & 2343 & 0 & 0 \\
    \texttt{printf.sage} & 228 & 321 & 1 & 0 & 321 & 1 & 0 \\
    \texttt{regexp.sage} & 113 & 391 & 2 & 0 & 391 & 2 & 0 \\
    \texttt{stlc.sage} & 227 & 677 & 11 & 0 & 677 & 11 & 0 \\
    \hline
    Total & 1221 & 4962 & 125 & 0 & 5073 & 14 & 0 \\
  \end{tabular}
  \caption{Subtyping Algorithm Statistics}
  \label{tab:decisions}
\end{table}


\chapter{Contemporaneous Related Work}

\label{chapter:relatedwork}

Subsequent to and simultaneous with the work presented here, many of
the fields with which this dissertation is in conversation have
experienced advances from other research. This chapter discusses this
related work in four categories, roughly parallel to those from
\hhref{Chapter}{chapter:background}: Dynamic contracts
(\hhref{Section}{section:relatedwork:contracts}), static contract
checking (\hhref{Section}{section:relatedwork:esc}),
dependent/refinement types
(\hhref{Section}{section:relatedwork:refinementtypes}), and gradual
types (\hhref{Section}{section:relatedwork:gradualtypes}). 

\section{Dynamic Contracts}
\label{section:relatedwork:contracts}

The higher-order contracts of \citet{findler_contracts_2002} have seen
many advances since the research presented in this dissertation was
undertaken. Let us consider them in three broad categories:
foundations for higher-order contracts and blame, explorations of
contracts in typed languages, and static checking of contracts.

\subsection{Foundations for untyped higher-order contracts and blame}

\citet{findler_contracts_2002} does not directly address the question
of what it means for a program to satisfy a contract.  Instead, the
meaning of contract satisfaction is left implicit in the semantics of
contract checking: If the contract does not blame a piece of code for
a violation, then that code obeys the contract. Subsequent examination
of the meaning of contract satisfaction and blame is closely related
to the dynamic checking of executable refinement types. In turn, the
correspondence between refined notions of correctness for dynamic
checking and the firmly established definition of type soundness is a
primary contribution of variants of $\lambdah$, especially
$\lambdahtc$.

\citet{blume_sound_2006} and \citet{findler_contracts_2006}
explore denotational models of contracts in an untyped setting,
proposing non-tautological, implementation-independent, definitions of
contract satisfaction. In the process, they correct some missed
contract violations from \citet{findler_contracts_2002}. The
denotational semantics for executable refinement types of
\hhref{Chapter}{chapter:lambdah} are inspired by these denotational
models, but made simpler by leveraging the simply-typed
$\lambda$-calculus.

In all of the above research on contracts, a contract is annotated
with two blame labels: One for the term it is meant to describe, and
one for its context. If the term violates its contract, then its label
receives the blame. For example, the term $3$ does not satisfy the
contract $\predty x \Int {x=4}$ so it will be blamed if the contract
is applied. On the other hand, if the context violates the contract of
a term, then the context's label receives blame. For example, if
$\f$ has the contract $\arrowt {\predty x \Int {x>0}} \Int$ and the
context attempts to evaluate $f~0$, then the context is to blame for
the resulting contract violation.

\citet{gronski_unifying_2007} question whether both blame labels are
necessary, and answer ambivalently. They exhibit a mapping from a
language $\lambda_C$ with dynamic contracts that are independent of
the type system to a language like $\lambdahtc$ with type casts that
are tightly integrated into the type system. In $\lambda_C$, contract
checks are of the form ${\cast T}^{l,l'}$ -- they have \emph{one}
contract and \emph{two} blame labels. In their variant of
$\lambdahtc$, type casts are of the form ${\casttwo S T}^l$ -- they
have \emph{two} types per cast but just \emph{one} blame label.  A
single blame label suffices per cast, but each contract check from
$\lambda_C$ is interpreted as two composed casts carrying each
label. In essence, the contract check
\[
{\cast T}^{l,l'}
\]
is translated into the composed type casts
\[
{\casttwo {\Base T} T}^{l'} \circ {\casttwo T {\Base T}}^l
\]
where one cast only ever blames the term to which the contract is
applied and one cast only every blames its context. This decomposition
corresponds with the reasoning of \citet{hinze_typed_2006} and
\citet{wadler_well-typed_2009}, each of whom disentangle the
``positive'' and ``negative'' components of a contract.

\citet{greenberg_contracts_2012}
follows up \citet{gronski_unifying_2007} with an examination of
whether a domain contract is rechecked when the input to a function
appears in the codomain contract, yielding distinct contract
disciplines.  When the domain contract is not rechecked, the system is
called ``lax''. A function contract $(\cast {\lamt x {T_1} {T_2}} ~
\f)~\e$ in in the lax system reduces to
\[
\cast {\subst x \e {} \, {T_2}}
~~
(f ~~(\cast {T_1} ~ \e))
\]
When function contracts are reechecked, the system is called ``picky''.
In the picky system, the same contract check reduces to
\[
\cast {\subst x {\cast {T_1} ~ \e} {} \, {T_2}}
~~
(f ~~(\cast {T_1} ~ \e))
\]
The lax discipline misses violations of $T_1$ in the codomain $T_2$,
thus may lead to erroneous crashes; it is incorrect regardless of
blame. \citet{dimoulas_contract_2011}, \citet{dimoulas_correct_2011}
and \citet{dimoulas_complete_2012} further examine the two-label
contract disciplines and expose that while the lax discipline misses
some violations, the picky system may blame the wrong module. This is
resolved by introducing a third blame label for the contract itself,
treating it as an independent party mediating the interaction between
a term and its context. Notably, the lax/picky distinction is
impossible in $\lambdahtc$. Supposing $\f$ has original type $\lamt x
{S_1} {S_2}$, the analogous term $(\casttwo {\lamt x {S_1} {S_2}}
{\lamt x {T_1} {T_2}} ~ \f)~\e$ reduces to
\[
\casttwo {\subst x {\casttwo {T_1} {S_1}\, \e} {}\, S_2} {\subst x \e {}\, T_2} 
~~
(f ~~(\casttwo {T_1} {S_1} ~ \e))
\]
This enforcement of a function contract is entirely derived from the
types; there is no choice of where to recheck the domain contract.
The domain contract $T_1$ is \emph{not} rechecked in $T_2$, because
the type system makes any violation impossible. Conversely, it
\emph{must} be rechecked in $S_2$, otherwise this term would be ill-typed.
If $S_2$ and $T_2$ are function types, then this does result in checks
that are not present in the lax system, but fewer than in the picky
system. The type system has yielded an integral insight into exactly
which checks are needed, and why.

\subsection{Higher-order contracts for lazy, typed languages}

The foundational work for contracts took place in an untyped setting,
but this dissertation is rooted firmly in type theory. The theory of
executable refinement types provides one possible foundation for
contracts in a typed setting. But alongside this theoretical work,
other research on the foundations of typed contracts has been driven
by implementations of typed higher-order contracts, generally as a
library rather than as a language construct.

\citet{hinze_typed_2006} presents an implementation of contracts as a
library for Haskell. Unfortunately, applying a contract to a term can
change the semantics of the term by altering its strictness.
\citet{chitil_lazy_2004},
\citet{chitil_monadic_2007,chitil_pattern_2007}, and
\citet{chitil_semantics_2011} develop a sound semantics for lazy
assertions (which do not include blame or higher-order functions) but
\citet{degen_true_2009,degen_interaction_2012} demonstrates some cases
in which lazy assertions are not suitable for use in contracts due to
incomplete contract monitoring. \citet{chitil_practical_2012} presents
an implementation of contracts including blame and ``picky'' semantics
in Haskell and briefly claims that it is easy to incorporate the
``indy'' monitoring of \citet{dimoulas_correct_2011}.

The feasibility of implementing typed contracts as a library is a
testament to the power of modern type systems and languages, but
executable refinement relate the correctness of the contract system to
the correctness of the type system in a way that is not possible when
contracts are implemented as a library. Hybrid type checking, for
example, would add safety while improving performance, but requires
contracts and types to share their specification language.

Executable refinement types are related to the discussion of laziness
only to the extent that laziness is inextricably related to concerns
of complete monitoring, but some of the properties promoted in arguing
for the correctness of the implementations may be interesting to
consider both for the semantics of contract enforcement in
$\lambdahtc$ and for the denotational interpretation of executable
refinement types.

\subsection{Parametric polymorphism in contracts}

The languages in this dissertation lack parametric polymorphism.
While functions in \sage{} may take a type as a parameter, the
property of relational parametricity is certainly not ensured.
Parametric polymorphism is common in modern typed programming
languages, but it is generally a feature of a static type system. The
foundation of \emph{dynamic} enforcement of parametric polymorphism is
more recent.

\citet{pierce_relating_2000} and \citet{sumii_bisimulation_2004}
connect the dynamic sealing of
\citet{morris_protection_1973,morris_types_1973} with parametricity
using bisimulation, and this technique was refined by
\citet{sangiorgi_logical_2007}. In these systems, polymorphic values
are essentially encrypted so that it is not possible to violate
parametricity.

\citet{guha_relationally-parametric_2007} present initial approaches
to using blame instead of encryption to enforce parametricity, then
\citet{matthews_parametric_2008} and \citet{ahmed_blame_2009} combine
the technique with gradual typing (discussed below in
\hhref{Section}{section:relatedwork:gradualtyping}).

\citet{belo_polymorphic_2011} supports a combination of executable
refinement types and polymorphic contracts, including refinements of
any type. Unfortunately their refinement system is considerably
weakened to only include closed refinements, and their type
system is noncompositional in a fundamental way, so the result
is not directly applicable nor comparable.

Combining the techniques presented in this dissertation and advances
in dynamic enforcement of parametric polymorphism is a promising
approach both to the theoretical goal of building a unified
theory of static and dynamic polymorphism and towards the
pragmatic goal of an effective language with both hybrid
types and guarantees of parametricity.

\section{Static Contract Checking}
\label{section:relatedwork:esc}

The field of static contract checking is broad and not all directly
relevant, so this section reviews only the most closely related work,
and some highlights from the rest of the field of ``SMT as a
platform''~\citep{bierman_semantic_2010}.  In particular, ESC/Haskell
and Liquid Types are in direct conversation with work on hybrid type
checking and executable refinement types

\subsection{ESC/Haskell}

\citet{xu_extended_2006} introduces ESC/Haskell, named by analogy with
ESC/Java, that statically checks contract annotations in Haskell
programs. ESC/Haskell verifies specifications by first inserting the
symbol \t{BAD} where a contract failure would occur, then attempting
to eliminate all occurrences of \t{BAD} with inlining, symbolic
evaluations, and counterexample-guided unrolling. If the symbol
remains in the program, then the program is rejected. 

In order to discuss the correctness of ESC/Haskell,
\citet{xu_static_2009} gives a declarative definition of what it means
for a term to satisfy a contract, similar to those discussed in
\hhref{Section}{section:relatedwork:contracts} and the denotational
model of \hhref{Section}{section:lambdah:denotation}, and proves that
if ESC/Haskell eliminates all occurrences of \t{BAD} and the term is
well-typed (like us, leveraging the underlying type system in a
crucial way) then the term satisfies the contract.

\citet{xu_hybrid_2012} then changes context to O'Caml, a strict
language, and makes two significant changes: It leverages an SMT
solver to determine reachability of code and allows \t{BAD} to be left
in the program optionally and checked at runtime. In the latter
situation no program is rejected statically, unlike HTC. 

\subsection{Liquid Types}
\label{section:relatedwork:liquidtypes}

\citet{rondon_liquid_2008} introduces \emph{logically qualified
  types}, or ``liquid types'' for short, which are comparable to
executable refinement types in that they are based upon predicates
written in the programming language at hand, but given meaning by
translation to SMT\footnote{\emph{Satifiability Modulo Theories}, or
  SMT, is a now-standard technique for automated theorem provers.}
terms.  These predicates are combined with the abstract interpretation
technique of \emph{predicate abstraction} to automatically infer
refinements including loop invariants, which are verified statically.
A programmer omits all refinements from their program but provides a
set of template \emph{logical qualifiers} such as $\{x > \star, \star
> x, x \geq \star\}$ which are instantiated by replacing $\star$ with
arbitrary program variables until constraints are satisfied or the
program is proven to be ill-typed.\footnote{ The term ``liquid'' type
  plays two roles: both as name for the whole system, but also for
  types whose refinements fall within the language of logical
  qualifiers.  } In addition to supporting nearly full type
reconstruction for programs whose predicates fall within the provided
template language, \citet{rondon_liquid_2008} includes an
implementation for O'Caml and hence support ML-style
polymorphism. \citet{terauchi_dependent_2010} demonstrates how to omit
the set of template logical qualifiers by using counterexample-guided
refinement and an interpolating theorem prover to generate new
templates until either a program is found to not be typable (vs the
templates simply being too coarse) or a suitably fine-grained
refinement is discovered.  \citet{rondon_low-level_2010} applies
liquid types to C, demonstrating the generality of the approach.

\citet{kawaguchi_type-based_2009} incorporates a modified form of
dependent sum type to use the liquid types technology for verifying
recursively-defined data types. They also automatically convert
``measures'', designated functions defined via catamorphism, into
refinements on the constructors of a data type, the initial refinement
design of \citet{freeman_refinement_1991}. 

\citet{jhala_hmc:_2011} explains the underlying technology in another
way by reframing type reconstruction steps analogous to those from
\citet{knowles_type_2007} and \citet{rondon_liquid_2008} as a
translation from a higher-order functional program to a first-order
imperative program whose verification implies the verification of the
source program. \citet{chugh_nested_2012} embeds the typing and
subtyping relations themselves into the SMT logic, in order to reason
about them \emph{within} refinement predicates in a dynamically typed
language.

\citet{vazou_abstract_2013} adds the ability for a function to be
polymorphic over refinement predicates, so that even monomorphic
functions such as $\t{max} : [\Int] \rightarrow \Int$ may be given
types such as $[\predty x \Int {isPrime~x}] \rightarrow {\predty x
  \Int {isPrime~x}}$. While one could already write max as a function
of type $\lamt p {(\Int \rightarrow \Int)} {\arrowt {\predty x \Int
    {p~x}} {\predty x \Int {p~x}}}$ with equivalent proof obligations,
this is extremely cumbersome and requires anticipation of all places
where a refinement may be added.

\subsection{Other work}

\citet{bengtson_refinement_2008} augments the F\# language with
refinement predicates written directly in an SMT-compatible constraint
language to statically verify security properties of programs such as
encryption protocols and access control implementations.

\citet{bierman_semantic_2010} embeds the type relation of a small
functional calculus into the logic of SMT and shows that subtyping
can then be implemented as implication between predicate interpretations
of types.

\section{Dependent/Refinement Types}
\label{section:relatedwork:refinementtypes}
\label{section:relatedwork:dependenttypes}

This section briefly discusses the continuation of the work on
refinement types \emph{other than} those expressed as logical
predicates ornamenting structural types. Programming directly in
powerful languages based on Type Theory is now also becoming more
common, and each of these allows the direct expression of many of the
ideas promoted in refinement type systems.

The Stardust language \citep{dunfield_refined_2007} serves as a good
summary of the field, as well as a novel system, by combining
\emph{datasort refinements} \citep{freeman_refinement_1991} and
\emph{index refinements} \citep{xi_dependent_1999} into single
language that also includes \emph{intersection types}
\citep{davies_intersection_2000} and \emph{union types}
\citep{dunfield_type_2003, dunfield_tridirectional_2004}.

Epigram~\citep{mcbride_view_2004},
Agda~\citep{norell_dependently_2009}, and
Coq~\citep{the_coq_development_team_coq_2012} are each languages based
on Type Theory where via propositions-as-types correspondence
programmers may explicitly prove correctness of their programs (or any
other property they may choose). Proofs are generally quite
challenging and manual, but proponents often profess a great
appreciation of the process of structuring their programs to be
provable. Each of these languages supports and promotes programming
with indexed types.

\citet{atkey_when_2011} and \citet{dagand_transporting_2012} examine
the foundations of when a particular recursive function over a
recursively-defined type may be expressed as an index to its recursive
formulation, offering a foundational look at what
\citet{kawaguchi_type-based_2009} calls a ``measure''.

$\Omega$mega~\citep{sheard_putting_2005, sheard_type-level_2007} is a
proposal that combines indexed types and user-defined kinds with
common extensions of Haskell, specifically generalized algebraic data
types (GADTs, also a feature of all systems built on Type Theory)
and type-level functions, to provide many of the benefits of 
programming in Type Theory while retaining a familiar Haskell-esque flavor.

Concoqtion~\citep{fogarty_concoqtion:_2007} layers Coq proofs atop
MetaOCaml to allow manual proofs of precise specifications. Hoare Type
Theory and its language YNot \citep{nanevski_polymorphism_2006,
  nanevski_abstract_2007, nanevski_ynot:_2008,
  nanevski_verification_2011, nanevski_dependent_2013} also blend Coq
proofs into a programming language but specifically add monadic
side-effecting computations which may be reasoned about with
Hoare-style axiomatic semantics that can be related to facts proven
about program values.

The balance between manual proof, automation and assurance is at stake
in all such systems. A program, viewed as a document explicating the
problem it solves, may be enhanced by explicit proofs, or its meaning
may be obscured. One may also ask whether it is necessary for the
program to be written in the same language used for carrying out
proofs. Could a checker of executable refinement types generate proofs
in Coq or another metatheoretical framework? In what manner can manual
proof or other forms of automation be integrated into hybrid type
checking?

\section{Gradual Typing}
\label{section:relatedwork:gradualtyping}
\label{section:relatedwork:gradualtypes}

Gradual typing \citep{siek_gradual_2006} is so named because it
addresses the problem of mixing typed and untyped program text while
being able to recover the root cause of errors. The formal techniques
of gradual typing resemble those of hybrid type checking, while the
use of blame is derived directly from research on higher-order
contracts. Thus gradual typing is more closely related than it may
appear at first.

Unlike refinement types, gradual typing is based not on subtyping but
on an idea of type consistency. Two types are consistent if they can
be unified by replacing occurrences of \t{Dynamic} in their syntax
with arbitrary other types. For example $\arrowt \Int \t{Dynamic}$ is
consistent when $\arrowt \Int \Int$ but also with simply $\t{Dynamic}$.
Unlike subtyping, compatibility cannot be transitive or else all types
would be compatible. Gradual typing formalizes an idea that is
implicit is \sage{}'s algorithmic treatment of $\t{Dynamic}$.

\citet{siek_gradual_2007} applies gradual typing to an object-oriented
language, adding subtyping alongside
consistency. \citet{siek_gradual_2008} adds unification-based type
inference. Notably, many na\"{\i}ve approaches to adding unification
fail to maintain the ``gradual'' flavor. \citet{rastogi_ins_2012}
explores dataflow-based type inference for gradual typing in the
context of ActionScript.  \citet{sagonas_gradual_2008} applies gradual
typing to existing Erlang codebases, while \citet{ren_ruby_2013} does
the same for Ruby.  \citet{wadler_well-typed_2009} examines the use of
blame in a language blending statically typed code and dynamically
typed code, with contracts as an intermediary, and proves that only
the dynamically typed portion can ever receive blame. By a careful
tracking of blame labels, a new syntactic technique is used to prove
soundness of a gradual typing system somewhat simpler than $\lambdah$.
\citet{ahmed_blame_2009} incorporates recent research on polymorphic
contracts into a gradually typed language.  \citet{ina_towards_2009,
  ina_gradual_2011} apply this to a more mainstream object-oriented
setting where polymorphism is referred to as ``generics''.

\citet{swamy_theory_2009} attempts a unified theory of type-directed
coercion insertion. This would, if successful, apply to both hybrid
type checking and gradual typing. Subtyping, where it appears in HTC,
is replaced by \emph{coercion generation}. Instead of a subsumption
rule, the theory features coercion generation at those same points, in
a reorganization of the same reasoning. Consider this rule as an
illustration:
\[
\inferrule*{
  \judgec \env d e S
  \and
  \judgect \env {\casttwo S T} c
}{
  \judgecc \env d {\langle\langle c\rangle\rangle(e)} T
}
\]
where $\judgect \env {\casttwo S T} c$ states that in order to coerce
a term of type $S$ to type $T$, the coercion $c$ must be applied. In
the event that $c$ is the identity, $\langle\langle c
\rangle\rangle(e)$ elides its application, resulting in just
$e$.\footnote{The operation $\langle\langle c\rangle\rangle$ is a
  metafunction carried out on the syntax, not a term of the language.}
This framework can generate a system equivalent to hybrid type
checking via the following two rules for generating coercions
\begin{mathpar}
\inferrule*{
  \judgesubalg \pmaybe \env S T
}{
  \judgect \env {\casttwo S T} {\casttwo S T}
}

\inferrule*{
  \judgesubalg \pyes \env S T
}{
  \judgect \env {\casttwo S T} {id}
}
\end{mathpar}
where $id$ is a designated symbol and
\begin{eqnarray*}
  \langle\langle {\,\casttwo S T\,} \rangle\rangle (e) & \defeq & \casttwo S T ~ e \\
  \langle\langle id\rangle\rangle(e) & \defeq & e
\end{eqnarray*}
Gradual typing may be similarly generated by using consistency checks
to determine the coercion.

Implementation of gradual typing faces performance challenges from the
use of higher-order coercions and also simply having dynamically typed
portions.  \citet{siek_threesomes_2009} and
\citet{herman_space-efficient_2010} address the problem of the
proliferation of function contract cast wrappers, which pose a serious
problem to adoption of higher-order constracts due to memory
consumption. \citet{allende_application_2011} proposes some initial
steps towards recovering type-based optimizations for programs using
gradual typing.

\citet{bayne_always-available_2011} allows a programmer to work while
their program is in an ill-typed state by optionally deferring type
errors until run time. Their proposal works by replacing ill-typed
terms by $\perp$, which may be assigned any type. The Glasgow Haskell
Compiler now incorporates this feature. In a hybrid type checking
system, this may be viewed as optionally using a theorem prover that
converts all ``no'' answers to ``maybe'' during development.

\citet{wolff_gradual_2011} blends gradual typing with
\emph{typestate}, where imperative update may change the type of an
imperative reference cell. In the context of executable refinement
types, this may present a way to temporarily abandon refinements and
then dynamically check that they have been restored.

\citet{politz_progressive_2012} takes a complementary approach to
gradual typing: Instead of offering a single form of soundness and
allowing the program to omit types here and there to control how much
they take advantage of the guarantee, their system offers control over
what \emph{guarantee} the programmer wants and then enforces types
appropriately. This obviously applies to HTC, and especially
\sage{}. Any HTC system incorporating may present itself as different
subtyping algorithms that are more or less confident in their
rejections of programs.

\chapter{Conclusion}
\label{chapter:conclusion}

\section{Summary}

Types are a proven and widespread language for describing programs
rigorously, but most type system express only very coarse
specifications. Refinement types increase the expressiveness of types,
but remain limited in order to ensure decidable and efficient
checking. Dynamic contracts and extended static checking each support
more expressive specification languages that structural type system,
but suffer from incomplete checking and excessive false positives,
respectively.

Executable refinement types enrich traditional type systems with
precise specifications as expressive as those of dynamic contracts and
extended static checking. This expressiveness renders traditional forms
of type checking and type reconstruction undecidable.

This dissertation has introduced and explored executable refinement
types from theoretical and practical perspectives, yielding the
following contributions:

\begin{itemize}


\item[-]
Executable refinement types may be interpreted via a lightweight
denotational semantics as sets of terms of the simply-typed
$\lambda$-calculus. Based on this denotational semantics, it follows
that implication and subtyping are proper preorders and that the type
system of $\lambdah$ is sound.  In addition to type safety, this
dissertation presents a new definition of extensional equivalence for
dependently typed calculi and proves that for $\lambdah$ extensional
equivalence implies contextual equivalence.


\item[-]
\hhref{Chapter}{chapter:htc} defines hybrid type checking, a technique
combining static type checking with dynamic contract checking, and
applies it to executable refinement types via the calculus
$\lambdahtc$.  By employing hybrid type checking, we may ensure that
all specification violations are caught either statically or
dynamically.  Furthermore, for any decidable approximation of the
$\lambdahtc$ type system, there is some program which hybrid type
checking rejects \emph{statically} that the decidable approximation
fails to reject.  Using our new proof extensional equivalence, we have
quickly proved that whenever a program is well-typed, all inserted
run-time checks succeed.


\item[-]
Type reconstruction as traditionally defined conflates the
reconstruction of omitted types with the checking of well-typedness of
a program. This conflation obscures useful distinctions between the
difficulty of the type reconstruction problem and the type checking
problem, and renders the definition useless for undecidable type
systems.  This dissertation proposes a novel definition of type
reconstruction that applies to undecidable type systems while
coincides with the old definition for decidable type systems.  Using
this new definition, a type reconstruction algorithm for executable
refinement types is presented.


\item[-]
Standard formulations of dependent types rely on non-compositional
reasoning, violating the abstraction that types represent. We have
shown how to restore compositionality by a new use of existential
types.  Even with this formulation that is restricted to compositional
reasoning, type checking can still captures the complete semantics of
a term -- the precision of the system is entirely parameterized by the
precision of the types of primitives.  When all predicates are written
in a decidable language, compositional type checking of executable
refinement types is decidable.


\item[-]
\sage{} is a prototype implementation of hybrid type checking for
executable refinement types. Experiments show that type checking time
-- including use of an SMT solver -- is short and that the number of
run-time checks inserted for many examples is small.  \sage{} also
demonstrates an architecture whereby any run-time failure is fed back
into the static portion of the hybrid type checker and becomes a
static type error: Every run-time failure can happen at most once.

\end{itemize}

\section{Open Avenues}

In addition to laying a foundation for future work that has yet
to be conceived, this dissertation suggests concrete directions
for research related to executable refinement types.

\subsection{Parametric Polymorphism}

Parametric polymorphism is standard in modern type theories, and
combines well with static refinement type systems such as Dependent ML
and Liquid Types. But parametricity for dynamic contracts is still
advancing. 

The addition of parametric polymorphism demands a new development
parallel to this dissertation: While types of $\lambdah$ are easily
interpreted as sets of $STLC$ terms, does this approach immediately
generalize to System F? How is extensional equivalence affected? What
are the consequences for hybrid type checking when the semantics of
relationally parametric contracts must be related to the semantics of
polymorphic types?

\subsection{Multiple Representations For Specifications}

Executable refinement types are easy for certain purposes, such as
translation to an SMT prover, or communicating with other humans. But
indexed types enforce many interesting specifications by carefully
architecture types for data constructors. \citet{atkey_when_2011},
\citet{dagand_transporting_2012}, and \citet{vazou_abstract_2013} all
investigate how to relate refinements on data constructors to
global properties of a data structure. This problem is far from solved,
and integral to making executable refinement types more usable.
What specifications do we wish to write and which specifications do
we hope for an implementation to derive from them? Can this
be automated sufficiently that we can always write simply the global
property of interest?

More generally, the cited works study equivalences between
representations that are best for different purposes: Some
representations are more effective for static checking, others more
effective for human comprehension. The practice of hybrid type
checking introduces a third goal to start to consider: \emph{effective
  run-time checking}.  This means using state-of-the-art data
structures, potentially mutable data, and probably much more obscure
code. Dynamic contract systems support arbitrary specifications, but
connecting this representation to the former two remains incomplete.

\subsection{Contract Properties}

This dissertation has focused on establishing that executable
refinement types enjoy the desirable properties expected of a type
system. However, contract systems have other correctness properties,
such as blame-correctness and complete monitoring. When a
counterexample is reported back to an implementation such as \sage{},
how does correct blame affect \eg the error message to report?

\subsection{Compositionality}

Compositionality of the sort analyzed in
\hhref{Chapter}{chapter:compositionality} is a property that
denotational semantics generally enjoys, and closely parallels
homomorphisms between semantics (where a type system is a morphism
between the syntactic identity semantics and the semantics that maps
each term to its type). Especially considering this in light of
abstract interpretation may yield insight into the relationship
between type systems and abstract interpretation, further leading to
advances in connections between syntactic and semantic approaches to
programming language theory.

\subsection{Improving hybrid type checking}

Our presentation of hybrid type checking is only proven to insert
sufficiently many casts to ensure safety. But, in fact, some
components of casts are extraneous: In a function subtyping question,
if only the domain type needs a cast, the function will nonetheless be
fully wrapped.  How to define the smallest number of casts, and prove
that an HTC system is optimal, is a moderately interesting theoretical
question. More realistically, an HTC implementation may simply
informally develop techniques for inserting casts that do not perform
needless checks.

\subsection{Side Effects}

Side effects in contracts are a huge issue not directly part of this
research thread, but are urgent in this context due to the connection
of dynamic contracts (where side effects could be used on a ``buyer
beware'' basis) and static semantics (where side effects do not
readily translate into simple specifications). JML actually had this
issue as well, because you want to use existing libraries.  An
alternative is to develop \emph{logical} libraries that do not depend
on side effects. But these will generally not be efficient to execute,
again raising the problem of connecting efficient execution with
effective verification.

\subsection{Error messages, Testing, and Counterexamples}

Type error message in modern languages can already be difficult to
interpret. When a type incompatibility results from the inability to
prove an implication between refinement predicates -- which may be
very large, especially when they are generated via type reconstruction
-- the problem is exacerbated.  Mathematically, a constructive way to
refute a proposition is by providing a counterexample, which would
also yield a simple way to communicate an error message. But SMT is
nonconstructive by design, so new theorem proving techniques may need
to be applied or developed in order to generate readable error
messages.

Counterexamples have a larger role to play. In \sage{}, they
strengthen the type checker by refuting more subtyping queries
statically. Another good source of counterexamples aside from a
counterexample-producing theorem prover is a testing
process. Randomized testing tools such as
QuickCheck~\citep{claessen_quickcheck:_2000} and
DART~\citep{godefroid_dart:_2005} may be effective for enhancing the
static capabilities of a hybrid type checking system.

\subsection{Large-Scale Implementation for Empirical Work}

\sage{} demonstrates that executable refinement types can be
incorporated into a realistic programming language, but this has not
yet been attempted.  How powerful does the counterexample database
become when it receives crash reports from all users of a system?  How
fully are proposed benefits of precise specification realized?  For
what sorts of programs are executable refinement types most useful?

\section{Closing Remarks}

The theory of executable refinement types is sound and versatile.  It
combines the strengths of type theory and dynamic contracts,
benefiting from advances in each. This exploration of executable
refinement types has yielded not only new algorithms to address the
standard problems of type systems, but also new definitions of the
problems. Through these new definitions have come new approaches to
type checking, type reconstruction, compositionality, and system
architecture.

However, most of the executable refinement types of this dissertation
are based upon the simply-typed $\lambda$-calculus; their significance
should not be overstated. They lack parametric polymorphism,
side-effects, and type classes, to name a few other features we may
want in a programming language. They are a foundation upon which to
build systems that more closely resemble today's programming practice.

Conversely, programming practice should be influenced to move towards
concepts with well-developed foundations. Concepts such as lexical
scoping, higher-order functions, structural types, and careful
management of side effects all open the door from common practice to
the state of the art in programming languages, including executable
refinement types.

Just as this dissertation could take for granted the simply-typed
$\lambda$-calculus as a basis, future work may take for granted that
executable refinement types are sound in theory and effective in
practice. New theories may build up from this point, and practice may
build towards it.

\addcontentsline{toc}{chapter}{Bibliography}
\renewcommand{\bibsection}{\chapter*{Bibliography}}
\bibliographystyle{myabbrvnat}
\bibliography{main.bib}

\begin{thebibliography}{159}
\expandafter\ifx\csname natexlab\endcsname\relax\def\natexlab#1{#1}\fi
\expandafter\ifx\csname url\endcsname\relax
  \def\url#1{{\tt #1}}\fi

\bibitem[Abadi et~al.(1990)Abadi, Cardelli, Curien, and
  Levy]{abadi_explicit_1990}
M.~Abadi, L.~Cardelli, P.-L. Curien, and J.-J. Levy.
\newblock Explicit substitutions.
\newblock In {\em Proceedings of the 17th {ACM} {SIGPLAN-SIGACT} Symposium on
  Principles of Programming Languages}, {POPL} '90, page 31{\textendash}46, New
  York, {NY}, {USA}, 1990. {ACM}.

\bibitem[Abadi et~al.(1991)Abadi, Cardelli, Pierce, and
  Plotkin]{abadi_dynamic_1991}
M.~Abadi, L.~Cardelli, B.~Pierce, and G.~Plotkin.
\newblock Dynamic typing in a statically typed language.
\newblock {\em {ACM} Transactions on Programming Languages and Systems},
  13\penalty0 (2):\penalty0 237{\textendash}268, Apr. 1991.

\bibitem[Ahmed et~al.(2009)Ahmed, Findler, Matthews, and
  Wadler]{ahmed_blame_2009}
A.~Ahmed, R.~B. Findler, J.~Matthews, and P.~Wadler.
\newblock Blame for all.
\newblock In {\em Proceedings for the 1st workshop on Script to Program
  Evolution}, {STOP} '09, page 1{\textendash}13, New York, {NY}, {USA}, 2009.
  {ACM}.

\bibitem[Aiken and Wimmers(1993)]{aiken_type_1993}
A.~Aiken and E.~L. Wimmers.
\newblock Type inclusion constraints and type inference.
\newblock In {\em Proceedings of the conference on Functional programming
  languages and computer architecture}, {FPCA} '93, page 31{\textendash}41, New
  York, {NY}, {USA}, 1993. {ACM}.

\bibitem[Aiken et~al.(1994)Aiken, Wimmers, and Lakshman]{aiken_soft_1994}
A.~Aiken, E.~L. Wimmers, and T.~K. Lakshman.
\newblock Soft typing with conditional types.
\newblock In {\em Proceedings of the 21st {ACM} {SIGPLAN-SIGACT} symposium on
  Principles of programming languages}, {POPL} '94, page 163{\textendash}173,
  New York, {NY}, {USA}, 1994. {ACM}.

\bibitem[Allende and Fabry(2011)]{allende_application_2011}
E.~Allende and J.~Fabry.
\newblock Application optimization when using gradual typing.
\newblock In {\em Proceedings of the 6th Workshop on Implementation,
  Compilation, Optimization of Object-Oriented Languages, Programs and
  Systems}, {ICOOOLPS} '11, page 3:1{\textendash}3:6, New York, {NY}, {USA},
  2011. {ACM}.

\bibitem[Atkey et~al.(2011)Atkey, Johann, and Ghani]{atkey_when_2011}
R.~Atkey, P.~Johann, and N.~Ghani.
\newblock When is a type refinement an inductive type?
\newblock In {\em Proceedings of the 14th international conference on
  Foundations of software science and computational structures: part of the
  joint European conferences on theory and practice of software},
  {FOSSACS'11/ETAPS'11}, page 72{\textendash}87, Berlin, Heidelberg, 2011.
  Springer-Verlag.

\bibitem[Augustsson(1998)]{augustsson_cayenne_1998}
L.~Augustsson.
\newblock Cayenne - a language with dependent types.
\newblock In {\em Proceedings of the third {ACM} {SIGPLAN} international
  conference on Functional programming}, {ICFP} '98, page 239{\textendash}250,
  New York, {NY}, {USA}, 1998. {ACM}.

\bibitem[Back(1988)]{back_calculus_1988}
R.~J.~R. Back.
\newblock A calculus of refinements for program derivations.
\newblock {\em Acta Inf.}, 25\penalty0 (6):\penalty0 593{\textendash}624, Aug.
  1988.

\bibitem[Barendregt(1991)]{barendregt_introduction_1991}
H.~Barendregt.
\newblock An introduction to generalized type systems.
\newblock {\em Journal of Functional Programming}, 1\penalty0 (2):\penalty0
  125--154, Apr. 1991.

\bibitem[Barnett et~al.(2005)Barnett, Leino, and Schulte]{barnett_spec_2005}
M.~Barnett, K.~R.~M. Leino, and W.~Schulte.
\newblock The spec\# programming system: an overview.
\newblock In {\em Proceedings of the 2004 international conference on
  Construction and Analysis of Safe, Secure, and Interoperable Smart Devices},
  {CASSIS'04}, page 49{\textendash}69, Berlin, Heidelberg, 2005.
  Springer-Verlag.

\bibitem[Bayne et~al.(2011)Bayne, Cook, and Ernst]{bayne_always-available_2011}
M.~Bayne, R.~Cook, and M.~D. Ernst.
\newblock Always-available static and dynamic feedback.
\newblock In {\em Proceedings of the 33rd International Conference on Software
  Engineering}, {ICSE} '11, page 521{\textendash}530, New York, {NY}, {USA},
  2011. {ACM}.

\bibitem[Belo et~al.(2011)Belo, Greenberg, Igarashi, and
  Pierce]{belo_polymorphic_2011}
J.~F. Belo, M.~Greenberg, A.~Igarashi, and B.~C. Pierce.
\newblock Polymorphic contracts.
\newblock In G.~Barthe, editor, {\em Programming Languages and Systems}, number
  6602 in Lecture Notes in Computer Science, pages 18--37. Springer Berlin
  Heidelberg, Jan. 2011.

\bibitem[Bengtson et~al.(2008)Bengtson, Bhargavan, Fournet, Gordon, and
  Maffeis]{bengtson_refinement_2008}
J.~Bengtson, K.~Bhargavan, C.~Fournet, A.~D. Gordon, and S.~Maffeis.
\newblock Refinement types for secure implementations.
\newblock In {\em Proceedings of the 2008 21st {IEEE} Computer Security
  Foundations Symposium}, {CSF} '08, page 17{\textendash}32, Washington, {DC},
  {USA}, 2008. {IEEE} Computer Society.

\bibitem[Bierman et~al.(2010)Bierman, Gordon, Hri{\textbackslash}ctcu, and
  Langworthy]{bierman_semantic_2010}
G.~M. Bierman, A.~D. Gordon, C.~Hri{\textbackslash}ctcu, and D.~Langworthy.
\newblock Semantic subtyping with an {SMT} solver.
\newblock In {\em Proceedings of the 15th {ACM} {SIGPLAN} international
  conference on Functional programming}, {ICFP} '10, page 105{\textendash}116,
  New York, {NY}, {USA}, 2010. {ACM}.

\bibitem[Blume and {McAllester}(2006)]{blume_sound_2006}
M.~Blume and D.~{McAllester}.
\newblock Sound and complete models of contracts.
\newblock {\em J. Funct. Program.}, 16\penalty0 (4-5):\penalty0
  375{\textendash}414, July 2006.

\bibitem[Breazu-Tannen et~al.(1991)Breazu-Tannen, Coquand, Gunter, and
  Scedrov]{breazu-tannen_inheritance_1991}
V.~Breazu-Tannen, T.~Coquand, C.~A. Gunter, and A.~Scedrov.
\newblock Inheritance as implicit coercion.
\newblock {\em Information and Computation}, 93\penalty0 (1):\penalty0
  172{\textendash}221, July 1991.

\bibitem[Bruce et~al.(1999)Bruce, Cardelli, and Pierce]{bruce_comparing_1999}
K.~B. Bruce, L.~Cardelli, and B.~C. Pierce.
\newblock Comparing object encodings.
\newblock {\em Information and Computation}, 155\penalty0 (1-2):\penalty0
  108{\textendash}133, Nov. 1999.

\bibitem[Burdy et~al.(2005)Burdy, Cheon, Cok, Ernst, Kiniry, Leavens, Leino,
  and Poll]{burdy_overview_2005}
L.~Burdy, Y.~Cheon, D.~R. Cok, M.~D. Ernst, J.~R. Kiniry, G.~T. Leavens,
  K.~R.~M. Leino, and E.~Poll.
\newblock An overview of {JML} tools and applications.
\newblock {\em International Journal of Software Tools for Technology
  Transfer}, 7\penalty0 (3):\penalty0 212{\textendash}232, June 2005.

\bibitem[Cardelli(1988)]{cardelli_phase_1988}
L.~Cardelli.
\newblock Phase distinctions in type theory.
\newblock Technical report, 1988.

\bibitem[Chitil(2011)]{chitil_semantics_2011}
O.~Chitil.
\newblock A semantics for lazy assertions.
\newblock In {\em Proceedings of the 20th {ACM} {SIGPLAN} workshop on Partial
  evaluation and program manipulation}, {PEPM} '11, page 141{\textendash}150,
  New York, {NY}, {USA}, 2011. {ACM}.

\bibitem[Chitil(2012)]{chitil_practical_2012}
O.~Chitil.
\newblock Practical typed lazy contracts.
\newblock {\em {SIGPLAN} Not.}, 47\penalty0 (9):\penalty0 67{\textendash}76,
  Sept. 2012.

\bibitem[Chitil and Huch(2007{\natexlab{a}})]{chitil_monadic_2007}
O.~Chitil and F.~Huch.
\newblock Monadic, prompt lazy assertions in haskell.
\newblock In {\em Proceedings of the 5th Asian Conference on Programming
  Languages and Systems}, {APLAS'07}, page 38{\textendash}53, Berlin,
  Heidelberg, 2007{\natexlab{a}}. Springer-Verlag.

\bibitem[Chitil and Huch(2007{\natexlab{b}})]{chitil_pattern_2007}
O.~Chitil and F.~Huch.
\newblock A pattern logic for prompt lazy assertions in haskell.
\newblock In {\em Proceedings of the 18th International Conference on
  Implementation and Application of Functional Languages}, {IFL'06}, page
  126{\textendash}144, Berlin, Heidelberg, 2007{\natexlab{b}}. Springer-Verlag.

\bibitem[Chitil et~al.(2004)Chitil, {McNeill}, and Runciman]{chitil_lazy_2004}
O.~Chitil, D.~{McNeill}, and C.~Runciman.
\newblock Lazy assertions.
\newblock In {\em Proceedings of the 15th International Conference on
  Implementation of Functional Languages}, {IFL'03}, page 1{\textendash}19,
  Berlin, Heidelberg, 2004. Springer-Verlag.

\bibitem[Chugh et~al.(2012{\natexlab{a}})Chugh, Herman, and
  Jhala]{chugh_dependent_2012}
R.~Chugh, D.~Herman, and R.~Jhala.
\newblock Dependent types for {JavaScript}.
\newblock In {\em Proceedings of the {ACM} international conference on Object
  oriented programming systems languages and applications}, {OOPSLA} '12, page
  587{\textendash}606, New York, {NY}, {USA}, 2012{\natexlab{a}}. {ACM}.

\bibitem[Chugh et~al.(2012{\natexlab{b}})Chugh, Rondon, and
  Jhala]{chugh_nested_2012}
R.~Chugh, P.~M. Rondon, and R.~Jhala.
\newblock Nested refinements: a logic for duck typing.
\newblock {\em {SIGPLAN} Not.}, 47\penalty0 (1):\penalty0 231{\textendash}244,
  Jan. 2012{\natexlab{b}}.

\bibitem[Church(1940)]{church_formulation_1940}
A.~Church.
\newblock A formulation of the simple theory of types.
\newblock {\em The Journal of Symbolic Logic}, 5\penalty0 (2):\penalty0 56,
  June 1940.

\bibitem[Claessen and Hughes(2000)]{claessen_quickcheck:_2000}
K.~Claessen and J.~Hughes.
\newblock {QuickCheck:} a lightweight tool for random testing of haskell
  programs.
\newblock In {\em Proceedings of the Fifth {ACM} {SIGPLAN} International
  Conference on Functional Programming}, {ICFP} '00, page 268{\textendash}279,
  New York, {NY}, {USA}, 2000. {ACM}.

\bibitem[Coquand(1986)]{coquand_analysis_1986}
T.~Coquand.
\newblock An analysis of girard's paradox.
\newblock In {\em In Symposium on Logic in Computer Science}, page
  227{\textendash}236. {IEEE} Computer Society Press, 1986.

\bibitem[Cousot and Cousot(1995)]{cousot_formal_1995}
P.~Cousot and R.~Cousot.
\newblock Formal language, grammar and set-constraint-based program analysis by
  abstract interpretation.
\newblock In {\em Proceedings of the seventh international conference on
  Functional programming languages and computer architecture}, {FPCA} '95, page
  170{\textendash}181, New York, {NY}, {USA}, 1995. {ACM}.

\bibitem[Cui et~al.(2005)Cui, Donnelly, and Xi]{cui_ats:_2005}
S.~Cui, K.~Donnelly, and H.~Xi.
\newblock {ATS:} a language that combines programming with theorem proving.
\newblock In {\em Proceedings of the 5th international conference on Frontiers
  of Combining Systems}, {FroCoS'05}, page 310{\textendash}320, Berlin,
  Heidelberg, 2005. Springer-Verlag.

\bibitem[Curry and Feys(1958)]{curry_combinatory_1958}
H.~Curry and R.~Feys.
\newblock {\em Combinatory Logic}, volume~1.
\newblock North-Holland, 1958.

\bibitem[Curry(1934)]{curry_functionality_1934}
H.~B. Curry.
\newblock Functionality in combinatory logic.
\newblock {\em Proceedings of the National Academy of Sciences of the United
  States of America}, 20\penalty0 (11):\penalty0 584--590, Nov. 1934.
\newblock {PMID:} 16577644 {PMCID:} {PMC1076489}.

\bibitem[Dagand and {McBride}(2012)]{dagand_transporting_2012}
P.-E. Dagand and C.~{McBride}.
\newblock Transporting functions across ornaments.
\newblock In {\em Proceedings of the 17th {ACM} {SIGPLAN} International
  Conference on Functional Programming}, {ICFP} '12, page 103{\textendash}114,
  New York, {NY}, {USA}, 2012. {ACM}.

\bibitem[Davies and Pfenning(2000)]{davies_intersection_2000}
R.~Davies and F.~Pfenning.
\newblock Intersection types and computational effects.
\newblock In {\em Proceedings of the Fifth {ACM} {SIGPLAN} International
  Conference on Functional Programming}, {ICFP} '00, page 198{\textendash}208,
  New York, {NY}, {USA}, 2000. {ACM}.

\bibitem[Degen et~al.(2009)Degen, Thiemann, and Wehr]{degen_true_2009}
M.~Degen, P.~Thiemann, and S.~Wehr.
\newblock True lies: Lazy contracts for lazy languages.
\newblock In {\em Arbeitstagung Programmiersprachen ({ATPS'09)}}, 2009.

\bibitem[Degen et~al.(2012)Degen, Thiemann, and Wehr]{degen_interaction_2012}
M.~Degen, P.~Thiemann, and S.~Wehr.
\newblock The interaction of contracts and laziness.
\newblock In {\em Proceedings of the {ACM} {SIGPLAN} 2012 workshop on Partial
  evaluation and program manipulation}, {PEPM} '12, page 97{\textendash}106,
  New York, {NY}, {USA}, 2012. {ACM}.

\bibitem[Denney(1998)]{denney_refinement_1998}
E.~Denney.
\newblock Refinement types for specification.
\newblock In {\em Proceedings of the {IFIP} {TC2/WG2.2},2.3 International
  Conference on Programming Concepts and Methods}, {PROCOMET} '98, page
  148{\textendash}166, London, {UK}, {UK}, 1998. Chapman \&amp; Hall, Ltd.

\bibitem[Detlefs et~al.(2005)Detlefs, Nelson, and Saxe]{detlefs_simplify:_2005}
D.~Detlefs, G.~Nelson, and J.~B. Saxe.
\newblock Simplify: a theorem prover for program checking.
\newblock {\em Journal of the {ACM}}, 52\penalty0 (3):\penalty0
  365{\textendash}473, May 2005.

\bibitem[Dijkstra(1976)]{dijkstra_discipline_1976}
E.~W. Dijkstra.
\newblock {\em A Discipline of Programming}.
\newblock Prentice Hall {PTR}, Upper Saddle River, {NJ}, {USA}, 1st edition,
  1976.

\bibitem[Dimoulas and Felleisen(2011)]{dimoulas_contract_2011}
C.~Dimoulas and M.~Felleisen.
\newblock On contract satisfaction in a higher-order world.
\newblock {\em {ACM} Transactions on Programming Languages and Systems},
  33\penalty0 (5):\penalty0 16:1{\textendash}16:29, Nov. 2011.

\bibitem[Dimoulas et~al.(2011)Dimoulas, Findler, Flanagan, and
  Felleisen]{dimoulas_correct_2011}
C.~Dimoulas, R.~B. Findler, C.~Flanagan, and M.~Felleisen.
\newblock Correct blame for contracts: No more scapegoating.
\newblock In {\em Proceedings of the 38th Annual {ACM} {SIGPLAN-SIGACT}
  Symposium on Principles of Programming Languages}, {POPL} '11, page
  215{\textendash}226, New York, {NY}, {USA}, 2011. {ACM}.

\bibitem[Dimoulas et~al.(2012)Dimoulas, Tobin-Hochstadt, and
  Felleisen]{dimoulas_complete_2012}
C.~Dimoulas, S.~Tobin-Hochstadt, and M.~Felleisen.
\newblock Complete monitors for behavioral contracts.
\newblock In {\em Proceedings of the 21st European conference on Programming
  Languages and Systems}, {ESOP'12}, page 214{\textendash}233, Berlin,
  Heidelberg, 2012. Springer-Verlag.

\bibitem[Dreyer et~al.(2003)Dreyer, Crary, and Harper]{dreyer_type_2003}
D.~Dreyer, K.~Crary, and R.~Harper.
\newblock A type system for higher-order modules.
\newblock In {\em Proceedings of the 30th {ACM} {SIGPLAN-SIGACT} Symposium on
  Principles of Programming Languages}, {POPL} '03, page 236{\textendash}249,
  New York, {NY}, {USA}, 2003. {ACM}.

\bibitem[Dunfield(2007)]{dunfield_refined_2007}
J.~Dunfield.
\newblock Refined typechecking with stardust.
\newblock In {\em Proceedings of the 2007 workshop on Programming languages
  meets program verification}, {PLPV} '07, page 21{\textendash}32, New York,
  {NY}, {USA}, 2007. {ACM}.

\bibitem[Dunfield and Pfenning(2003)]{dunfield_type_2003}
J.~Dunfield and F.~Pfenning.
\newblock Type assignment for intersections and unions in call-by-value
  languages.
\newblock In {\em Proceedings of the 6th International conference on
  Foundations of Software Science and Computation Structures and joint European
  conference on Theory and practice of software}, {FOSSACS'03/ETAPS'03}, page
  250{\textendash}266, Berlin, Heidelberg, 2003. Springer-Verlag.

\bibitem[Dunfield and Pfenning(2004)]{dunfield_tridirectional_2004}
J.~Dunfield and F.~Pfenning.
\newblock Tridirectional typechecking.
\newblock In {\em Proceedings of the 31st {ACM} {SIGPLAN-SIGACT} Symposium on
  Principles of Programming Languages}, {POPL} '04, page 281{\textendash}292,
  New York, {NY}, {USA}, 2004. {ACM}.

\bibitem[Escard{\'o}(2004)]{escardo_synthetic_2004}
M.~Escard{\'o}.
\newblock Synthetic topology of data types and classical spaces.
\newblock In {\em Electronic Notes in Theoretical Computer Science}, page 2004.
  Elsevier, 2004.

\bibitem[Fagan(1991)]{fagan_soft_1991}
M.~Fagan.
\newblock {\em Soft typing: an approach to type checking for dynamically typed
  languages}.
\newblock PhD thesis, Rice University, Houston, {TX}, {USA}, 1991.
\newblock {UMI} Order No. {GAX91-36021}.

\bibitem[Fahndrich and Aiken(1996)]{fahndrich_making_1996}
M.~Fahndrich and A.~Aiken.
\newblock Making set-constraint program analyses scale.
\newblock Technical report, University of California at Berkeley, Berkeley,
  {CA}, {USA}, 1996.

\bibitem[Findler and Blume(2006)]{findler_contracts_2006}
R.~B. Findler and M.~Blume.
\newblock Contracts as pairs of projections.
\newblock In {\em Proceedings of the 8th international conference on Functional
  and Logic Programming}, {FLOPS'06}, page 226{\textendash}241, Berlin,
  Heidelberg, 2006. Springer-Verlag.

\bibitem[Findler and Felleisen(2002)]{findler_contracts_2002}
R.~B. Findler and M.~Felleisen.
\newblock Contracts for higher-order functions.
\newblock In {\em Proceedings of the seventh {ACM} {SIGPLAN} international
  conference on Functional programming}, {ICFP} '02, page 48{\textendash}59,
  New York, {NY}, {USA}, 2002. {ACM}.

\bibitem[Flanagan(2006)]{flanagan_hybrid_2006}
C.~Flanagan.
\newblock Hybrid type checking.
\newblock In {\em Conference Record of the 33rd {ACM} {SIGPLAN-SIGACT}
  Symposium on Principles of Programming Languages}, {POPL} '06, page
  245{\textendash}256, New York, {NY}, {USA}, 2006. {ACM}.

\bibitem[Flanagan and Felleisen(1997)]{flanagan_componential_1997}
C.~Flanagan and M.~Felleisen.
\newblock Componential set-based analysis.
\newblock In {\em Proceedings of the {ACM} {SIGPLAN} 1997 conference on
  Programming language design and implementation}, {PLDI} '97, page
  235{\textendash}248, New York, {NY}, {USA}, 1997. {ACM}.

\bibitem[Flanagan et~al.(1996)Flanagan, Flatt, Krishnamurthi, Weirich, and
  Felleisen]{flanagan_catching_1996}
C.~Flanagan, M.~Flatt, S.~Krishnamurthi, S.~Weirich, and M.~Felleisen.
\newblock Catching bugs in the web of program invariants.
\newblock In {\em Proceedings of the {ACM} {SIGPLAN} 1996 Conference on
  Programming Language Design and Implementation}, {PLDI} '96, page
  23{\textendash}32, New York, {NY}, {USA}, 1996. {ACM}.

\bibitem[Flanagan et~al.(2002)Flanagan, Leino, Lillibridge, Nelson, Saxe, and
  Stata]{flanagan_extended_2002}
C.~Flanagan, K.~R.~M. Leino, M.~Lillibridge, G.~Nelson, J.~B. Saxe, and
  R.~Stata.
\newblock Extended static checking for java.
\newblock In {\em Proceedings of the {ACM} {SIGPLAN} 2002 Conference on
  Programming language design and implementation}, {PLDI} '02, page
  234{\textendash}245, New York, {NY}, {USA}, 2002. {ACM}.

\bibitem[Floyd(1967)]{floyd_assigning_1967}
R.~Floyd.
\newblock Assigning meaning to programs.
\newblock In {\em Proceedings of the Symposium in Applied Mathematics:
  Mathematical Aspects of Computer Science}, pages 19--32, 1967.

\bibitem[Fogarty et~al.(2007)Fogarty, Pasalic, Siek, and
  Taha]{fogarty_concoqtion:_2007}
S.~Fogarty, E.~Pasalic, J.~Siek, and W.~Taha.
\newblock Concoqtion: indexed types now!
\newblock In {\em Proceedings of the 2007 {ACM} {SIGPLAN} symposium on Partial
  evaluation and semantics-based program manipulation}, {PEPM} '07, page
  112{\textendash}121, New York, {NY}, {USA}, 2007. {ACM}.

\bibitem[Freeman and Pfenning(1991)]{freeman_refinement_1991}
T.~Freeman and F.~Pfenning.
\newblock Refinement types for {ML}.
\newblock In {\em Proceedings of the {ACM} {SIGPLAN} 1991 conference on
  Programming language design and implementation}, {PLDI} '91, page
  268{\textendash}277, New York, {NY}, {USA}, 1991. {ACM}.

\bibitem[Frisch et~al.(2008)Frisch, Castagna, and
  Benzaken]{frisch_semantic_2008}
A.~Frisch, G.~Castagna, and V.~Benzaken.
\newblock Semantic subtyping: Dealing set-theoretically with function, union,
  intersection, and negation types.
\newblock {\em Journal of the {ACM}}, 55\penalty0 (4):\penalty0
  19:1{\textendash}19:64, Sept. 2008.

\bibitem[Fuh and Mishra(1988)]{fuh_type_1988}
Y.-C. Fuh and P.~Mishra.
\newblock Type inference with subtypes.
\newblock {\em Theoretical Computer Science}, 73\penalty0 (2):\penalty0
  155{\textendash}175, Jan. 1988.

\bibitem[Girard(1972)]{girard_interpretation_1972}
J.-Y. Girard.
\newblock Interpr{\'e}tation fonctionnelle et {\'e}limination des coupures de
  l{\textquoteright}arithm{\'e}tique d{\textquoteright}ordre sup{\'e}rieur.
\newblock 1972.

\bibitem[Godefroid et~al.(2005)Godefroid, Klarlund, and
  Sen]{godefroid_dart:_2005}
P.~Godefroid, N.~Klarlund, and K.~Sen.
\newblock {DART:} directed automated random testing.
\newblock In {\em Proceedings of the 2005 {ACM} {SIGPLAN} conference on
  Programming language design and implementation}, {PLDI} '05, page
  213{\textendash}223, New York, {NY}, {USA}, 2005. {ACM}.

\bibitem[Greenberg et~al.(2012)Greenberg, Pierce, and
  Weirich]{greenberg_contracts_2012}
M.~Greenberg, B.~c. Pierce, and S.~Weirich.
\newblock Contracts made manifest.
\newblock {\em Journal of Functional Programming}, 22\penalty0 (3):\penalty0
  225{\textendash}274, May 2012.

\bibitem[Gronski and Flanagan(2007)]{gronski_unifying_2007}
J.~Gronski and C.~Flanagan.
\newblock Unifying hybrid types and contracts.
\newblock In {\em In Eighth Symposium on Trends in Functional Programming},
  2007.

\bibitem[Gronski et~al.(2006)Gronski, Knowles, Tomb, Freund, and
  Flanagan]{gronski_sage:_2006}
J.~Gronski, K.~Knowles, A.~Tomb, S.~N. Freund, and C.~Flanagan.
\newblock Sage: Hybrid checking for flexible specifications.
\newblock In {\em Proceedings of the 7th workshop on Scheme and functional
  programming}, Portland, {OR}, 2006. {ACM}.

\bibitem[Grossman(2002)]{grossman_existential_2002}
D.~Grossman.
\newblock Existential types for imperative languages.
\newblock In {\em Proceedings of the 11th European Symposium on Programming
  Languages and Systems}, {ESOP} '02, page 21{\textendash}35, London, {UK},
  {UK}, 2002. Springer-Verlag.

\bibitem[Guha et~al.(2007)Guha, Matthews, Findler, and
  Krishnamurthi]{guha_relationally-parametric_2007}
A.~Guha, J.~Matthews, R.~B. Findler, and S.~Krishnamurthi.
\newblock Relationally-parametric polymorphic contracts.
\newblock In {\em Proceedings of the 2007 symposium on Dynamic languages},
  {DLS} '07, page 29{\textendash}40, New York, {NY}, {USA}, 2007. {ACM}.

\bibitem[Harper and Lillibridge(1994)]{harper_type-theoretic_1994}
R.~Harper and M.~Lillibridge.
\newblock A type-theoretic approach to higher-order modules with sharing.
\newblock In {\em Proceedings of the 21st {ACM} {SIGPLAN-SIGACT} symposium on
  Principles of programming languages}, {POPL} '94, page 123{\textendash}137,
  New York, {NY}, {USA}, 1994. {ACM}.

\bibitem[Hayashi(1993)]{hayashi_logic_1993}
S.~Hayashi.
\newblock Logic of refinement types.
\newblock In {\em Proceedings of the international workshop on Types for proofs
  and programs}, {TYPES} '93, page 108{\textendash}126, Secaucus, {NJ}, {USA},
  1993. Springer-Verlag New York, Inc.

\bibitem[Heintze(1992)]{heintze_set_1992}
N.~C. Heintze.
\newblock {\em Set based program analysis}.
\newblock PhD thesis, Carnegie Mellon University, Pittsburgh, {PA}, {USA},
  1992.
\newblock {UMI} Order No. {GAX93-22866}.

\bibitem[Henglein(1994)]{henglein_dynamic_1994}
F.~Henglein.
\newblock Dynamic typing: syntax and proof theory.
\newblock In {\em Selected papers of the symposium on Fourth European symposium
  on programming}, {ESOP'92}, page 197{\textendash}230, Amsterdam, The
  Netherlands, The Netherlands, 1994. Elsevier Science Publishers B. V.

\bibitem[Herman et~al.(2010)Herman, Tomb, and
  Flanagan]{herman_space-efficient_2010}
D.~Herman, A.~Tomb, and C.~Flanagan.
\newblock Space-efficient gradual typing.
\newblock {\em Higher Order Symbolic Compututation}, 23\penalty0 (2):\penalty0
  167{\textendash}189, June 2010.

\bibitem[Hinze et~al.(2006)Hinze, Jeuring, and L{\"o}h]{hinze_typed_2006}
R.~Hinze, J.~Jeuring, and A.~L{\"o}h.
\newblock Typed contracts for functional programming.
\newblock In {\em In {FLOPS} {\textquoteright}06: Functional and Logic
  Programming: 8th International Symposium}, page 208{\textendash}225.
  Springer-Verlag, 2006.

\bibitem[Hoang and Mitchell(1995)]{hoang_lower_1995}
M.~Hoang and J.~C. Mitchell.
\newblock Lower bounds on type inference with subtypes.
\newblock In {\em Proceedings of the 22nd {ACM} {SIGPLAN-SIGACT} symposium on
  Principles of programming languages}, {POPL} '95, page 176{\textendash}185,
  New York, {NY}, {USA}, 1995. {ACM}.

\bibitem[Hoare(1969)]{hoare_axiomatic_1969}
C.~A.~R. Hoare.
\newblock An axiomatic basis for computer programming.
\newblock {\em Communications of the {ACM}}, 12\penalty0 (10):\penalty0
  576{\textendash}580, Oct. 1969.

\bibitem[Holt and Cordy(1988)]{holt_turing_1988}
R.~C. Holt and J.~R. Cordy.
\newblock The turing programming language.
\newblock {\em Communications of the {ACM}}, 31\penalty0 (12):\penalty0
  1410{\textendash}1423, Dec. 1988.

\bibitem[Howard(1980)]{howard_formulae-as-types_1980}
W.~A. Howard.
\newblock The formulae-as-types notion of construction.
\newblock {\em To {HB} Curry: essays on combinatory logic, lambda calculus and
  formalism}, 44:\penalty0 479---490, 1980.

\bibitem[Hurkens(1995)]{hurkens_simplification_1995}
A.~J.~C. Hurkens.
\newblock A simplification of girard's paradox.
\newblock In {\em Proceedings of the Second International Conference on Typed
  Lambda Calculi and Applications}, {TLCA} '95, page 266{\textendash}278,
  London, {UK}, {UK}, 1995. Springer-Verlag.

\bibitem[Ina and Igarashi(2009)]{ina_towards_2009}
L.~Ina and A.~Igarashi.
\newblock Towards gradual typing for generics.
\newblock In {\em Proceedings for the 1st workshop on Script to Program
  Evolution}, {STOP} '09, page 17{\textendash}29, New York, {NY}, {USA}, 2009.
  {ACM}.

\bibitem[Ina and Igarashi(2011)]{ina_gradual_2011}
L.~Ina and A.~Igarashi.
\newblock Gradual typing for generics.
\newblock In {\em Proceedings of the 2011 {ACM} international conference on
  Object oriented programming systems languages and applications}, {OOPSLA}
  '11, page 609{\textendash}624, New York, {NY}, {USA}, 2011. {ACM}.

\bibitem[Jhala et~al.(2011)Jhala, Majumdar, and Rybalchenko]{jhala_hmc:_2011}
R.~Jhala, R.~Majumdar, and A.~Rybalchenko.
\newblock {HMC:} verifying functional programs using abstract interpreters.
\newblock In {\em Proceedings of the 23rd international conference on Computer
  aided verification}, {CAV'11}, page 470{\textendash}485, Berlin, Heidelberg,
  2011. Springer-Verlag.

\bibitem[Kawaguchi et~al.(2009)Kawaguchi, Rondon, and
  Jhala]{kawaguchi_type-based_2009}
M.~Kawaguchi, P.~Rondon, and R.~Jhala.
\newblock Type-based data structure verification.
\newblock In {\em Proceedings of the 2009 {ACM} {SIGPLAN} conference on
  Programming language design and implementation}, {PLDI} '09, page
  304{\textendash}315, New York, {NY}, {USA}, 2009. {ACM}.

\bibitem[Knowles and Flanagan(2007)]{knowles_type_2007}
K.~Knowles and C.~Flanagan.
\newblock Type reconstruction for general refinement types.
\newblock In {\em Proceedings of the 16th European conference on Programming},
  {ESOP'07}, page 505{\textendash}519, Berlin, Heidelberg, 2007.
  Springer-Verlag.

\bibitem[Knowles and Flanagan(2009)]{knowles_compositional_2009}
K.~Knowles and C.~Flanagan.
\newblock Compositional reasoning and decidable checking for dependent contract
  types.
\newblock In {\em Proceedings of the 3rd workshop on Programming languages
  meets program verification}, {PLPV} '09, page 27{\textendash}38, New York,
  {NY}, {USA}, 2009. {ACM}.

\bibitem[Knowles and Flanagan(2010)]{knowles_hybrid_2010}
K.~Knowles and C.~Flanagan.
\newblock Hybrid type checking.
\newblock {\em {ACM} Transactions on Programming Languages and Systems},
  32\penalty0 (2):\penalty0 6:1{\textendash}6:34, Feb. 2010.

\bibitem[K{\"o}lling and Rosenberg(1996)]{kolling_blue_1996}
M.~K{\"o}lling and J.~Rosenberg.
\newblock Blue - a language for teaching object-oriented programming.
\newblock {\em {SIGCSE} Bulletins}, 28\penalty0 (1):\penalty0
  190{\textendash}194, Mar. 1996.

\bibitem[Lambek and Scott(1986)]{lambek_introduction_1986}
J.~Lambek and P.~J. Scott.
\newblock {\em Introduction to higher order categorical logic}.
\newblock Cambridge University Press, New York, {NY}, {USA}, 1986.

\bibitem[Leavens and Cheon(2006)]{leavens_design_2006}
G.~T. Leavens and Y.~Cheon.
\newblock {\em Design by Contract with {JML}}.
\newblock 2006.

\bibitem[Leroy et~al.(2013)Leroy, Doligez, Frisch, Garrigue, R{\'e}my, and
  Vouillon]{leroy_ocaml_2013}
X.~Leroy, D.~Doligez, A.~Frisch, J.~Garrigue, D.~R{\'e}my, and J.~Vouillon.
\newblock The {OCaml} system release 4.01.
\newblock Technical report, Sept. 2013.

\bibitem[Lincoln and Mitchell(1992)]{lincoln_algorithmic_1992}
P.~Lincoln and J.~C. Mitchell.
\newblock Algorithmic aspects of type inference with subtypes.
\newblock In {\em Proceedings of the 19th {ACM} {SIGPLAN-SIGACT} symposium on
  Principles of programming languages}, {POPL} '92, page 293{\textendash}304,
  New York, {NY}, {USA}, 1992. {ACM}.

\bibitem[Luckham(1990)]{luckham_programming_1990}
D.~C. Luckham.
\newblock {\em Programming with Specifications: An Introduction to Anna, a
  Language for Specifying {ADA} Programs}.
\newblock Springer-Verlag New York, Inc., Secaucus, {NJ}, {USA}, 1990.

\bibitem[Mandelbaum et~al.(2003)Mandelbaum, Walker, and
  Harper]{mandelbaum_effective_2003}
Y.~Mandelbaum, D.~Walker, and R.~Harper.
\newblock An effective theory of type refinements.
\newblock In {\em Proceedings of the eighth {ACM} {SIGPLAN} international
  conference on Functional programming}, {ICFP} '03, page 213{\textendash}225,
  New York, {NY}, {USA}, 2003. {ACM}.

\bibitem[Marlow(2010)]{marlow_haskell_2010}
S.~Marlow.
\newblock Haskell 2010 language report, 2010.

\bibitem[Martin-L{\"o}f(1975)]{martin-lof_intuitionistic_1975}
P.~Martin-L{\"o}f.
\newblock An intuitionistic theory of types: Predicative part.
\newblock In {{H.E.} Rose and {J.C.} Shepherdson}, editor, {\em Studies in
  Logic and the Foundations of Mathematics}, volume Volume 80 of {\em Logic
  Colloquium '73 Proceedings of the Logic Colloquium}, pages 73--118. Elsevier,
  1975.

\bibitem[Matthews and Ahmed(2008)]{matthews_parametric_2008}
J.~Matthews and A.~Ahmed.
\newblock Parametric polymorphism through run-time sealing or, theorems for
  low, low prices!
\newblock In {\em Proceedings of the Theory and Practice of Software, 17th
  European Conference on Programming Languages and Systems},
  {ESOP'08/ETAPS'08}, page 16{\textendash}31, Berlin, Heidelberg, 2008.
  Springer-Verlag.

\bibitem[{McBride}(2002)]{mcbride_faking_2002}
C.~{McBride}.
\newblock Faking it (simulating dependent types in haskell).
\newblock {\em Journal of Functional Programming}, 12\penalty0 (5):\penalty0
  375{\textendash}392, July 2002.

\bibitem[{McBride} and {McKinna}(2004)]{mcbride_view_2004}
C.~{McBride} and J.~{McKinna}.
\newblock The view from the left.
\newblock {\em Journal of Functional Programming}, 14\penalty0 (1):\penalty0
  69{\textendash}111, Jan. 2004.

\bibitem[Meyer(1988)]{meyer_object-oriented_1988}
B.~Meyer.
\newblock {\em Object-Oriented Software Construction}.
\newblock Prentice-Hall, Inc., Upper Saddle River, {NJ}, {USA}, 1st edition,
  1988.

\bibitem[Meyer(1992)]{meyer_eiffel:_1992}
B.~Meyer.
\newblock {\em Eiffel: the language}.
\newblock Prentice-Hall, Inc., Upper Saddle River, {NJ}, {USA}, 1992.

\bibitem[Mitchell(1984)]{mitchell_coercion_1984}
J.~C. Mitchell.
\newblock Coercion and type inference.
\newblock In {\em Proceedings of the 11th {ACM} {SIGACT-SIGPLAN} symposium on
  Principles of programming languages}, {POPL} '84, page 175{\textendash}185,
  New York, {NY}, {USA}, 1984. {ACM}.

\bibitem[Mitchell and Plotkin(1988)]{mitchell_abstract_1988}
J.~C. Mitchell and G.~D. Plotkin.
\newblock Abstract types have existential type.
\newblock {\em {ACM} Transactions on Programming Languages and Systems},
  10\penalty0 (3):\penalty0 470{\textendash}502, July 1988.

\bibitem[Morris(1973{\natexlab{a}})]{morris_protection_1973}
J.~H. Morris.
\newblock Protection in programming languages.
\newblock {\em Commun. {ACM}}, 16\penalty0 (1):\penalty0 15{\textendash}21,
  Jan. 1973{\natexlab{a}}.

\bibitem[Morris(1973{\natexlab{b}})]{morris_types_1973}
J.~H. Morris.
\newblock Types are not sets.
\newblock In {\em Proceedings of the 1st annual {ACM} {SIGACT-SIGPLAN}
  symposium on Principles of programming languages}, {POPL} '73, page
  120{\textendash}124, New York, {NY}, {USA}, 1973{\natexlab{b}}. {ACM}.

\bibitem[Morrisett et~al.(1999)Morrisett, Walker, Crary, and
  Glew]{morrisett_system_1999}
G.~Morrisett, D.~Walker, K.~Crary, and N.~Glew.
\newblock From system f to typed assembly language.
\newblock {\em {ACM} Transactions on Programming Languages and Systems},
  21\penalty0 (3):\penalty0 527{\textendash}568, May 1999.

\bibitem[Nanevski et~al.(2007)Nanevski, Ahmed, Morrisett, and
  Birkedal]{nanevski_abstract_2007}
A.~Nanevski, A.~Ahmed, G.~Morrisett, and L.~Birkedal.
\newblock Abstract predicates and mutable adts in hoare type theory.
\newblock In {\em Proceedings of the 16th European conference on Programming},
  {ESOP'07}, page 189{\textendash}204, Berlin, Heidelberg, 2007.
  Springer-Verlag.

\bibitem[Nanevski et~al.(2011)Nanevski, Banerjee, and
  Garg]{nanevski_verification_2011}
A.~Nanevski, A.~Banerjee, and D.~Garg.
\newblock Verification of information flow and access control policies with
  dependent types.
\newblock In {\em Proceedings of the 2011 {IEEE} Symposium on Security and
  Privacy}, {SP} '11, page 165{\textendash}179, Washington, {DC}, {USA}, 2011.
  {IEEE} Computer Society.

\bibitem[Nanevski et~al.(2013)Nanevski, Banerjee, and
  Garg]{nanevski_dependent_2013}
A.~Nanevski, A.~Banerjee, and D.~Garg.
\newblock Dependent type theory for verification of information flow and access
  control policies.
\newblock {\em {ACM} Transactions on Programming Languages and Systems},
  35\penalty0 (2):\penalty0 6:1{\textendash}6:41, July 2013.

\bibitem[Nanevski et~al.(2006)Nanevski, Morrisett, and
  Birkedal]{nanevski_polymorphism_2006}
A.~Nanevski, G.~Morrisett, and L.~Birkedal.
\newblock Polymorphism and separation in hoare type theory.
\newblock In {\em Proceedings of the eleventh {ACM} {SIGPLAN} international
  conference on Functional programming}, {ICFP} '06, page 62{\textendash}73,
  New York, {NY}, {USA}, 2006. {ACM}.

\bibitem[Nanevski et~al.(2008)Nanevski, Morrisett, Shinnar, Govereau, and
  Birkedal]{nanevski_ynot:_2008}
A.~Nanevski, G.~Morrisett, A.~Shinnar, P.~Govereau, and L.~Birkedal.
\newblock Ynot: dependent types for imperative programs.
\newblock In {\em Proceedings of the 13th {ACM} {SIGPLAN} international
  conference on Functional programming}, {ICFP} '08, page 229{\textendash}240,
  New York, {NY}, {USA}, 2008. {ACM}.

\bibitem[Necula et~al.(2005)Necula, Condit, Harren, {McPeak}, and
  Weimer]{necula_ccured:_2005}
G.~C. Necula, J.~Condit, M.~Harren, S.~{McPeak}, and W.~Weimer.
\newblock {CCured:} type-safe retrofitting of legacy software.
\newblock {\em {ACM} Transactions on Programming Languages and Systems},
  27\penalty0 (3):\penalty0 477{\textendash}526, May 2005.

\bibitem[Norell(2009)]{norell_dependently_2009}
U.~Norell.
\newblock Dependently typed programming in agda.
\newblock In {\em Proceedings of the 4th international workshop on Types in
  language design and implementation}, {TLDI} '09, page 1{\textendash}2, New
  York, {NY}, {USA}, 2009. {ACM}.

\bibitem[Odersky et~al.(1999)Odersky, Sulzmann, and Wehr]{odersky_type_1999}
M.~Odersky, M.~Sulzmann, and M.~Wehr.
\newblock Type inference with constrained types.
\newblock {\em Theory and Practice of Object Systems}, 5\penalty0 (1):\penalty0
  35{\textendash}55, Jan. 1999.

\bibitem[Ou et~al.(2004)Ou, Tan, Mandelbaum, and Walker]{ou_dynamic_2004}
X.~Ou, G.~Tan, Y.~Mandelbaum, and D.~Walker.
\newblock Dynamic typing with dependent types.
\newblock In J.-J. Levy, E.~W. Mayr, and J.~C. Mitchell, editors, {\em
  Exploring New Frontiers of Theoretical Informatics}, number 155 in {IFIP}
  International Federation for Information Processing, pages 437--450. Springer
  {US}, Jan. 2004.

\bibitem[Parnas(1972{\natexlab{a}})]{parnas_criteria_1972}
D.~L. Parnas.
\newblock On the criteria to be used in decomposing systems into modules.
\newblock {\em Communications of the {ACM}}, 15\penalty0 (12):\penalty0
  1053{\textendash}1058, Dec. 1972{\natexlab{a}}.

\bibitem[Parnas(1972{\natexlab{b}})]{parnas_technique_1972}
D.~L. Parnas.
\newblock A technique for software module specification with examples.
\newblock {\em Communications of the {ACM}}, 15\penalty0 (5):\penalty0
  330{\textendash}336, May 1972{\natexlab{b}}.

\bibitem[Pierce and Sumii(2000)]{pierce_relating_2000}
B.~Pierce and E.~Sumii.
\newblock Relating cryptography and polymorphism.
\newblock 2000.

\bibitem[Pierce(2002)]{pierce_types_2002}
B.~C. Pierce.
\newblock {\em Types and programming languages}.
\newblock {MIT} Press, Cambridge, {MA}, {USA}, 2002.

\bibitem[Pierce(2004)]{pierce_advanced_2004}
B.~C. Pierce.
\newblock {\em Advanced Topics in Types and Programming Languages}.
\newblock The {MIT} Press, 2004.

\bibitem[Pierce and Turner(1998)]{pierce_local_1998}
B.~C. Pierce and D.~N. Turner.
\newblock Local type inference.
\newblock In {\em Proceedings of the 25th {ACM} {SIGPLAN-SIGACT} symposium on
  Principles of programming languages}, {POPL} '98, page 252{\textendash}265,
  New York, {NY}, {USA}, 1998. {ACM}.

\bibitem[Politz et~al.(2012)Politz, Quay-de~la Vallee, and
  Krishnamurthi]{politz_progressive_2012}
J.~G. Politz, H.~Quay-de~la Vallee, and S.~Krishnamurthi.
\newblock Progressive types.
\newblock In {\em Proceedings of the {ACM} international symposium on New
  ideas, new paradigms, and reflections on programming and software}, Onward!
  '12, page 55{\textendash}66, New York, {NY}, {USA}, 2012. {ACM}.

\bibitem[Pottier(1996)]{pottier_simplifying_1996}
F.~Pottier.
\newblock Simplifying subtyping constraints.
\newblock In {\em Proceedings of the First {ACM} {SIGPLAN} International
  Conference on Functional Programming}, {ICFP} '96, page 122{\textendash}133,
  New York, {NY}, {USA}, 1996. {ACM}.

\bibitem[Rastogi et~al.(2012)Rastogi, Chaudhuri, and Hosmer]{rastogi_ins_2012}
A.~Rastogi, A.~Chaudhuri, and B.~Hosmer.
\newblock The ins and outs of gradual type inference.
\newblock In {\em Proceedings of the 39th annual {ACM} {SIGPLAN-SIGACT}
  symposium on Principles of programming languages}, {POPL} '12, page
  481{\textendash}494, New York, {NY}, {USA}, 2012. {ACM}.

\bibitem[Ren et~al.(2013)Ren, Toman, Strickland, and Foster]{ren_ruby_2013}
B.~M. Ren, J.~Toman, T.~S. Strickland, and J.~S. Foster.
\newblock The ruby type checker.
\newblock In {\em Proceedings of the 28th Annual {ACM} Symposium on Applied
  Computing}, {SAC} '13, page 1565{\textendash}1572, New York, {NY}, {USA},
  2013. {ACM}.

\bibitem[Rondon et~al.(2012)Rondon, Bakst, Kawaguchi, and
  Jhala]{rondon_csolve:_2012}
P.~Rondon, A.~Bakst, M.~Kawaguchi, and R.~Jhala.
\newblock {CSolve:} verifying c with liquid types.
\newblock In {\em Proceedings of the 24th international conference on Computer
  Aided Verification}, {CAV'12}, page 744{\textendash}750, Berlin, Heidelberg,
  2012. Springer-Verlag.

\bibitem[Rondon et~al.(2010)Rondon, Kawaguchi, and
  Jhala]{rondon_low-level_2010}
P.~M. Rondon, M.~Kawaguchi, and R.~Jhala.
\newblock Low-level liquid types.
\newblock In {\em Proceedings of the 37th Annual {ACM} {SIGPLAN-SIGACT}
  Symposium on Principles of Programming Languages}, {POPL} '10, page
  131{\textendash}144, New York, {NY}, {USA}, 2010. {ACM}.

\bibitem[Rondon et~al.(2008)Rondon, Kawaguci, and Jhala]{rondon_liquid_2008}
P.~M. Rondon, M.~Kawaguci, and R.~Jhala.
\newblock Liquid types.
\newblock In {\em Proceedings of the 2008 {ACM} {SIGPLAN} conference on
  Programming language design and implementation}, {PLDI} '08, page
  159{\textendash}169, New York, {NY}, {USA}, 2008. {ACM}.

\bibitem[Rushby et~al.(1998)Rushby, Owre, and Shankar]{rushby_subtypes_1998}
J.~Rushby, S.~Owre, and N.~Shankar.
\newblock Subtypes for specifications: Predicate subtyping in {PVS}.
\newblock {\em {IEEE} Transactions on Software Engineering}, 24\penalty0
  (9):\penalty0 709{\textendash}720, Sept. 1998.

\bibitem[Sagonas and Luna(2008)]{sagonas_gradual_2008}
K.~Sagonas and D.~Luna.
\newblock Gradual typing of erlang programs: a wrangler experience.
\newblock In {\em Proceedings of the 7th {ACM} {SIGPLAN} workshop on {ERLANG}},
  {ERLANG} '08, page 73{\textendash}82, New York, {NY}, {USA}, 2008. {ACM}.

\bibitem[Sangiorgi et~al.(2007)Sangiorgi, Kobayashi, and
  Sumii]{sangiorgi_logical_2007}
D.~Sangiorgi, N.~Kobayashi, and E.~Sumii.
\newblock Logical bisimulations and functional languages.
\newblock In {\em Proceedings of the 2007 international conference on
  Fundamentals of software engineering}, {FSEN'07}, page 364{\textendash}379,
  Berlin, Heidelberg, 2007. Springer-Verlag.

\bibitem[Scott(1969)]{scott_type-theoretical_1969}
D.~S. Scott.
\newblock A type-theoretical alternative to {ISWIM}, {CUCH}, {OWHY}.
\newblock 1969.

\bibitem[Sheard(2005)]{sheard_putting_2005}
T.~Sheard.
\newblock Putting curry-howard to work.
\newblock In {\em Proceedings of the 2005 {ACM} {SIGPLAN} workshop on Haskell},
  Haskell '05, page 74{\textendash}85, New York, {NY}, {USA}, 2005. {ACM}.

\bibitem[Sheard(2007)]{sheard_type-level_2007}
T.~Sheard.
\newblock Type-level computation using narrowing in \&\#937;mega.
\newblock {\em Electron. Notes Theor. Comput. Sci.}, 174\penalty0 (7):\penalty0
  105{\textendash}128, June 2007.

\bibitem[Siek and Taha(2006)]{siek_gradual_2006}
J.~Siek and W.~Taha.
\newblock Gradual typing for functional languages.
\newblock In {\em Proceedings of the 7th workshop on Scheme and functional
  programmin}, 2006.

\bibitem[Siek and Taha(2007)]{siek_gradual_2007}
J.~Siek and W.~Taha.
\newblock Gradual typing for objects.
\newblock In {\em Proceedings of the 21st European conference on
  Object-Oriented Programming}, {ECOOP'07}, page 2{\textendash}27, Berlin,
  Heidelberg, 2007. Springer-Verlag.

\bibitem[Siek and Vachharajani(2008)]{siek_gradual_2008}
J.~G. Siek and M.~Vachharajani.
\newblock Gradual typing with unification-based inference.
\newblock In {\em Proceedings of the 2008 symposium on Dynamic languages},
  {DLS} '08, page 7:1{\textendash}7:12, New York, {NY}, {USA}, 2008. {ACM}.

\bibitem[Siek and Wadler(2009)]{siek_threesomes_2009}
J.~G. Siek and P.~Wadler.
\newblock Threesomes, with and without blame.
\newblock In {\em Proceedings for the 1st workshop on Script to Program
  Evolution}, {STOP} '09, page 34{\textendash}46, New York, {NY}, {USA}, 2009.
  {ACM}.

\bibitem[Sperber et~al.(2007)Sperber, Dybvig, Flatt, Straaten, Kelsey, Clinger,
  Rees, Findler, and Matthews]{sperber_revised_2007}
M.~Sperber, R.~K. Dybvig, M.~Flatt, A.~V. Straaten, R.~Kelsey, W.~Clinger,
  J.~Rees, R.~B. Findler, and J.~Matthews.
\newblock The revised report on the algorithmic language scheme.
\newblock Technical report, Sept. 2007.

\bibitem[Statman(1985)]{statman_logical_1985}
R.~Statman.
\newblock Logical relations and the typed lambda-calculus.
\newblock pages 85--97, 1985.

\bibitem[Stone and Harper(2000)]{stone_deciding_2000}
C.~A. Stone and R.~Harper.
\newblock Deciding type equivalence in a language with singleton kinds.
\newblock In {\em Proceedings of the 27th {ACM} {SIGPLAN-SIGACT} symposium on
  Principles of programming languages}, {POPL} '00, page 214{\textendash}227,
  New York, {NY}, {USA}, 2000. {ACM}.

\bibitem[Stoutamire and Omohundro(1996)]{stoutamire_sather_1996}
D.~Stoutamire and S.~Omohundro.
\newblock The sather 1.1 specification.
\newblock 1996.

\bibitem[Sumii and Pierce(2004)]{sumii_bisimulation_2004}
E.~Sumii and B.~C. Pierce.
\newblock A bisimulation for dynamic sealing.
\newblock {\em {SIGPLAN} Notices}, 39\penalty0 (1):\penalty0
  161{\textendash}172, Jan. 2004.

\bibitem[Swamy et~al.(2009)Swamy, Hicks, and Bierman]{swamy_theory_2009}
N.~Swamy, M.~Hicks, and G.~M. Bierman.
\newblock A theory of typed coercions and its applications.
\newblock In {\em Proceedings of the 14th {ACM} {SIGPLAN} international
  conference on Functional programming}, {ICFP} '09, page 329{\textendash}340,
  New York, {NY}, {USA}, 2009. {ACM}.

\bibitem[Tarditi et~al.(1996)Tarditi, Morrisett, Cheng, Stone, Harper, and
  Lee]{tarditi_til:_1996}
D.~Tarditi, G.~Morrisett, P.~Cheng, C.~Stone, R.~Harper, and P.~Lee.
\newblock {TIL:} a type-directed optimizing compiler for {ML}.
\newblock In {\em Proceedings of the {ACM} {SIGPLAN} 1996 conference on
  Programming language design and implementation}, {PLDI} '96, page
  181{\textendash}192, New York, {NY}, {USA}, 1996. {ACM}.

\bibitem[Terauchi(2010)]{terauchi_dependent_2010}
T.~Terauchi.
\newblock Dependent types from counterexamples.
\newblock In {\em Proceedings of the 37th Annual {ACM} {SIGPLAN-SIGACT}
  Symposium on Principles of Programming Languages}, {POPL} '10, page
  119{\textendash}130, New York, {NY}, {USA}, 2010. {ACM}.

\bibitem[Thatte(1990)]{thatte_quasi-static_1990}
S.~Thatte.
\newblock Quasi-static typing.
\newblock In {\em Proceedings of the 17th {ACM} {SIGPLAN-SIGACT} symposium on
  Principles of programming languages}, {POPL} '90, page 367{\textendash}381,
  New York, {NY}, {USA}, 1990. {ACM}.

\bibitem[{The Coq development team}(2012)]{the_coq_development_team_coq_2012}
{The Coq development team}.
\newblock The coq proof assistant reference manual.
\newblock Technical report, 2012.

\bibitem[Vazou et~al.(2013)Vazou, Rondon, and Jhala]{vazou_abstract_2013}
N.~Vazou, P.~M. Rondon, and R.~Jhala.
\newblock Abstract refinement types.
\newblock In {\em Proceedings of the 22nd European conference on Programming
  Languages and Systems}, {ESOP'13}, page 209{\textendash}228, Berlin,
  Heidelberg, 2013. Springer-Verlag.

\bibitem[Wadler and Findler(2009)]{wadler_well-typed_2009}
P.~Wadler and R.~B. Findler.
\newblock Well-typed programs can't be blamed.
\newblock In {\em Proceedings of the 18th European Symposium on Programming
  Languages and Systems: Held as Part of the Joint European Conferences on
  Theory and Practice of Software, {ETAPS} 2009}, {ESOP} '09, page
  1{\textendash}16, Berlin, Heidelberg, 2009. Springer-Verlag.

\bibitem[Winskel(1993)]{winskel_formal_1993}
G.~Winskel.
\newblock {\em The formal semantics of programming languages: an introduction}.
\newblock {MIT} Press, Cambridge, {MA}, {USA}, 1993.

\bibitem[Wolff et~al.(2011)Wolff, Garcia, Tanter, and
  Aldrich]{wolff_gradual_2011}
R.~Wolff, R.~Garcia, {\'E}.~Tanter, and J.~Aldrich.
\newblock Gradual typestate.
\newblock In {\em Proceedings of the 25th European conference on
  Object-oriented programming}, {ECOOP'11}, page 459{\textendash}483, Berlin,
  Heidelberg, 2011. Springer-Verlag.

\bibitem[Wright and Cartwright(1994)]{wright_practical_1994}
A.~K. Wright and R.~Cartwright.
\newblock A practical soft type system for scheme.
\newblock {\em {SIGPLAN} Lisp Pointers}, {VII}\penalty0 (3):\penalty0
  250{\textendash}262, July 1994.

\bibitem[Xi(2000)]{xi_imperative_2000}
H.~Xi.
\newblock Imperative programming with dependent types.
\newblock In {\em Proceedings of the 15th Annual {IEEE} Symposium on Logic in
  Computer Science}, {LICS} '00, page 375{\textendash}, Washington, {DC},
  {USA}, 2000. {IEEE} Computer Society.

\bibitem[Xi and Pfenning(1999)]{xi_dependent_1999}
H.~Xi and F.~Pfenning.
\newblock Dependent types in practical programming.
\newblock In {\em Proceedings of the 26th {ACM} {SIGPLAN-SIGACT} symposium on
  Principles of programming languages}, {POPL} '99, page 214{\textendash}227,
  New York, {NY}, {USA}, 1999. {ACM}.

\bibitem[Xu(2006)]{xu_extended_2006}
D.~N. Xu.
\newblock Extended static checking for haskell.
\newblock In {\em Proceedings of the 2006 {ACM} {SIGPLAN} workshop on Haskell},
  Haskell '06, page 48{\textendash}59, New York, {NY}, {USA}, 2006. {ACM}.

\bibitem[Xu(2012)]{xu_hybrid_2012}
D.~N. Xu.
\newblock Hybrid contract checking via symbolic simplification.
\newblock In {\em Proceedings of the {ACM} {SIGPLAN} 2012 workshop on Partial
  evaluation and program manipulation}, {PEPM} '12, page 107{\textendash}116,
  New York, {NY}, {USA}, 2012. {ACM}.

\bibitem[Xu et~al.(2009)Xu, Peyton~Jones, and Claessen]{xu_static_2009}
D.~N. Xu, S.~Peyton~Jones, and K.~Claessen.
\newblock Static contract checking for haskell.
\newblock In {\em Proceedings of the 36th Annual {ACM} {SIGPLAN-SIGACT}
  Symposium on Principles of Programming Languages}, {POPL} '09, page
  41{\textendash}52, New York, {NY}, {USA}, 2009. {ACM}.

\bibitem[Zenger(1997)]{zenger_indexed_1997}
C.~Zenger.
\newblock Indexed types.
\newblock {\em Theoretical Compututer Science}, 187\penalty0 (1-2):\penalty0
  147{\textendash}165, Nov. 1997.

\end{thebibliography}


\begin{thebibliography}{10}

\bibitem{des:tex}
Jacques D{\'e}sarm{\'e}nien.
\newblock How to run {\TeX} in french.
\newblock Technical Report SATN-CS-1013, Computer Science Department, Stanford
  University, Stanford, California, August 1984.

\bibitem{fuchs:dvi0}
David Fuchs.
\newblock The format of {\TeX}'s {DVI} files version 1.
\newblock {\em TUGboat}, 2(2):12--16, July 1981.

\bibitem{fuchs:dvi}
David Fuchs.
\newblock Device independent file format.
\newblock {\em TUGboat}, 3(2):14--19, October 1982.

\bibitem{furuta:pctex}
Richard~K. Furuta and Pierre~A. MacKay.
\newblock Two {\TeX} implementations for the {IBM PC}.
\newblock {\em Dr. Dobb's Journal}, 10(9):80--91, September 1985.

\bibitem{knuth:web}
Donald~E. Knuth.
\newblock The {WEB} system for structured documentation, version 2.3.
\newblock Technical Report STAN-CS-83-980, Computer Science Department,
  Stanford University, Stanford, California, September 1983.

\bibitem{knuth:tex}
Donald~E. Knuth.
\newblock {\em The {\TeX} Book}.
\newblock Addison-Wesley, Reading, Massachusetts, 1984.
\newblock Reprinted as Vol. A of {\it Computers \& Typesetting\/}, 1986.

\bibitem{knuth:lp}
Donald~E. Knuth.
\newblock Literate programming.
\newblock {\em The Computer Journal}, 27(2):97--111, May 1984.

\bibitem{knuth:tor}
Donald~E. Knuth.
\newblock A torture test for {\TeX}, version 1.3.
\newblock Technical Report STAN-CS-84-1027, Computer Science Department,
  Stanford University, Stanford, California, November 1984.

\bibitem{knuth:pgm}
Donald~E. Knuth.
\newblock {\em {\TeX}: The Program}, volume~B of {\em Computers \&
  Typesetting}.
\newblock Addison-Wesley, Reading, Massachusetts, 1986.

\bibitem{lamport:latex}
Leslie Lamport.
\newblock {\em {\LaTeX}: A Document Preparation System. User's Guide and
  Reference Manual}.
\newblock Addison-Wesley, Reading, Massachusetts, 1986.

\bibitem{patashnik:bibtex}
Oren Patashnik.
\newblock {\em Bib{\TeX}ing}.
\newblock Computer Science Department, Stanford University, Stanford,
  California, January 1988.
\newblock Available in the Bib{\TeX} release.

\bibitem{patashnik:bibhax}
Oren Patashnik.
\newblock {\em Designing Bib{\TeX} Styles}.
\newblock Computer Science Department, Stanford University, January 1988.

\bibitem{samuel:tex}
Arthur~L. Samuel.
\newblock First grade {\TeX}: A beginner's {\TeX} manual.
\newblock Technical Report SATN-CS-83-985, Computer Science Department,
  Stanford University, Stanford, California, November 1983.

\bibitem{spivak:ams}
Michael~D. Spivak.
\newblock {\em The Joy of {\TeX}}.
\newblock American Mathematical Society, 1985.

\end{thebibliography}

\end{document}